\documentclass[twocolumn]{aastex63}
\usepackage[caption=false]{subfig}
\usepackage{grffile}
\usepackage{amsmath}
\usepackage{bm}
\usepackage{booktabs}
\usepackage{morefloats}
\newcommand{\mic}{\mu{\rm m}}
\newcommand{\tprime}{\hbox{$t^{\prime}$}}
\newcommand{\ism}{\textsc{ism}}

\submitjournal{ApJS}

\shorttitle{BayeSED-GALAXIES I.}
\shortauthors{Han et al. 2023}

\begin{document}

\title{BayeSED-GALAXIES I. Performance test for simultaneous photometric redshift and stellar population parameter estimation of galaxies in the CSST wide-field multiband imaging survey}

\correspondingauthor{Yunkun Han}
\email{hanyk@ynao.ac.cn}

\author[0000-0002-2547-0434]{Yunkun Han}
\affiliation{Yunnan Observatories, Chinese Academy of Sciences, 396 Yangfangwang, Guandu District, Kunming, 650216, P. R. China}
\affiliation{Center for Astronomical Mega-Science, Chinese Academy of Sciences, 20A Datun Road, Chaoyang District, Beijing, 100012, P. R. China}
\affiliation{Key Laboratory for the Structure and Evolution of Celestial Objects, Chinese Academy of Sciences, 396 Yangfangwang, Guandu District, Kunming, 650216, P. R. China}
\affiliation{International Centre of Supernovae, Yunnan Key Laboratory, Kunming 650216, P. R. China}

\author[0000-0003-4200-4432]{Lulu Fan}
\affiliation{Deep Space Exploration Laboratory / Department of Astronomy, University of Science and Technology of China, Hefei 230026, China}
\affiliation{School of Astronomy and Space Science, University of Science and Technology of China, Hefei 230026, China}

\author[0000-0003-3728-9912]{Xian Zhong Zheng}
\affiliation{Purple Mountain Observatory, Chinese Academy of Sciences, 10 Yuanhua Road, Nanjing 210023, China}

\author{Jin-Ming Bai}
\affiliation{Yunnan Observatories, Chinese Academy of Sciences, 396 Yangfangwang, Guandu District, Kunming, 650216, P. R. China}
\affiliation{Center for Astronomical Mega-Science, Chinese Academy of Sciences, 20A Datun Road, Chaoyang District, Beijing, 100012, P. R. China}
\affiliation{Key Laboratory for the Structure and Evolution of Celestial Objects, Chinese Academy of Sciences, 396 Yangfangwang, Guandu District, Kunming, 650216, P. R. China}

\author[0000-0001-9204-7778]{Zhanwen Han}
\affiliation{Yunnan Observatories, Chinese Academy of Sciences, 396 Yangfangwang, Guandu District, Kunming, 650216, P. R. China}
\affiliation{Center for Astronomical Mega-Science, Chinese Academy of Sciences, 20A Datun Road, Chaoyang District, Beijing, 100012, P. R. China}
\affiliation{Key Laboratory for the Structure and Evolution of Celestial Objects, Chinese Academy of Sciences, 396 Yangfangwang, Guandu District, Kunming, 650216, P. R. China}
\affiliation{International Centre of Supernovae, Yunnan Key Laboratory, Kunming 650216, P. R. China}



\begin{abstract}
The forthcoming CSST wide-field multiband imaging survey will produce seven-band photometric spectral energy distributions (SEDs) for billions of galaxies.
The effective extraction of astronomical information from these massive datasets of SEDs relies on the techniques of both SED synthesis (or modeling) and analysis (or fitting).
We evaluate the performance of the latest version of BayeSED code combined with SED models with increasing complexity for simultaneously determining the photometric redshifts and stellar population parameters of galaxies in this survey.
By using an empirical statistics-based mock galaxy sample without SED modeling errors, we show finding that the random observational errors in photometries are more important sources of errors than the parameter degeneracies and Bayesian analysis method and tool.
By using a Horizon-AGN hydrodynamical simulation-based mock galaxy sample with SED modeling errors about the star formation histories (SFHs) and dust attenuation laws (DALs), the simple typical assumptions lead to significantly worse parameter estimation with CSST photometries only.
The SED models with more flexible (or complicated) forms of SFH/DAL do not necessarily lead to better estimation of redshift and stellar population parameters.
We discuss the selection of the best SED model by means of Bayesian model comparison in different surveys.
Our results reveal that the Bayesian model comparison with Bayesian evidence may favor SED models with different complexities when using photometries from different surveys.
Meanwhile, the SED model with the largest Bayesian evidence tends to give the best performance of parameter estimation, which is more clear for photometries with larger discriminative power.


\end{abstract}



\section{Introduction} \label{sec:intro}
Understanding the complex ecosystem of stars, interstellar gas and dust, and supermassive black holes in galaxies is one of the most important challenges in modern astrophysics \citep{National-Academies-of-SciencesE2021g}.
The new generation of space and ground telescopes and the corresponding large surveys will provide vast amounts of multi-band data for understanding the cosmic ecosystems and all the complex physical processes involved.
For example, the James Webb Space Telescope \citep[JWST,][]{RiekeM2005p,GardnerJ2006l,BeichmanC2012r} is able to detect the earliest stages of galaxies from infrared at unprecedented depths, and is expected to provide decisive observations of the first generation of stars and galaxies \citep{BeichmanC2012r,RobertsonB2022b}.
Meanwhile, forthcoming deep and wide field surveys with the Chinese Space Station Telescope \citep[CSST,][]{ZhanH2011i,ZhanH2018y,ZhanH2021j}, the Euclid Space Telescope \citep{LaureijsR2011a,JoachimiB2016b}, the Vera C. Rubin Observatory Legacy Survey of Space and Time \citep[LSST,][]{IvezicZ2019a,BreivikK2022a}, and the Nancy Grace Roman Space Telescope \cite[NGRST,][]{GreenJ2012c,DoreO2019p} will provide multi-band photometric and spectroscopic information for billions of galaxies.
Especially, the CSST wide-field multiband imaging survey is set to image approximately 17,500 square degrees of the sky using NUV, u, g, r, i, z, and y bands in about 10 years of orbital time, which aims to achieve a $5\sigma$ limiting magnitude of 26 (AB mag) or higher for point sources in the g and r bands.
How to effectively and reliably measure the redshift and the properties of various physical components of galaxies from the obtained huge amount of photometric spectral energy distributions (SEDs) data has become an urgent task to be done.
A new generation of SED synthesis and analysis methods and tools are strongly demanded to effectively extract physical information from those massive datasets of observational SEDs.

The SED synthesis and analysis of galaxies are two aspects that are both opposite and unified in nature.
The reliability and efficiency of the SED synthesis and analysis methods and tools will directly determine the reliability and efficiency of physical information extraction from the massive multi-wavelength datasets.
In terms of SED synthesis of galaxies, the evolutionary synthesis technique of stellar population has become the core method from the pioneering works of \cite{TinsleyB1976a,TinsleyB1978a}.
Nowadays, the stellar population synthesis models of BC03 \citep{BruzualG2003a}, M05 \citep{MarastonC2005a}, FSPS \citep{ConroyC2009a}, BPASS \citep{EldridgeJ2009a}, among others are widely used in the study of the formation and evolution of galaxies.
However, in the SED synthesis models of galaxies, many important uncertainties remain in almost all the model ingredients \citep{ConroyC2009a,ConroyC2010b,ConroyC2010a,ConroyC2013b}, such as the initial stellar mass function (IMF) \citep{PadoanP1997a,HoverstenE2008a,van-DokkumP2008a,BastianN2010a,CappellariM2012b,FerrerasI2013a,GennaroM2018d}, the physics of stellar evolution \citep{ThomasD2003a,ZhangF2005c,MarastonC2006a,HanZ2007a,MarigoP2008a,BertelliG2008a,BrottI2011a,Hernandez-PerezF2013a}, stellar spectral libraries \citep{CoelhoP2009a,ChoiJ2019a,KnowlesA2019t,CoelhoP2020k,KnowlesA2021d,YanR2019a}, the complex star formation and metallicity enrichment histories (SFHs and MEHs) \citep{DebsarmaS2016a,IyerK2019a,CarnallA2019b,LejaJ2019a,AufortG2020a,WangE2020a,IyerK2020a,CoteB2016c,MaiolinoR2019a,ValentiniM2019a}, the reprocessing by the interstellar gas and dust \citep{DraineB2003a,DraineB2010a,GallianoF2018b,KewleyL2019a,SalimS2020a,TacconiL2020b}, and the possible contribution from active galactic nuclei (AGNs) at the center of galaxies \citep{AntonucciR1993a,AntonucciR2012l,NetzerH2015a,HickoxR2018i,BrownA2019a,BrownM2019a,LyuJ2022c}.
Different choices of these model ingredients will lead to very different estimation of the redshifts and physical parameters of galaxies, as well as different and even conflicting conclusions about the formation and evolution of galaxies.
Therefore, the proper selection of these model ingredients is an essential step in any SED analysis work of galaxies \citep{HanY2019a,HanY2020a}.

In terms of SED analysis of galaxies, the Bayesian method has been widely adopted in the last decade.
For example, the widely used and actively developing SED fitting codes, such as MAGPHYS \citep{da-CunhaE2008a}, CIGALE \citep{NollS2009b,BoquienM2019a}, GalMC \citep{AcquavivaV2011a}, BayeSED \citep{HanY2012b,HanY2014a}, BEAGLE \citep{ChevallardJ2016a}, Prospector \citep{LejaJ2017a}, BAGPIPES \citep{CarnallA2018a}, and ProSpect \citep{RobothamA2020a} are all based on the Bayesian methods.
Besides, a long list of new SED fitting codes, such as MCSED \citep{BowmanW2020a}, piXedfit \citep{Abdurrouf2021j},  gsf \citep{MorishitaT2022a}, and Lightning \citep{DooreK2023a} among others, are build along this way.
The application of Bayesian methods implies that the SED analysis of galaxies is considered as a more general Bayesian inference problem instead of the previous Chi-square minimization-based optimization problem known as SED fitting.
For the parameter estimation of a give SED model, the Bayesian approach provides the complete posterior probability distribution of parameters as the solution to the SED analysis problem, which is computationally more demanding but allows a more formal and simultaneous estimation of parameter values and their uncertainties.
More importantly, for the selection of model ingredients, the Bayesian approach also provides the very useful Bayesian evidence which can be considered as a quantified Occam's razor for effective model selection.

A noteworthy difference among the Bayesian SED analysis tools is that the earlier tools (e.g. MAGPHYS and CIGALE) are based on irregular or regular grid search, while the newer generation of tools (e.g. GalMC and BayeSED) are based on more advanced random sampling techniques such as Markov Chain Monte Carlo \citep[MCMC,][]{SharmaS2017b,HoggD2018a} and Nested Sampling \citep[NS,][]{SkillingJ2006a,BuchnerJ2021w,AshtonG2022w}.
The advantage of the grid-based Bayesian approach is that an SED library with regular or irregular model grids can be built in advance for only once.
Besides, the prior probabilities can be set more freely during this procedure.
Then it can be used in the analysis of a large sample of galaxies of any size without the generation of new SEDs.
However, the size of SED library needs to be very large to allow a reasonable parameter estimation for all galaxies in the sample, especially in the case with regular grids where the number of required grids will increase dramatically with the number of free parameters.
In contrast, a sampling-based Bayesian approach allows for a more detailed and efficient sampling of the parameter space for each galaxy and allows for a finer reconstruction of the posterior, leading to more reliable parameter estimates.
However, the theoretical SED synthesis needs to be done in real-time and repeated many times, which could be very computationally expensive for the analysis of very large samples of galaxies.
Fortunately, a much more efficient synthesis SED models can be achieved with the help of machine learning techniques.
For example, in \cite{HanY2014a} we have employed the artificial neural network (ANN) and K-nearest neighbor (KNN) searching techniques to speed up the sampling-based Bayesian approach.
The combination of sampling-based Bayesian inference and machine learning techniques enables the detailed Bayesian SED analysis of very large samples of galaxies \citep{HanY2019u}.
Although the training phase of machine-learning-based SED synthesis method could be very time-consuming, especially for very complex SED models with many free parameters and the accurate synthesis of high-resolution SED, it is very promising with more advanced training techniques \citep{AlsingJ2020a,GildaS2021m,QiuY2022x,HahnC2022x} and worthwhile to carry out further exploration in this direction.

For the study of galaxy formation and evolution, the ideal SED synthesis and analysis tool should be able to simultaneously account for the contributions of stars, interstellar gas and dust, and AGN components, and to provide accurate and efficient estimates of the redshift and the physical properties of all components.
However, in practice, it is very difficult, if not impossible, to fully satisfy all of these requirements.
Therefore, a good SED synthesis and analysis tool should attempt to achieve a reasonable balance among these requirements as much as possible.
This is what we are trying to achieve during the development of the BayeSED code \citep{HanY2012b,HanY2014a,HanY2019a,HanY2019u,HanY2020a}.
In this work, we will rigorously test the performance of the latest version of BayeSED code combined with SED models with increasing complexity for simultaneous photometric redshift and stellar population parameter estimation of galaxies, so as to be ready for the analysis of the forthcoming massive datasets from the CSST wide-field multiband imaging survey and others.

We begin in \S \ref{sec:bayesed_enc} by introducing the methods we have employed for the generation of empirical statistics-based (\S \ref{ss:mock_sed}) and hydrodynamical simulation-based (\S \ref{ss:mock_hyd}) mock catalog of galaxies, observational error modeling (\S \ref{ss:noise}) and the selection of samples (\S \ref{sec:sample}) that will be used for the performance test.
In \S \ref{sec:bayesed_dec}, we briefly describe the Bayesian approach of photometric SED analysis methods, including parameter estimation (\S \ref{ss:estimation}) and model selection (\S \ref{ss:model_select}).
Especially, in \S \ref{sss:mutinest}, we will introduce some runtime parameters of MultiNest algorithm which is the core engine of BayeSED.
They need to be properly tuned to improve the performance of BayeSED.
We present the results of performance test for the case without SED modeling errors using an empirical statistics-based mock galaxy sample in \S \ref{sec:pf_sed}.
In \S \ref{sec:pf_hyd}, by employing the simplest SED model, we present the results of performance test for the case with SED modeling errors about the SFH and DAL of galaxies using a Horizon-AGN hydrodynamical simulation-based mock galaxy sample for CSST-like imaging survey.
In \S \ref{sec:disc}, we discuss the effectiveness of more flexible (or complex) forms of SFH and DAL of galaxies for improving the performance of simultaneous redshift and stellar parameter estimation in CSST-like (\S \ref{ss:disc_csst}), CSST+Euclid-like (\S \ref{ss:disc_csst_euclid}), and COSMOS-like (\S \ref{ss:disc_cosmos}) surveys with increasing discriminative power, respectively.
Especially, we also discuss the relation between the metrics of the quality of  parameter estimation and Bayesian evidence, as well as how they depend on the different surveys.
Finally, a summary of our results and conclusions is presented in \S \ref{sec:summary}.

Throughout this work, we assume a flat $\Lambda$-cold dark matter cosmology with $H_{0}=70$\,km\,s$^{-1}$\,Mpc$^{-1}$, $\Omega_{m}=0.3$, $\Omega_{\Lambda}=0.7$ \cite[][WMAP-5]{SpergelD2003a}, and the \cite{ChabrierG2003a} IMF.
All the presented magnitudes are in the AB system \citep{OkeJ1974a}.

\section{Bayesian photometric SED synthesis with BayeSED} \label{sec:bayesed_enc}
The SED synthesis (or modeling) module is an essential part in any Bayesian SED fitting code.
In BayeSED-V3, we have added more functions for SED synthesis, especially for the simulation of mock observation of galaxies in a Bayesian way.
This is not just crucial for the current work, but also lays the foundation for future applications of machine learning and simulation-based Bayesian inference methods in Bayesian SED fitting \citep{HahnC2022x,HahnC2022d}.
For this work, we use the empirical statistics-based (\S \ref{ss:mock_sed}) and hydrodynamical simulation-based (\S \ref{ss:mock_hyd}) methods to generate mock photometric catalog, add noise with a simple magnitude limit-based approach (\ref{ss:noise}), and select a sample (\ref{sec:sample}) similar to previous works for the performance test in the next two sections.
In the following, we introduce them in more detail.

\subsection{Empirical statistics-based photometric mock catalog} \label{ss:mock_sed}
The first method to generate mock photometric catalog is built by randomly draw samples from the parameter space of a particular SED model while under the constraints of some empirical statistical properties of galaxies.
The sampling is performed with the same nested sampler MultiNest as in the Bayesian SED analysis mode of BayeSED.
A sample from this catalog will be used in \S \ref{sec:pf_sed} to test the performance of redshift and stellar population parameter estimation in the case where the SED modeling is perfect, since exactly the same SED modeling method will be used in the Bayesian SED analysis of it.
\subsubsection{SED modeling} \label{sss:sedm}
As in \cite{HanY2019a}, the SED of a galaxy is modeled as the luminosity of starlight from stellar populations of varying ages and metallicities, transmitted through the Interstellar Medium (ISM) and the Intergalactic Medium (IGM) to the observer.
Specifically, the luminosity emitted at wavelength $\lambda$ by a galaxy with  $age=t$ can be given as:
\begin{align}
    L_\lambda(t) &= \int_0^t {\rm d}\tprime \, \psi(t-\tprime) \, S_\lambda[\tprime,{Z}(t-\tprime)] \, T^\ism_\lambda(t,t^\prime) \label{eq:csp1}\\
                 &= T^\ism_\lambda \,\int_0^t {\rm d}\tprime \, \psi(t-\tprime) \, S_\lambda[\tprime,{Z}(t-\tprime)] \label{eq:csp2}\,  ,
\end{align}
where $\psi(t-\tprime)$ is the star formation history (SFH) describing the SFR as a function of the time $t-\tprime$, and $S_\lambda[\tprime,{Z}(t-t^\prime)]$ is the luminosity emitted per unit wavelength per unit mass by a simple stellar population (SSP) of age $t^\prime$ and metallicity ${Z}(t-t^\prime)$.

$T^\ism_\lambda(t,t^\prime)$ is the transmission function of the ISM \citep{CharlotS2001a}, which is contributed by two components:
\begin{equation}
T^\ism_\lambda(t,t^\prime)=T^+_\lambda(t,t^\prime) T^0_\lambda(t,t^\prime),
\end{equation}
where $T^+_\lambda(t,t^\prime)$ and $T^0_\lambda(t,t^\prime)$ are the transmission functions of the ionized gas and the neutral ISM, respectively.
The transmission through ionized gas can be modeled with photoionization code such as CLOUDY.
However, we set $T^+_\lambda(t,t^\prime)=1$ in this work to be consistent with the hydrodynamical-simulation based catalog (\S \ref{ss:mock_hyd}).
A detailed modeling of $T^+_\lambda(t,t^\prime)$ with CLOUDY to account for the combined effects of starlight absorption, nebular line emission, ionized continuum emission, and possible emission from warm dust within HII regions will be presented in a companion paper.
Meanwhile, the transmission functions of the neutral ISM is considered with a simple time-independent dust attenuation law (DAL) and uniformly applied to the whole galaxy.

${Z}(t-t^\prime)$ is the stellar metallicity as a function of the time $t-t^\prime$, which describe the chemical evolution history of the galaxy.
In previous works, we assume a time-independent metallicity, i.e. ${Z}(t-t^\prime)=Z_0$, as in many SED fitting codes of galaxy.
To properly consider the evolution of stellar metallicity, we additionally employ a linear SFH-to-metallicity mapping model\citep{DriverS2013a,RobothamA2020a,ThorneJ2021n,AlsingJ2023g}:
\begin{equation}
    Z(t-t^\prime)=\left(Z(t_{\mathrm{age}})-Z_{\min }\right) \frac{1}{M} \int_0^t \psi(t-\tprime) {\rm d}\tprime+Z_{\min }
\end{equation}

Generally, the main ingredients for our SED modeling of galaxies are the SSP model, SFH, CEH and DAL.
In this work,as the construction of the Horizon-AGN hydrodynamical-simulation based catalog (\S \ref{ss:mock_hyd}), we use the SSP model assuming a \cite{ChabrierG2003a} stellar IMF from the widely used stellar population synthesis model of \cite{BruzualG2003a}.
The SFH of galaxies is typically parameterized as the exponentially declining form: ${\rm SFR}(t) \propto e^{-t/\tau}\,$(hereafter $\tau$ model).
The $\tau$ model only describes the SFH of galaxies in a closed box without inflow of pristine gas and outflow of processed gas, where the gases are converted to stars at a rate proportional to the remaining gas and with a fixed efficiency \citep{SchmidtM1959a,TinsleyB1980n}.
It is widely discussed in the literature that this simple assumption may lead to systematically biased estimation of stellar population parameters, especially for galaxies at $z\gtrsim2$ \citep{LeeS2009a,LeeS2010a,ReddyN2012a,CieslaL2017a,CarnallA2018a}. 
Therefore, some more flexible and physically inspired form of models have been suggested to improve the measurement of SFHs of galaxies and the estimation of their stellar population parameters and photometric redshift \citep{PacificiC2012a,CieslaL2017a,IyerK2017a,CarnallA2019b,LejaJ2019a,IyerK2019a,LowerS2020u,SuessK2022v}.

In the present work, we employ three extension of the $\tau$ model with different complexity.
The first one is described as:
\begin{equation}
    \psi(t)\propto t^{\beta}\times \exp({-t/\tau}),
    \label{eq:sfh_beta_tau}
\end{equation}
which is just an extended form of the delayed-$\tau$ model \citep{LeeS2010a}.
Apparently, the typical $\tau$ model and delayed-$\tau$ model are just two special cases of this model (hereafter $\beta$-$\tau$ model)  with $\beta=0$ and $\beta=1$, respectively.
The second one is the $\beta$-$\tau$ model combined with a quenching (or rejuvenation) component which is described as \citep{CieslaL2016a}:
\begin{equation}
    \psi(t)\propto  \begin{cases}t^{\beta} \times \exp (-t / \tau), & \text { when } t<=t_{\text {trunc }} \\ r_{\mathrm{SFR}} \times \psi\left(t=t_{\text {trunc}}\right), & \text { when } t>t_{\text {trunc }}\end{cases}
\end{equation}
where $t_{\text {trunc}}$ is the time when the star formation is quenched ($r_{\mathrm{SFR}}<1$) or rejuvenated ($r_{\mathrm{SFR}}>1$), and $r_{\mathrm{SFR}}$ is the ratio between $\psi(t>t_{\text {trunc }})$ and $\psi(t=t_{\text {trunc }})$:
\begin{equation}
    r_{\mathrm{SFR}}=\frac{\psi\left(t>t_{\text {trunc }}\right)}{\psi\left(t_{\text {trunc }}\right)}.
\end{equation}
This model (hereafter $\beta$-$\tau$-r model) is a further extension of the $\beta$-$\tau$ model with the latter being the special case with $r_{\mathrm{SFR}}=1$.
The third one is the double power-law model \citep{DiemerB2017a,CarnallA2018a,AlsingJ2023g} combined with a quenching (or rejuvenation) component which is described as:
\begin{equation}
    \psi(t)\propto  \begin{cases}\frac{1.0}{{(\frac{t}{t^*})^\alpha + (\frac{t}{t^*})^{-\beta}}}, & \text { when } t<=t_{\text {trunc }} \\ r_{\mathrm{SFR}} \times \psi\left(t=t_{\text {trunc}}\right), & \text { when } t>t_{\text {trunc }}\end{cases}
\end{equation}
where $\alpha$ and $\beta$ are the falling and rising slopes, respectively, and $t^*$ is related to the time at which star formation peaks, which is defined as $t^* \equiv \tau t_{\mathrm{age}}$ for the age of the galaxy $t_{\mathrm{age}}$.
A major advantage of the double-power-law model is the decoupling of the rising and falling parts of the SFH.
Therefore, this model (hereafter $\alpha$-$\beta$-$\tau$-r model) is even more flexible than the $\beta$-$\tau$-r model.

The dust attenuation law (DAL) is another very important ingredient for the SED modeling of galaxies \citep{WalcherJ2011a,ConroyC2013b}.
When deriving the photometric redshift and physical properties of galaxies from the analysis of their photometric or spectroscopic observations, a universal DAL as a simple uniform screen is commonly assumed. 
However, different choices of the universal law may lead to very different estimation of photometric redshift and physical parameters of galaxies \citep{PforrJ2012a,PforrJ2013a,SalimS2020a}.
Especially, many studies show that the dust attenuation curve of different galaxies are very different \citep{KriekM2013a,ReddyN2015a,SalmonB2016a,SalimS2019a,ShivaeiI2020k}, and therefore there is no universal DAL as expected on theoretical grounds \citep{WittA2000a,SeonK2016c,NarayananD2018o,LowerS2022g}.
By a detailed study of the dust attenuation curves of about 230,000 individual galaxies in the local universe using photometric data covering from UV to IR bands, \cite{SalimS2018a} presented new forms of attenuation laws that are suitable for normal star-forming galaxies, high-z analogs, and quiescent galaxies \citep[See also][]{NollS2009b}.
In this work, we additionally employ this new form of DAL which is parameterized as following:
\begin{equation}
    \frac{A_{\rm \lambda,{\rm mod}}}{A_{\rm V}}=\frac{k_{\lambda}}{R_{\rm V}}=
    \frac{k_{\lambda,{\rm Cal}}}{R_{V,{\rm Cal}}}
    \left (\frac{\lambda}{5500\text{\AA}}\right)^{\delta}+\frac{D_{\lambda}}{R_{V,{\rm mod}}},
    \label{eq:dal_salim2018}
\end{equation}
where ${k_{\lambda,{\rm Cal}}}/{R_{V,{\rm Cal}}}$ is the \cite{CalzettiD2000a} DAL with ${R_{V,{\rm Cal}}}=4.05$. The power law term with an exponent $\delta$ is introduced to deviate from the slope of the \cite{CalzettiD2000a} DAL. $R_{V,{\rm mod}}$ is the $\delta$-dependent ratio of total to selective extinction for the modified law. The term $D_{\lambda}$ is introduced to add a UV bump.
The relationship between $R_{V,{\rm mod}}$ and $\delta$ is given by:
\begin{equation}
    R_{V,{\rm mod}} = \frac{R_{V,{\rm Cal}}}{(R_{V,{\rm
    Cal}}+1)(0.44/0.55)^{\delta}-R_{V,{\rm Cal}}}.
    \label{eqn:RV_mod}
\end{equation}
The UV bump following a Drude profile \citep{FitzpatrickE1986f} is represented as:
\begin{equation}
    D_{\lambda}(E_{\rm b}) = \frac{E_{\rm b}\lambda^2\gamma^2}{[\lambda^2-(\lambda_0)^2]^2+\lambda^2\gamma^2},
    \label{eq:D_bump}
\end{equation}
with the amplitude $E_{\rm b}$, fixed central wavelength $\lambda_0=0.2175\mic$ and width $\gamma=0.35\mic$.

In total, we have considered six different combinations of SFH, CEH, and DAL with increasing complexity.
A summary of these models, their parameters and priors are shown in Table \ref{tab:models}.
\begin{table}
    \begin{center}
        \caption{Summary of SED models, parameters and priors.}
        \begin{tabular}{lllll}
            \hline
            \textbf{Model}&\textbf{Parameter} & \textbf{Range} & \textbf{Prior} & \\
                          &$z$ & $[0,4]$ & U & \\
                          &${\rm log}(M_*/{\rm M_{\odot}})$ & $[8,12]$ & U & \\
                          &${\rm log}(age/\rm{yr})$\tablenotemark{c} & $[8,10.14]$ & U& \\
                          &$A_{\rm v}$ & $[0,4]$ & U & \\
                          \hline
            SFH=$\tau$,CEH=no &${\rm log}(Z_0/Z_{\odot})$ & $[-2.30,0.70]$ & U & \\
            DAL=Cal+00\tablenotemark{a} &${\rm log}(\tau/\rm{yr})$& $[6,10]$ & U & \\
            \hline
            SFH=$\tau$,CEH=yes &${\rm log}(Z(t_{\mathrm{age}})/Z_{\odot})$ & $[-2.30,0.70]$ & U & \\
            DAL=Cal+00\tablenotemark{a} &${\rm log}(\tau/\rm{yr})$& $[6,10]$ & U & \\
            \hline
            SFH=$\tau$,CEH=yes &${\rm log}(Z(t_{\mathrm{age}})/Z_{\odot})$ & $[-2.30,0.70]$ & U & \\
            DAL=Sal+18\tablenotemark{b} &${\rm log}(\tau/\rm{yr})$& $[6,10]$ & U & \\
                                        &$E_{\rm b}$ & $[-2,6]$ & U & \\
                                        &$\delta$ & $[-1.2,0.4]$ & U & \\
                                        \hline
            SFH=$\beta$-$\tau$          &${\rm log}(Z(t_{\mathrm{age}})/Z_{\odot})$ & $[-2.30,0.70]$ & U & \\
            CEH=yes                     &${\rm log}(\tau/\rm{yr})$& $[6,10]$ & U & \\
            DAL=Sal+18\tablenotemark{b} &$\beta$ & $[0,1]$ & U & \\
                                        &$E_{\rm b}$ & $[-2,6]$ & U & \\
                                        &$\delta$ & $[-1.2,0.4]$ & U & \\
                                        \hline
            SFH=$\beta$-$\tau$-r        &${\rm log}(Z(t_{\mathrm{age}})/Z_{\odot})$ & $[-2.30,0.70]$ & U & \\
            CEH=yes                     &${\rm log}(\tau/\rm{yr})$& $[6,10]$ & U & \\
            DAL=Sal+18\tablenotemark{b} &rSFR & $[1e-6,1e6]$ & LU\tablenotemark{d} & \\
                                        &$t_{\rm trunc}/t_{\mathrm{age}}$ & $[0,1]$ & U & \\
                                        &$\beta$ & $[0,1]$ & U & \\
                                        &$E_{\rm b}$ & $[-2,6]$ & U & \\
                                        &$\delta$ & $[-1.2,0.4]$ & U & \\
                                        \hline
            SFH=$\alpha$-$\beta$-$\tau$-r &${\rm log}(Z(t_{\mathrm{age}})/Z_{\odot})$ & $[-2.30,0.70]$ & U & \\
            CEH=yes                       &$\tau=t^*/t_{\mathrm{age}}$& $[0.007,1]$ & U & \\
            DAL=Sal+18\tablenotemark{b}   &$\alpha$ & $[0.01,1000]$ & LU & \\
                                          &$\beta$ & $[0.01,1000]$ & LU & \\
                                          &rSFR & $[1e-6,1e6]$ & LU\tablenotemark{d} & \\
                                          &$t_{\rm trunc}/t_{\mathrm{age}}$ & $[0,1]$ & U & \\
                                          &$E_{\rm b}$ & $[-2,6]$ & U & \\
                                          &$\delta$ & $[-1.2,0.4]$ & U & \\
                                          \hline
        \end{tabular}
        \tablecomments
        {
            \tablenotetext{a}{\cite{CalzettiD2000a}}
            \tablenotetext{b}{\cite{SalimS2018a}}
            \tablenotetext{c}{We also apply the constraint that the age of galaxy is less than the age of Universe at z.}
            \tablenotetext{d}{Uniform in log space.}
        }
        \label{tab:models}
    \end{center}
\end{table}
Finally, we also include the effect of IGM absorption with the description of \cite{MadauP1995h}.
Other other more recent consideration of IGM absorption are also available in BayeSED.
However, the exploration of the effects of different choices of IGM absorption models on the redshift and stellar population parameter estimation is beyond the scope of this work, which will not change the conclusions given here.

\subsubsection{Galaxy population modeling} \label{sss:galaxy_pop}
To model the galaxy population, we need to set the joint probability distribution that characterizes the statistical properties of the galaxy population.
The statistical properties of the galaxy population are the results of the complex physical procedures happened during the formation and evolution of galaxies.
In this work, we employ some widely discussed empirical statistical properties of galaxies to model the galaxy population phenomenologically.
It should be mentioned that there are large uncertainties in these statistical properties, and we do not attempt to use the most up-to-date results for all of them in this work.
The other choices of statistical properties of galaxies will not change the conclusions of this work.
\begin{figure*}[]
    \centering
    \includegraphics[scale=0.55]{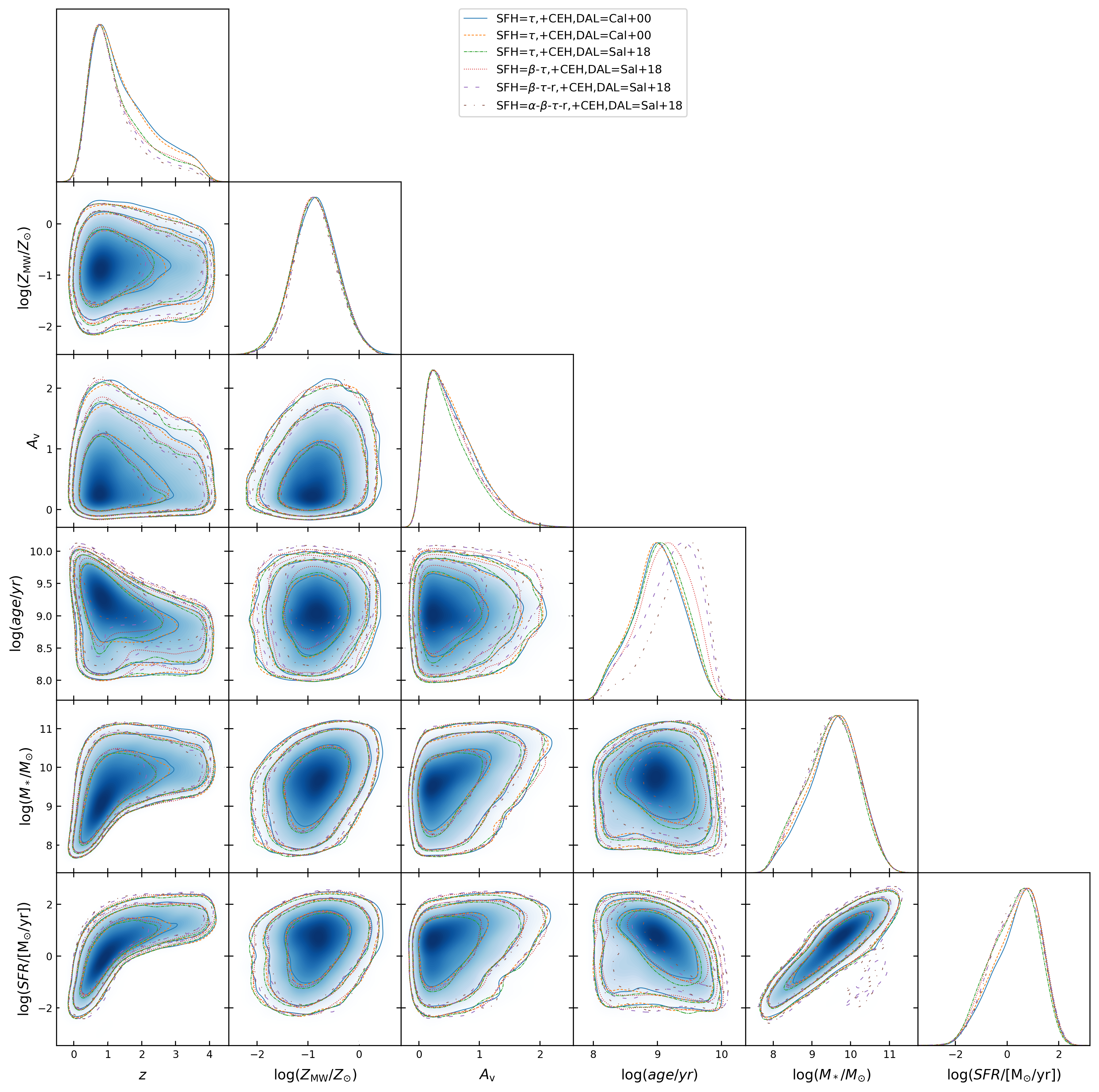}
    \caption{The joint distributions of redshift and physical parameters of the empirical statistics-based mock galaxy population produced with BayeSED combined with SED models of different complexity. With the same set of empirical statistics, different SED models lead to slightly different redshift and age distribution, which is likely due to different mapping relations from free parameters to derived parameters. Meanwhile, only the two SED models with a quenching component produce a clear region of quiescent galaxies below the star-forming main sequence.}
    \label{fig:gal_pop}
\end{figure*}

Similar to \cite{TanakaM2015a}\citep[See also][]{AlsingJ2023g}, we assume that the joint probability distribution of the stellar population parameters and redshift of galaxy population can be factorized as
\begin{equation}
    \begin{aligned}
& P(\boldsymbol{G}, z)=P(M_*, z) \\
& \quad \times P(SFR \mid M_*,z) \\
& \quad \times P(A_{\rm V} \mid SFR,z) \\
& \quad \times P(age \mid M_*, z) \\
& \quad \times P(Z_{\rm MW} \mid M_*, z).
    \end{aligned}
    \label{eq:P_G}
\end{equation}
The joint distribution of stellar mass and redshift is defined as
\begin{equation}
    P(M_*, z) \propto \Phi(M_*, z) d V(z),
\end{equation}
where $\Phi(M_*, z)$ is the unnormalized stellar mass function and $d V(z)$ is the differential comoving volume element. 
We employ the recent measurement of stellar mass function and its redshift evolution from \cite{LejaJ2020a}, while a WMAP-5 \citep{SpergelD2003a} cosmology for the comoving volume element.

Following \cite{TanakaM2015a}, we assume that $P(SFR \mid M_*,z)$ can be expressed as the sum of two Gaussians to represent two distinct sequences formed by star forming and quiescent galaxies:
\begin{equation}
    \begin{gathered}
        P\left(\mathrm{SFR} \mid M_*, z\right) \propto \\
        \frac{(1-f_{\mathrm{Q}})}{\sigma_{\mathrm{SF}}} \exp \left[-\frac{1}{2}\left(\frac{\log \mathrm{SFR}-\log \mathrm{SFR}_{\mathrm{SF}}\left(M_*, z\right)}{\sigma_{\mathrm{SF}}}\right)^2\right] \\
        \quad+\frac{f_{\mathrm{Q}}}{\sigma_{\mathrm{Q}}} \exp \left[-\frac{1}{2}\left(\frac{\log \mathrm{SFR}-\log \mathrm{SFR}_{\mathrm{SF}}\left(M_*, z\right)+2}{\sigma_{\mathrm{Q}}}\right)^2\right],
    \end{gathered}
\end{equation}
where $\operatorname{SFR}_{\mathrm{SF}}\left(M_*, z\right)$ is the mean SFR of star forming galaxies  given by
\begin{equation}
    \operatorname{SFR}_{\mathrm{SF}}\left(M_*, z\right)=\operatorname{SFR}^*(z) \times \frac{M_*}{10^{11} M_{\odot}} M_{\odot} \mathrm{yr}^{-1},
\end{equation}
with
\begin{equation}
    \operatorname{SFR}^*(z)= \begin{cases}10 \times(1+z)^{2.1} & (z<2) \\ 19 \times(1+z)^{1.5} & (z \geqslant 2)\end{cases},
\end{equation}
and the fraction of quiescent galaxies is given as a function of stellar mass and redshift \citep{BehrooziP2013a}:
\begin{equation}
    f_{\mathrm{Q}}\left(M_*, z\right)=\left[\left(\frac{M_*}{10^{10.2+0.5 z} M_{\odot}}\right)^{-1.3}+1\right]^{-1}.
\end{equation}

As in \cite{TanakaM2015a}, the dust attenuation is considered to positively correlate with SFR:
\begin{equation}
    P\left(\tau_V \mid \mathrm{SFR}, z\right) \propto \frac{1}{{\sigma_{\tau_V}}}\exp \left[-\frac{1}{2}\left(\frac{\tau_V-\left\langle\tau_V\right\rangle}{\sigma_{\tau_V}}\right)^2\right],
\end{equation}
where  $\sigma_{\tau_V}=0.5$, and
\begin{equation}
    \left\langle\tau_V\right\rangle= \begin{cases}0.2 & \left(\mathrm{SFR}_0<1\right) \\ 0.2+0.5 \log \mathrm{SFR}_0 & \left(\mathrm{SFR}_0>1\right)\end{cases},
\end{equation}
\begin{equation}
    \mathrm{SFR}_0=100 \frac{\mathrm{SFR}}{\mathrm{SFR}^*(z)},
\end{equation}
Then, we use the relation between $\tau_V$ and  $A_{\rm V}$ to obtain $P(A_{\rm V} \mid SFR,z)$.

The probability of the age of a galaxy is described conditionally on the stellar mass and redshift:
\begin{equation}
    P\left(\operatorname{age} \mid M_*, z\right) \propto \exp \left[-\frac{\langle\mathrm{SFR}\rangle}{\operatorname{SFR}^*(z)}\right],
\end{equation}
where
\begin{equation}
    \langle\mathrm{SFR}\rangle=M_* / \text { age. }
\end{equation}
This lead to a low probability for a massive galaxy with young age, while a high probability for a low-mass galaxy with the same age. 

Finally, the probability of mass weighted stellar metallicity is modeled  as:
\begin{equation}
    \begin{array}{l}
        P\left(Z_{\rm MW} \mid M_*, z\right) \propto \\
        \frac{1}{{\sigma_{\log(Z_{\rm MW})}}}\exp \left[-\frac{1}{2}\left(\frac{\log(Z_{\rm MW})-\left\langle\log(Z_{\rm MW})\right\rangle}{\sigma_{\log(Z_{\rm MW})}}\right)^2\right],
    \end{array}
\end{equation}
where ${\sigma_{\log(Z_{\rm MW})}}=0.1$, and
\begin{equation}
    \begin{array}{l}
        \left\langle\log(Z_{\rm MW})\right\rangle =\\
        0.40\left[\log M_*-\right.10] +0.67 \exp (-0.50 z)-1.04
    \end{array}
\end{equation}
is the redshift dependent stellar mass and metallicity relation \citep{MaX2016o} which is predicted by using the high-resolution cosmological zoom-in simulations from the Feedback in Realistic Environment (FIRE) project \citep{HopkinsP2014b}.

To generate empirical statistics-based mock catalog of galaxies, we employ the MultiNest algorithm to draw samples from the joint probability distribution of the stellar population parameters and redshift of galaxy population by setting $P(\boldsymbol{G}, z)$ in Equation \ref{eq:P_G} to be the likelihood function.
Besides, to simulate a magnitude-limited sample, we can additionally set the likelihood function to be $0$ when the magnitude in a given band is larger than a given value.
Since the sampling points with likelihood to be $0$ will be ignored by MultiNest, the obtained posterior sample can be used to buid a magnitude-limited sample of mock galaxies with some physical constraints from the empirical statistical properties of galaxies.
More details about the selection of magnitude-limited sample is presented in \S \ref{sec:sample}.
The mock catalog can be build with the posterior sample of redshift and all physical parameters given by MultiNest.
However, this is a weighted sample \citep{YallupD2022q}, which can not be directly used as a mock sample of galaxies.
To build a more realistic mock sample of galaxies, we use bootstrap resampling method to obtain an unweighted sample.

In total, we have build six mock catalog of galaxies by employing SED models with different combinations of SFH, CEH, and DAL and increasing complexity as shown in Table \ref{tab:models}, respectively.
The employed priors of redshift and stellar population parameters are listed in the same Table for each model, respectively.
In Figure \ref{fig:gal_pop}, we show the joint distributions of redshift and physical parameters of the six empirical statistics-based mock galaxy population.
Although the sampe set of empirical are employed, different SED models lead to slightly different distribution of parameters, especially for redshift and galaxy age.
This is likely due to different mapping relations from free parameters to derived parameters.
For example, different forms of SFH may lead to different relations between the age of galaxy and its recent SFR.


\subsection{Hydrodynamical simulation-based photometric mock catalog} \label{ss:mock_hyd}
The second method to generate mock photometric catalog is based on an SED library which is built by the post-processing of galaxies from a hydrodynamical simulation.
This catalog will be used in \S \ref{sec:pf_hyd} to test the performance of redshift and stellar population parameter estimation in the case where the SED modeling is imperfect, since the SED modeling method employed in the Bayesian SED analysis will be very different from the one used to built it.

We start from the rest-frame spectra of galaxies which are produced using the light-cone from the cosmological hydrodynamical simulation Horizon-AGN \citep{DuboisY2014a}.
The computation of these spectra, which accounts for the complex star formation history and metal enrichment of Horizon-AGN galaxies, and consistently includes dust attenuation, is described in details by \cite{LaigleC2019a} and \cite{DavidzonI2019a}.
The dust attenuation of galaxies in Horizon-AGN simulation is modelled for each stellar particle, assumed to be a SSP, by using the gas metal mass distribution as an approximation of the dust mass distribution, assuming a constant dust-to-metal mass ratio \citep{LaigleC2019a}.
Besides, to obtain the amount of extinction at a given wavelength, the \cite{WeingartnerJ2001a} model of Milky Way dust grain with $R_{\rm V}=3.1$ and the prominent $2175{\rm \AA}$-graphite bump is employed for post-processing the simulated galaxies. 
As mentioned in \cite{LaigleC2019a}, the overall attenuation curve becomes less steep and the bump tends to be reduced when summing up the contribution of all the SSPs to obtain the resulting galaxy spectrum.
They also noticed that the averaged attenuation curve in Horizon-AGN simulation can not be well reproduced by either the model of \cite{CalzettiD2000a} or \cite{ArnoutsS2013a}.
The more flexible form of DAL as given by Equation \ref{eq:dal_salim2018} is more likely to reproduce the attenuation curves of galaxies in the Horizon-AGN simulation.
In order to isolate the possible differences in the convolution with filter response function, observational error modeling, and consideration of IGM absorption, we choose to convert their rest-frame spectra of mock galaxies to corresponding mock photometries with BayeSED, instead of using their virtual photometries directly\footnote{We have compared the two set of photometries careful and found that they are actually very similar, only with some differences at the very faint end.}.
The consideration for the effects of IGM absorption is the same as in \S \ref{sss:sedm}.
Therefore, the difference between the empirical statistics-based (\S \ref{ss:mock_sed}) and hydrodynamical simulation-based photometric mock catalog are only driven by the different SFH, CEH, DAL of mock galaxies and their different distribution of redshift and physical parameters.

\subsection{Observational error modeling} \label{ss:noise}
The modeling of realistic errors on the flux is crucial for a meaningful performance test of redshift and stellar population parameter estimation.
Here, we introduce the method we have employed to compute flux errors of mock galaxies and perturb their fluxes accordingly.
The flux error for a wavelength band $i$ with N$\sigma$ AB magnitude limit $m_{{\rm lim},i}=-2.5*{\rm log}(F_{\rm lim,i})+23.9$ is given by:
\begin{equation}
    \sigma_{F,i}=\sqrt{(F_{\rm lim,i}/N)^2+\sigma^2_{F,i,{\rm sys}}},
\end{equation}
where the flux limit $F_{\rm lim,i}$  and the systematic flux error $\sigma_{F,i,{\rm sys}}$ are in unit of $\mu{\rm Jy}$.
From the relation that magnitude error $\sigma_m \approx 1.08574/SNR$ and signal-to-noise ratio $SNR=F/\sigma_F$, we can obtain:
\begin{equation}
\sigma_{F,i,{\rm sys}}= \sigma_{m,{\rm sys}}*F_i/1.08574.
\end{equation}

\begin{figure*}
    \gridline{\leftfig{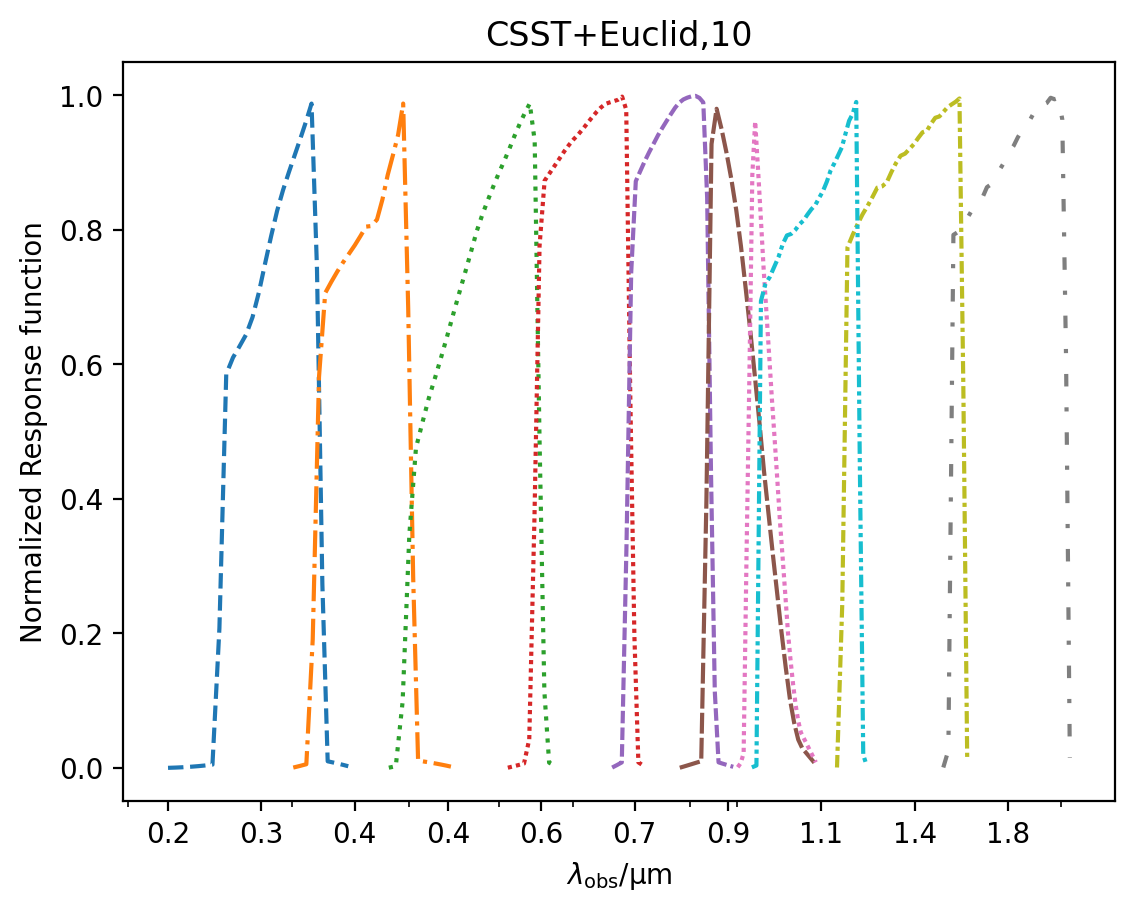}{0.48\textwidth}{(a)}
        \rightfig{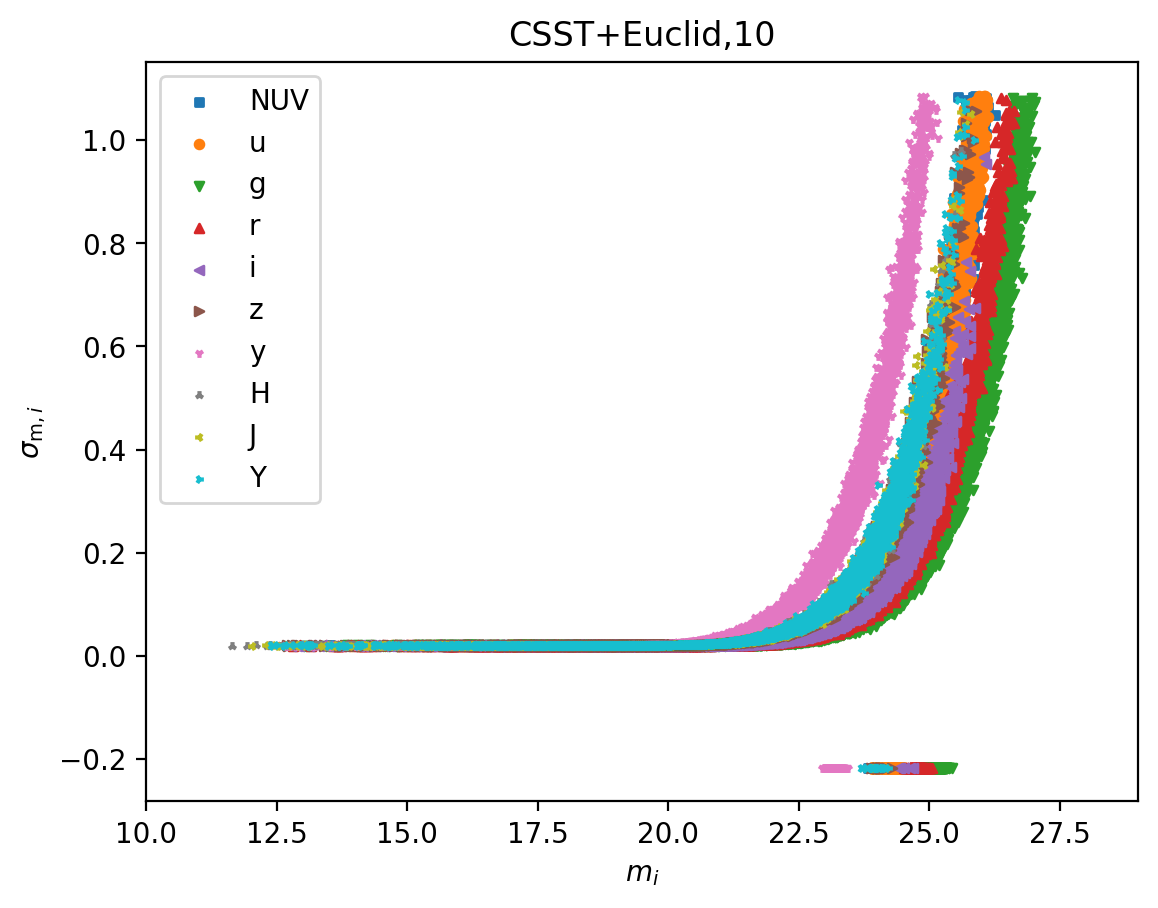}{0.48\textwidth}{(b)}
    }
    \gridline{\leftfig{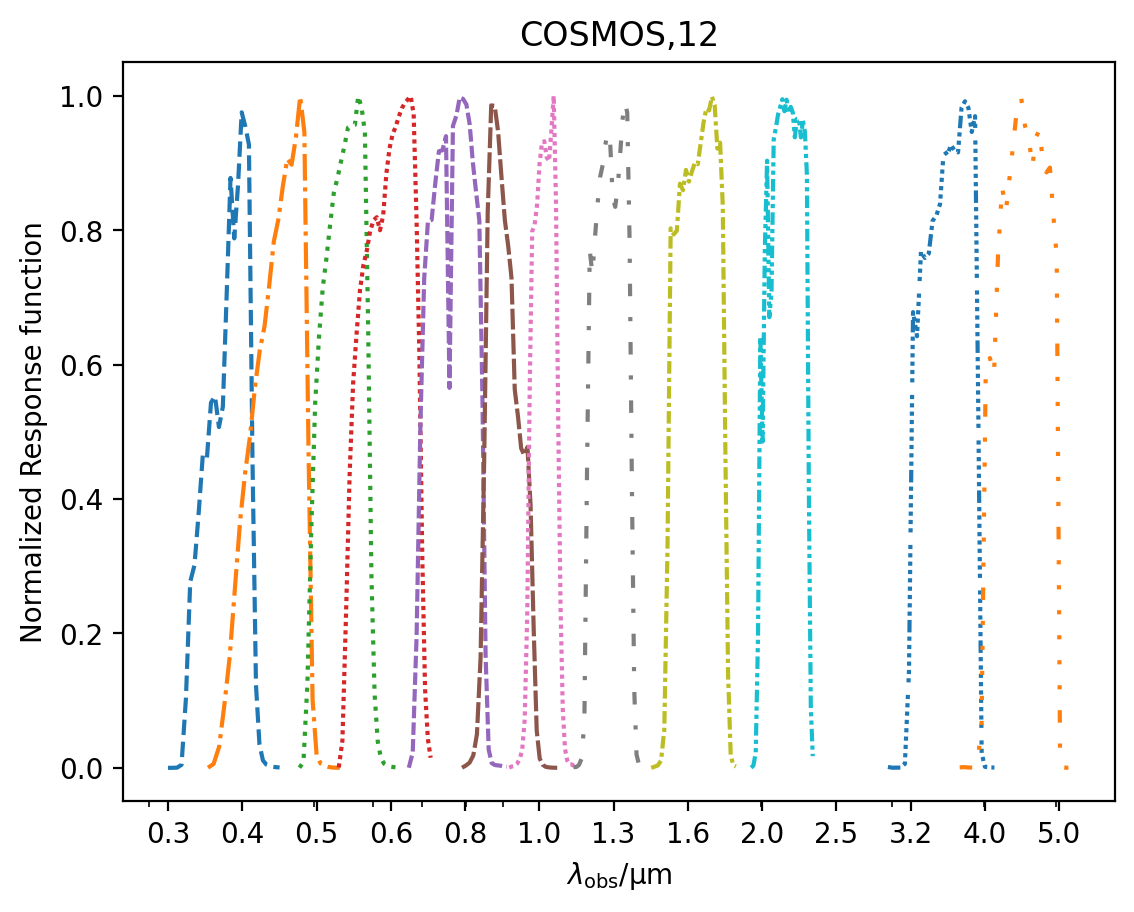}{0.48\textwidth}{(c)}
        \rightfig{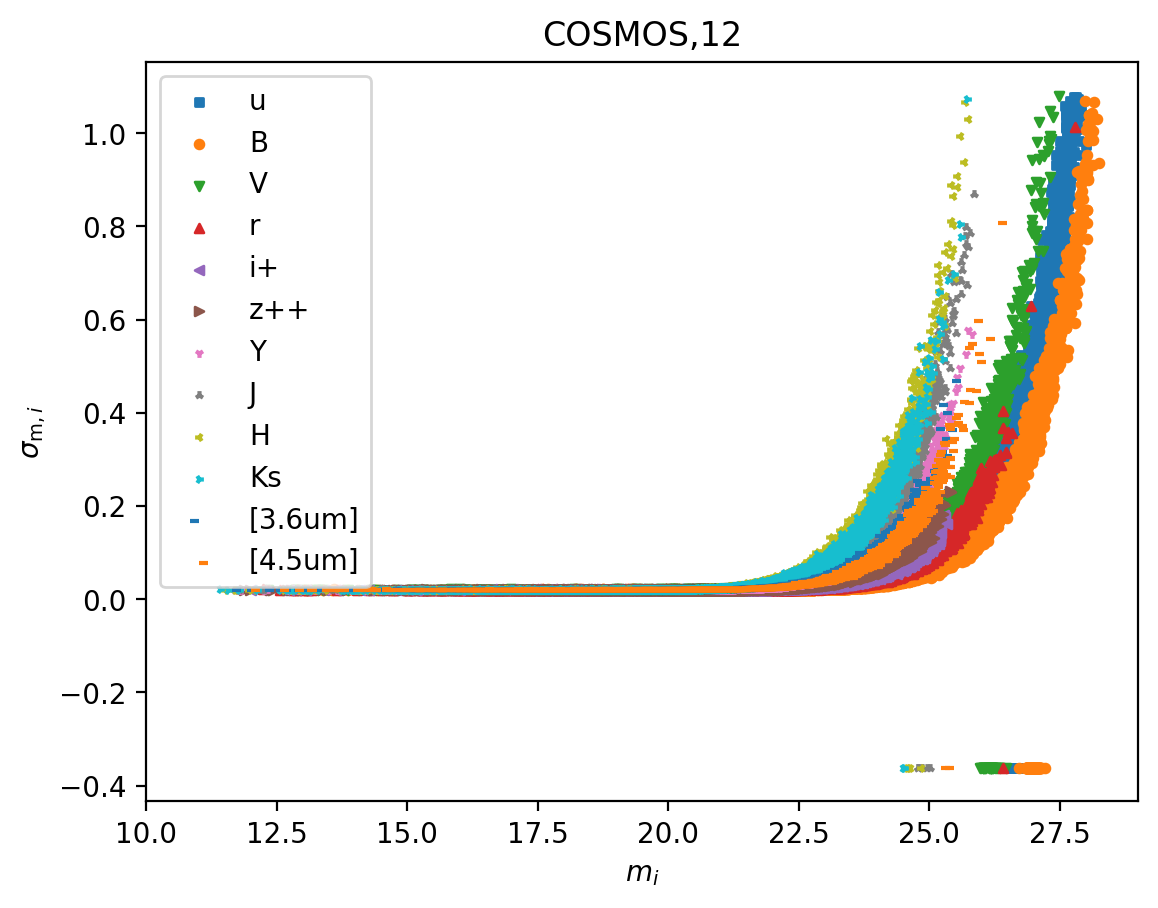}{0.48\textwidth}{(d)}
    }
    \caption{\textbf{(a)} Response functions for CSST and Euclid bands. \textbf{(b)} The modeled relation between magnitude and magnitude error for CSST bands. Sources with $SNR<1$ (i.e. $\sigma_{\rm m,i}>1.08574$) are considered as non-detections. The non-detections in a wavelength band $i$ with N$\sigma$ flux limit $F_{\rm lim,i}$  and magnitude limit $m_{{\rm lim},i}$ are represented as sources with $F_i=F_{\rm lim,i}$ and $\sigma_{F,i}=-F_{\rm lim,i}/N$ (the flux case), or $m_i=m_{{\rm lim},i}$ and $\sigma_{\rm m,i}=-1.08574/N$ (the magnitude case). These conventions make sure the consistent conversion between flux data and magnitude data in the input file of BayeSED. \textbf{(c)}, \textbf{(d)}: Same as in \textbf{(a)}, \textbf{(b)}, but for COSMOS bands. For clarity, the twelve intermediate bands (IBs) and two narrow bands (NBs) are not shown. 
    \label{fig:filters}}
\end{figure*}

As in \cite{CaoY2018a}, we assume a systematic magnitude error $\sigma_{m,{\rm sys}}=0.02$.
The final mock flux is obtained by the original flux perturbed by a Gaussian noise $\epsilon \sim \mathcal{N}\left(0, \sigma_{F,i}^{2}\right)$.
In practice, the magnitude limit may have a dispersion $\sigma_{m,{\rm lim},i}$ for galaxies with different sizes.
So, the actually used magnitude limit is drawn from the Gaussian distribution $\mathcal{N}\left(m_{{\rm lim},i}, \sigma_{m,{\rm lim},i}^{2}\right)$.
In this work, we set $\sigma_{m,{\rm lim},i}=0.1$.
We have generated three sets of mock catalog for CSST-like, Euclid-like, and COSMOS-like surveys, respectively.
A summary of the adopted depths in all bands of the three surveys is shown in Table \ref{tab:depths}.
The response functions and modeled relation between magnitude and magnitude error are shown in the panels of Figure \ref{fig:filters} for the 7 CSST bands, 3 Euclid bands and 26 COSMOS bands, respectively.
To separate the effects of observational errors on the accuracy of parameter estimation, we also generated another two sets of mock catalog without adding observational errors.
In this case (the no noise case in Figure \ref{fig:filters}), the magnitude errors are all fixed to be $0.01$, but the photometries have not been perturbed accordingly.

\begin{table}
    \begin{center}
        \begin{tabular}{ccc}
            \textbf{Survey} & \textbf{band} & \textbf{depth}  \\
                            \hline
                            & $NUV$ & $24.2\pm0.1$   \\
                            & $u$ & $24.2\pm0.1$   \\
            CSST-like       & $g$ & $25.1\pm0.1$   \\
            $5\sigma$ depth\tablenotemark{b} & $r$ & $24.8\pm0.1$   \\
                            & $i$ & $24.6\pm0.1$   \\
                            & $z$ & $24.1\pm0.1$   \\
                            & $y$ & $23.2\pm0.1$   \\
                            \hline
            Euclid-like     & $J$ & $24.0\pm0.1$   \\
            $5\sigma$ depth\tablenotemark{a} & $H$ & $24.0\pm0.1$   \\
                            & $Y$ & $24.0\pm0.1$   \\
                            \hline
                            & $u$ & $26.0\pm0.1$   \\
                            & $B$ & $26.4\pm0.1$   \\
                            & $V$ & $25.6\pm0.1$   \\
                            & $r$ & $25.9\pm0.1$   \\
                            & $i^{+}$ & $25.6\pm0.1$   \\
                            & $z^{++}$ &$25.3\pm0.1$   \\
                            & $Y$ & $24.7\pm0.1$  \\
COSMOS-like                 & $J$ & $24.3\pm0.1$   \\
$5\sigma$ depth \tablenotemark{a}& $H$ & $24.0\pm0.1$   \\
                            & $K_{\rm s}$ & $24.1\pm0.1$  \\
                            & $IB$ & $24-25\pm0.1$  \\
                            & $NB711$ & $24.5\pm0.1$  \\
                            & $NB816$ & $24.6\pm0.1$  \\
                            & $3.6\mic$ & $24.9\pm0.1$  \\
                            & $4.5\mic$ & $24.9\pm0.1$  \\
                            \hline
        \end{tabular}
        \tablecomments
        {
            \tablenotetext{a}{The $3\sigma$ depths for COSMOS-like survey provided by \cite{LaigleC2016a} have been converted to the $5\sigma$ depths.}
            \tablenotetext{b}{The $5\sigma$ depths for extended sources in the CSST wide-field multiband imaging survey \citep{GongY2019a}. The CSST deep survey can be at least 1 mag deeper . The results of performance tests for the latter will be presented in future works.}
        }
    \end{center}
    \caption{A summary of the adopted depths in all bands for CSST-like, Euclid-like, and  COSMOS-like mock observations.}
    \label{tab:depths}
\end{table}
\subsection{Sample selection} \label{sec:sample}
\begin{figure}[]
\centering
\includegraphics[scale=0.55]{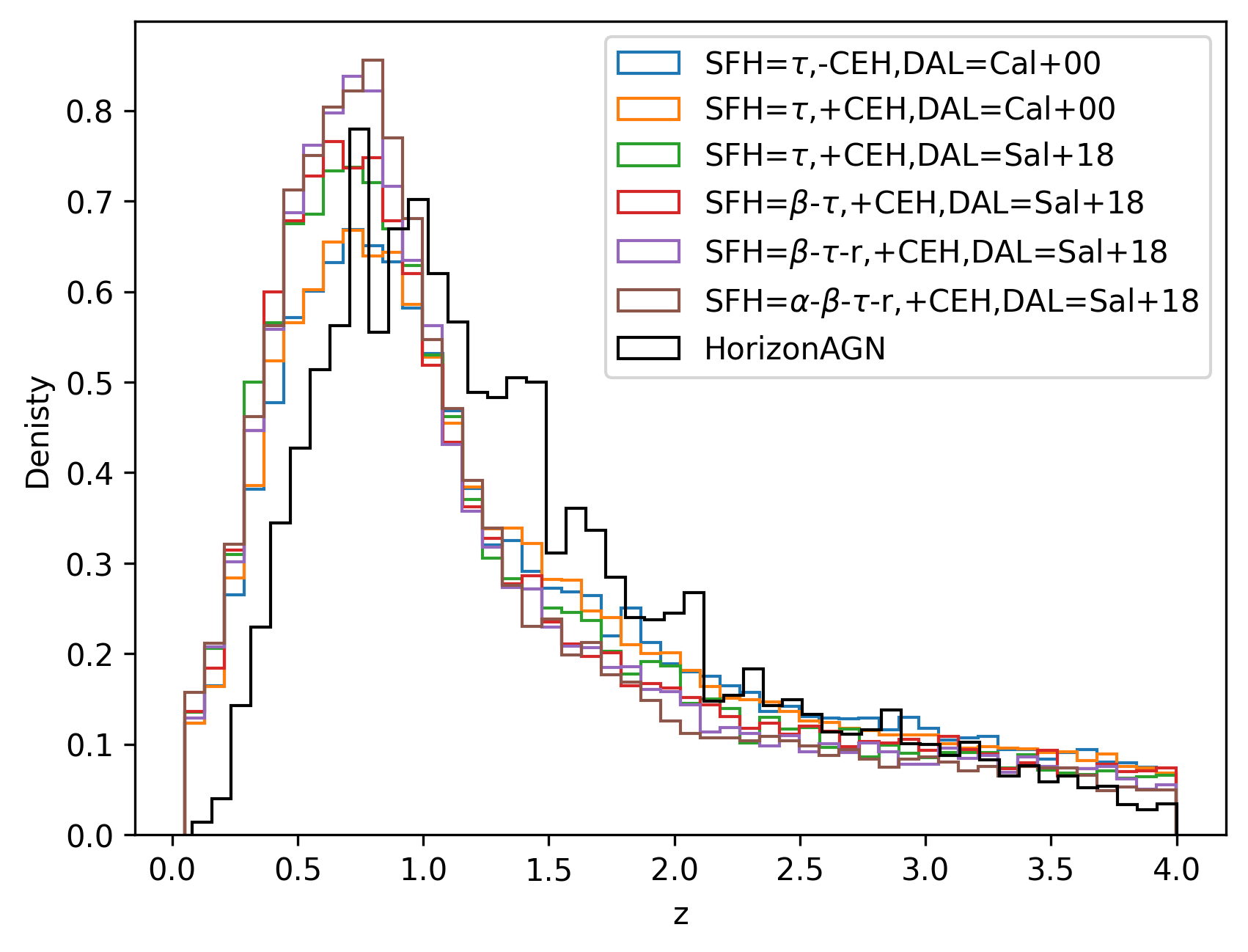}
\includegraphics[scale=0.55]{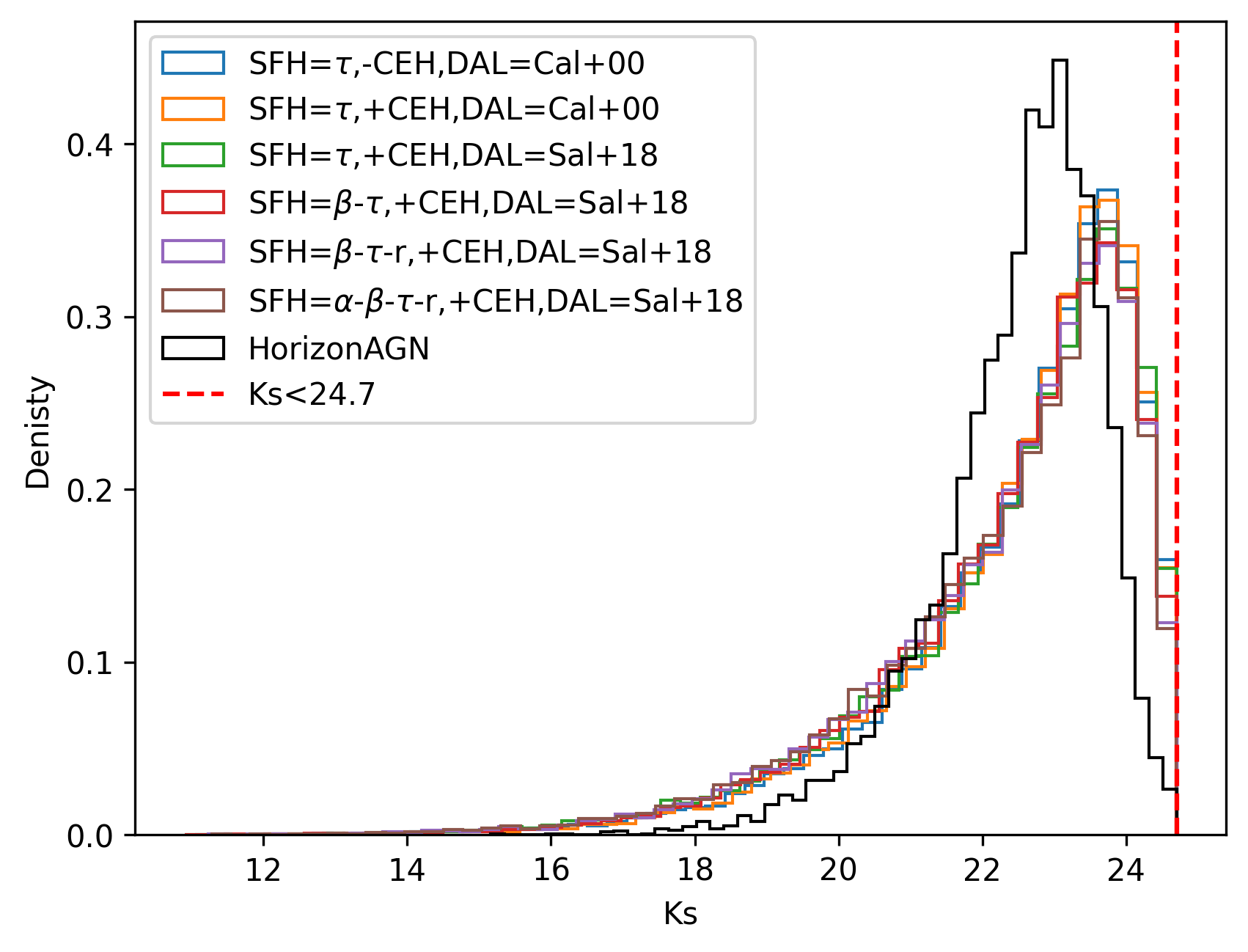}
\includegraphics[scale=0.55]{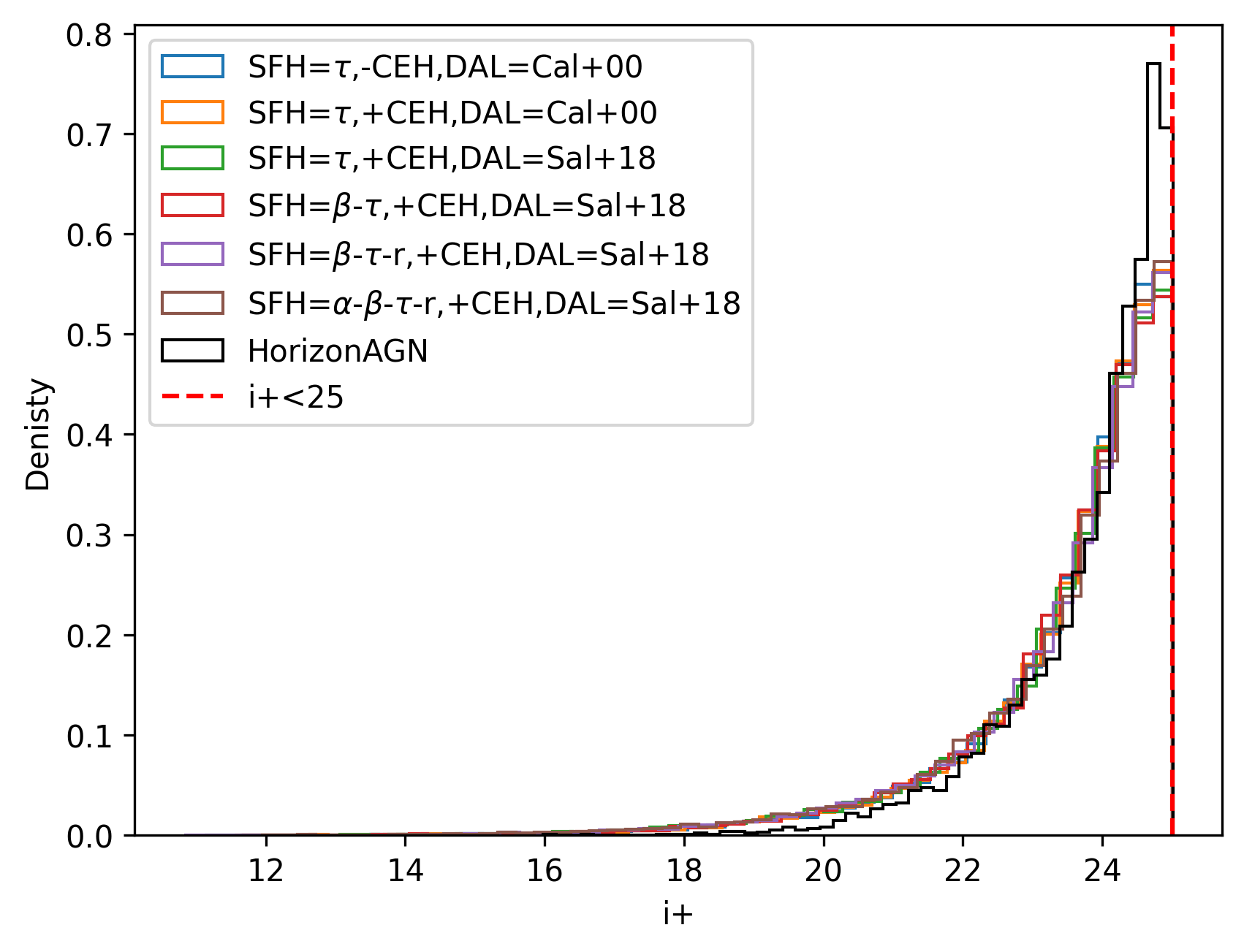}
\caption{The comparison of redshift and magnitude distributions of the empirical statistics-based and the Horizon-AGN hydrodynamical simulation-based mock galaxy samples.}
\label{fig:BayeSED2HorizonAGN}
\end{figure}

To test the performance of BayeSED, we selected two sets of samples of galaxies with $K_s<24.7$ \citep{LaigleC2019a} and $i+<25$ \citep{CaoY2018a}\footnote{In \cite{CaoY2018a} and \cite{ZhouX2022s,ZhouX2022v}, only the high-quality sources with the $SNR\geq10$ in g or i band were selected to test the performance of photometric redshift estimation. This selection is not used in this work to obtain a more full picture for the performance of the CSST wide-field multiband imaging survey.} from the empirical statistics-based mock catalog (\S \ref{ss:mock_sed}) and the hydrodynamical simulation-based mock catalog (\S \ref{ss:mock_hyd}), respectively.
The first set of samples are obtained directly with BayeSED combined with SED models with different complexity by using the method present in \S \ref{sss:galaxy_pop}.
The second sample is selected from the Horizon-AGN hydrodynamical simulation-based photometric catalogs for COSMOS-like configuration\footnote{\url{https://www.horizon-simulation.org/PHOTOCAT/HorizonAGN_LAIGLE-DAVIDZON+2019_COSMOS_v1.6.fits}} which contains $789,354$ galaxies.
We find that a sample with $10,000$ galaxies is large enough to obtain stable results for the performance tests as presented in \S \ref{sec:pf_sed} and \S \ref{sec:pf_hyd}.
The redshift and magnitude distributions of the two samples are presented in Figure \ref{fig:BayeSED2HorizonAGN}.
When employing different SED models with different complexity and the same set of empirical statistics, the empirical statistics-based samples show some differences, especially the redshift distribution.
This is likely due to the different mapping relations from physical parameters to photometries and from free parameters to derived parameters for different SED models.
Generally, the hydrodynamical simulation-based sample is consistent with  the empirical statistics-based samples.
We attribute the difference between the two set of samples to the different modeling of the SFH, CEH and  DAL of galaxies and their different distribution of redshift and physical parameters.

\section{Bayesian photometric SED analysis with BayeSED} \label{sec:bayesed_dec}
The general method for the application of Bayesian inference to photometric SED analysis of galaxies is the same as in \cite{HanY2014a,HanY2019a}.
In this section, we introduce some special aspects of Bayesian parameter estimation (\S \ref{ss:estimation}) and model selection (\S \ref{ss:model_select}) which are relevant to the current work.

\subsection{Bayesian parameter estimation} \label{ss:estimation}
For the Bayesian analysis of the mock data generated in the last section, we employ the same SED modeling procedure and setting of priors for free parameters as in \S \ref{sss:sedm}, while the commonly used Gaussian form of likelihood function is employed.
The performance of this Bayesian analysis, including its speed and quality, is crucial for the analysis of large sample of galaxies in the big data era.
We need some metrics to quantify the performance of parameter estimation which is the main subject of this work.

While the speed of parameter estimation can be easily quantified by the running time, some metrics for the quality of parameter estimation are required.
Similar to \cite{AcquavivaV2015a}, we use three metrics to quantify the quality of parameter estimation.
Bias, which characterizes the median separation between the predicted and the true values, is defined as:
\begin{equation}
    BIA={\rm Median}(\Delta x),
    \label{eq:bia}
\end{equation}
while the precision, which describes the scatter between predicted and the true values, is defined as:
\begin{equation}
    \sigma_{\rm NMAD}=1.4826*{\rm Median}(\lvert \Delta x \rvert),
    \label{eq:mad}
\end{equation}
where  $\Delta x =(x_{\mathrm{phot}}-x_{\mathrm{true}})|/(1+x_{\mathrm{true}})$ for redshift \citep{IlbertO2009a,DahlenT2013a,SalvatoM2018a}, and $ \Delta x =(x_{\mathrm{phot}}-x_{\mathrm{true}})/(x_{\mathrm{true}}^{\mathrm{max}}-x_{\mathrm{true}}^{\mathrm{min}})$ for other parameters.
The median-base definition makes them to be less sensitive to outliers (sources with unexpectedly large errors).
The fraction of outliers is defined as:
\begin{equation}
     OLF=\frac{1}{N_{\mathrm{obj}}} \times N(|\Delta x|>0.15).
    \label{eq:olf}
\end{equation}
\subsection{Bayesian model selection} \label{ss:model_select}
An important advantage of nested sampling-based algorithm, such as MultiNest, over MCMC-based method is the ability to carry out a simultaneous parameter estimation and model selection.
While the main subject of this work is parameter estimation, it is also interesting to explore the effects of sampling parameters (namingly, $nlive$ and $efr$) of MultiNest on the computation of Bayesian evidence, the quantity which is crucial for Bayesian model selection.

In \cite{HanY2019a}, we presented a mathematical framework to discriminate the different assumptions about SSP, DAL and SFH in the SED modeling of galaxies based on the Bayesian evidence for a sample of galaxies.
In this work, since the SSP model employed in the generation of mock data is the same as that employed in their Bayesian analysis, we do not need to consider the different choices of SSP.
So, the problem is significantly simplified.
In this work, we focus on the computation of the Bayesian evidence for the SED modeling of a sample of galaxies with SSP, SFH, and DAL all being assumed to be universal (i.e. ${{\bm M}({ssp}_{0},{sfh}_{0},{dal}_{0})}$-like model \cite[See Section 5.1 of][]{HanY2019a}.)
The sample Bayesian evidence in this case (as Equation 33 of \cite{HanY2019a}) is: 
\begin{flalign}
&p(\bm d_1,\bm d_2,\dots,\bm d_N|{{\bm M}({ssp}_{0},{sfh}_{0},{dal}_{0})},\bm I)=&\notag\\
&\prod\limits_{g = 1}^N\int p(\bm d_g|{\bm \theta_g}, {{\bm M}({ssp}_{0},{sfh}_{0},{dal}_{0})},\bm I)&\notag\\
&p({\bm \theta_g}|{{\bm M}({ssp}_{0},{sfh}_{0},{dal}_{0})},\bm I) {\mathrm{d}\bm \theta_g}&\notag\\
&=\prod\limits_{g = 1}^N p(\bm d_g|{{\bm M}({ssp}_{0},{sfh}_{0},{dal}_{0})},\bm I).
\label{eq:bayes_ev_NSED_UUU_000}
\end{flalign}
Although the detailed SFH and DAL of different galaxies can vary significantly, the sample Bayesian evidence computed in this manner remains valuable for identifying the most efficient combination of SFH and DAL for analyzing a vast sample of galaxies, such as the one provided by the CSST wide-field imaging survey.

In practice, the natural logarithmic of Bayesian evidence is commonly used for Bayesian model selection.
Therefore, Equation \ref{eq:bayes_ev_NSED_UUU_000} can be rewritten as:
\begin{flalign}
&\ln(BE)\equiv \ln(p(\bm d_1,\bm d_2,\dots,\bm d_N|{{\bm M}({ssp}_{0},{sfh}_{0},{dal}_{0})},\bm I))&\notag\\
&=\sum\limits_{g = 1}^N \ln(p(\bm d_g|{{\bm M}({ssp}_{0},{sfh}_{0},{dal}_{0})},\bm I)),
\label{eq:lnBE}
\end{flalign}
where $\ln(p(\bm d_g|{{\bm M}({ssp}_{0},{sfh}_{0},{dal}_{0})},\bm I))$, the Bayesian evidence for an individual galaxy, can be directly obtained in BayeSED with MultiNest.
However, the individual Bayesian evidences estimated with MultiNest contain errors.
A more strict Bayesian model selection should consider the effects of error propagation.
In our case, the error of the sample Bayesian evidence $\ln(BE)$ is simply the sum of errors for individual galaxies which is provided by MultiNest as well.

The minimum $\chi^2$ method is also widely used for model selection.
For the case with Gaussian observational errors, there is only a constant difference between the minimum $\chi^2$ and the natural logarithmic of maximum likelihood.
The sample maximum likelihood (as Equation 32 of \cite{HanY2019a}) is: 
\begin{flalign}
&\mathcal{L}_{\rm max}(\hat{\bm \theta}_1,\hat{\bm \theta}_2,\dots,\hat{\bm \theta}_N)\notag\\
&\equiv{\mathop {\rm max} \limits_ {\bm \theta_1,\bm \theta_2,\dots,\bm \theta_N}}[p(\bm d_1,\bm d_2,\dots,\bm d_N|{\bm \theta_1,\bm \theta_2,\dots,\bm \theta_N},\notag\\
&{{\bm M}({ssp}_{0},{sfh}_{0},{dal}_{0})},\bm I)]\notag\\
&=\prod\limits_{g = 1}^N{\mathop {\rm max} \limits_ {\bm \theta_g}}[p(\bm d_g|{\bm \theta_g}, {{\bm M}({ssp}_{0},{sfh}_{0},{dal}_{0})},\bm I)].
\label{eq:ML}
\end{flalign}
Then, the natural logarithmic of sample maximum likelihood is:
\begin{flalign}
&\ln(ML)\equiv \ln(\mathcal{L}_{\rm max}(\hat{\bm \theta}_1,\hat{\bm \theta}_2,\dots,\hat{\bm \theta}_N))\notag\\
&=\sum\limits_{g = 1}^N \ln({\mathop {\rm max} \limits_ {\bm \theta_g}}[p(\bm d_g|{\bm \theta_g}, {{\bm M}({ssp}_{0},{sfh}_{0},{dal}_{0})},\bm I)]),
\label{eq:lnML}
\end{flalign}
where $\ln({\mathop {\rm max} \limits_ {\bm \theta_g}}[p(\bm d_g|{\bm \theta_g}, {{\bm M}({ssp}_{0},{sfh}_{0},{dal}_{0})},\bm I)])$, the natural logarithmic of maximum likelihood for an individual galaxy, can be directly obtained in BayeSED with MultiNest.
Similar to the model selection with Bayesian evidence, only the difference of $\ln(ML)$ between different models is useful for the model selection.
Therefore, the model selection with $\ln(ML)$ is equivalent to that with minimum $\chi^2$.
In \S \ref{sec:disc}, we will discuss the difference between the two model selection methods.

\subsection{Runtime parameters of MultiNest algorithm} \label{sss:mutinest}
As the Bayesian inference engine of BayeSED, MultiNest has some runtime parameters.
The values of these runtime parameters have very important effects on the performance of BayeSED for redshift and stellar population parameter estimation of galaxies.
Here, we briefly introduce the meaning of these runtime parameters of MultiNest algorithm.

Nested sampling (NS) \citep{SkillingJ2004a,SkillingJ2006a}, as a Monte Carlo (MC) method primarily designed for the efficient computation of the Bayesian evidence, allows posterior inference as a by-product at the same time.
So, it provides a way to carry out simultaneous Bayesian parameter estimation and model selection.
As an algorithm built on the NS framework, MultiNest \citep{FerozF2008b,FerozF2009b} is special for its efficiency in sampling from posteriors that may contain several modes and/or degeneracies.
It has been improved further by the implementation of importance nested sampling (INS) \citep{CameronE2014f,FerozF2019a} to increase the efficiency for evidence computation.
In the latest version of BayeSED, the V3.12 version of MultiNest, which includes the implementation of INS, is employed.

Similar to most nested sampling algorithms, MultiNest explores the posterior distribution by maintaining a fixed number \cite[See also][for new methods using variable number]{HigsonE2019b,SpeagleJ2020a} of samples drawn from the prior distribution, called live points, and iteratively replaces the point with the lowest likelihood value (the dead point) with another point drawn from the prior but has a higher value of likelihood.
While there are many runtime parameters of MultiNest which can be set in BayeSED, only two of them are of particular importance.
They largely determined the accuracy and computational cost for the running of MultiNest algorithm and therefore BayeSED.
The first one is the total number of live points ($nlive$), which determines the effective sampling resolution.
The second one is the target sampling efficiency ($efr$), which determines the ratio of points accepted to those sampled.
Generally, the larger $nlive$ and lower $efr$ lead to more accurate posteriors and evidence values but higher computational cost.
The optimal value of $nlive$ and $efr$ should be problem-dependent, although $efr$ equals to $0.8$ and $0.3$ are recommended by the authors of MultiNest for parameter estimation and evidence evalutaion, respectively.

In this work, we will explore the effects of $nlive$ and $efr$ on the estimation of photometric redshift and stellar population parameters.
The results are presented in \S \ref{ss:pf_sed_z} and \S \ref{ss:pf_sed_p}, respectively.

\section{Results of performance tests using empirical statistics-based mock galaxy sample} \label{sec:pf_sed}
In this section, we present the results of performance tests of photometric redshift and stellar population parameter estimation by using empirical statistics-based mock galaxy sample for CSST wide-field multiband imaging survey.
Since the SED model employed in the Bayesian parameter estimation is exactly the same as that used in the generation of mock observations, the error of parameter estimation is mainly contributed by the random error in the data, parameter degeneracies, the stochastic nature of the employed MultiNest sampling algorithm and other potential errors in the BayeSED code.
To separate out the effects of the random photometric error in the data, we will consider the two cases with and without adding random noise to the photometric data.
Besides, to find out the optimal run parameters, we have considered six different choices of the target sampling efficiency ($efr$) and eight choices of the number of live points ($nlive$) for the MultiNest sampling algorithm.
Furthermore, we compare the performance of different SED models with increasing complexity in terms of running time and quality of parameter estimation.

\begin{figure*}
    \gridline
    {
        \leftfig{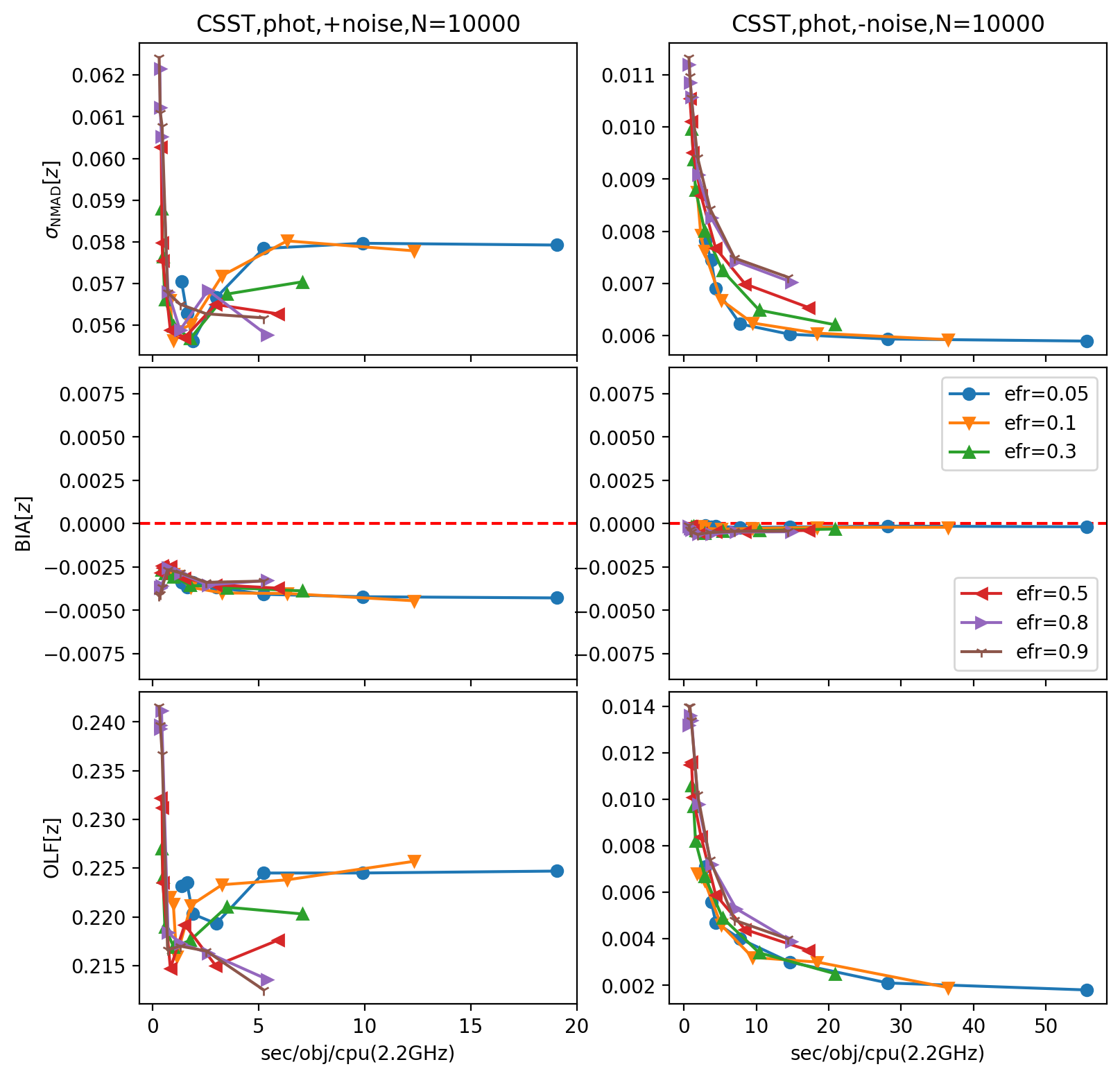}{0.48\textwidth}{(a)}
        \rightfig{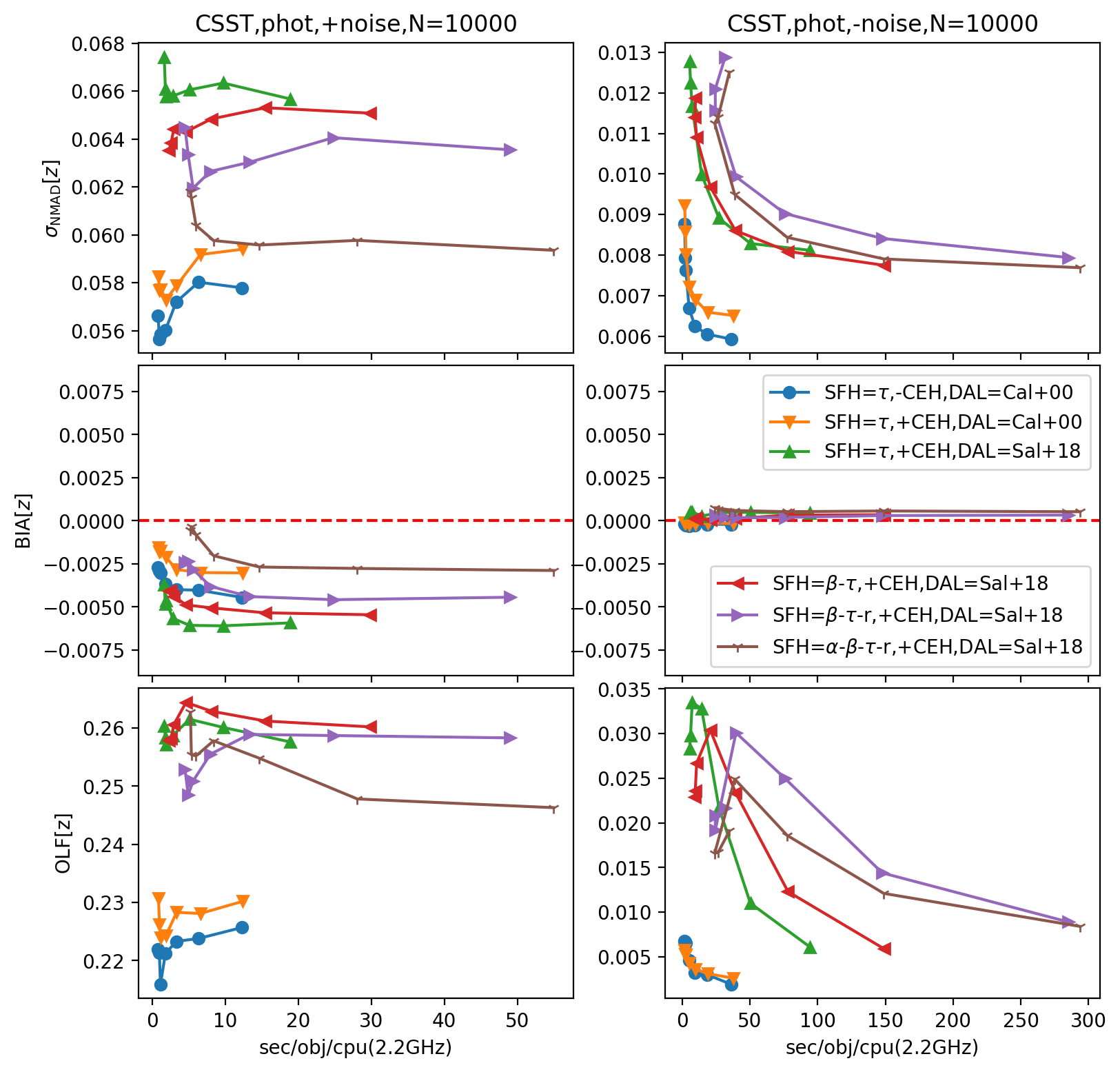}{0.48\textwidth}{(b)}
    }
    \caption{ 
         Performance test with empirical statistics-based mock galaxy sample for the photometric redshift estimation of galaxies in the CSST wide-field multiband imaging survey.
        \textbf{(a)}
        The results for only the simplest SED model (SFH=$\tau$,-CEH,DAL=Cal+00) employed in this work.
        We have considered six different choices of $efr$ (target sampling efficiency) as shown by different symbols.
        For each $efr$, we have considered eight cases with the number of live points ($nlive$), which determines the effective sampling resolution, equals to $10$, $15$, $20$, $25$, $50$, $100$, $200$, and $400$, respectively.
        The relations between the computation time (in sec/obj/cpu, by employing one core of a 2.2 GHz cpu) and the performance metrics $\sigma_{\rm NMAD}$ (top panels), BIA (middle panels), OLF (bottom panels) are shown respectively. 
        The results for the two cases with (left panels) and without (right panels) observational noise in the mock data are shown respectively.
        In general, larger value of $nlive$ and smaller value of $efr$ lead to better quality of redshift estimation, but with the cost of longer running time.
        \textbf{(b)}
        The results for six different SED models with increasing complexity.
        Only the results with $efr=0.1$ are shown.
        In general, more complicated SED models require longer running time.
        They also lead to worse quality of photometric redshift estimation, which is likely due to more severe parameter degeneracies.
        Actually, for the last four more complicated SED models with the DAL of \cite{SalimS2018a}, the number of free parameter is greater than the number of photometric data points (7 for CSST imaging survey), as shown in Table \ref{tab:models}.
    \label{fig:perf_phot_multinest_z}}
\end{figure*}
\subsection{Photometric redshift estimation} \label{ss:pf_sed_z}
The results of performance tests for photometric redshift estimation are shown in Figure \ref{fig:perf_phot_multinest_z}.
In Figure \ref{fig:perf_phot_multinest_z}\textbf{(a)}, we show the results for only the simplest SED model (SFH=$\tau$,-CEH,DAL=Cal+00) employed in this work.
As shown in the top right panel of this figure, in the case without noise, there is a clear anti-correlation between the computation time (or the sampling resolution $nlive$) and the error $\sigma_{\rm NMAD}$ (defined in Equation \ref{eq:mad}) of photometric redshift estimation.
Meanwhile, a smaller $efr$ makes the anti-correlation converge faster with the increasing of $nlive$.
There is a clear lower limit for the value of $\sigma_{\rm NMAD}$, which is about $0.006$.
As shown in the top left panel of Figure \ref{fig:perf_phot_multinest_z}\textbf{(a)}, in the case with noise, the error of photometric redshift estimation does not always decrease with the sampling resolution $nlive$.
When we set $efr=0.1$, the lowest error ($\lesssim0.056$) of redshift estimation is obtained when $nlive$ is about $25$.
When $nlive>25$, the error of redshift estimation start to increase with $nlive$ and finally converge to $\sim 0.058$.
This is most likely due to the overfitting to the noise added to the mock data.

The middle two panels of Figure \ref{fig:perf_phot_multinest_z}\textbf{(a)} show the relation between the computation time (or $nlive$) and the bias (defined in Equation \ref{eq:bia}) of photometric redshift estimation.
In the case with noise, the relation between the computation time (or $nlive$) and bias has almost the opposite profile of that of the error $\sigma_{\rm NMAD}$.
However, the bias of photometric redshift estimation is generally very small, which is almost zero in the noise-free case.

The bottom two panels of Figure \ref{fig:perf_phot_multinest_z}\textbf{(a)} show the relation between the computation time (or $nlive$) and the fraction of outliers OLF (defined in Equation \ref{eq:olf}) of photometric redshift estimation.
Similar to that for $\sigma_{\rm NMAD}$, in the noise-free case, there is a clear anti-correlation between the computation time (or $nlive$) and OLF.
In this case, the lower limit for the value of OLF is about $0.002$.
In the case with noise, the relation between the computation time (or $nlive$) and OLF has the same profile as that of the error $\sigma_{\rm NMAD}$.
When we set $efr=0.1$, the lowest OLF ($\lesssim0.215$) of redshift estimation is also obtained when $nlive$ is about $25$.
When $nlive>25$, the OLF of redshift estimation start to increase with $nlive$ and finally converge to $\sim 0.225$.

In Figure \ref{fig:perf_phot_multinest_z}\textbf{(b)}, we show the results for all of the six SED models with increasing complexity.
Here, only the results with $efr=0.1$ are shown.
In the case without noise, as shown in the top right panel of this figure, the error $\sigma_{\rm NMAD}$ of photometric redshift estimation tends to converge to a larger value when more complicated SED model is employed.
This is not strange, since more complicated SED models have more free parameters and thus suffer from more severe parameter degeneracies.
Besides, more complicated SED models apparently require longer running time.
The bias of redshift estimation is always very small no matter which SED is employed.
In general, when more complicated SED model is employed, the OLF of redshift estimation also increases significantly, and decreases much slower with the increasing of $nlive$.

In the case with noise, as shown in the left panels of Figure \ref{fig:perf_phot_multinest_z}\textbf{(b)}, the results are a little more complicated.
For the first three simplest SED models, the error $\sigma_{\rm NMAD}$ of photometric redshift estimation apparently increases with the increasing of model complexity.  
However, when more complicated forms of SFH is considered, $\sigma_{\rm NMAD}$ start to decreases with the increasing of model complexity, although not very significantly.
Meanwhile, the most complicated SED model (SFH=$\alpha$-$\beta$-$\tau$-r,+CEH,DAL=Sal+18) lead to the smallest absolute value of bias, although the bias is actually very small in all cases.
The situation for OLF is somewhat similar to that of $\sigma_{\rm NMAD}$.
No matter which SED model is employed, when $nlive\gtrsim25$, both $\sigma_{\rm NMAD}$ and OLF of redshift estimation start to increase, and then slowly decrease to a stable value.

\subsection{Stellar population parameter estimation} \label{ss:pf_sed_p}
In this subsection, we show the results of performance tests for the photometric stellar population parameter estimation.
While the estimates of many stellar population parameter are available, we only show the results for stellar mass and SFR, which are two of the most important physical parameters for the study of the formation and evolution of galaxies.
\begin{figure*}
    \gridline
    {
        \leftfig{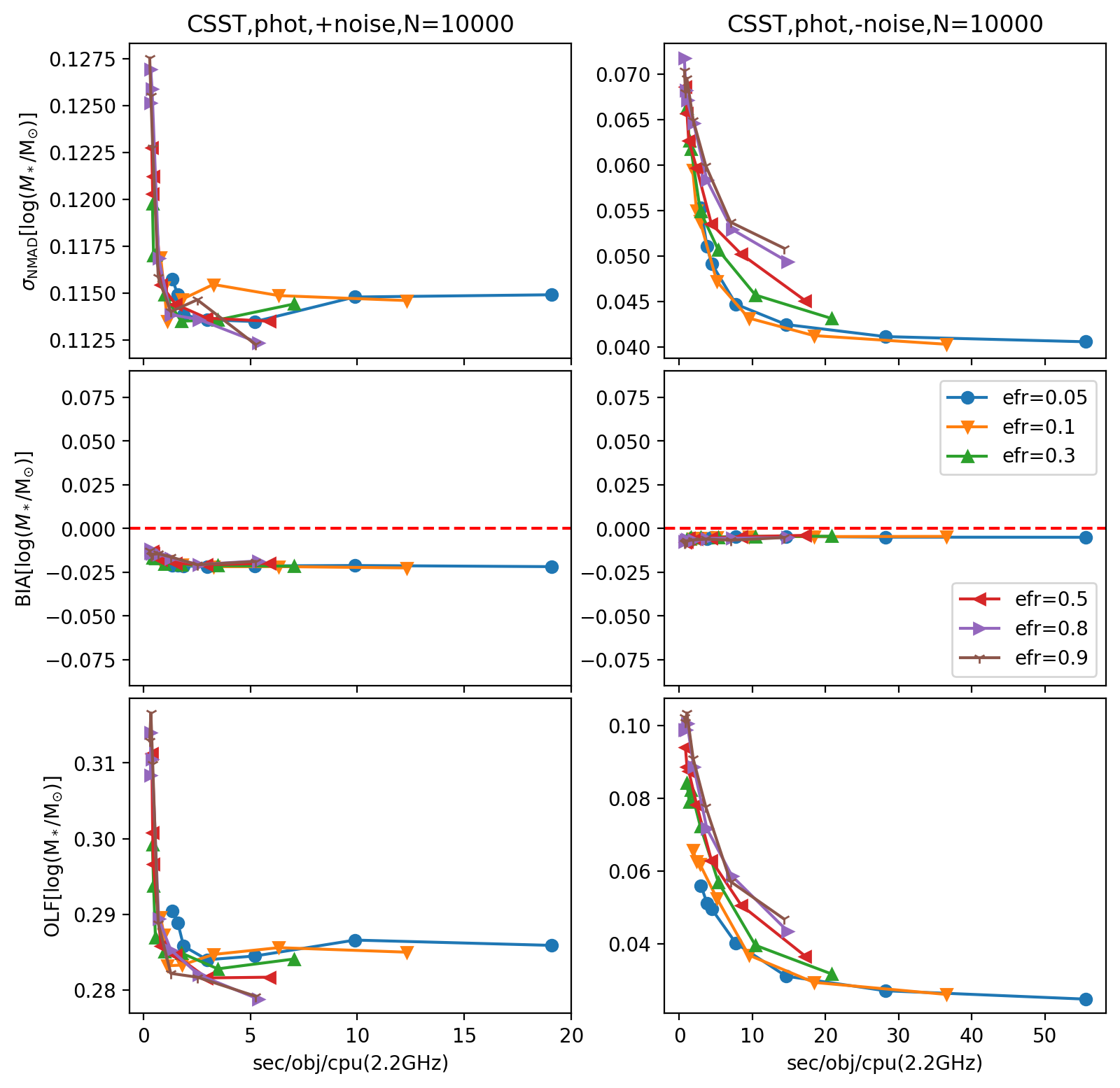}{0.48\textwidth}{(a)}
        \rightfig{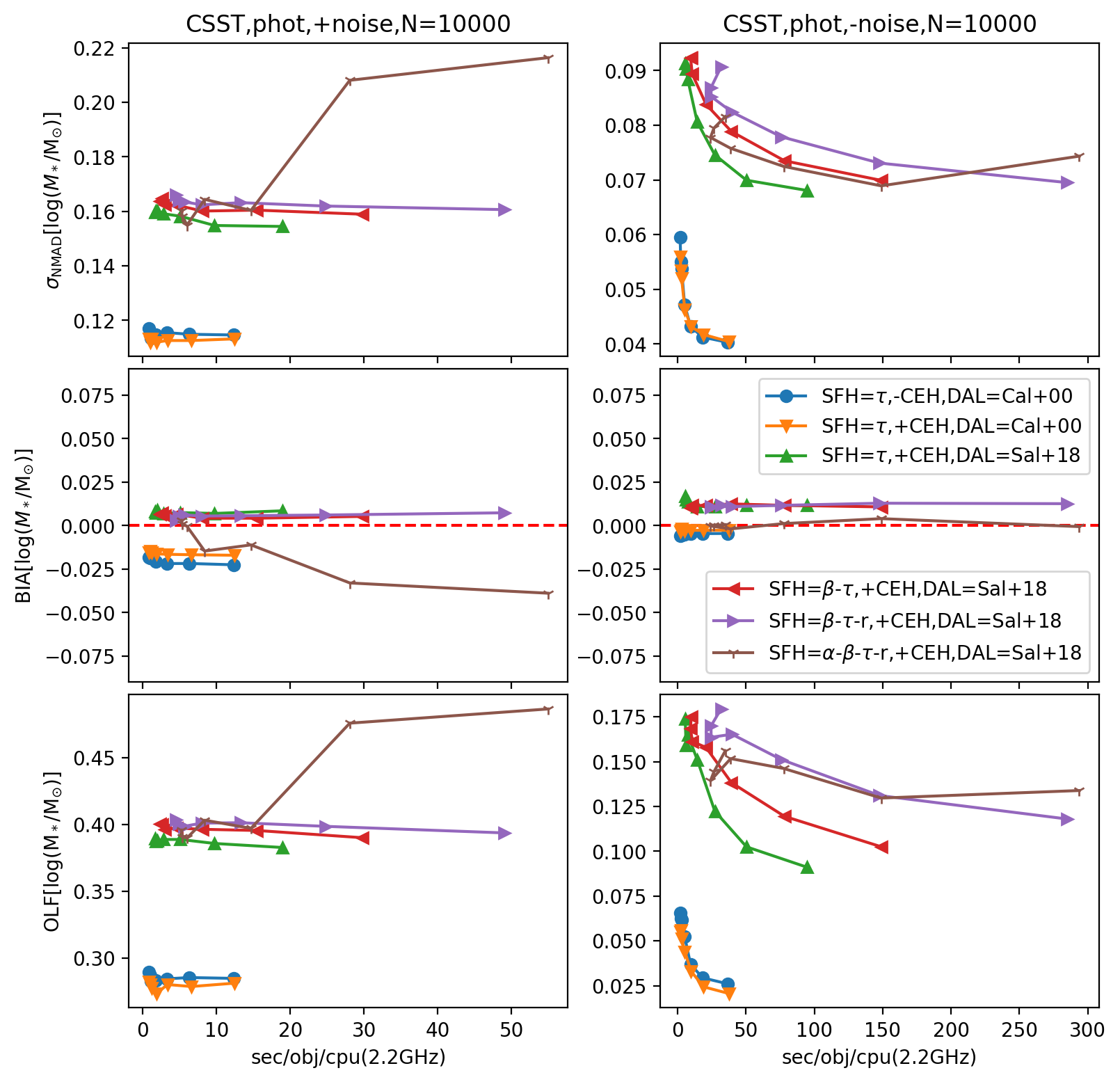}{0.48\textwidth}{(b)}
    }
    \caption
    {
        As in Figure \ref{fig:perf_phot_multinest_z}, but for the stellar mass estimation.
        The quality of stellar mass estimation, in terms of $\sigma_{\rm NMAD}$, BIA and OLF, is worse than that of redshift estimation and more sensitive to the selection of SED models.
        In the case with noise, for the most complicated SED model (SFH=$\alpha$-$\beta$-$\tau$-r,+CEH,DAL=Sal+18) used in this work, the $\sigma_{\rm NMAD}$, bias and OLF of stellar mass estimation increases significantly when $nlive>100$.
        This should be a clear indication of overfitting to the noise in the data.
        In general, more complicated SED models lead to worse quality of stellar mass estimation.
        \label{fig:perf_phot_multinest_mass}
    }
\end{figure*}
\subsubsection{Stellar mass} \label{ss:pf_sed_m}
The results of performance tests for stellar mass estimation are shown in Figure \ref{fig:perf_phot_multinest_mass}.
In Figure \ref{fig:perf_phot_multinest_mass}\textbf{(a)}, we show the results for only the simplest SED model (SFH=$\tau$,-CEH,DAL=Cal+00) employed in this work.
As shown in the top right panel of this figure, in the case without noise, there is also a clear anti-correlation between the computation time (or the sampling resolution $nlive$) and the error $\sigma_{\rm NMAD}$ of photometric stellar mass estimation.
The behavior of error $\sigma_{\rm NMAD}$ of stellar mass with respect to the change of efr is similar to that of photometric redshift.
There is also a clear lower limit for the value of $\sigma_{\rm NMAD}$, which is about $0.04$.
As shown in the top left panel of Figure \ref{fig:perf_phot_multinest_mass}\textbf{(a)}, in the case with noise, the error of stellar mass estimation does not always decrease with the sampling resolution $nlive$ as well.
When we set $efr=0.1$, the lowest error ($\sim0.1130$) of stellar mass estimation is also obtained when $nlive$ is about $25$.
When $nlive>25$, the error of stellar mass estimation only slightly increase with $nlive$.
The error of stellar mass is about two times larger than that of photometric redshift estimation.
The bias of stellar mass estimation is also larger, but still very small when comparing with $\sigma_{\rm NMAD}$.
In the noise-free case, there is also a clear anti-correlation between the computation time (or $nlive$) and the OLF of stellar mass estimation, where the lower limit for the value of OLF is about $0.03$.
In the case with noise, when we set $efr=0.1$, the lowest OLF ($\lesssim0.285$) of stellar mass estimation is also obtained when $nlive$ is about $25$.
When $nlive>25$, the OLF of stellar mass estimation only slightly increase with $nlive$ as well.
The OLF of stellar mass estimation is slightly larger than that of photometric redshift estimation.

In Figure \ref{fig:perf_phot_multinest_mass}\textbf{(b)}, we show the results for all of the six SED models with increasing complexity, where only the results with $efr=0.1$ are shown.
In the two cases with or without noise, as shown in the top right panel of this figure, the error $\sigma_{\rm NMAD}$ of photometric stellar mass estimation tends to converge to a larger value when more complicated SED model is employed.
The same is true for the OLF of photometric stellar mass estimation.
The behavior of bias is somewhat different, but it is generally very small when comparing with $\sigma_{\rm NMAD}$.
Besides, in the case with noise, for the most complicated SED model (SFH=$\alpha$-$\beta$-$\tau$-r,+CEH,DAL=Sal+18) used in this work, the $\sigma_{\rm NMAD}$, bias and OLF of stellar mass estimation increases significantly when $nlive>100$.
This should be a very clear indication of overfitting to the noise in the data.
In general, more complicated SED models lead to worse quality of stellar mass estimation.

\begin{figure*}
    \gridline
    {
        \leftfig{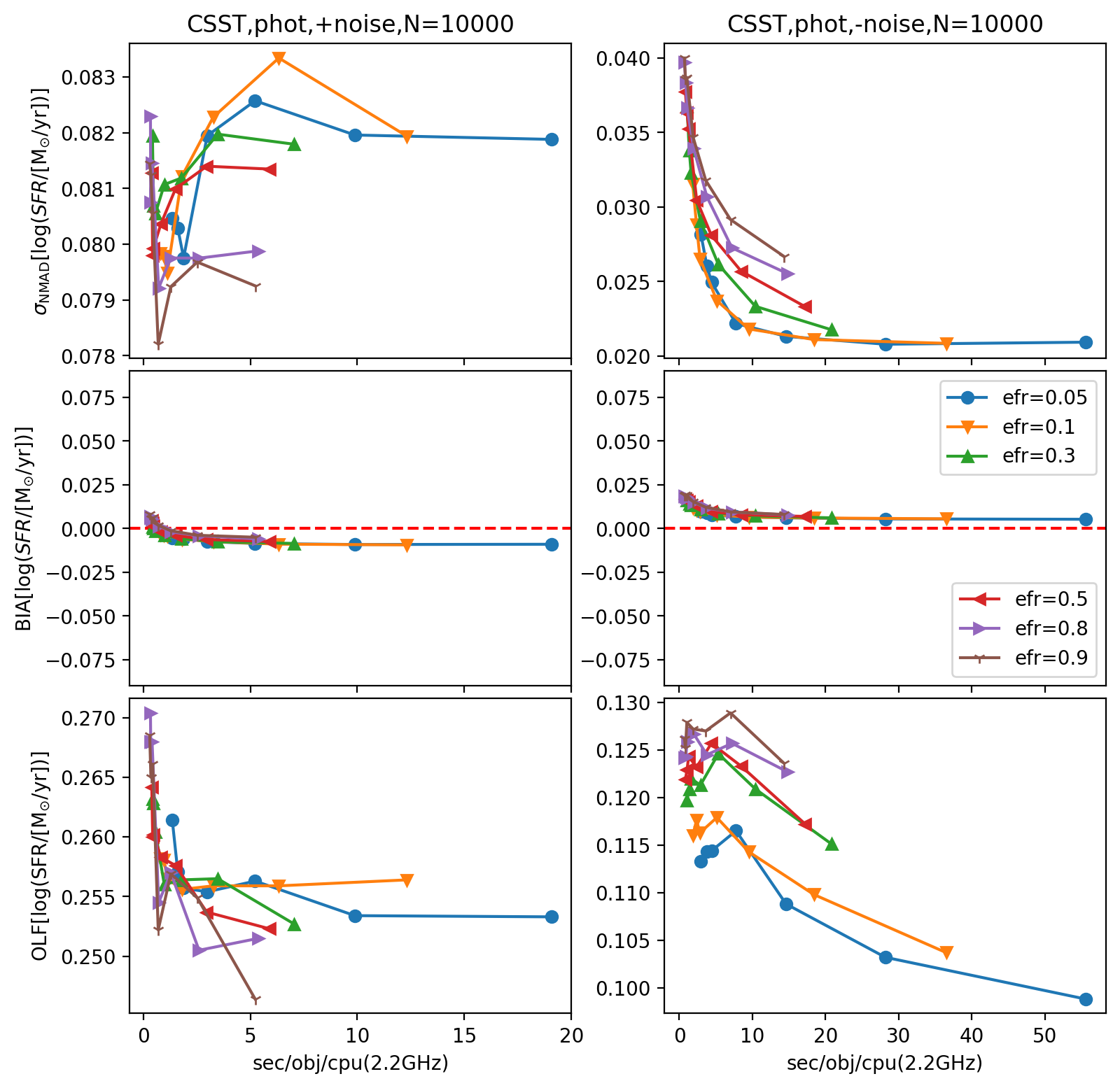}{0.48\textwidth}{(a)}
        \rightfig{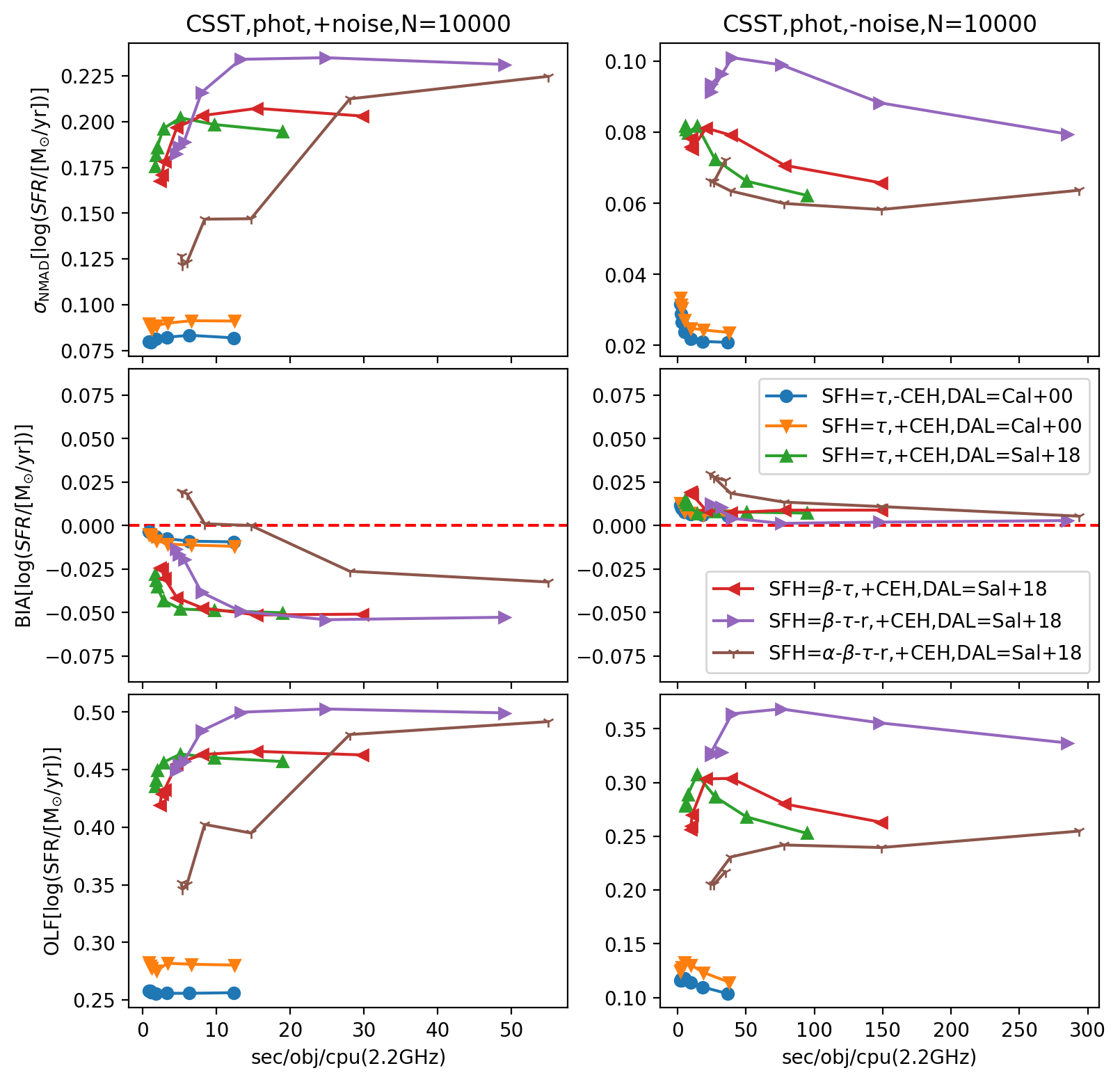}{0.48\textwidth}{(b)}
    }
    \caption
    {
        As in Figure \ref{fig:perf_phot_multinest_z}, but for the star-formation rate estimation.
        The quality of star-formation rate estimation, in terms of $\sigma_{\rm NMAD}$, BIA and OLF, is slightly better than that of stellar mass estimation, but even more sensitive to the selection of SED models.
        In the case with noise, for the four most complicated SED models used in this work, the $\sigma_{\rm NMAD}$, bias and OLF of SFR estimation significantly increases with $nlive$.
        This is another even more clear indication of overfitting to the noise in the data.
        In general, more complicated SED models lead to worse quality of SFR estimation.
        \label{fig:perf_phot_multinest_sfr}
    }
\end{figure*}
\subsubsection{Star-formation rate} \label{ss:pf_sed_s}
The results of performance tests for SFR estimation are shown in Figure \ref{fig:perf_phot_multinest_sfr}.
In Figure \ref{fig:perf_phot_multinest_sfr}\textbf{(a)}, we show the results for only the simplest SED model (SFH=$\tau$,-CEH,DAL=Cal+00) employed in this work.
Similar to the results for photometric redshift and stellar mass estimation, in the case without noise, there is also a clear anti-correlation between the computation time (or the sampling resolution $nlive$) and the error $\sigma_{\rm NMAD}$ of photometric stellar mass estimation.
The behavior of error $\sigma_{\rm NMAD}$ of SFR with respect to the change of efr is similar to that of photometric redshift and stellar mass.
There is also a clear lower limit for the value of $\sigma_{\rm NMAD}$, which is about $0.02$.
As shown in the top left panel of Figure \ref{fig:perf_phot_multinest_mass}\textbf{(a)}, in the case with noise, the error of SFR estimation increases apparently when the sampling resolution $nlive\gtrsim25$.
Generally, the error of SFR estimation is slightly smaller than that of stellar mass estimation.
The bias of SFR estimation is also slightly smaller, and ignorable with respect to $\sigma_{\rm NMAD}$.
In the noise-free case, the relation between the computation time (or $nlive$) and the OLF of SFR estimation is somewhat different from that of photometric redshift and stellar mass.
Even with $nlive=500$, the OLF of SFR estimation still does not seem to converge.
The lower limit for the value of OLF seems near $0.1$.
In the case with noise, the OLF of SFR estimation converge much faster to about $0.255$ when we set $efr=0.1$.
This is slightly smaller than that of stellar mass estimation.

In Figure \ref{fig:perf_phot_multinest_sfr}\textbf{(b)}, we show the results for all of the six SED models with increasing complexity, where only the results with $efr=0.1$ are shown.
In the two cases with or without noise, as shown in the top right panel of this figure, the error $\sigma_{\rm NMAD}$ of SFR estimation tends to converge to a larger value when more complicated SED model is employed, and is more sensitive to the selection of SED model than that of stellar mass.
The same is true for the OLF of SFR estimation.
The behavior of bias is somewhat different, but it is generally very small when comparing with $\sigma_{\rm NMAD}$.
Besides, in the case with noise, for the four most complicated SED models used in this work, the $\sigma_{\rm NMAD}$, bias and OLF of SFR estimation significantly increases with $nlive$.
This is another even more clear indication of overfitting to the noise in the data.
In general, more complicated SED models lead to worse quality of SFR estimation.

\subsection{Computation of Bayesian evidence} \label{ss:pf_sed_BE}
In this subsection, we present the results of performance test for the computation of Bayesian evidence, a quantity which is crucial for Bayesian model selection.

\begin{figure*}
    \gridline
    {
        \leftfig{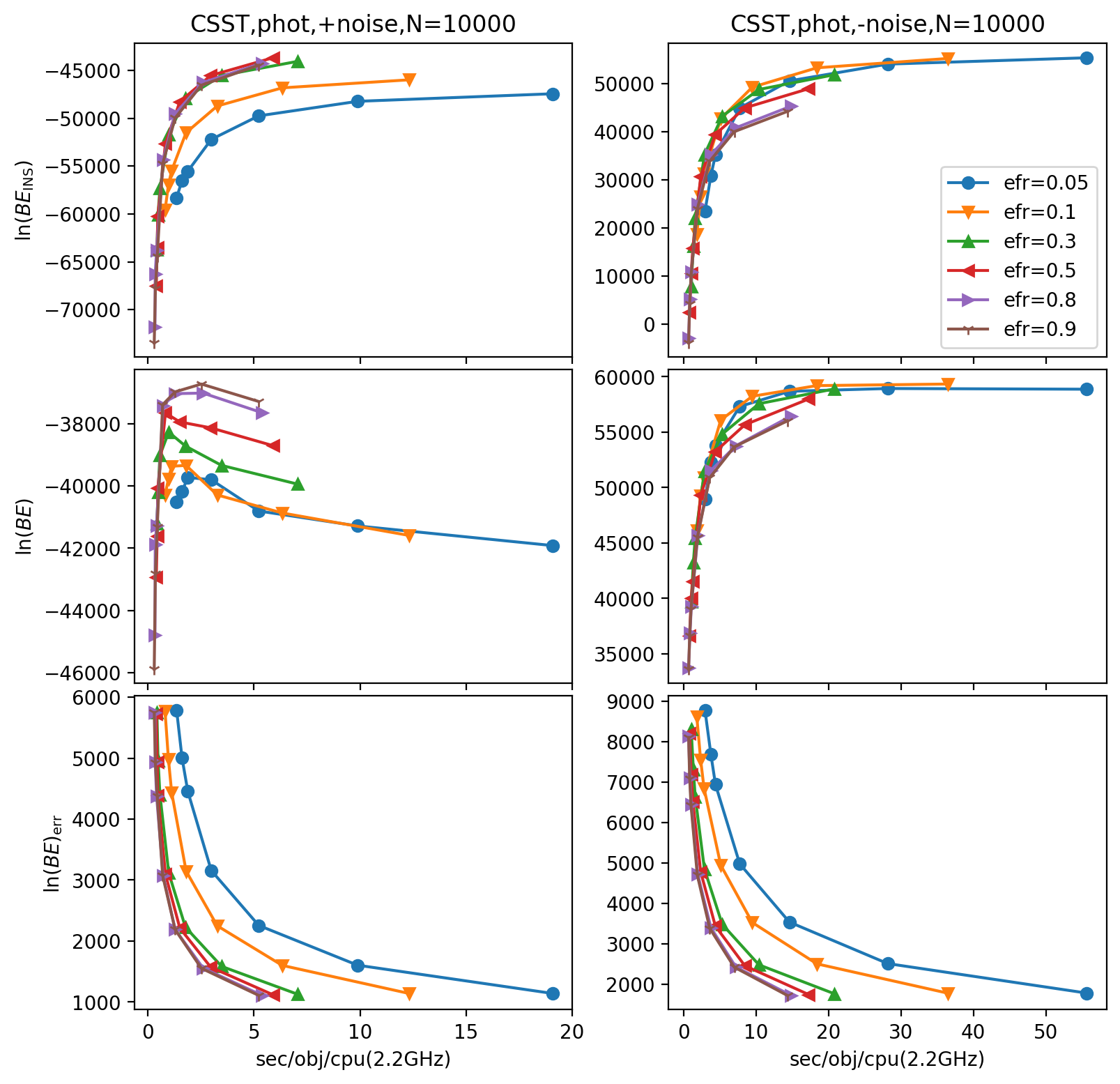}{0.48\textwidth}{(a)}
        \rightfig{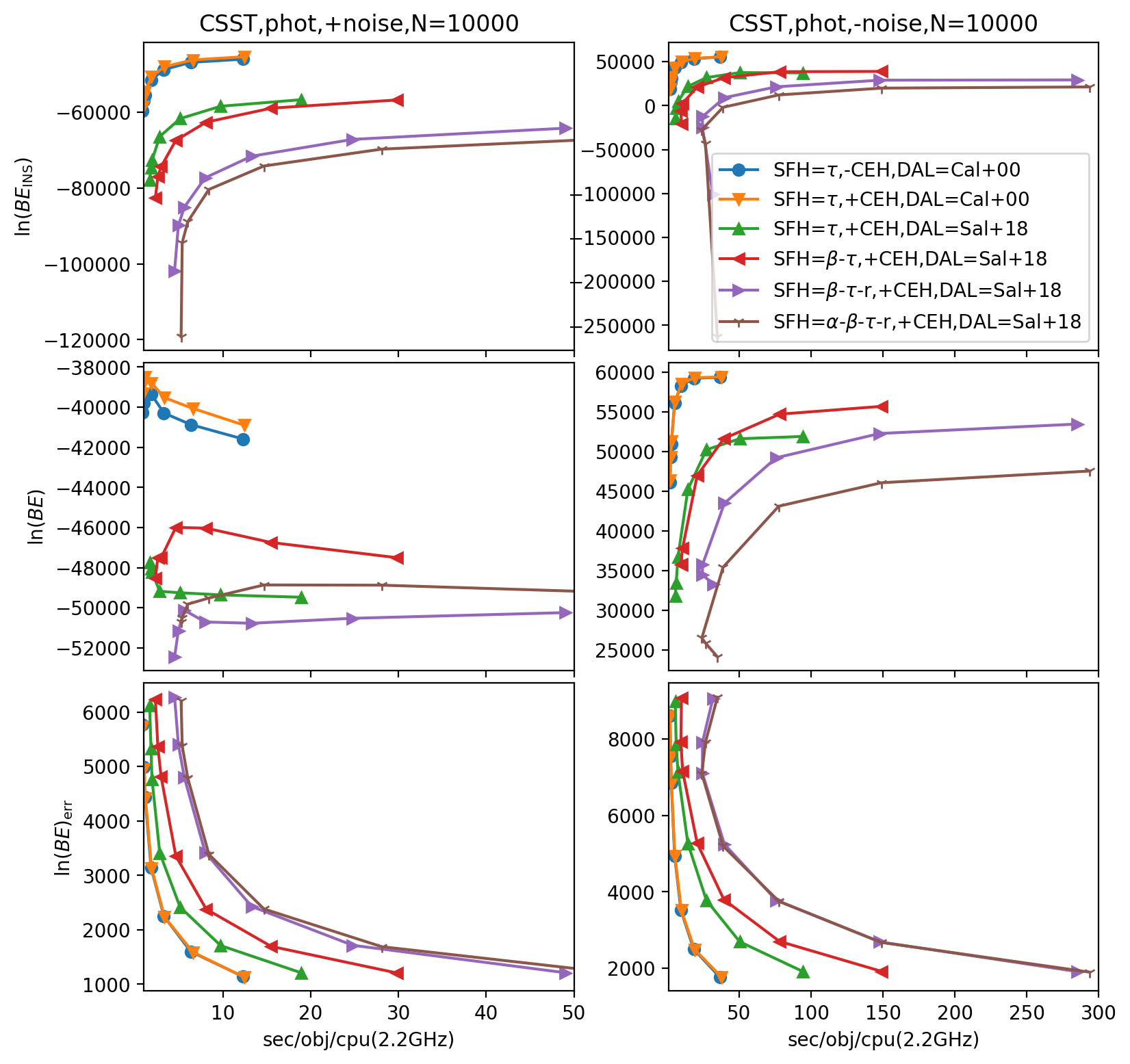}{0.48\textwidth}{(b)}
    }
    \caption
    {
        Performance test of Bayesian evidence estimation with empirical statistics-based mock galaxy sample for CSST imaging survey. The relations between the computation time (in sec/obj by employing a single 2.2 GHz cpu core) and the value of the natural logarithm of Bayesian evidence (as computed with Equation \ref{eq:lnBE}) and its error (bottom panels) for the whole galaxy sample are shown. 
        The results for two versions of Bayesian evidence with (top panels) or without (middle panels) importance sampling have been shown.
        \textbf{(a)} 
        The results for only the simplest SED model (SFH=$\tau$,-CEH,DAL=Cal+00) employed in this work.
        We have considered six different choices of $efr$ (target sampling efficiency) as shown by different symbols.
        The results for seven cases with $nlive$ equals to $15$, $20$, $25$, $50$, $100$, $200$, and $400$ are shown respectively.
        The left panels show the results with noisy data while the right panels show the results with noise-free data.
        \textbf{(b)} 
        The results for six different SED models with increasing complexity.
        Only the results with $efr=0.1$ are shown.
        The value of Bayesian evidence clearly decreases with the increasing of the complexity of the SED model.
        \label{fig:perf_phot_multinest_logZ}
    }
\end{figure*}

In Figure \ref{fig:perf_phot_multinest_logZ}\textbf{(a)}, we show the results for only the simplest SED model (SFH=$\tau$,-CEH,DAL=Cal+00) employed in this work.
As shown in the top and middle panels of this figure, the Bayesian evidence computed with importance sampling is more stable than that without importance sampling, especially in the case with noise.
So, hereafter and especially in \S \ref{sec:pf_hyd} and \ref{sec:disc}, all mentioned Bayesian evidences are computed with importance sampling.
The value of Bayesian evidence increases with the number of live points ($nlive$) which determines the effective sampling resolution.
In all cases, it eventually converges to a stable value when $nlive$ is very large, while a smaller sampling efficiency ($efr$) leads to faster convergence rate.
A good balance between the speed and quality of Bayesian evidence estimation can be achieved when the MultiNest runtime parameters $efr$ equals to $0.1$ and $nlive$ equals to $50$.

On the other hand, as shown in the bottom panels of Figure \ref{fig:perf_phot_multinest_logZ}\textbf{(a)}, the error of Bayesian evidence decreases with $nlive$, while a larger sampling efficiency ($efr$) also leads to faster convergence rate.
However, unlike the value of Bayesian evidence, the error of Bayesian evidence converges slower with the increasing of $nlive$ in all cases.
As a result, if we set $efr=0.1$ and $nlive=50$, the error of Bayesian evidence would be overestimated.
A much larger value of $nlive$ seems required to obtain a more reliable estimation for the error of Bayesian evidence with MultiNest, which would be very computationally expansive and not suitable for the analysis of massive photometric data.
However, in practice, this may not be a serious issue, since an overestimated error of Bayesian evidence only leads to a more conservative conclusion about model comparison.
We just need to keep this in mind.

In Figure \ref{fig:perf_phot_multinest_logZ}\textbf{(b)}, we show the results for all SED models with increasing complexity, where only the results with $efr=0.1$ are shown.
Although the data used for the computation of Bayesian evidence is different for different SED models, the value of Bayesian evidence clearly decreases with the increasing of the complexity of SED model.
This is reasonable.
Since the same SED model is employed for the generation of mock data and their Bayesian analysis, the mock data can always be interpreted well.
However, a more complicated SED model is penalized for being distributed over a larger space, of which only a smaller fraction is useful for the given mock data.

\section{Results of performance tests using hydrodynamical simulation-based mock galaxy sample} \label{sec:pf_hyd}
In this section, we present the results of performance tests of photometric redshift and stellar population parameter estimation  by using hydrodynamical simulation-based mock galaxy sample for CSST-like imaging survey.
Only the results obtained with the simplest SED model are shown.
In \S \ref{sec:disc}, we will discuss the effect of more flexible SFH and DAL for CSST-like, CSST+Euclid-like and COSMOS-like surveys, respectively.

As mentioned in \S \ref{ss:mock_hyd}, the generation of this mock galaxy sample accounts for the complex SFH and metal enrichment of Horizon-AGN galaxies, and consistently includes dust attenuation.
However, for the Bayesian analysis of this more theoretical mock galaxy sample, we firstly employ the widely used assumptions about SFH (exponentially declining), metal enrichment history (constant but free), and dust attenuation (uniform foreground dust screen with \cite{CalzettiD2000a} DAL).
The results in this section will help us to quantify the systematic errors resulting from these simplified assumptions.
Besides, as mentioned in \S \ref{sec:sample}, galaxies in the mock sample used here are selected with $K_s<24.7$ and $i_+<25$.
In the literature, it is quite common to exclude some pathological cases  with a kind of $\chi^2$ selection \citep{DavidzonI2017a,CaputiK2015f,LaigleC2019a} before presenting the results of performance test.
However, no such cut was made here because the pathological cases are precisely what we want to investigate.

\begin{figure*}
    \gridline
    {
        \leftfig{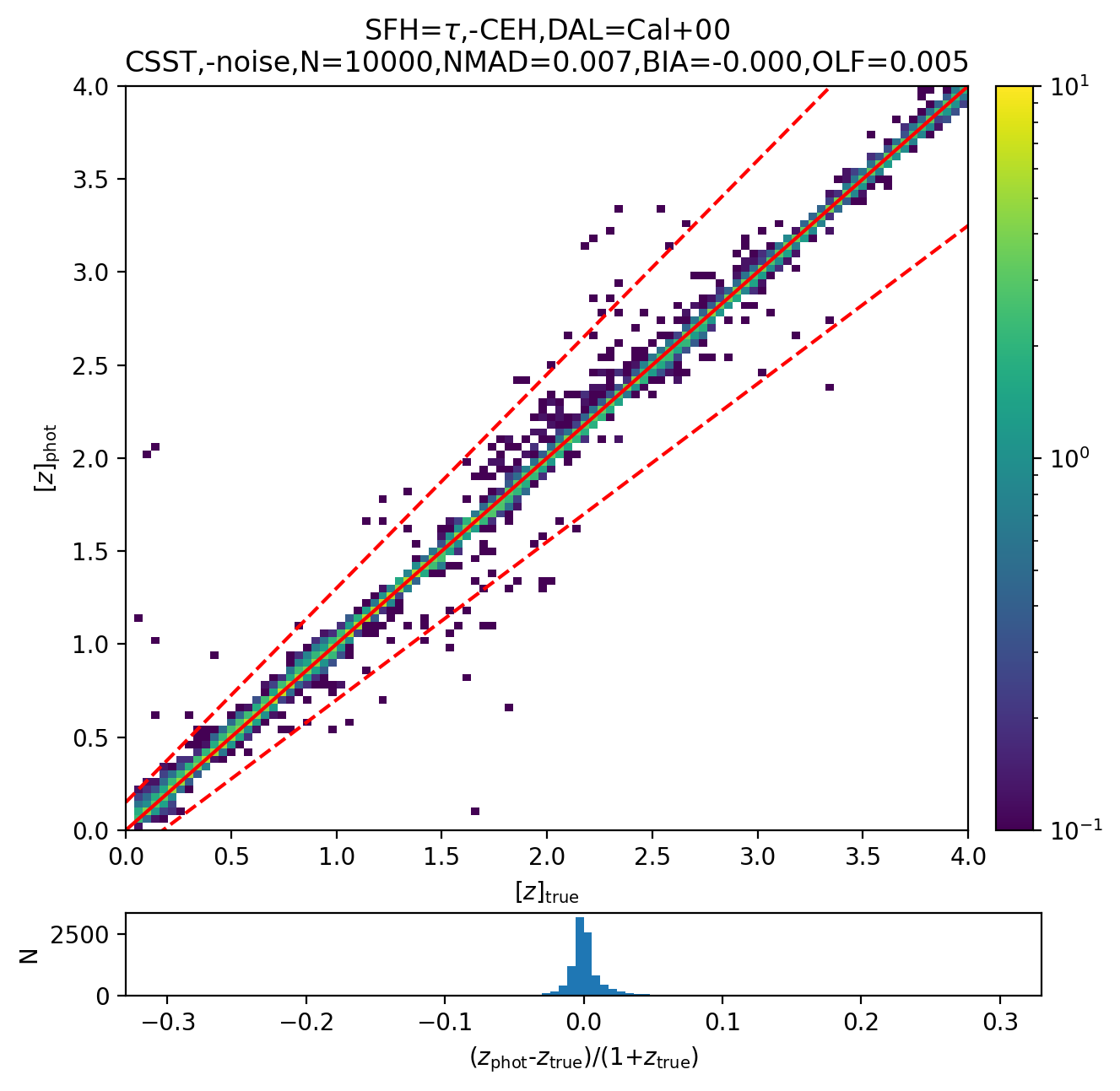}{0.48\textwidth}{(a)}
        \rightfig{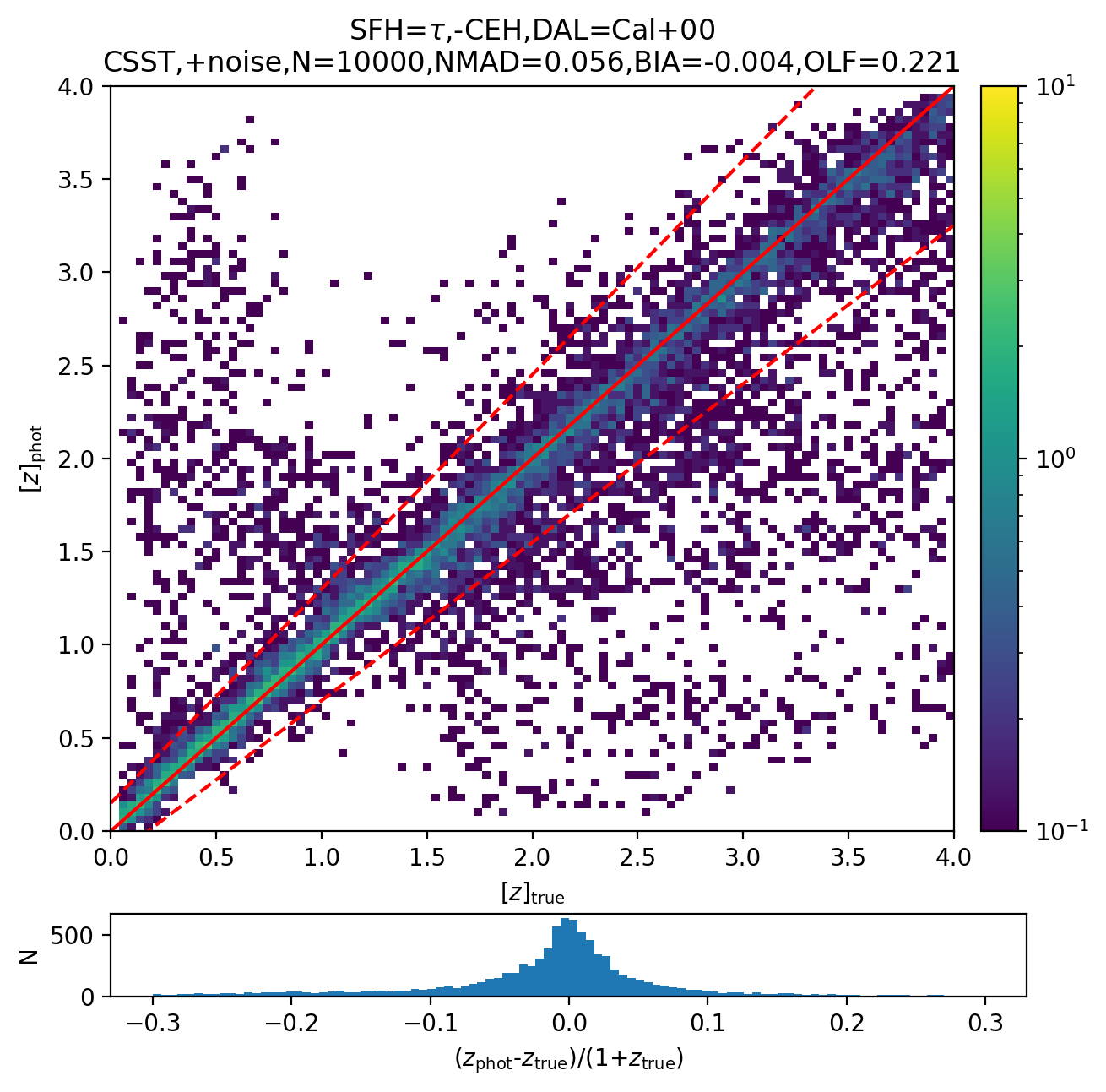}{0.48\textwidth}{(b)}
    }
    \gridline
    {
        \leftfig{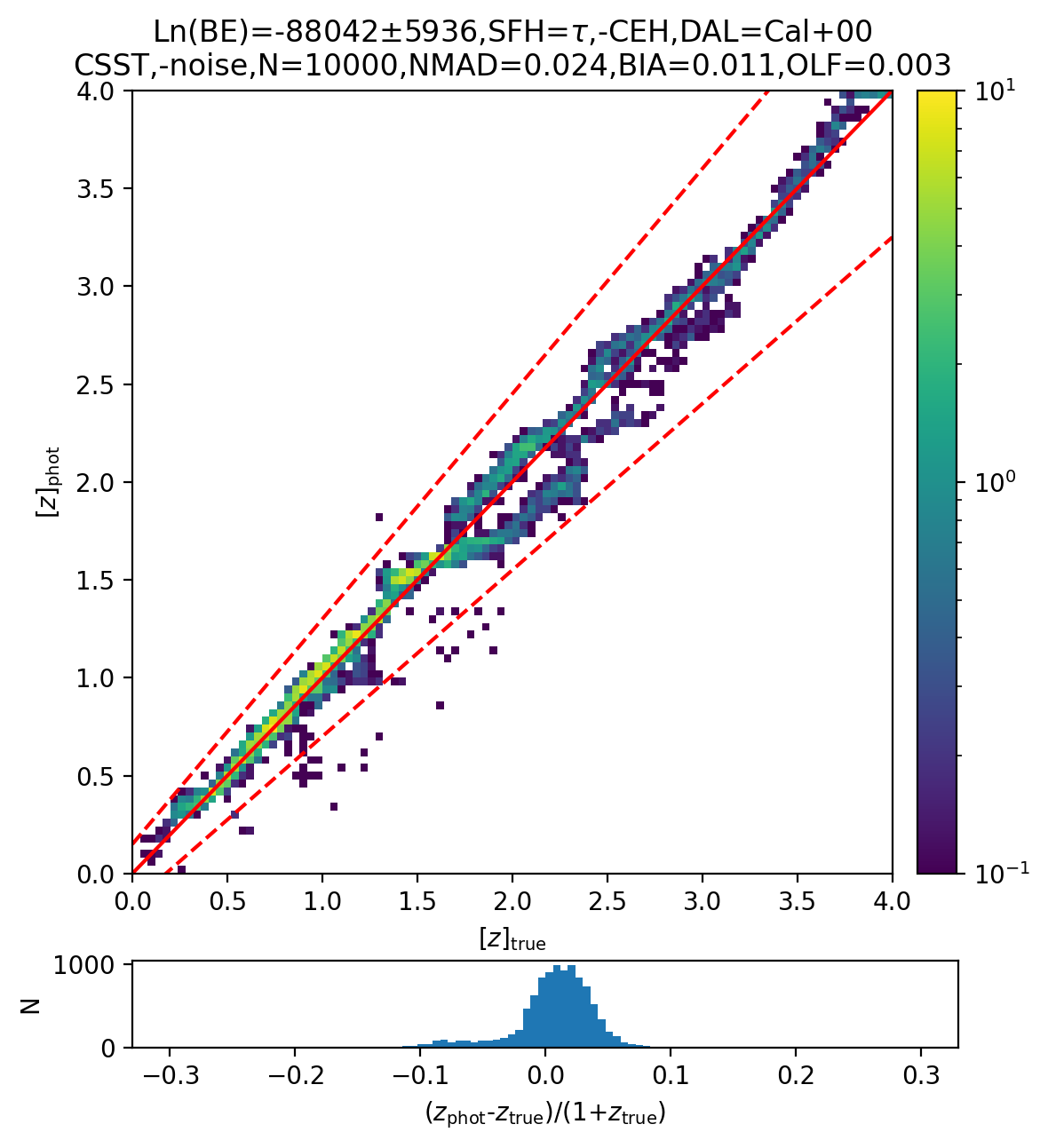}{0.48\textwidth}{(c)}
        \rightfig{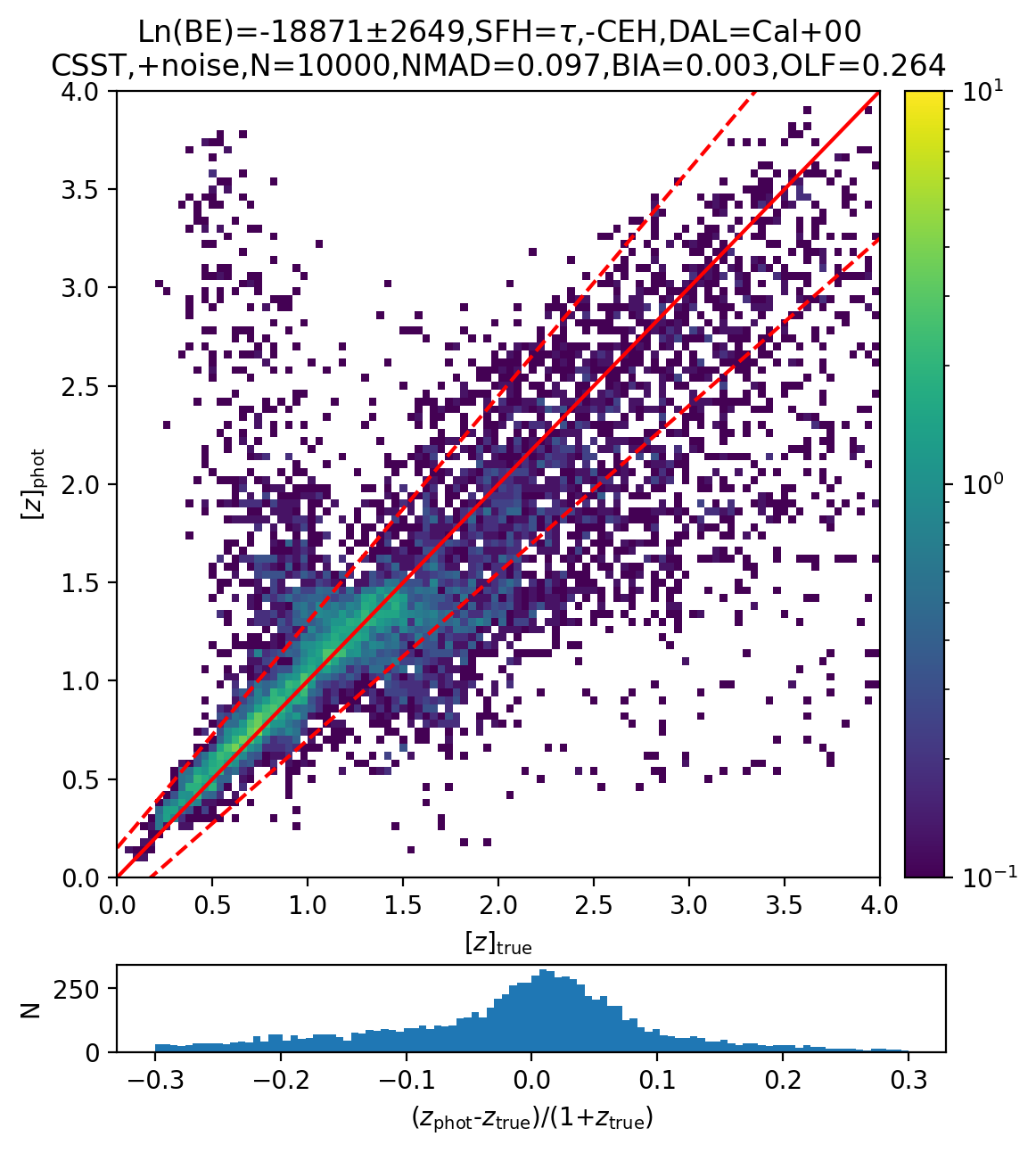}{0.48\textwidth}{(d)}
    }
    \caption
    {
        The results of photometric redshift estimation with (righ panels) and without (left panels) noise, and for the analysis of the empirical statistics-based (top panels) and hydrodynamical simulation-based (bottom panels) mock galaxy sample, respectively.
        The error from imperfect SED modeling will only present for the analysis of the hydrodynamical simulation-based mock galaxy sample.
        The photometric redshifts ($z_{\rm phot}$) are estimated by employing the $\tau$ model of SFH without consideration of metallicity evolution and the \cite{CalzettiD2000a} model of DAL.
        The red solid line indicate the identity while the red dotted lines indicate the outlier limits, i.e. $\left|z_{\mathrm{phot}}-z_{\mathrm{true}}\right|/(1+z_{\mathrm{true}})>0.15$.
        Here, we show the results obtained with the MultiNest runtime parameters $efr$ equals to $0.1$ and $nlive$ equals to $50$.
        In general, the observational noise is the more important source of error for the photometric redshift estimation of galaxies, and the contribution from imperfect SED modeling is also very important.
    \label{fig:perf_phot_z}
}
\end{figure*}
\begin{figure*}
    \gridline
    {
        \leftfig{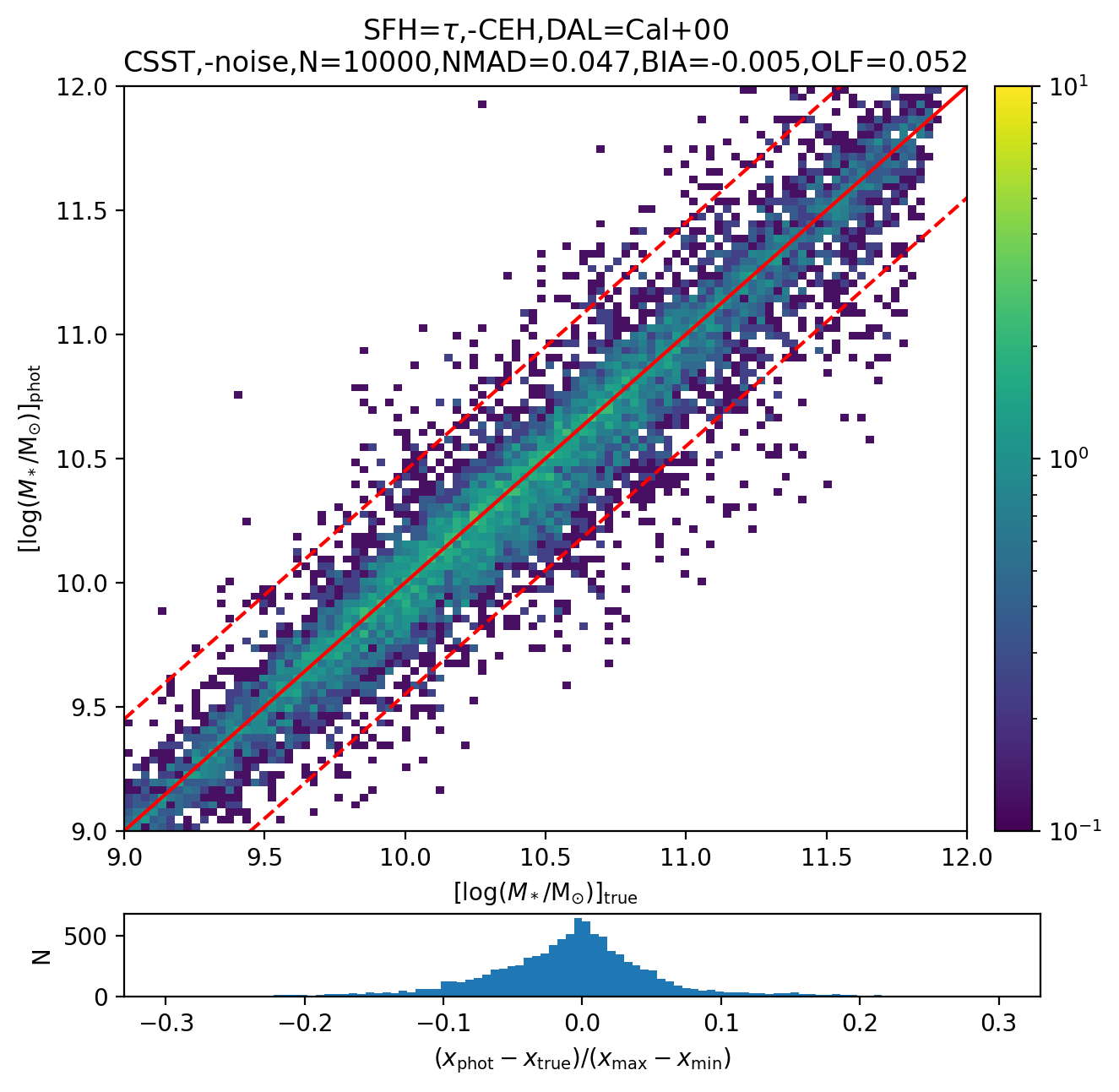}{0.48\textwidth}{(a)}
        \rightfig{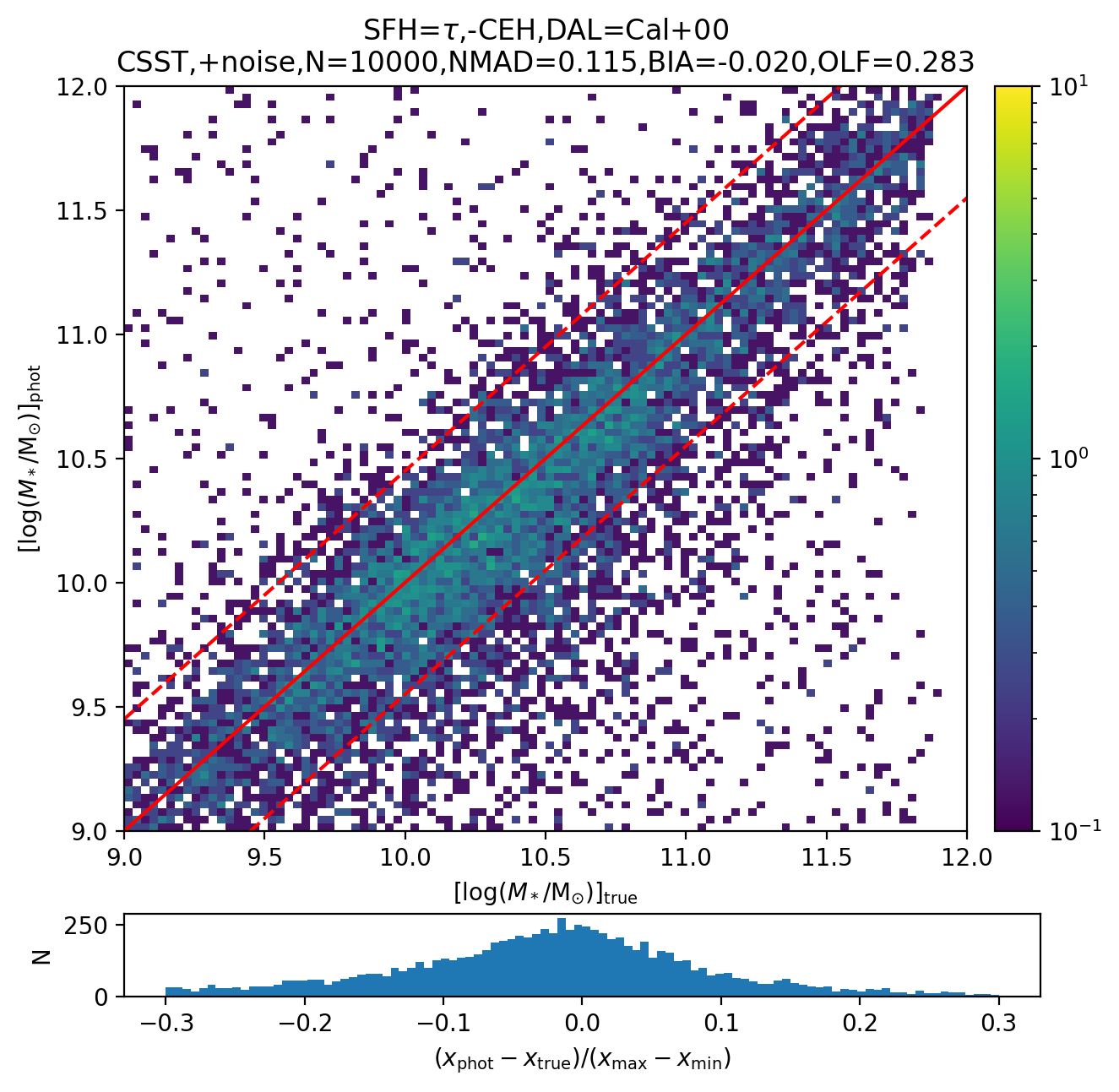}{0.48\textwidth}{(b)}
    }
    \gridline
    {
        \leftfig{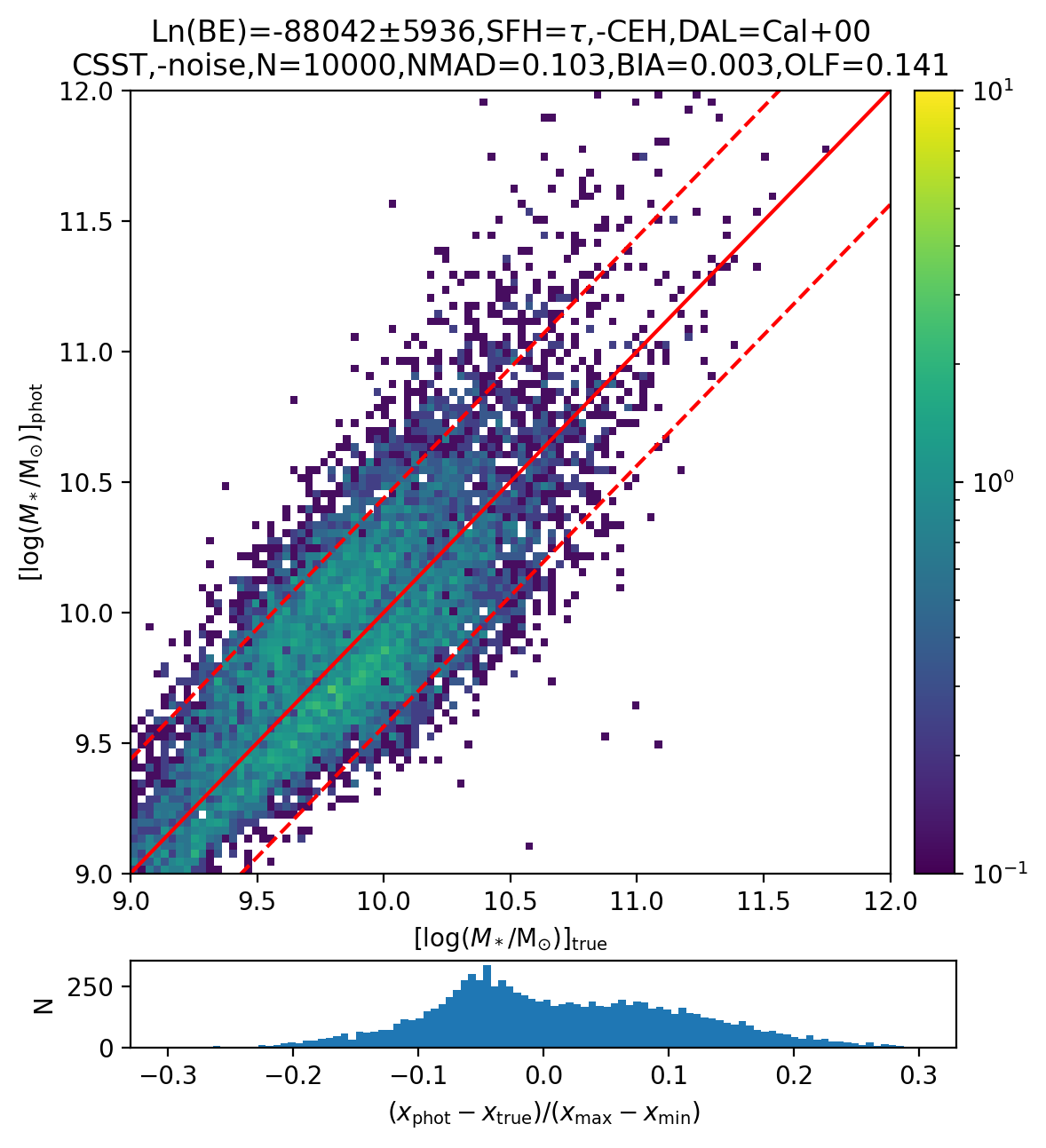}{0.48\textwidth}{(c)}
        \rightfig{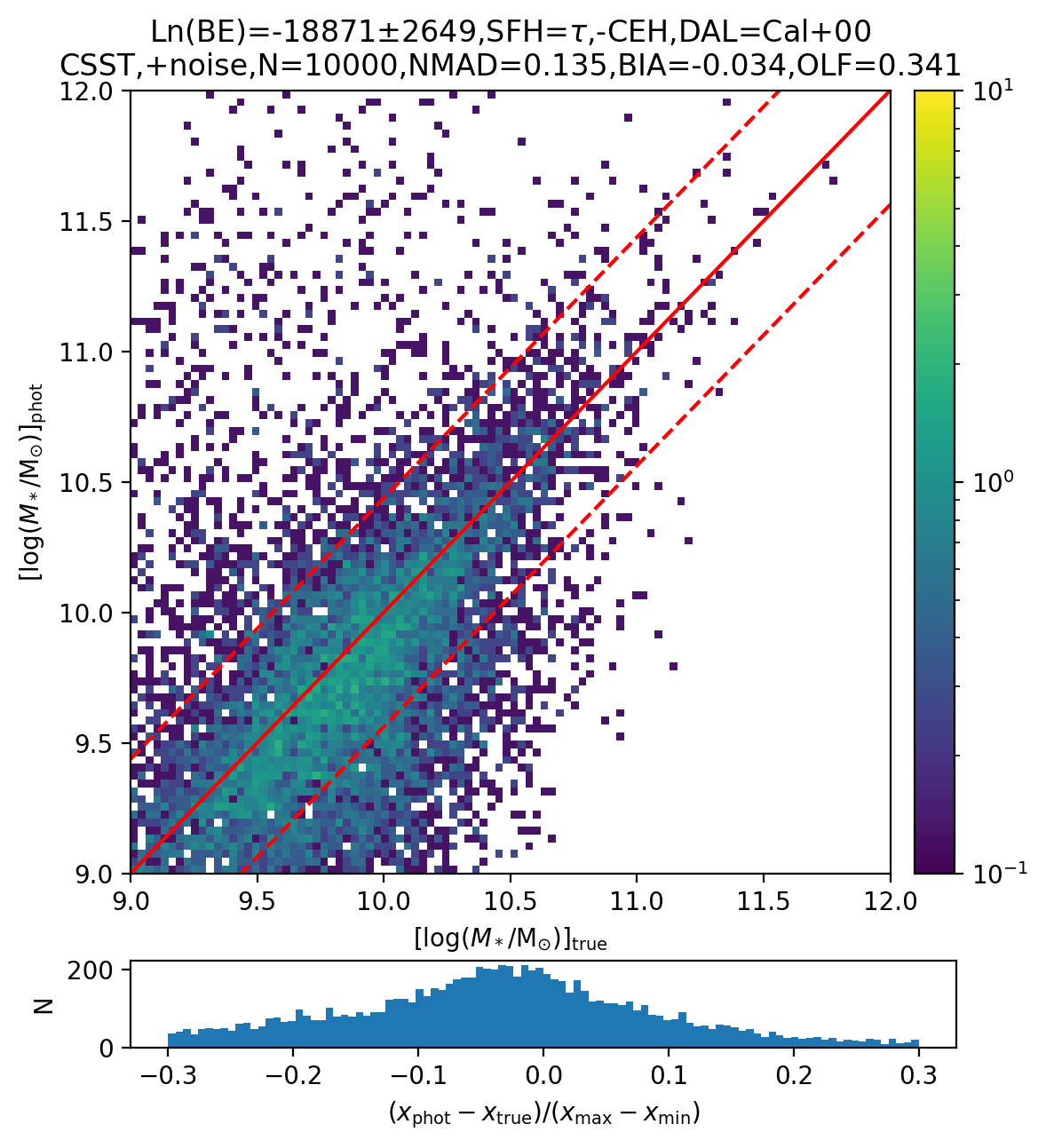}{0.48\textwidth}{(d)}
    }
    \caption
    {
        As Figure \ref{fig:perf_phot_z}, but for the stellar mass. 
        In general, the observational noise is the more important source of error for the photometric stellar mass estimation of galaxies, and the contribution from imperfect SED modeling is almost comparable.
        \label{fig:perf_phot_mass}
    }
\end{figure*}
\begin{figure*}
    \gridline
    {
        \leftfig{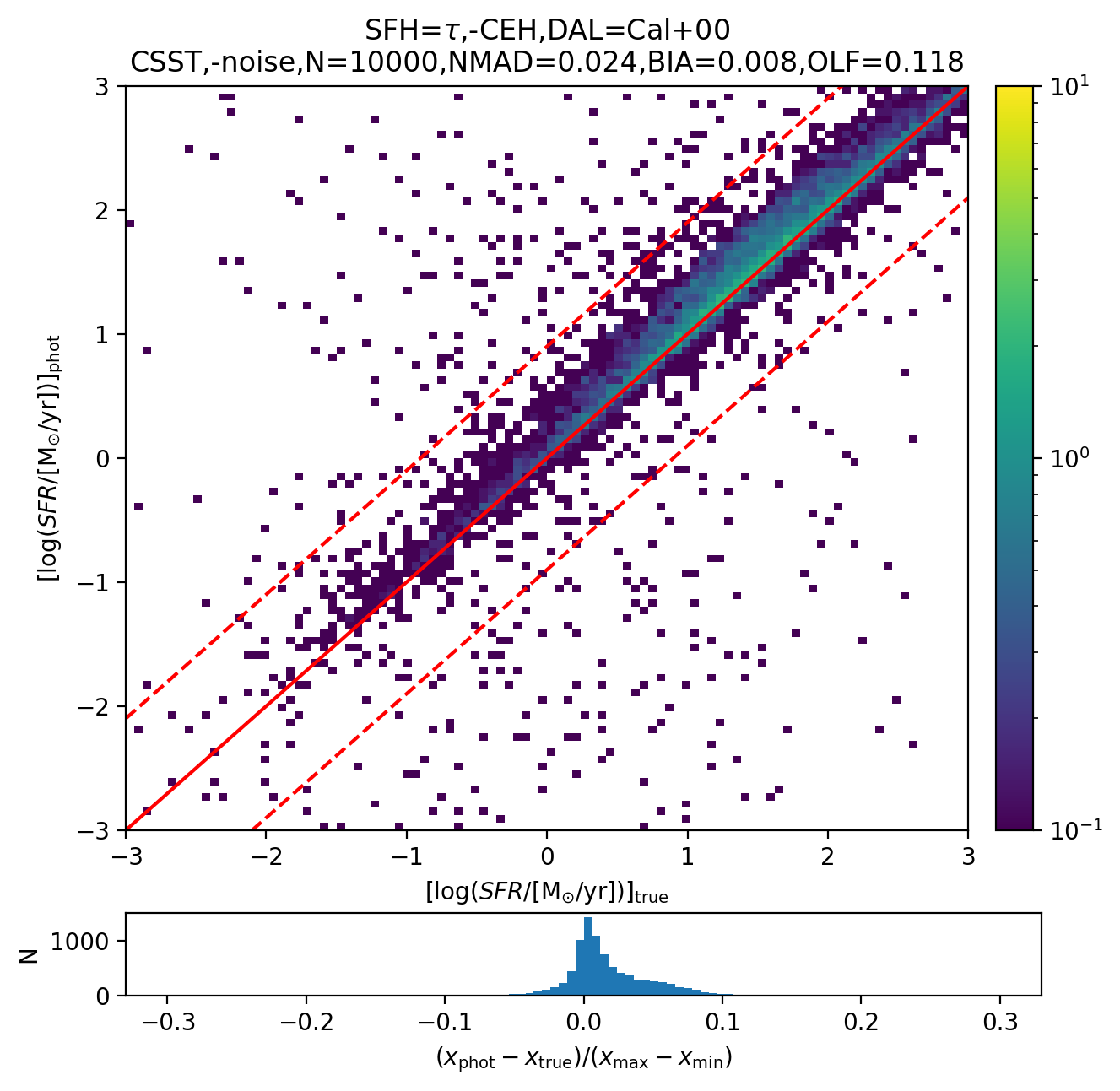}{0.48\textwidth}{(a)}
        \rightfig{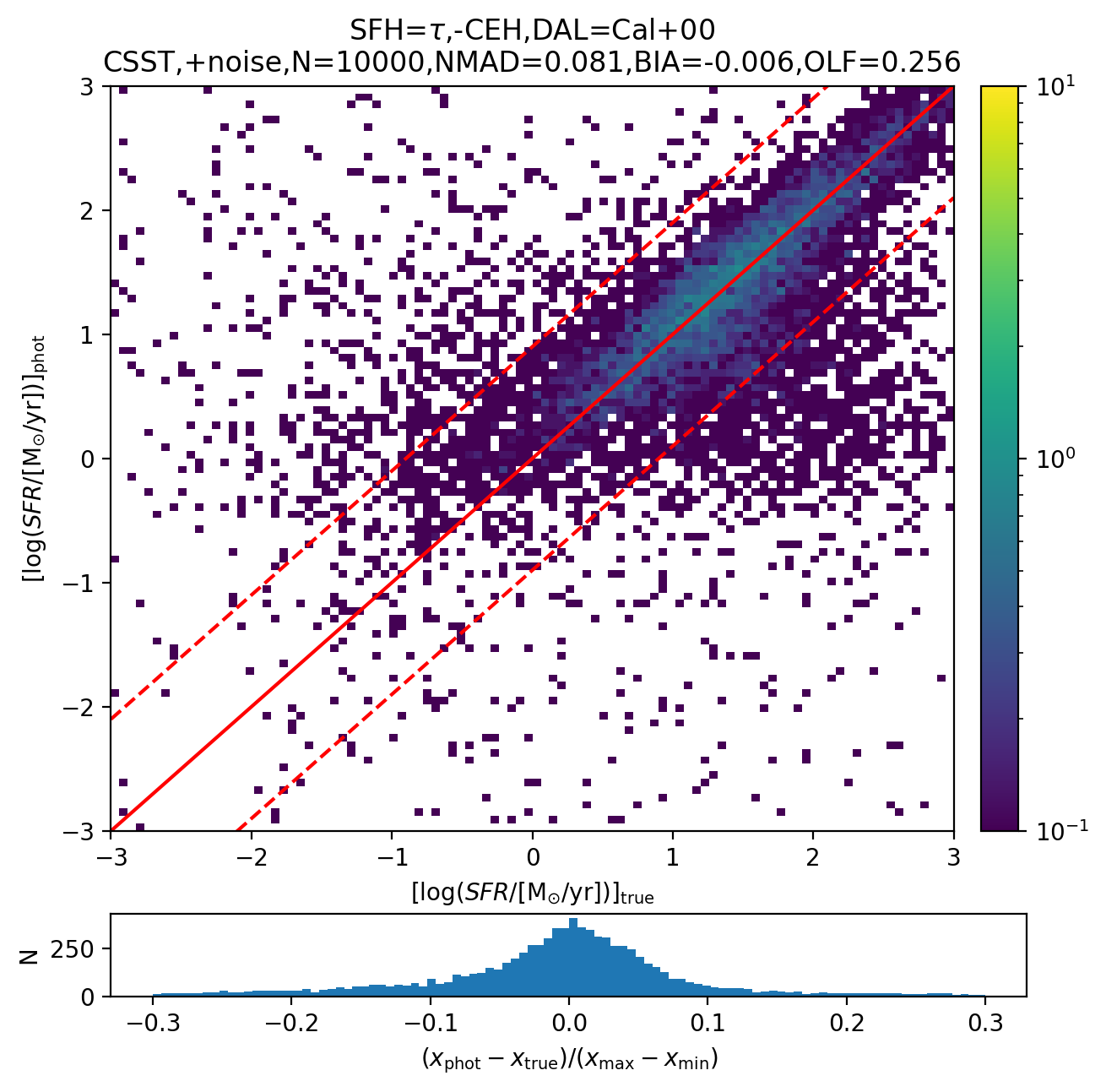}{0.48\textwidth}{(b)}
    }
    \gridline
    {
        \leftfig{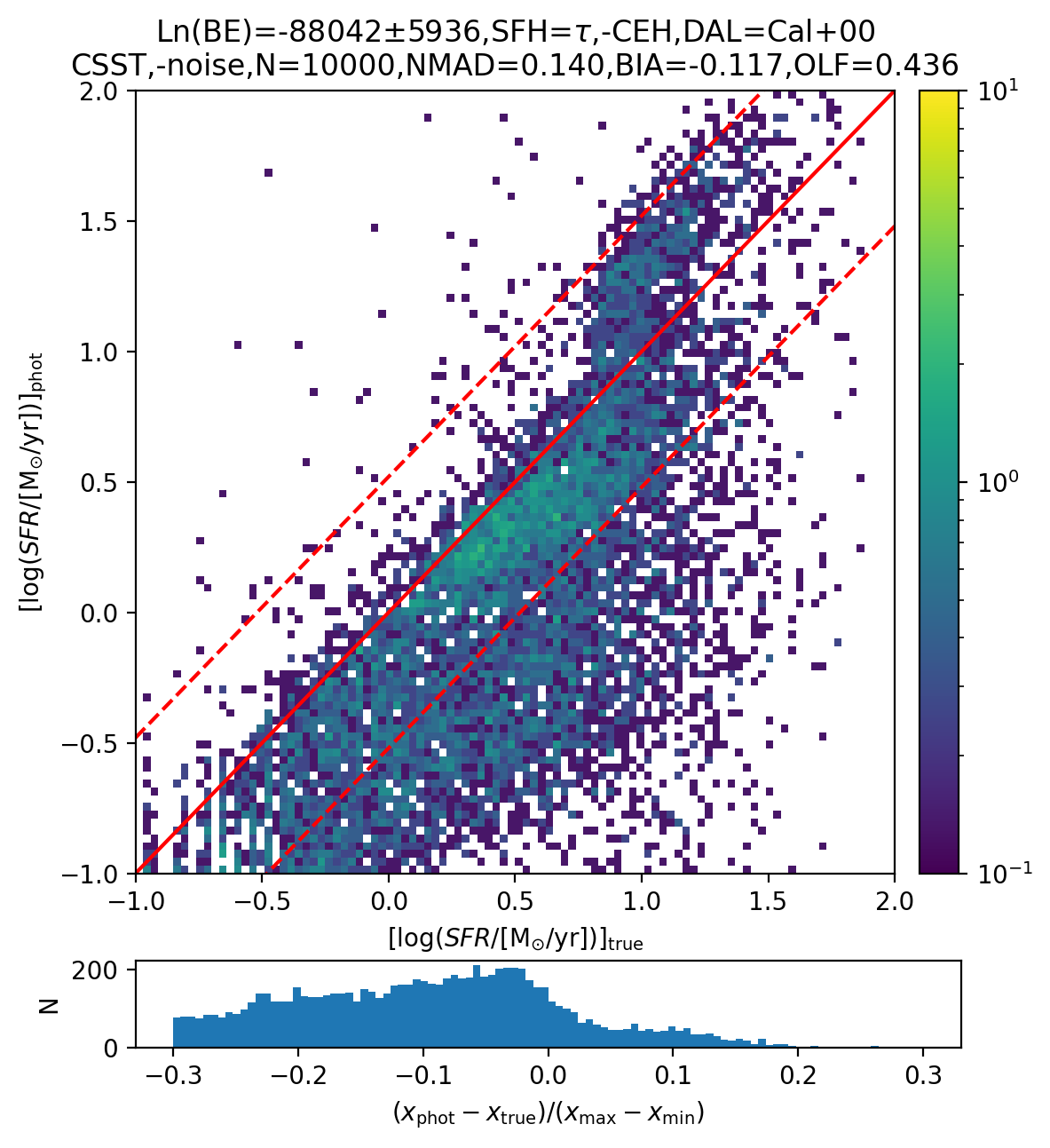}{0.48\textwidth}{(c)}
        \rightfig{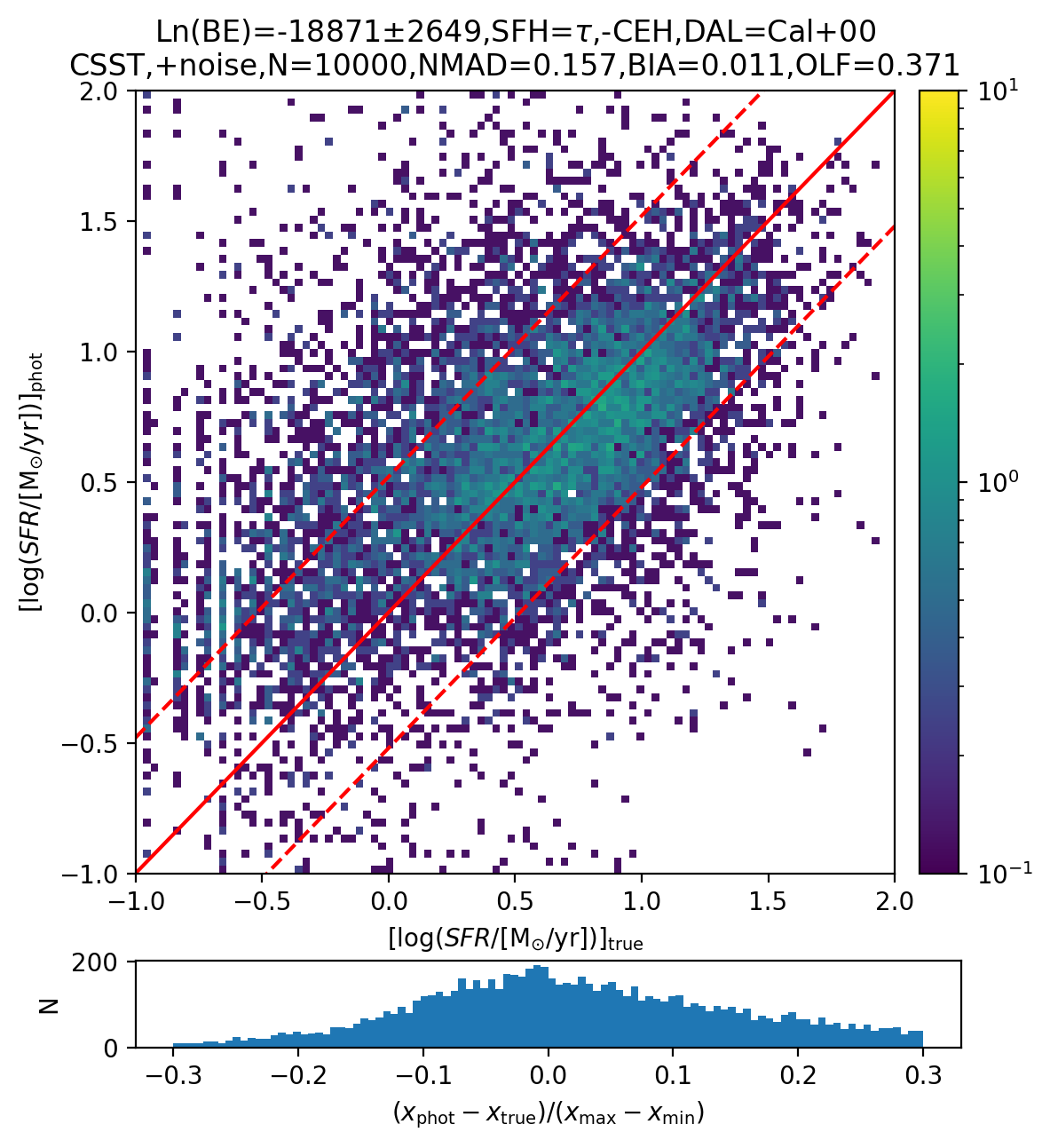}{0.48\textwidth}{(d)}
    }
    \caption
    {
        As Figure \ref{fig:perf_phot_z}, but for the SFR.
        In general, the imperfect SED modeling is more important source of error for the photometric SFR estimation of galaxies, and the contribution from imperfect SED modeling is also very important.
        Especially, in the case without noise, it is clear that the simple $\tau$ model of SFH and \cite{CalzettiD2000a} form of DAL lead to severely biased estimation of SFR for the Horizon-AGN simulation based mock sample of galaxies.
        \label{fig:perf_phot_sfr}
    }
\end{figure*}
\subsection{Photometric redshift estimation} \label{ss:pf_hyd_z}
In Figure \ref{fig:perf_phot_z}, we investigate the performance of BayeSED combined with the simplest SED model to estimate the photometric redshifts of hydrodynamical simulation-based mock galaxy sample for CSST-like imaging survey.
Aa a reference, the panel a of this figure show the ideal case without observational noise and SED modeling errors.
So, in this case, the effects of parameter degeneracies, the stochastic nature of MultiNest sampling algorithm and other potential errors in the BayeSED code are the sources of error.
As shown clearly, the total error from all of these sources is very small.
By comparing panels a and b of this figure, with only the observational noise added, the $\sigma_{\rm NMAD}$ of photometric redshift estimation increases by eight times and the OLF increases by more than forty times.
The bias also increases, but is ignorable with respect to $\sigma_{\rm NMAD}$.
By comparing panels a and c of this figure, with only the error from the imperfect SED modeling added, the $\sigma_{\rm NMAD}$ of photometric redshift estimation increases by more than three times and the OLF decreases slightly.
Besides, there are some additional systematic patterns in the relation between true and estimated values of photometric redshift.
The bias also increases and is comparable to $\sigma_{/rm NMAD}$.
In general, the observational noise is more important source of error for the photometric redshift estimation of galaxies, although the other one is also very important.

Finally, as shown in panel d of Figure \ref{fig:perf_phot_z}, when all sources of error are included, the $\sigma_{\rm NMAD}$ of photometric redshift estimation increases to $0.097$, the OLF increases to $0.264$, and the bias becomes $0.003$.
The systematic patterns  shown in panel c seems being hidden due to the added noise.
The algorithm seems to be struggling to estimate the photometric redshifts correctly by only using the seven-band photometries from CSST-like imaging survey, especially for galaxies with redshift larger than one.

\subsection{Stellar population parameter estimation} \label{ss:pf_hyd_p}
In Figure \ref{fig:perf_phot_mass}, we investigate the performance of BayeSED combined with the simplest SED model to estimate the stellar population parameters of hydrodynamical simulation-based mock galaxies for CSST-like imaging survey.
Aa a reference, the panel a of this figure show the ideal case without observational noise and SED modeling errors.
In this ideal case, the $\sigma_{\rm NMAD}$, bias and OLF of stellar mass estimation are $0.047$, $0.005$ and $0.052$, respectively.
As shown in panel b, with only the observational noise added, the results become $0.115$, $0.02$ and $0.283$, respectively.
As shown in panel c, with only the error from the imperfect SED modeling added, the results become $0.103$, $0.003$ and $0.141$, respectively.
By comparing panels a and c, the performance of stellar mass estimation is severely affected by the simplified assumptions in the SED modeling.
However, by comparing panels b and c, the observational noise is more important source of error for the photometric stellar mass estimation of galaxies, although the other one is also very important.
Finally, as shown in panel d of this figure, when all sources of error are included, the $\sigma_{\rm NMAD}$ of photometric stellar mass estimation increases to $0.135$, the bias increases to $0.034$, and the OLF increases to $0.341$.
The algorithm seems to be even more struggling to estimate the photometric stellar mass correctly by only using the seven-band photometries from CSST-like imaging survey.

Similarly, Figure \ref{fig:perf_phot_sfr} shows the performance of BayeSED combined with the simplest SED model to estimate the SFR of hydrodynamical simulation-based mock galaxy sample for CSST-like imaging survey.
Comparing with the results for stellar mass estimation in Figure \ref{fig:perf_phot_mass}, the photometric SFR estimation is even more severely affected by the simplified assumptions in the SED modeling.
Actually, by comparing panels b and c of this figure, it is clear that the error from the imperfect SED modeling is more important source of error for the photometric SFR estimation of galaxies, although the other one is also very important.
Finally, it becomes even more struggling to estimate the photometric SFR correctly by only using the seven-band photometries from CSST-like imaging survey.


\section{Discussion} \label{sec:disc}
By comparing the results of performance tests for simultaneous photometric redshift and stellar parameters estimation using empirical statistics-based mock galaxy sample (\S \ref{sec:pf_sed}) and hydrodynamical simulation-based mock galaxy sample(\S \ref{sec:pf_hyd}), especially those presented in Figures \ref{fig:perf_phot_z}, \ref{fig:perf_phot_mass} and \ref{fig:perf_phot_sfr}, it is clear that the simple typical assumptions about the SFH and DAL of galaxies have severe impact on the performance of photometric parameter estimation of galaxies for CSST-like imaging survey.
It is not very surprising, since the SFHs and MEHs of galaxies in the cosmic hydrodynamical simulation, such as Horizon-AGN \citep{VolonteriM2016a,BeckmannR2017v,KavirajS2017x,BeckmannR2017a}, are much more complex and diverse \citep[See also][]{IyerK2020a} than the simple assumptions that have been employed in the previous Bayesian analysis of photometric mock data.

In this section, we will discuss the effects of more flexible forms of SFH and DAL on the performance of simultaneous photometric redshift and stellar population parameter estimation of galaxies.
As in \cite{HanY2012b,HanY2014a,HanY2019a} \citep[See also][]{SalmonB2016a,DriesM2016a,DriesM2018a,LawlerA2021n}, we mainly employ the Bayesian model comparison method to compare six different combinations of these model ingredients with increasing complexity (See Table \ref{tab:models} for details).
In addition to the CSST-like survey (\S \ref{ss:disc_csst}) where only the photometries from seven broad-bands are available, we also discuss the results obtained by using mock data for CSST+Euclid-like (\S \ref{ss:disc_csst_euclid}) and COSMOS-like surveys (\S \ref{ss:disc_cosmos}) with increasing discriminative power, respectively.

\subsection{Effects of more flexible SFH and DAL for CSST-like survey} \label{ss:disc_csst}
\begin{table*}
	\begin{center}
		\begin{tabular}{llllllllll}
			\toprule
			survey & noise & SFH & DAL & ln(BE) & ln(ML) & parameter & NMAD & BIA & OLF \\
\midrule
			 &  &  &  &  &  & $z_{\rm phot}$ & 0.024 & 0.011 & 0.003 \\
			CSST & - & $\tau$,-CEH & Cal+00 & -88042$\pm$5936 & 127197 & ${\rm log}(M_*)_{\rm phot}$ & 0.103 & 0.003 & 0.141 \\
			 &  &  &  &  &  & ${\rm log}(SFR/[\rm M_{\odot}/yr])_{\rm phot}$ & 0.140 & -0.117 & 0.436 \\
\midrule
			 &  &  &  &  &  & $z_{\rm phot}$ & 0.023 & 0.010 & 0.002 \\
			CSST & - & $\tau$,+CEH & Cal+00 & -74128$\pm$5973 & 144380 & ${\rm log}(M_*)_{\rm phot}$ & 0.092 & -0.010 & 0.141 \\
			 &  &  &  &  &  & ${\rm log}(SFR/[\rm M_{\odot}/yr])_{\rm phot}$ & 0.148 & -0.141 & 0.491 \\
\midrule
			 &  &  &  &  &  & $z_{\rm phot}$ & 0.016 & 0.014 & 0.004 \\
			CSST & - & $\tau$,+CEH & Sal+18 & 58878$\pm$5823 & 282267 & ${\rm log}(M_*)_{\rm phot}$ & 0.086 & 0.041 & 0.170 \\
			 &  &  &  &  &  & ${\rm log}(SFR/[\rm M_{\odot}/yr])_{\rm phot}$ & 0.153 & -0.019 & 0.390 \\
\midrule
			 &  &  &  &  &  & $z_{\rm phot}$ & 0.015 & 0.014 & 0.004 \\
			CSST & - & $\beta$-$\tau$,+CEH & Sal+18 & 56343$\pm$5780 & 283410 & ${\rm log}(M_*)_{\rm phot}$ & 0.079 & 0.041 & 0.150 \\
			 &  &  &  &  &  & ${\rm log}(SFR/[\rm M_{\odot}/yr])_{\rm phot}$ & 0.133 & 0 & 0.370 \\
\midrule
			 &  &  &  &  &  & $z_{\rm phot}$ & 0.017 & 0.011 & 0.006 \\
			CSST & - & $\beta$-$\tau$-r,+CEH & Sal+18 & 34213$\pm$5867 & 275861 & ${\rm log}(M_*)_{\rm phot}$ & 0.097 & 0.052 & 0.227 \\
			 &  &  &  &  &  & ${\rm log}(SFR/[\rm M_{\odot}/yr])_{\rm phot}$ & 0.257 & 0.045 & 0.579 \\
\midrule
			 &  &  &  &  &  & $z_{\rm phot}$ & 0.016 & 0.010 & 0.005 \\
			CSST & - & $\alpha$-$\beta$-$\tau$-r,+CEH & Sal+18 & 33044$\pm$5800 & 275886 & ${\rm log}(M_*)_{\rm phot}$ & 0.098 & 0.051 & 0.239 \\
			 &  &  &  &  &  & ${\rm log}(SFR/[\rm M_{\odot}/yr])_{\rm phot}$ & 0.214 & 0.065 & 0.544 \\
\midrule
			 &  &  &  &  &  & $z_{\rm phot}$ & 0.097 & 0.003 & 0.264 \\
			CSST & + & $\tau$,-CEH & Cal+00 & -18871$\pm$2649 & 49807 & ${\rm log}(M_*)_{\rm phot}$ & 0.135 & -0.034 & 0.341 \\
			 &  &  &  &  &  & ${\rm log}(SFR/[\rm M_{\odot}/yr])_{\rm phot}$ & 0.157 & 0.011 & 0.371 \\
\midrule
			 &  &  &  &  &  & $z_{\rm phot}$ & 0.096 & 0.005 & 0.264 \\
			CSST & + & $\tau$,+CEH & Cal+00 & -18661$\pm$2649 & 49469 & ${\rm log}(M_*)_{\rm phot}$ & 0.135 & -0.034 & 0.344 \\
			 &  &  &  &  &  & ${\rm log}(SFR/[\rm M_{\odot}/yr])_{\rm phot}$ & 0.155 & 0 & 0.366 \\
\midrule
			 &  &  &  &  &  & $z_{\rm phot}$ & 0.098 & 0.012 & 0.263 \\
			CSST & + & $\tau$,+CEH & Sal+18 & -26098$\pm$2956 & 54493 & ${\rm log}(M_*)_{\rm phot}$ & 0.158 & -0.023 & 0.387 \\
			 &  &  &  &  &  & ${\rm log}(SFR/[\rm M_{\odot}/yr])_{\rm phot}$ & 0.201 & 0 & 0.468 \\
\midrule
			 &  &  &  &  &  & $z_{\rm phot}$ & 0.099 & 0.011 & 0.263 \\
			CSST & + & $\beta$-$\tau$,+CEH & Sal+18 & -28141$\pm$2904 & 54484 & ${\rm log}(M_*)_{\rm phot}$ & 0.161 & -0.029 & 0.396 \\
			 &  &  &  &  &  & ${\rm log}(SFR/[\rm M_{\odot}/yr])_{\rm phot}$ & 0.194 & 0.028 & 0.472 \\
\midrule
			 &  &  &  &  &  & $z_{\rm phot}$ & 0.100 & 0.013 & 0.268 \\
			CSST & + & $\beta$-$\tau$-r,+CEH & Sal+18 & -35048$\pm$3020 & 54444 & ${\rm log}(M_*)_{\rm phot}$ & 0.171 & -0.037 & 0.415 \\
			 &  &  &  &  &  & ${\rm log}(SFR/[\rm M_{\odot}/yr])_{\rm phot}$ & 0.232 & 0.041 & 0.544 \\
\midrule
			 &  &  &  &  &  & $z_{\rm phot}$ & 0.104 & 0.007 & 0.290 \\
			CSST & + & $\alpha$-$\beta$-$\tau$-r,+CEH & Sal+18 & -33983$\pm$2869 & 54440 & ${\rm log}(M_*)_{\rm phot}$ & 0.177 & -0.052 & 0.430 \\
			 &  &  &  &  &  & ${\rm log}(SFR/[\rm M_{\odot}/yr])_{\rm phot}$ & 0.224 & 0.080 & 0.570 \\
			\bottomrule
		\end{tabular}
	\end{center}
	\caption
    {
        A summary of Bayesian evidences (BE), maximum likelihoods (ML) and the metrics of the quality of parameter estimation from the Bayesian analysis of the hydrodynamical simulation-based mock galaxy sampe for CSST-like survey by employing six SED models with increasing complexity in the form of SFH and DAL, as well as for the cases with and without noise.
    }
	\label{tab:sum_CSST}
\end{table*}

\begin{figure*}
    \centering
    \includegraphics{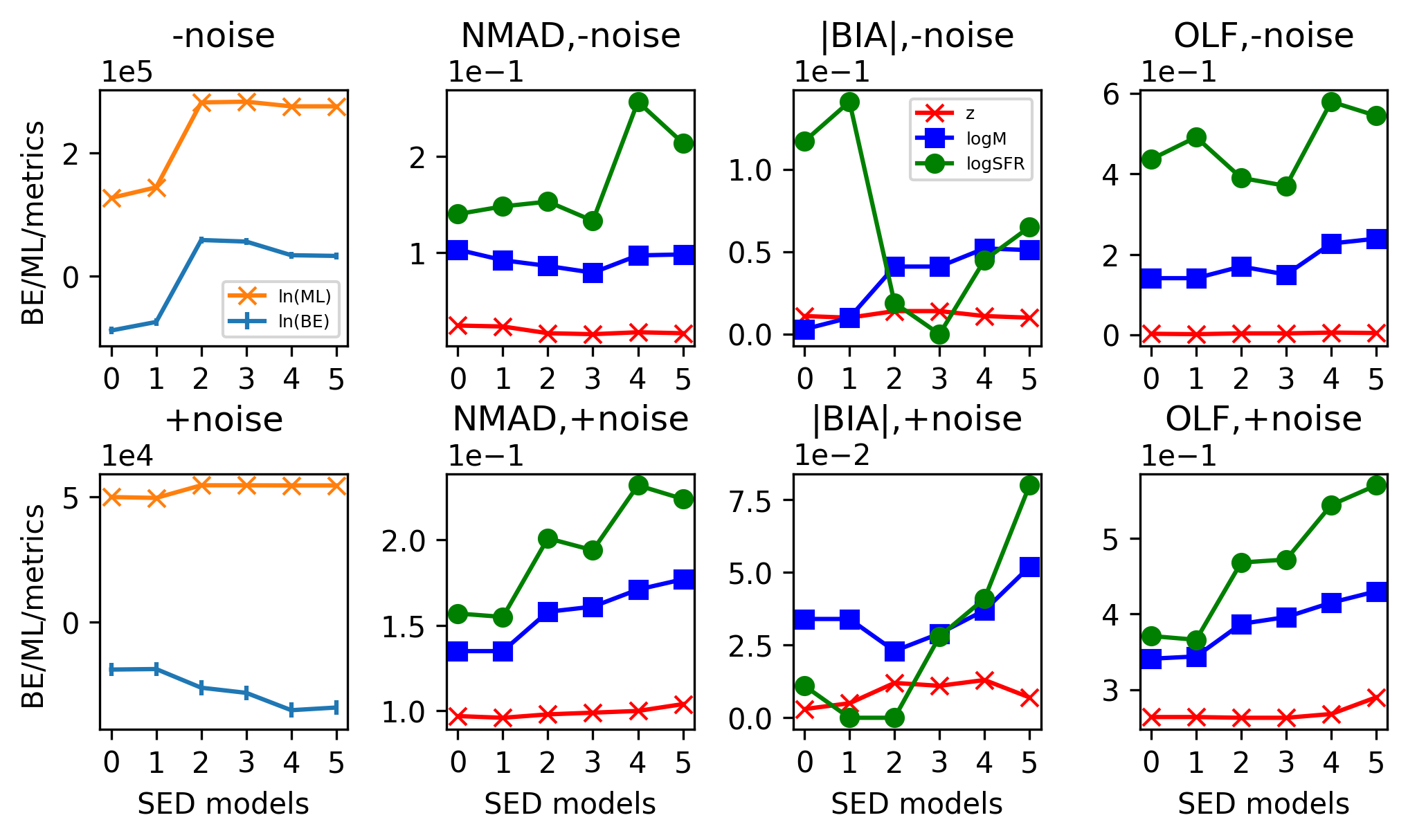}
    \caption
    {
        The Bayesian evidences (BE), maximum likelihood (ML) and metrics of photometric redshift (red star), stellar mass (blue square) and SFR (green circle) estimation from the Bayesian analysis of the hydrodynamical simulation-based mock galaxy sampe for CSST-like survey by employing six SED models (0:``SFH=$\tau$,-CEH,DAL=Cal+18'', 1:``SFH=$\tau$,+CEH,DAL=Cal+18'', 2:``SFH=$\tau$,+CEH,DAL=Sal+18'', 3:``SFH=$\beta$-$\tau$,+CEH,DAL=Sal+18'', 4:``SFH=$\beta$-$\tau$-r,+CEH,DAL=Sal+18'', 5:``SFH=$\alpha$-$\beta$-$\tau$-r,+CEH,DAL=Sal+18'') with increasing complexity in the forms of SFH and DAL , as well as for the cases with (bottom panels) and without noise (top panels).
        In the case without noise, the simplest model ``SFH=$\tau$,-CEH,DAL=Cal+18'' has the lowest Bayesian evidence of ${\rm ln}(BE)=-88042\pm5936$.
        Meanwhile, the SED models ``SFH=$\tau$,+CEH,DAL=Sal+18'' and ``SFH=$\beta$-$\tau$,+CEH,DAL=Sal+18'' which are neither the simplest nor the most complex models have the largest Bayesian evidences of ${\rm ln}(BE)=58878\pm5823$ and ${\rm ln}(BE)=56343\pm5780$ which are comparable within error bar.
        Interestingly, the same two models also give the highest quality parameter estimates.
        It is worth to mention that the model selection with maximum likelihood (or equivalently minimum $\chi^2$) lead to similar results.
        Contrary to the case without noise, in the case with noise, the two simplest SED models ``SFH=$\tau$,-CEH,DAL=Cal+00'' and ``SFH=$\tau$,+CEH,DAL=Cal+00'' have the largest Bayesian evidences of ${\rm ln}(BE)=-18871\pm2649$ and ${\rm ln}(BE)=-18661\pm2649$ which are comparable within error bar.
        It is very interesting to notice that the model selection with maximum likelihood (or equivalently minimum $\chi^2$) lead to exactly the opposite results, where the more complex (or flexible) models are always more favored.
        Similar to the case without noise, the two simplest SED models also give the highest quality parameter estimates.
        For photometric redshift, stellar mass and SFR, their accurate estimation becomes increasingly difficult, while the latter two are also more sensitive to the selection of SED models.
        Generally, the quality of parameter estimates is closely related to the level of Bayesian evidence, which is especially clear in the more realistic case with noise.
        Actually, the quality of parameter estimation, especially that of stellar mass and SFR estimation, significantly decrease with the increasing of SED model complexity, which is similar to the case with perfect SED modeling as shown in Figures \ref{fig:perf_phot_multinest_z}, \ref{fig:perf_phot_multinest_mass} and \ref{fig:perf_phot_multinest_sfr}, and should be caused by more severe parameter degeneracies suffered by the more flexible SED model.
        It is clear that, in the more realistic case with noise, the model selection with maximum likelihood (or equivalently minimum $\chi^2$) is not consistent with the measurements of the quality of parameter estimation.
        Since the direct measurements of the metrics such as NMAD, BIA and OLF are usually unavailable, the Bayesian model comparison with Bayesian evidence can be used to find the best SED model which is not only the most efficient but also give the best parameter estimation.
    }
    \label{fig:model_lnE_metrics_CSST}
\end{figure*}

In Table \ref{tab:sum_CSST}, we present a summary of the Bayesian evidences, maximum likelihoods and metrics of the quality of parameter estimation from the Bayesian analysis of the hydrodynamical simulation-based mock galaxy sampe for CSST-like survey by employing six different combinations of SFH and DAL with increasing complexity, as well as for the cases with and without noise.
The same results are also shown more clearly in Figure \ref{fig:model_lnE_metrics_CSST}.

\subsubsection{Model comparison} \label{ss:disc_csst_mod}
In the case without noise, as shown in the top left panel of Figure \ref{fig:model_lnE_metrics_CSST}, the simplest model ``SFH=$\tau$,-CEH,DAL=Cal+18'' has the lowest Bayesian evidence of ${\rm ln}(BE)=-88042\pm5936$.
With the additional consideration of metallicity evolution, the Bayesian evidence of the model ``SFH=$\tau$,+CEH,DAL=Cal+18'' increases to ${\rm ln}(BE)=-74128\pm5973$.
Then, with the adoption of the DAL of \cite{SalimS2018a}, the Bayesian evidence of the model ``SFH=$\tau$,+CEH,DAL=Sal+18'' increases significantly to ${\rm ln}(BE)=58878\pm5823$.
Apparently, the DAL of \cite{SalimS2018a} is a much better choice than that of \cite{CalzettiD2000a} for the hydrodynamical simulation-based mock galaxy sampe in CSST-like survey.
Furthermore, by employing a more complicated $\beta$-$\tau$ form of SFH, the Bayesian evidence of the model ``SFH=$\beta$-$\tau$,+CEH,DAL=Sal+18'' seems decreases a little to ${\rm ln}(BE)=56343\pm5780$.
Actually, the latter two SED models (``SFH=$\tau$,+CEH,DAL=Sal+18'' and ``SFH=$\beta$-$\tau$,+CEH,DAL=Sal+18'')  have the largest Bayesian evidences which are comparable within error bar.
They are neither the simplest nor the most complex models.
With a quenching (or rejuvenation) component added to the SFH, the Bayesian evidence of the model ``SFH=$\beta$-$\tau$-r,+CEH,DAL=Sal+18'' obviously decreases to ${\rm ln}(BE)=34213\pm5867$.
It seems that, while the rejuvenation or rapid quenching events may happen in some galaxies, this additional component of SFH is not very effective for most of the galaxies in the sample.
Finally, by employing a even more flexible double power-law form of SFH, the Bayesian evidence of the model ``SFH=$\alpha$-$\beta$-$\tau$-r,+CEH,DAL=Sal+18'' seems decreases a little to ${\rm ln}(BE)=33044\pm5800$.
However, the latter two SED models are actually comparable within error bar.
On the other hand, it is worth to mention that the model selection with maximum likelihood (or equivalently minimum $\chi^2$) lead to similar results.

In the case with noise, as shown in the bottom left panel of Figure \ref{fig:model_lnE_metrics_CSST}, the two simplest SED models ``SFH=$\tau$,-CEH,DAL=Cal+00'' and ``SFH=$\tau$,+CEH,DAL=Cal+00'' have the largest Bayesian evidences of ${\rm ln}(BE)=-18871\pm2649$ and ${\rm ln}(BE)=-18661\pm2649$, respectively.
Although the latter which has additionally considered the metallicity evolution seems better, their Bayesian evidences are actually comparable within error bar.
Then, with the adoption of the DAL of \cite{SalimS2018a}, the Bayesian evidence of the model ``SFH=$\tau$,+CEH,DAL=Sal+18'' decreases significantly to ${\rm ln}(BE)=-26098\pm2956$, exactly the opposite of the situation without noise.
It is likely that the more complicated form of DAL does not give a better fit to the noisier data.
By employing a more complicated $\beta$-$\tau$ form of SFH, the Bayesian evidence of the model ``SFH=$\beta$-$\tau$,+CEH,DAL=Sal+18'' seems decreases further to ${\rm ln}(BE)=-28141\pm2904$, although comparable with the former within error bar.
With a quenching (or rejuvenation) component added to the SFH, the Bayesian evidence of the model ``SFH=$\beta$-$\tau$-r,+CEH,DAL=Sal+18'' obviously decreases to ${\rm ln}(BE)=-35048\pm3020$, which is similar to the case without noise.
Finally, by employing a even more flexible double power-law form of SFH, the Bayesian evidence of the model ``SFH=$\alpha$-$\beta$-$\tau$-r,+CEH,DAL=Sal+18'' seems increases a little to ${\rm ln}(BE)=-33983\pm2869$, although comparable with the former within error bar.
On the other hand, it is very interesting to notice that the model selection with maximum likelihood (or equivalently minimum $\chi^2$) lead to exactly the opposite results, where the more complex (or flexible) models are always more favored.

\subsubsection{Parameter estimation} \label{ss:disc_csst_par}
In the right three panels of Figure \ref{fig:model_lnE_metrics_CSST}, we show the three metrics of the quality of photometric redshift, stellar mass and SFR estimation for different SED models, respectively.
In general, in the case without noise, the SED models ``SFH=$\tau$,+CEH,DAL=Sal+18'' and ``SFH=$\beta$-$\tau$,+CEH,DAL=Sal+18'' which are just the two with largest Bayesian evidence give the highest quality parameter estimates.
In the case with noise, the two simplest SED models ``SFH=$\tau$,-CEH,DAL=Cal+00'' and ``SFH=$\tau$,+CEH,DAL=Cal+00'' which are also the two with the largest Bayesian evidences give the highest quality parameter estimates.
In the following, we discuss these results in more detail.

\begin{figure*}
    \gridline
    {
        \leftfig{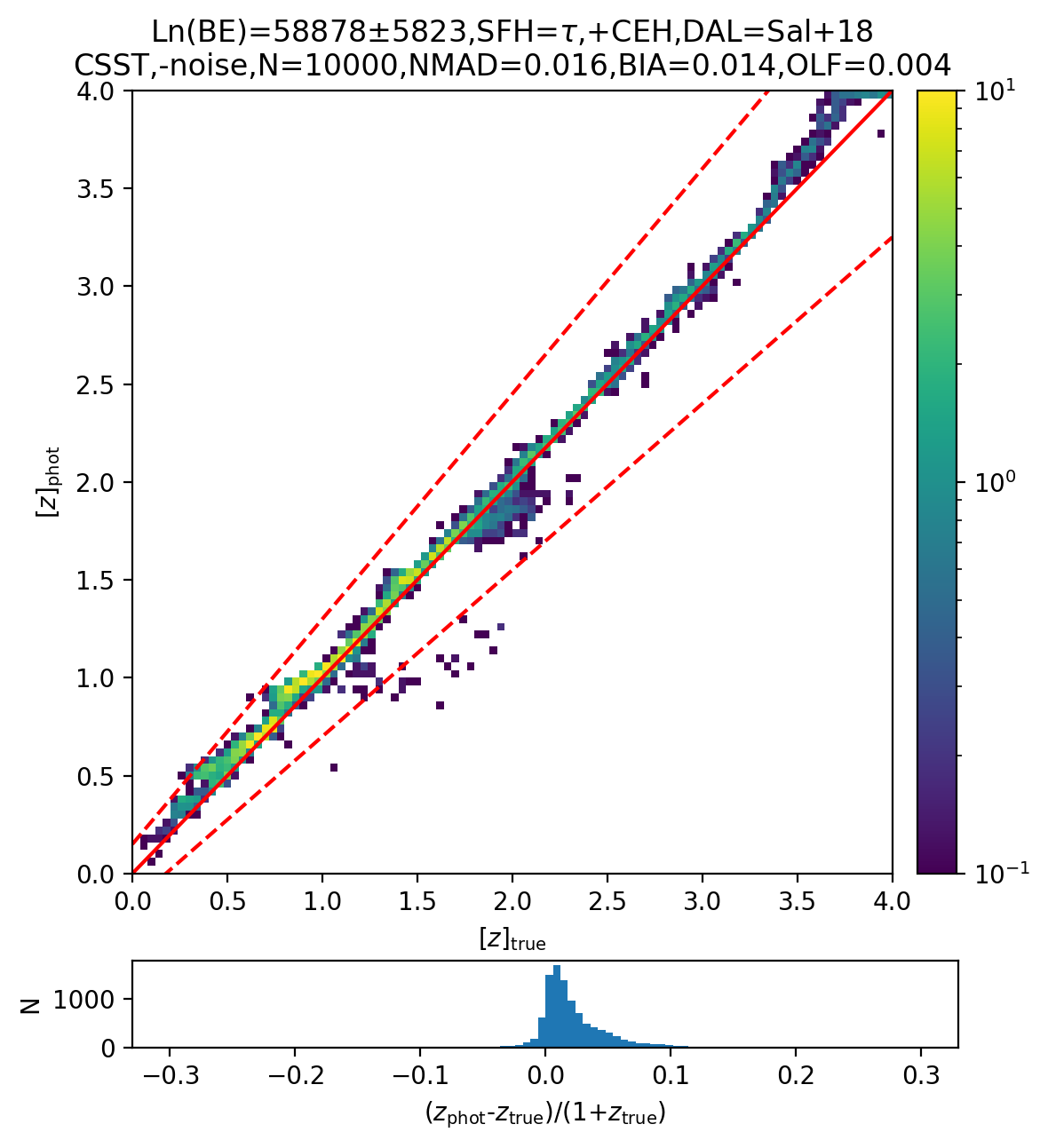}{0.48\textwidth}{(a)}
        \rightfig{perf_phot_0csp_sfh200_bc2003_lr_BaSeL_chab_i0000_2dal8_10_z_CSST_inoise2_par0}{0.48\textwidth}{(b)}
    }
    \gridline
    {
        \leftfig{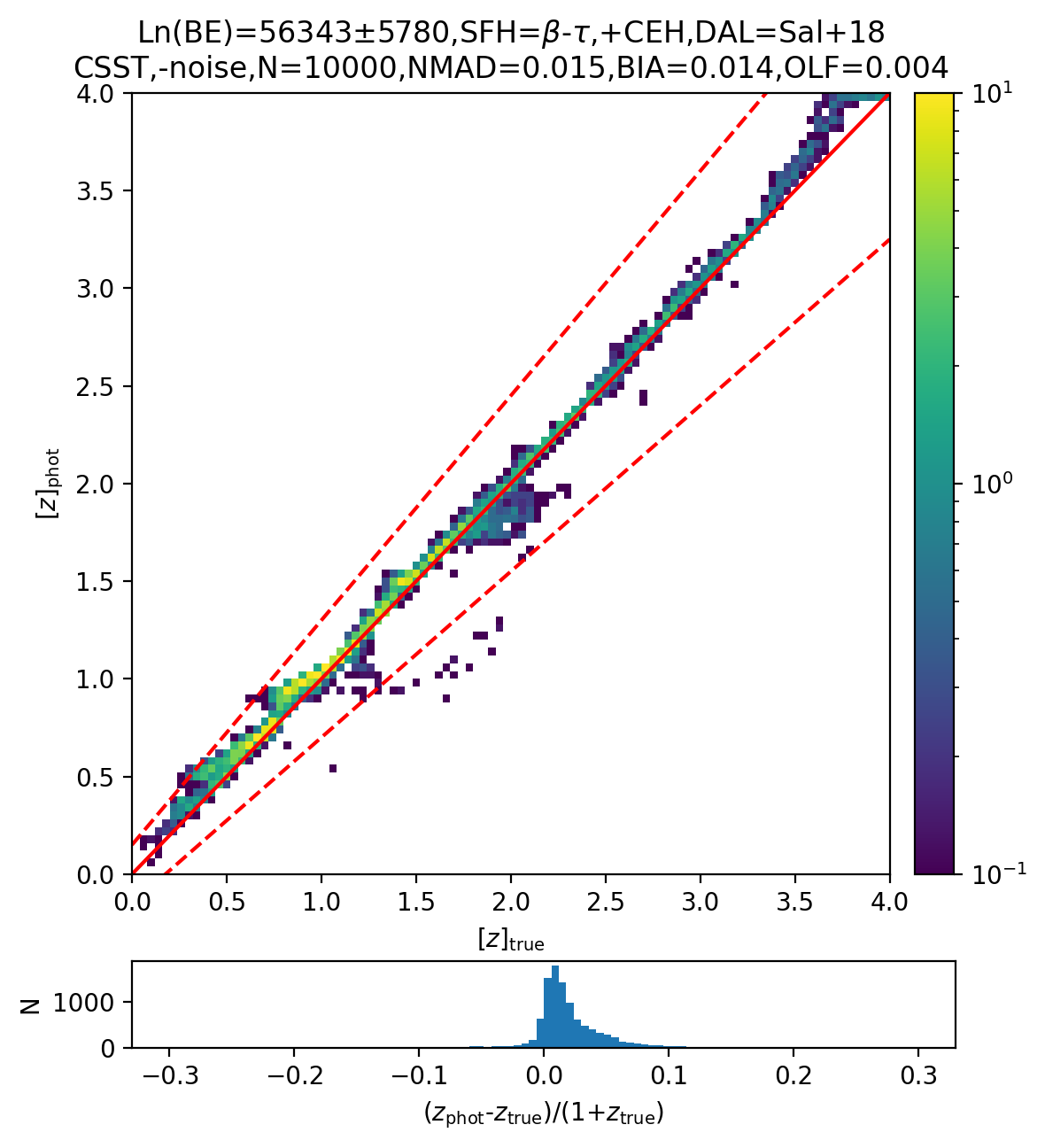}{0.48\textwidth}{(c)}
        \rightfig{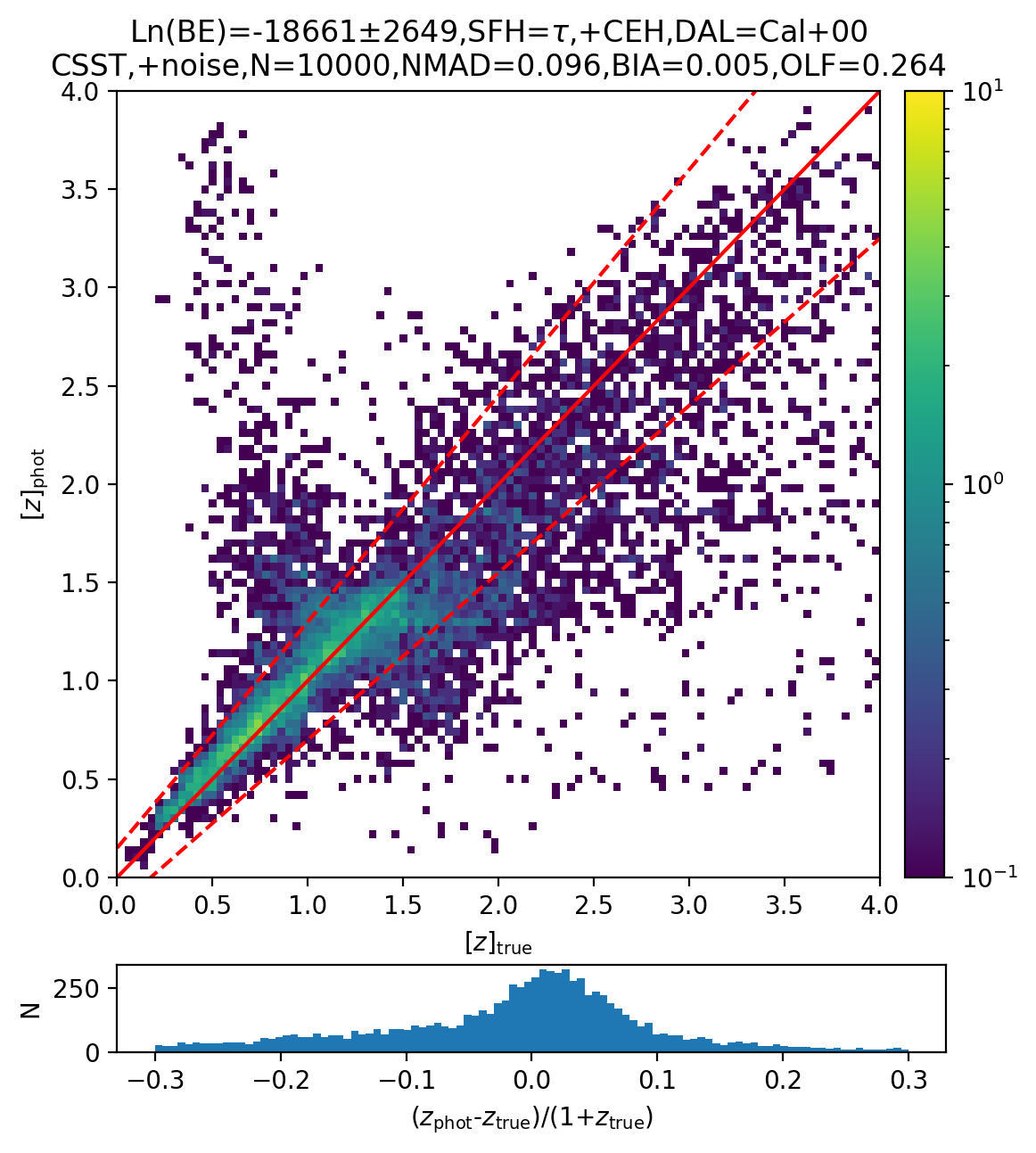}{0.48\textwidth}{(d)}
    }
    \caption
    {
        The results of photometric redshift estimation from the Bayesian analysis of the hydrodynamical simulation-based mock data for CSST-like imaging survey by employing the two SED models with the largest Bayesian evidence.
        \textbf{(a)} By comparing with the results in panel c of Figure \ref{fig:perf_phot_z}, it is clear that, with the additional consideration of metallicity evolution and the adoption of the DAL of \cite{SalimS2018a}, the $\sigma_{\rm NMAD}$ of photometric redshift estimation is obviously reduced, although the bias and OLF are slightly increased.
        Meanwhile, the systematic patterns in the former results are also largely reduced.
        \textbf{(c)} By additionally employing a more complicated $\beta$-$\tau$ form of SFH, the $\sigma_{\rm NMAD}$ of photometric redshift estimation is only slightly reduced while the bias and OLF are exactly the same.
        \textbf{(b, d)} In the case with noise, the best two models are quite similar in the quality of photometric redshift estimation.
        Besides, there are two clear branches of outliers caused by the mis-identification of Lyman and Balmer break features.
    }
    \label{fig:perf_phot_csst_z2}
\end{figure*}
The detailed results of photometric redshift estimation obtained by employing the two models with the largest Bayesian evidences are shown in Figure \ref{fig:perf_phot_csst_z2}.
In the case without noise, by comparing the results in panel c of Figure \ref{fig:perf_phot_z} with that in panel a of Figure \ref{fig:perf_phot_csst_z2}, it is clear that, with the additional consideration of metallicity evolution and the adoption of the DAL of \cite{SalimS2018a}, the $\sigma_{\rm NMAD}$ of photometric redshift estimation is obviously reduced while the bias and OLF are only slightly increased.
Meanwhile, the systematic patterns in the former results are also largely reduced.
However, as shown in panel c of Figure \ref{fig:perf_phot_csst_z2}, by additionally employing a more complicated $\beta$-$\tau$ form of SFH, the $\sigma_{\rm NMAD}$ of photometric redshift estimation is only slightly reduced while the bias and OLF are exactly the same.
Besides, as shown in Table \ref{tab:sum_CSST} and Figure \ref{fig:model_lnE_metrics_CSST}, the other two even more complicated forms of SFH lead to similar quality of photometric redshift estimation.
In the case with noise, the best two models are quite similar in the quality of photometric redshift estimation.
Besides, with the increasing of the complexity of SED models, the quality of photometric redshift estimation tend to decrease, although not very significantly.

\begin{figure*}
    \gridline
    {
        \leftfig{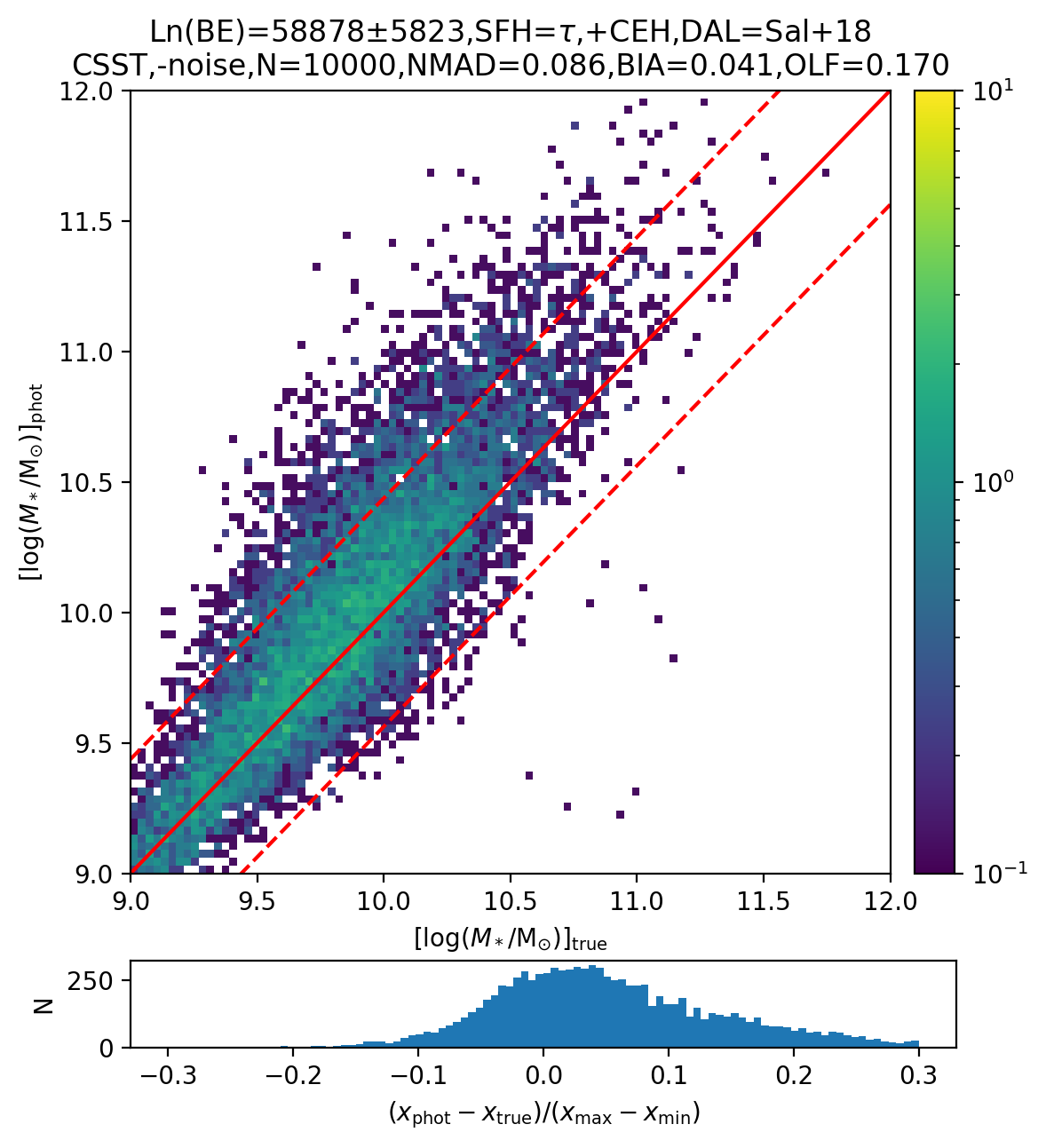}{0.48\textwidth}{(a)}
        \rightfig{perf_phot_0csp_sfh200_bc2003_lr_BaSeL_chab_i0000_2dal8_10_z_CSST_inoise2_par1}{0.48\textwidth}{(b)}
    }
    \gridline
    {
        \leftfig{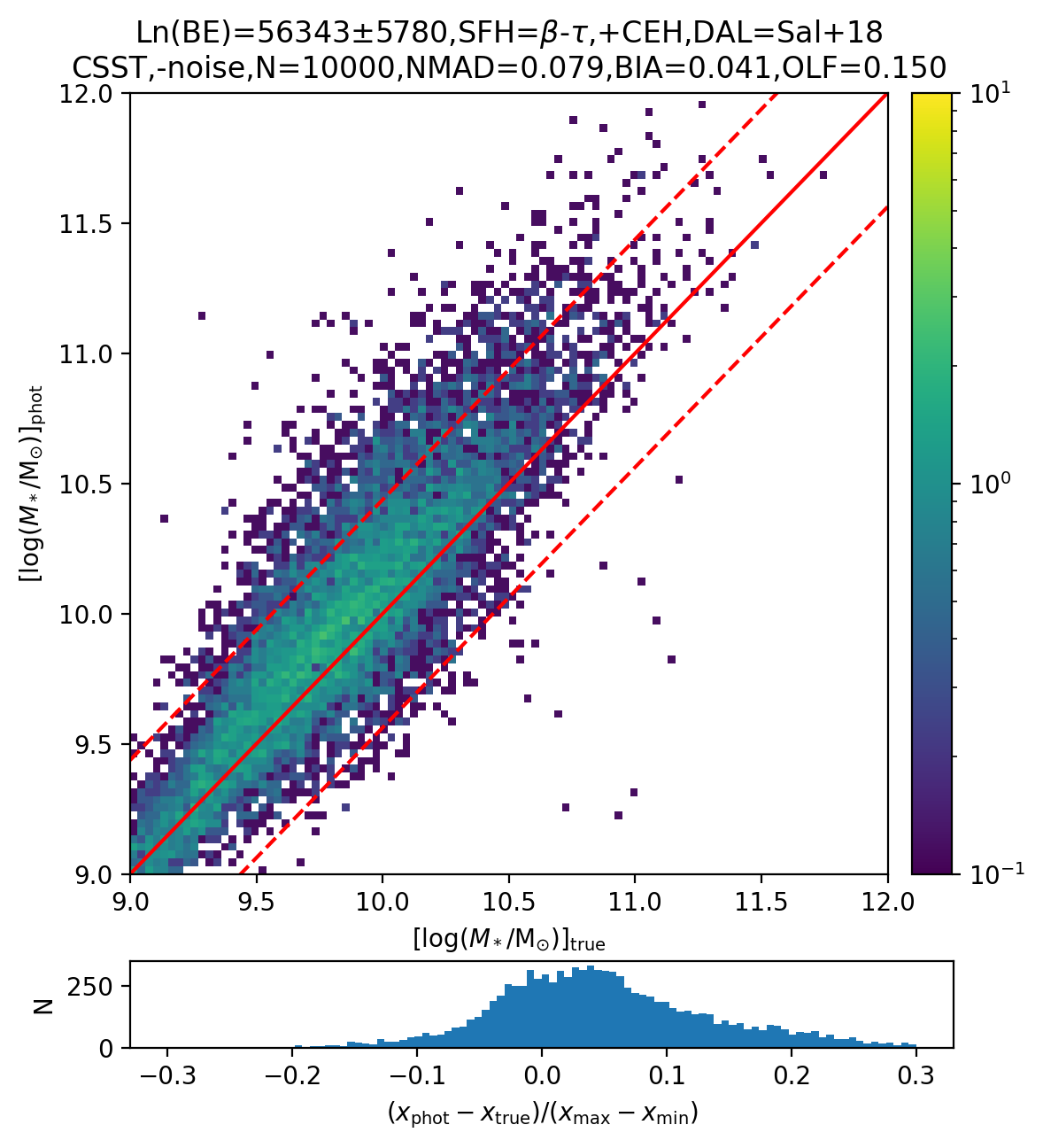}{0.48\textwidth}{(c)}
        \rightfig{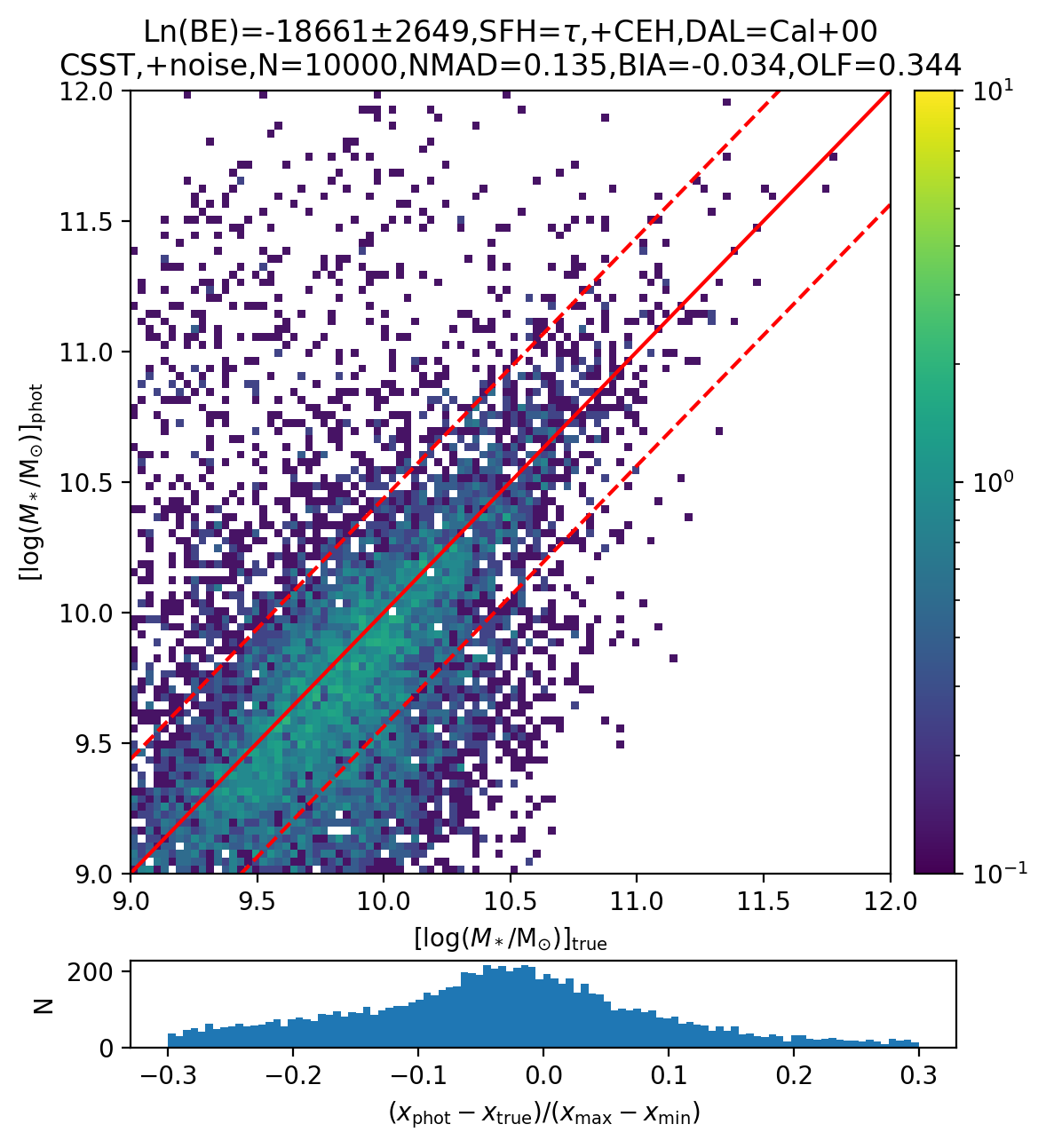}{0.48\textwidth}{(d)}
    }
    \caption
    {
        As in Figure \ref{fig:perf_phot_csst_z2}, but for the stellar mass.
        \textbf{(a)} By comparing with the results in panel c of Figure \ref{fig:perf_phot_mass}, it is clear that, with the additional consideration of metallicity evolution and the adoption of the DAL of \cite{SalimS2018a}, the $\sigma_{\rm NMAD}$ of photometric stellar mass estimation is reduced, although the bias and OLF are somewhat increased.
        \textbf{(c)} By additionally employing a more complicated $\beta$-$\tau$ form of SFH, the quality of photometric stellar mass estimation increase further.
        \textbf{(b, d)} In the case with noise, the best two models are exactly the same in the quality of photometric stellar mass estimation.
    }
    \label{fig:perf_phot_csst_mass2}
\end{figure*}
The detailed results of photometric stellar mass estimation obtained by employing the two models with the largest Bayesian evidences are shown in Figure \ref{fig:perf_phot_csst_mass2}.
In the case without noise, by comparing the results in panel c of Figure \ref{fig:perf_phot_mass} with that in panel a of Figure \ref{fig:perf_phot_csst_mass2}, it is clear that, with the additional consideration of metallicity evolution and the adoption of the DAL of \cite{SalimS2018a}, the $\sigma_{\rm NMAD}$ of photometric stellar mass estimation is reduced, although the bias and OLF are somewhat increased.
By additionally employing a more complicated $\beta$-$\tau$ form of SFH, as shown in panel c of Figure \ref{fig:perf_phot_csst_mass2}, the quality of photometric stellar mass estimation increase further.
However, as shown in Table \ref{tab:sum_CSST} and Figure \ref{fig:model_lnE_metrics_CSST}, with the increasing of the complexity of SED models, the quality of photometric stellar mass stimation decreases obviously. 
In the case with noise, the best two models are exactly the same in the quality of photometric stellar mass estimation.
Meanwhile, with the increasing of the complexity of SED models, the quality of photometric stellar mass stimation decreases even more obviously. 

\begin{figure*}
    \gridline
    {
        \leftfig{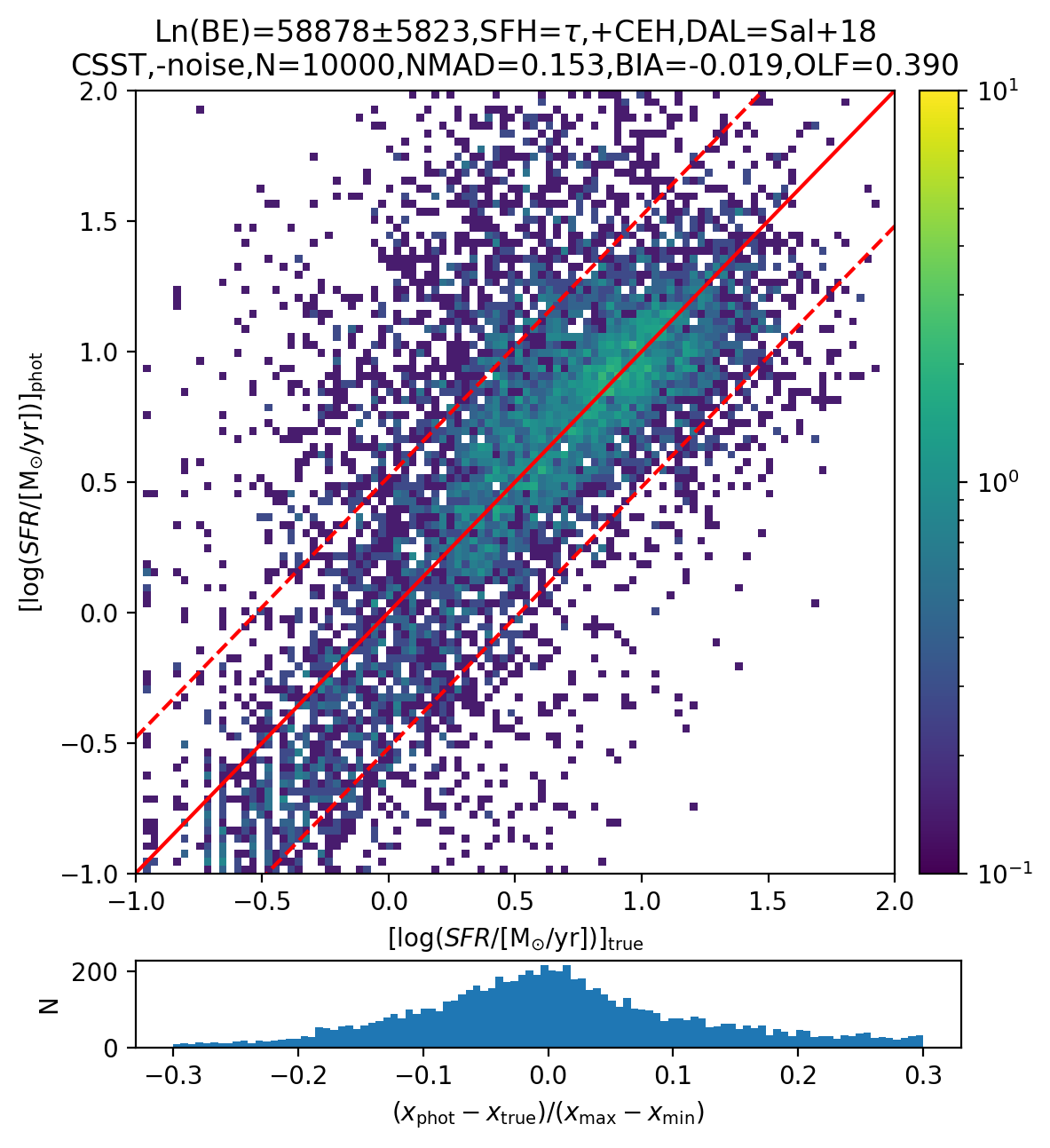}{0.48\textwidth}{(a)}
        \rightfig{perf_phot_0csp_sfh200_bc2003_lr_BaSeL_chab_i0000_2dal8_10_z_CSST_inoise2_par2}{0.48\textwidth}{(b)}
    }
    \gridline
    {
        \leftfig{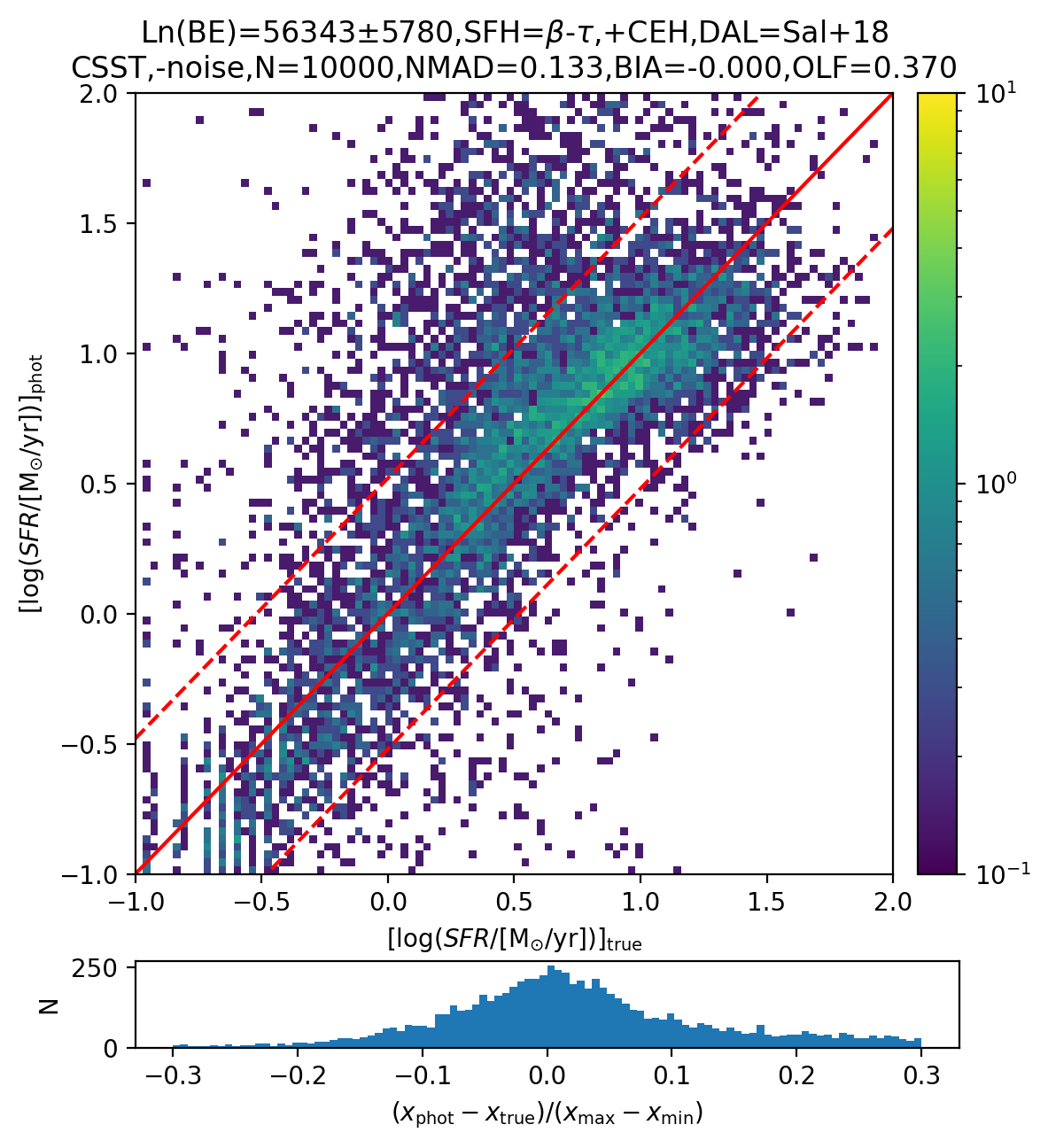}{0.48\textwidth}{(c)}
        \rightfig{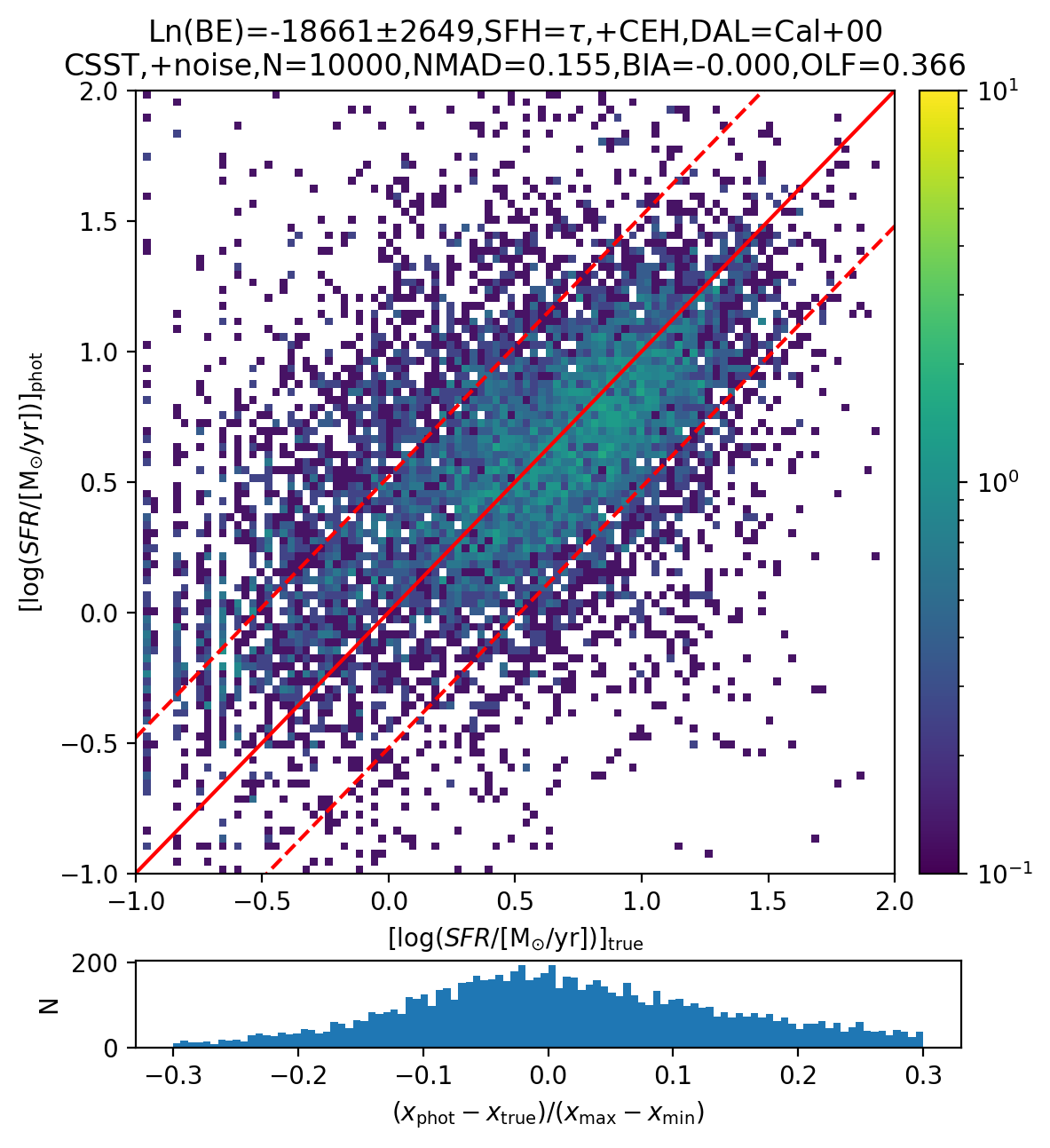}{0.48\textwidth}{(d)}
    }
    \caption
    {
        As in Figure \ref{fig:perf_phot_csst_z2}, but for the SFR.
        \textbf{(a)} By comparing with the results in panel c of Figure \ref{fig:perf_phot_sfr} with that in panel a of Figure \ref{fig:perf_phot_csst_sfr2}, it is clear that, with the additional consideration of metallicity evolution and the adoption of the DAL of \cite{SalimS2018a}, the systematic bias and OLF of photometric SFR are largely reduced, although the $\sigma_{\rm NMAD}$ is slightly increased.
        \textbf{(c)} By additionally employing a more complicated $\beta$-$\tau$ form of SFH the quality of photometric SFR estimation increase to the best.
        \textbf{(b, d)} In the case with noise, the best two models are also very similar in the quality of photometric stellar mass estimation.
    }
    \label{fig:perf_phot_csst_sfr2}
\end{figure*}
The detailed results of photometric SFR estimation obtained by employing the two models with the largest Bayesian evidences are shown in Figure \ref{fig:perf_phot_csst_sfr2}.
In the case without noise, by comparing with the results in panel c of Figure \ref{fig:perf_phot_sfr} with that in panel a of Figure \ref{fig:perf_phot_csst_sfr2}, it is clear that, with the additional consideration of metallicity evolution and the adoption of the DAL of \cite{SalimS2018a}, the systematic bias and OLF of photometric redshift estimation are largely reduced while the $\sigma_{\rm NMAD}$ is slightly increased.
By additionally employing a more complicated $\beta$-$\tau$ form of SFH, as shown in panel c of Figure \ref{fig:perf_phot_csst_sfr2}, the quality of photometric SFR estimation increase to the best.
However, as shown in Table \ref{tab:sum_CSST} and Figure \ref{fig:model_lnE_metrics_CSST}, with a additional quenching (or rejuvenation) component, the $\beta$-$\tau$-r form of SFH lead to a much worse quality of photometric SFR estimation.
Finally, the most complicated $\alpha$-$\beta$-$\tau$-r form of SFH lead to a slightly better SFR estimation.
In the case with noise, the best two models are also very similar in the quality of photometric stellar mass estimation.
Meanwhile, with the increasing of the complexity of SED models, the quality of photometric SFR stimation increases significantly. 

For photometric redshift, stellar mass and SFR, their accurate estimation becomes increasingly difficult.
Besides, the latter two are also more sensitive to the selection of SED models.
Generally, the quality of parameter estimates is closely related to the level of Bayesian evidence, which is especially clear in the more realistic case with noise.
Meanwhile, the model selection with maximum likelihood (or equivalently minimum $\chi^2$), where the more complex (or flexible) models are always more favored, is not consistent with the measurements of the quality of parameter estimation.
In practice, the direct measurements of the quality of parameter estimation as indicated by NMAD, BIA and OLF are usually unavailable.
So, the Bayesian model comparison with Bayesian evidence can be used to find the best SED model which is not only the most efficient but also give the best parameter estimation.

\begin{figure*}
    \centering
    \includegraphics[scale=0.55]{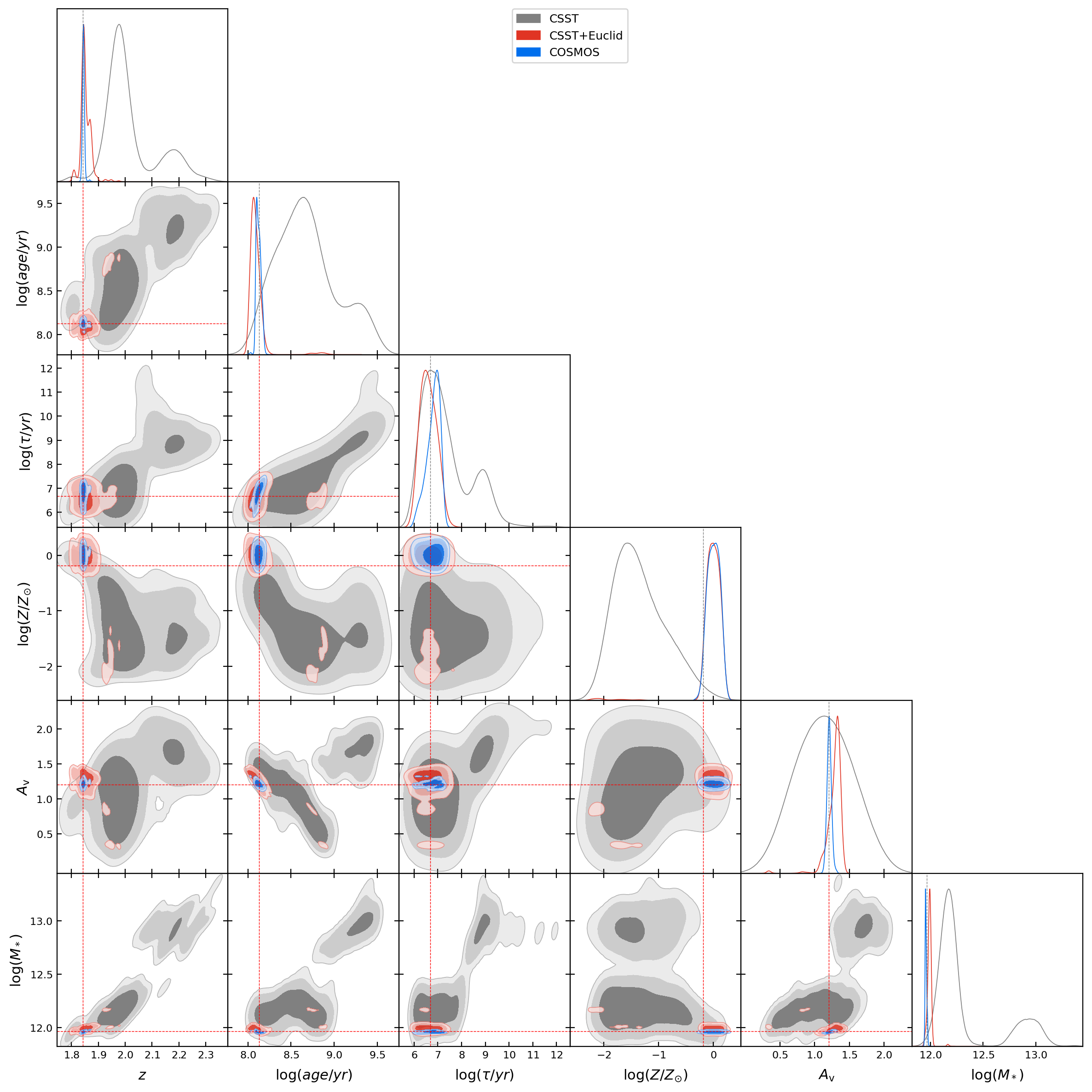}
    \caption
    {
        An example of 1D and 2D posterior probability distribution functions of free parameters obtained for the Bayesian analysis of a mock galaxy with CSST-like (grey), CSST+Euclid-like (red), and COSMOS-like (blue) photometric data, respectively.
        The contours show the $1\sigma$, $2\sigma$, and $3\sigma$ confidence regions, while the red dash lines show the ground truth values of each parameter.
        It is clear that the parameters are more tightly constrained and some degeneracies between them have been broken when using datasets with increasing discriminative powers.
    }
    \label{fig:pdftree}
\end{figure*}
\subsection{Effects of more flexible SFH and DAL for CSST+Euclid-like survey} \label{ss:disc_csst_euclid}
The results of both model comparison and parameter estimation are strongly dependent on the used datasets which may have very different discriminative powers.
In Figure \ref{fig:pdftree}, we show an example of 1D and 2D posterior probability distribution functions (PDFs) of free parameters obtained from the Bayesian analysis of the photometric data of a mock galaxy in the CSST-like, CSST-like+Euclid-like, and COSMOS-like surveys, respectively.
It is clear that different datasets lead to very different PDFs, due to their very different discriminative powers.
In this section and \S \ref{ss:disc_cosmos}, we discuss the effects of more flexible SFH and DAL for  CSST+Euclid-like and COSMOS-like surveys, respectively.
The addition of Euclid data extends the wavelength coverage of the data to the longer NIR band than with CSST-only data, which should be useful for enhancing the discriminative power of model comparison and the quality of the parameter estimation.

\begin{table*}
	\begin{center}
		\begin{tabular}{llllllllll}
			\toprule
			survey & noise & SFH & DAL & ln(BE) & ln(ML) & parameter & NMAD & BIA & OLF \\
\midrule
			 &  &  &  &  &  & $z_{\rm phot}$ & 0.020 & 0.002 & 0.001 \\
			CSST+Euclid & - & $\tau$,-CEH & Cal+00 & -243527$\pm$6307 & -3717 & ${\rm log}(M_*)_{\rm phot}$ & 0.055 & -0.023 & 0.045 \\
			 &  &  &  &  &  & ${\rm log}(SFR/[\rm M_{\odot}/yr])_{\rm phot}$ & 0.097 & -0.088 & 0.299 \\
\midrule
			 &  &  &  &  &  & $z_{\rm phot}$ & 0.021 & 0.001 & 0.001 \\
			CSST+Euclid & - & $\tau$,+CEH & Cal+00 & -216613$\pm$6390 & 30195 & ${\rm log}(M_*)_{\rm phot}$ & 0.057 & -0.014 & 0.037 \\
			 &  &  &  &  &  & ${\rm log}(SFR/[\rm M_{\odot}/yr])_{\rm phot}$ & 0.094 & -0.108 & 0.342 \\
\midrule
			 &  &  &  &  &  & $z_{\rm phot}$ & 0.013 & 0.010 & 0.005 \\
			CSST+Euclid & - & $\tau$,+CEH & Sal+18 & 72092$\pm$6475 & 340038 & ${\rm log}(M_*)_{\rm phot}$ & 0.029 & -0.015 & 0.009 \\
			 &  &  &  &  &  & ${\rm log}(SFR/[\rm M_{\odot}/yr])_{\rm phot}$ & 0.081 & -0.046 & 0.218 \\
\midrule
			 &  &  &  &  &  & $z_{\rm phot}$ & 0.012 & 0.010 & 0.004 \\
			CSST+Euclid & - & $\beta$-$\tau$,+CEH & Sal+18 & 70340$\pm$6409 & 339916 & ${\rm log}(M_*)_{\rm phot}$ & 0.027 & -0.010 & 0.007 \\
			 &  &  &  &  &  & ${\rm log}(SFR/[\rm M_{\odot}/yr])_{\rm phot}$ & 0.065 & -0.021 & 0.195 \\
\midrule
			 &  &  &  &  &  & $z_{\rm phot}$ & 0.017 & 0.012 & 0.010 \\
			CSST+Euclid & - & $\beta$-$\tau$-r,+CEH & Sal+18 & 25728$\pm$6566 & 315401 & ${\rm log}(M_*)_{\rm phot}$ & 0.032 & -0.005 & 0.017 \\
			 &  &  &  &  &  & ${\rm log}(SFR/[\rm M_{\odot}/yr])_{\rm phot}$ & 0.087 & 0.006 & 0.302 \\
\midrule
			 &  &  &  &  &  & $z_{\rm phot}$ & 0.016 & 0.012 & 0.016 \\
			CSST+Euclid & - & $\alpha$-$\beta$-$\tau$-r,+CEH & Sal+18 & 19560$\pm$6454 & 305020 & ${\rm log}(M_*)_{\rm phot}$ & 0.035 & -0.007 & 0.030 \\
			 &  &  &  &  &  & ${\rm log}(SFR/[\rm M_{\odot}/yr])_{\rm phot}$ & 0.087 & 0.011 & 0.286 \\
\midrule
			 &  &  &  &  &  & $z_{\rm phot}$ & 0.065 & -0.003 & 0.119 \\
			CSST+Euclid & + & $\tau$,-CEH & Cal+00 & -31783$\pm$3314 & 58861 & ${\rm log}(M_*)_{\rm phot}$ & 0.058 & -0.030 & 0.083 \\
			 &  &  &  &  &  & ${\rm log}(SFR/[\rm M_{\odot}/yr])_{\rm phot}$ & 0.164 & -0.020 & 0.379 \\
\midrule
			 &  &  &  &  &  & $z_{\rm phot}$ & 0.064 & -0.001 & 0.118 \\
			CSST+Euclid & + & $\tau$,+CEH & Cal+00 & -31347$\pm$3302 & 58766 & ${\rm log}(M_*)_{\rm phot}$ & 0.058 & -0.027 & 0.080 \\
			 &  &  &  &  &  & ${\rm log}(SFR/[\rm M_{\odot}/yr])_{\rm phot}$ & 0.171 & -0.027 & 0.392 \\
\midrule
			 &  &  &  &  &  & $z_{\rm phot}$ & 0.067 & 0.016 & 0.129 \\
			CSST+Euclid & + & $\tau$,+CEH & Sal+18 & -38081$\pm$3529 & 64421 & ${\rm log}(M_*)_{\rm phot}$ & 0.061 & -0.016 & 0.103 \\
			 &  &  &  &  &  & ${\rm log}(SFR/[\rm M_{\odot}/yr])_{\rm phot}$ & 0.186 & 0.011 & 0.449 \\
\midrule
			 &  &  &  &  &  & $z_{\rm phot}$ & 0.068 & 0.015 & 0.130 \\
			CSST+Euclid & + & $\beta$-$\tau$,+CEH & Sal+18 & -41164$\pm$3437 & 64349 & ${\rm log}(M_*)_{\rm phot}$ & 0.063 & -0.017 & 0.112 \\
			 &  &  &  &  &  & ${\rm log}(SFR/[\rm M_{\odot}/yr])_{\rm phot}$ & 0.177 & 0.042 & 0.476 \\
\midrule
			 &  &  &  &  &  & $z_{\rm phot}$ & 0.069 & 0.017 & 0.141 \\
			CSST+Euclid & + & $\beta$-$\tau$-r,+CEH & Sal+18 & -49796$\pm$3561 & 63873 & ${\rm log}(M_*)_{\rm phot}$ & 0.065 & -0.023 & 0.127 \\
			 &  &  &  &  &  & ${\rm log}(SFR/[\rm M_{\odot}/yr])_{\rm phot}$ & 0.213 & 0.059 & 0.571 \\
\midrule
			 &  &  &  &  &  & $z_{\rm phot}$ & 0.068 & 0.016 & 0.136 \\
			CSST+Euclid & + & $\alpha$-$\beta$-$\tau$-r,+CEH & Sal+18 & -49608$\pm$3426 & 64065 & ${\rm log}(M_*)_{\rm phot}$ & 0.073 & -0.025 & 0.169 \\
			 &  &  &  &  &  & ${\rm log}(SFR/[\rm M_{\odot}/yr])_{\rm phot}$ & 0.181 & 0.084 & 0.560 \\
			\bottomrule
		\end{tabular}
	\end{center}
	\caption
    {
        As in Table \ref{tab:sum_CSST}, but for CSST+Euclid-like survey.
    }
	\label{tab:sum_CSST_Euclid}
\end{table*}

\begin{figure*}
    \centering
    \includegraphics{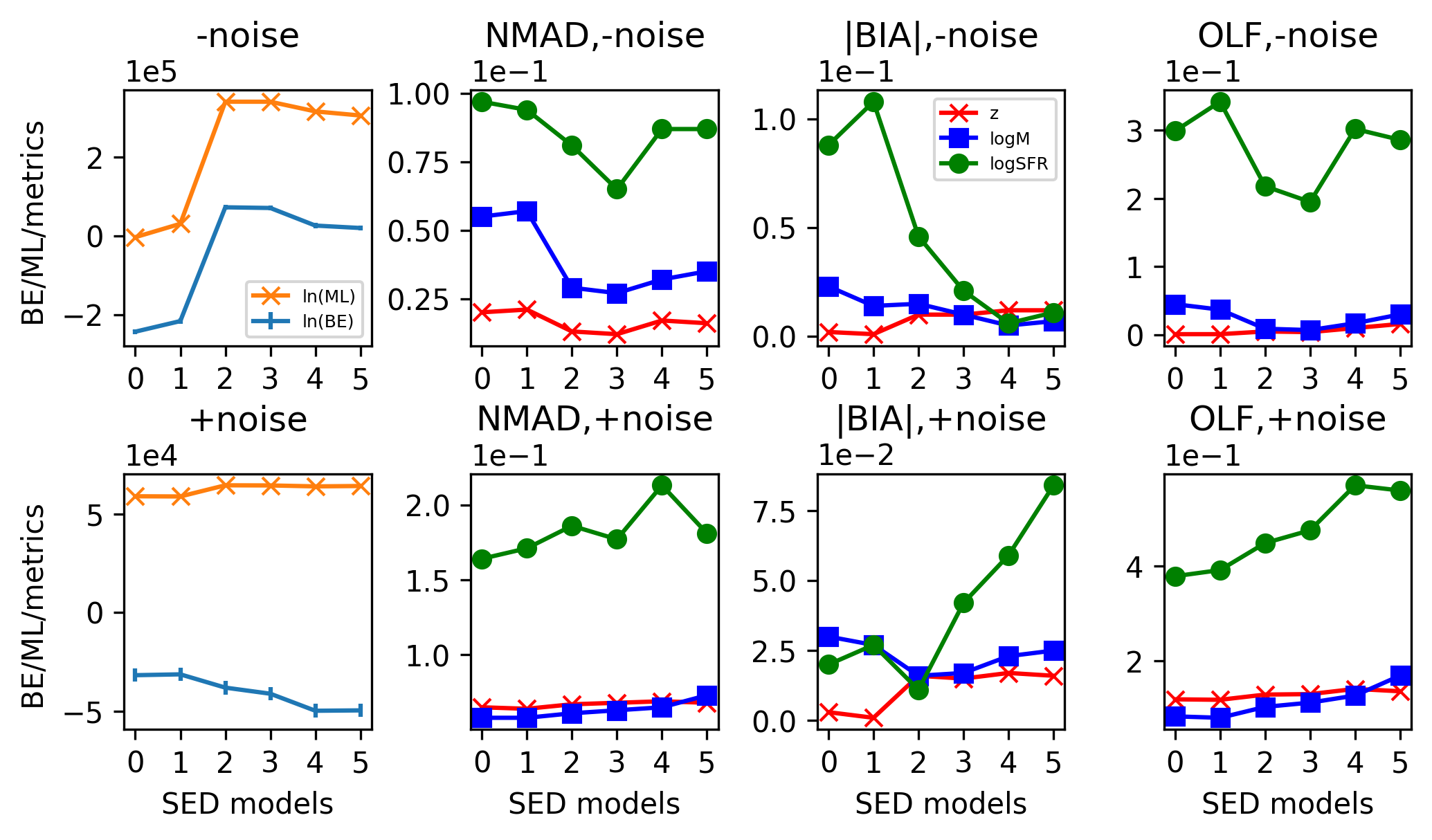}
    \caption
    {
        The Bayesian evidences (BE), maximum likelihood (ML) and metrics of photometric redshift (red star), stellar mass (blue square) and SFR (green circle) estimation from the Bayesian analysis of the hydrodynamical simulation-based mock galaxy sampe for CSST+Euclid-like survey by employing six SED models (0:``SFH=$\tau$,-CEH,DAL=Cal+18'', 1:``SFH=$\tau$,+CEH,DAL=Cal+18'', 2:``SFH=$\tau$,+CEH,DAL=Sal+18'', 3:``SFH=$\beta$-$\tau$,+CEH,DAL=Sal+18'', 4:``SFH=$\beta$-$\tau$-r,+CEH,DAL=Sal+18'', 5:``SFH=$\alpha$-$\beta$-$\tau$-r,+CEH,DAL=Sal+18'') with increasing complexity in the forms of SFH and DAL , as well as for the cases with (bottom panels) and without noise (top panels).
        In the case without noise, the simplest model ``SFH=$\tau$,-CEH,DAL=Cal+18'' has the lowest Bayesian evidence of ${\rm ln}(BE)=-243527\pm6307$.
        Meanwhile, the SED models ``SFH=$\tau$,+CEH,DAL=Sal+18'' and ``SFH=$\beta$-$\tau$,+CEH,DAL=Sal+18'' which are neither the simplest nor the most complex models have the largest Bayesian evidences of ${\rm ln}(BE)=72092\pm6475$ and ${\rm ln}(BE)=70340\pm6409$ which are comparable within error bar.
        As in the case for CSST-like survey, the model selection with maximum likelihood (or equivalently minimum $\chi^2$) lead to similar results.
        Contrary to the case without noise, in the case with noise, the two simplest SED models ``SFH=$\tau$,-CEH,DAL=Cal+00'' and ``SFH=$\tau$,+CEH,DAL=Cal+00'' have the largest Bayesian evidences of ${\rm ln}(BE)=-31783\pm3314$ and ${\rm ln}(BE)=-31347\pm3302$ which are comparable within error bar.
        However, the model selection with maximum likelihood (or equivalently minimum $\chi^2$) lead to exactly the opposite results, where the more complex (or flexible) models are always more favored.
        Generally, the quality of parameter estimates is closely related to the level of Bayesian evidence, which is especially clear in the more realistic case with noise.
        The model selection with Bayesian evidence is still more consistent with the measurements of the quality of parameter estimation than that with maximum likelihood (or equivalently minimum $\chi^2$).
        For photometric redshift, stellar mass and SFR, their accurate estimation becomes increasingly difficult, while the latter two are also more sensitive to the selection of SED models.
        In the case without noise, the SED models ``SFH=$\tau$,+CEH,DAL=Sal+18'' and ``SFH=$\beta$-$\tau$,+CEH,DAL=Sal+18'' which are just the two with the largest Bayesian evidence give the highest quality parameter estimates.
        In the case with noise, the two simplest SED models ``SFH=$\tau$,-CEH,DAL=Cal+00'' and ``SFH=$\tau$,+CEH,DAL=Cal+00'' which are also the two with the largest Bayesian evidences give the highest quality parameter estimates.
        All of these results are very similar to that for CSST-like survey.
        However, the relative error of Bayesian evidences have been reduced, especially in the case without noise.
        Besides, the quality of parameter estimation, especially that of stellar mass estimation, has been significantly improved.
        Furthermore, the quality of parameter estimation, especially that of stellar mass, increases more slowly with the increasing of SED model complexity. 
    }
    \label{fig:model_lnE_metrics_CSST_Euclid}
\end{figure*}

In Table \ref{tab:sum_CSST_Euclid}, we present a summary of the Bayesian evidences, maximum likelihoods and metrics of the quality of parameter estimation from the Bayesian analysis of the hydrodynamical simulation-based mock galaxy sampe for CSST+Euclid-like survey by employing six different combinations of SFH and DAL with increasing complexity, as well as for the cases with and without noise.
The same results are also shown more clearly in Figure \ref{fig:model_lnE_metrics_CSST_Euclid}.

\subsubsection{Model comparison} \label{ss:disc_csst_euclid_mod}
In the case without noise, as shown in the top left panel of Figure \ref{fig:model_lnE_metrics_CSST_Euclid}, the simplest model ``SFH=$\tau$,-CEH,DAL=Cal+18'' has the lowest Bayesian evidence of ${\rm ln}(BE)=-243527\pm6307$.
With the additional consideration of metallicity evolution, the Bayesian evidence of the model ``SFH=$\tau$,+CEH,DAL=Cal+18'' increases to ${\rm ln}(BE)=-216613\pm6390$.
Then, with the adoption of the DAL of \cite{SalimS2018a}, the Bayesian evidence of the model ``SFH=$\tau$,+CEH,DAL=Sal+18'' increases significantly to ${\rm ln}(BE)=72092\pm6475$.
Apparently, the DAL of \cite{SalimS2018a} is also a much better choice than that of \cite{CalzettiD2000a} for the hydrodynamical simulation-based mock galaxy sampe in CSST+Euclid-like survey.
Furthermore, by employing a more complicated $\beta$-$\tau$ form of SFH, the Bayesian evidence of the model ``SFH=$\beta$-$\tau$,+CEH,DAL=Sal+18'' seems decreases a little to ${\rm ln}(BE)=70340\pm6409$.
As in the case for CSST-like survey, the latter two SED models (``SFH=$\tau$,+CEH,DAL=Sal+18'' and ``SFH=$\beta$-$\tau$,+CEH,DAL=Sal+18'')  have the largest Bayesian evidences which are comparable within error bar.
With a quenching (or rejuvenation) component added to the SFH, the Bayesian evidence of the model ``SFH=$\beta$-$\tau$-r,+CEH,DAL=Sal+18'' decreases significantly to ${\rm ln}(BE)=25728\pm6566$.
Finally, by employing a even more flexible double power-law form of SFH, the Bayesian evidence of the model ``SFH=$\alpha$-$\beta$-$\tau$-r,+CEH,DAL=Sal+18'' seems decreases a little to ${\rm ln}(BE)=19560\pm6454$ which is comparable with the former within error bar.
On the other hand, the model selection with maximum likelihood (or equivalently minimum $\chi^2$) lead to similar results.

In the case with noise, the two simplest SED models ``SFH=$\tau$,-CEH,DAL=Cal+00'' and ``SFH=$\tau$,+CEH,DAL=Cal+00'' have the largest Bayesian evidences of ${\rm ln}(BE)=-31783\pm3314$ and ${\rm ln}(BE)=-31347\pm3302$, respectively.
Then, with the adoption of the DAL of \cite{SalimS2018a}, the Bayesian evidence of the model ``SFH=$\tau$,+CEH,DAL=Sal+18'' decreases significantly to ${\rm ln}(BE)=-38081\pm3529$.
By employing a more complicated $\beta$-$\tau$ form of SFH, the Bayesian evidence of the model ``SFH=$\beta$-$\tau$,+CEH,DAL=Sal+18'' seems decreases further to ${\rm ln}(BE)=-41164\pm3437$, although comparable with the former within error bar.
With a quenching (or rejuvenation) component added to the SFH, the Bayesian evidence of the model ``SFH=$\beta$-$\tau$-r,+CEH,DAL=Sal+18'' obviously decreases to ${\rm ln}(BE)=-49796\pm3561$, which is similar to the case without noise.
Finally, by employing a even more flexible double power-law form of SFH, the Bayesian evidence of the model ``SFH=$\alpha$-$\beta$-$\tau$-r,+CEH,DAL=Sal+18'' seems increases a little to ${\rm ln}(BE)=-49608\pm3426$, although comparable with the former within error bar.
However, the model selection with maximum likelihood (or equivalently minimum $\chi^2$) lead to exactly the opposite results, where the more complex (or flexible) models are always more favored.

In general, all of these results are very similar to that for CSST-like survey.
The model selection with Bayesian evidence is still more consistent with the measurements of the quality of parameter estimation.
However, the relative error of Bayesian evidences have been reduced, especially in the case without noise.

\subsubsection{Parameter estimation} \label{ss:disc_csst_euclid_par}
In the right three panels of Figure \ref{fig:model_lnE_metrics_CSST_Euclid}, we show the three metrics of the quality of photometric redshift, stellar mass and SFR estimation for different SED models, respectively.
In general, in the case without noise, the SED models ``SFH=$\tau$,+CEH,DAL=Sal+18'' and ``SFH=$\beta$-$\tau$,+CEH,DAL=Sal+18'' which are just the two with largest Bayesian evidence give the highest quality parameter estimates.
In the case with noise, the two simplest SED models ``SFH=$\tau$,-CEH,DAL=Cal+00'' and ``SFH=$\tau$,+CEH,DAL=Cal+00'' which are also the two with the largest Bayesian evidences give the highest quality parameter estimates.
In the following, we discuss these results in more detail.

\begin{figure*}
    \gridline
    {
        \leftfig{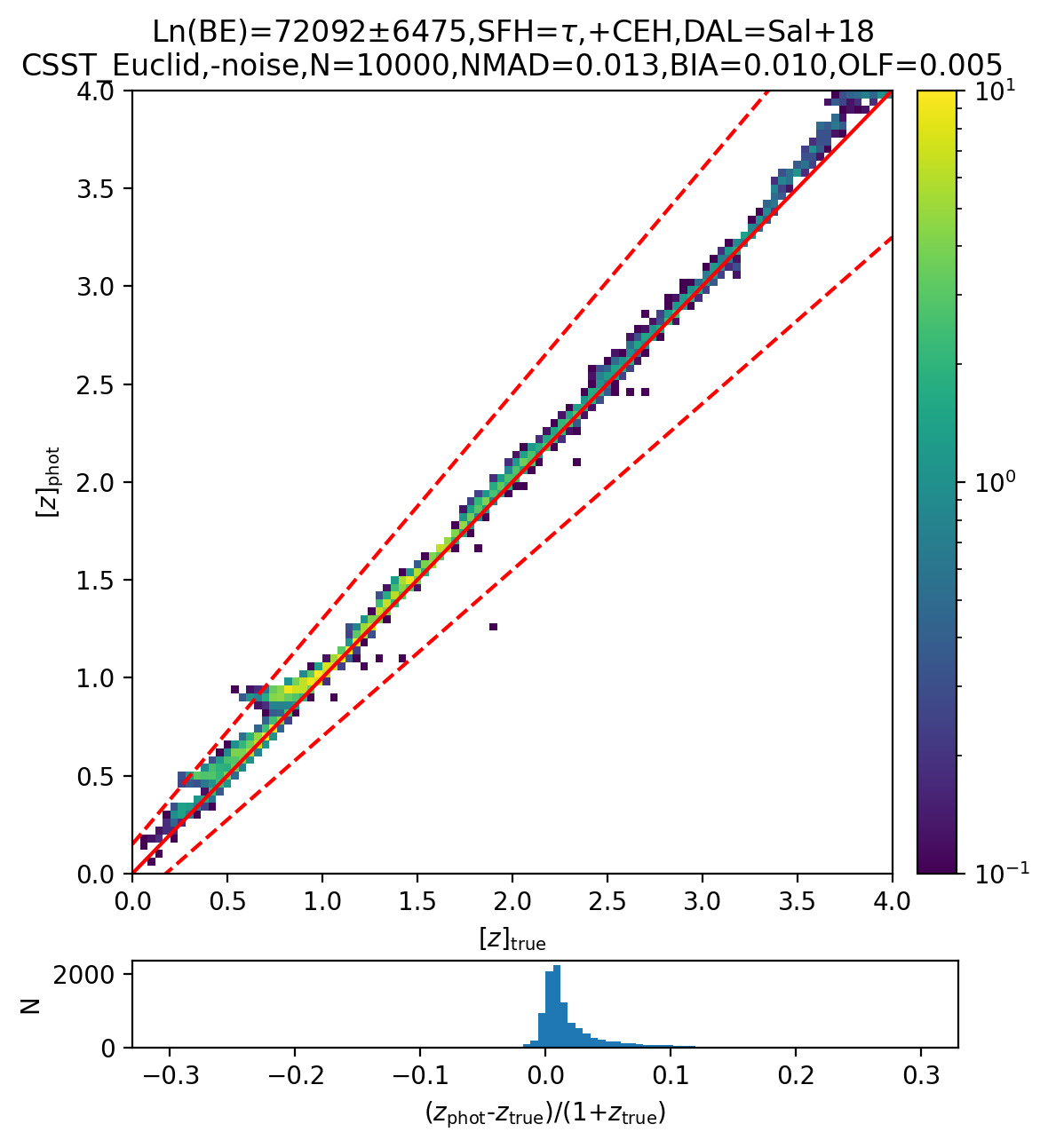}{0.48\textwidth}{(a)}
        \rightfig{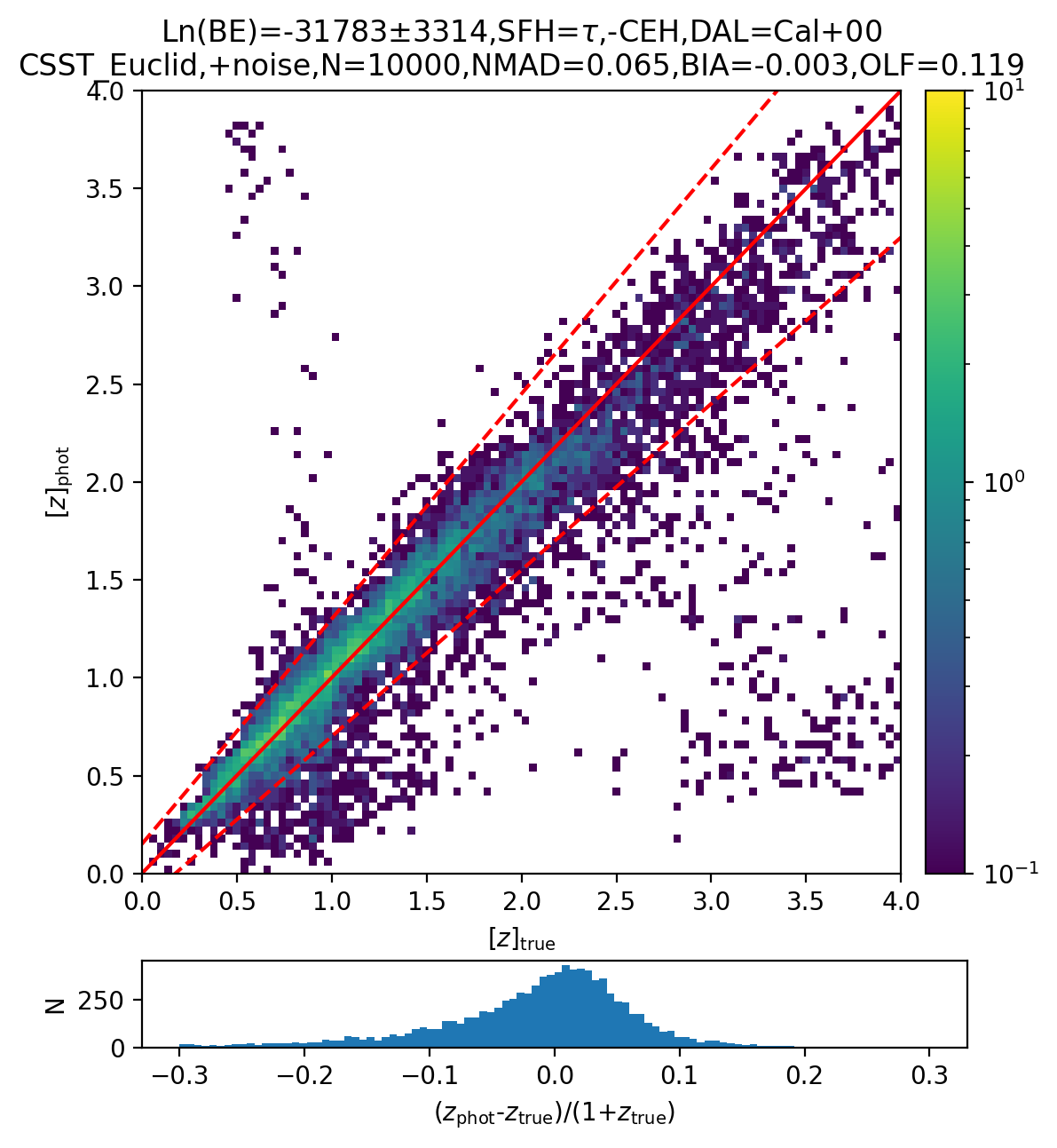}{0.48\textwidth}{(b)}
    }
    \gridline
    {
        \leftfig{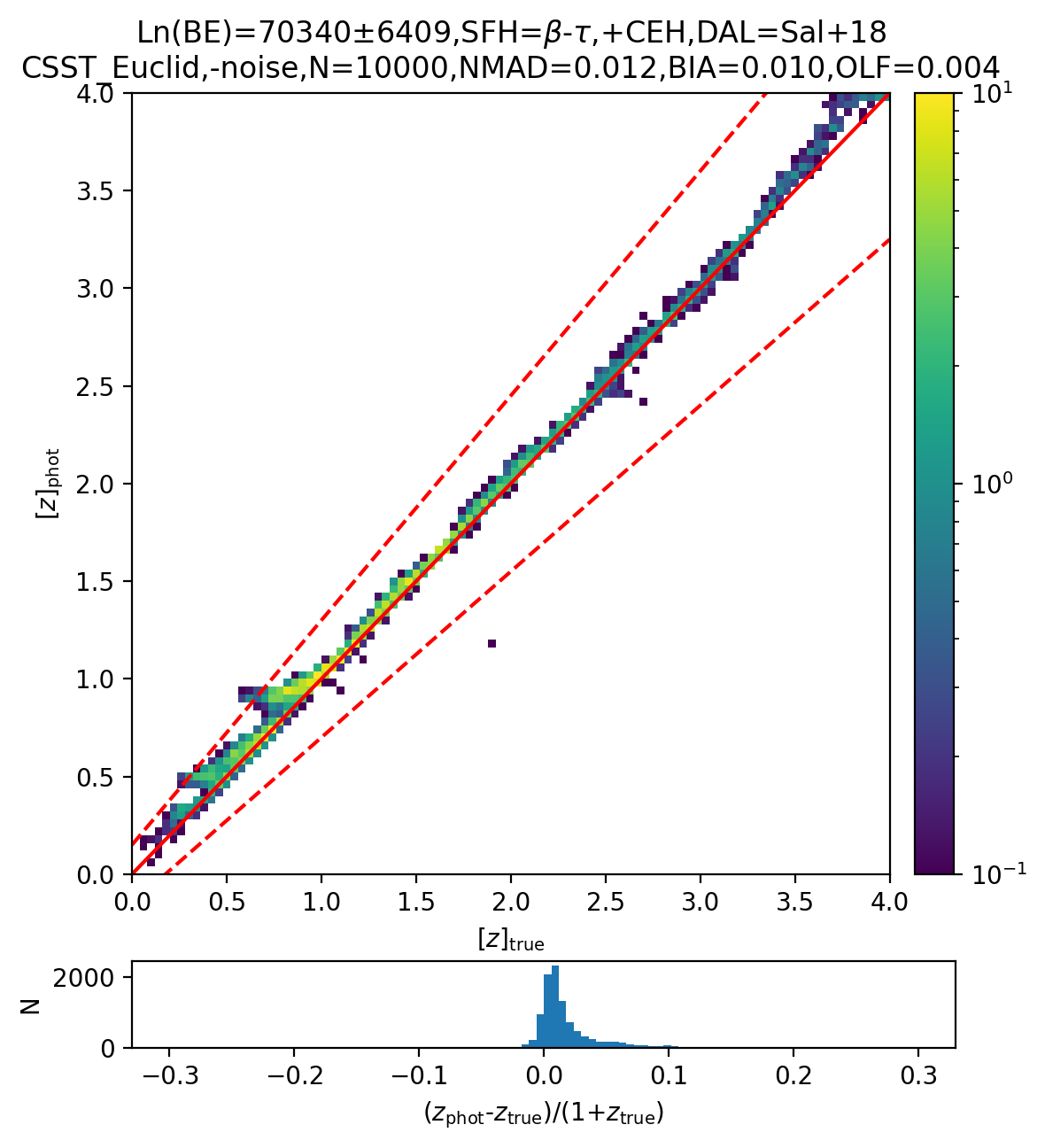}{0.48\textwidth}{(c)}
        \rightfig{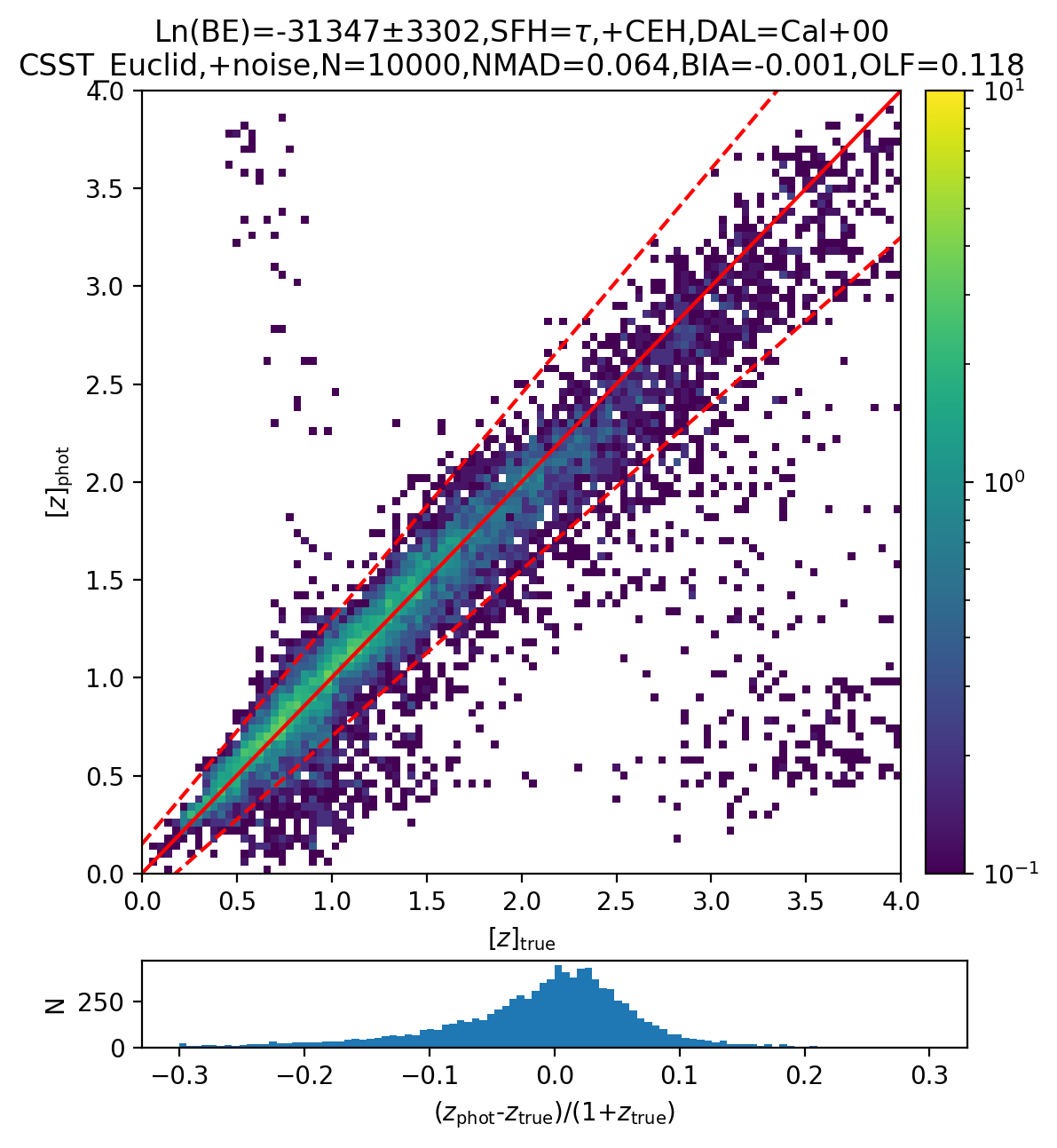}{0.48\textwidth}{(d)}
    }
    \caption
    {
        The results of photometric redshift estimation from the Bayesian analysis of the hydrodynamical simulation-based mock data for CSST+Euclid-like survey by employing the two SED models with the largest Bayesian evidence.
        By comparing with the results for CSST-like survey in Figure \ref{fig:perf_phot_csst_z2}, it is clear that the quality of photometric redshift estimation has been obviously increased in both the cases with and without noise.
        Especially, in the more realistic case with noise, the outliers caused by the mis-identification of Lyman and Balmer break features have been largely reduced.
        This suggest that, the inclusion of J, H and Y bands from Euclid is helpful for improving the photometric redshift estimation.
        However, the more complicated $\beta$-$\tau$ form of SFH is not very helpful for improving the quality of photometric redshift estimation.
    }
    \label{fig:perf_phot_csst_euclid_z2}
\end{figure*}
The detailed results of photometric redshift estimation obtained by employing the two models with the largest Bayesian evidences are shown in Figure \ref{fig:perf_phot_csst_euclid_z2}.
By comparing with the results for CSST-like survey in Figure \ref{fig:perf_phot_csst_z2}, it is clear that the quality of photometric redshift estimation has been obviously increased in both the cases with and without noise.
Especially, in the more realistic case with noise, the outliers caused by the mis-identification of Lyman and Balmer break features have been largely reduced.
This suggest that, the inclusion of J, H and Y bands from Euclid is helpful for improving the photometric redshift estimation.
However, the more complicated $\beta$-$\tau$ form of SFH is not very helpful for improving the quality of photometric redshift estimation.

\begin{figure*}
    \gridline
    {
        \leftfig{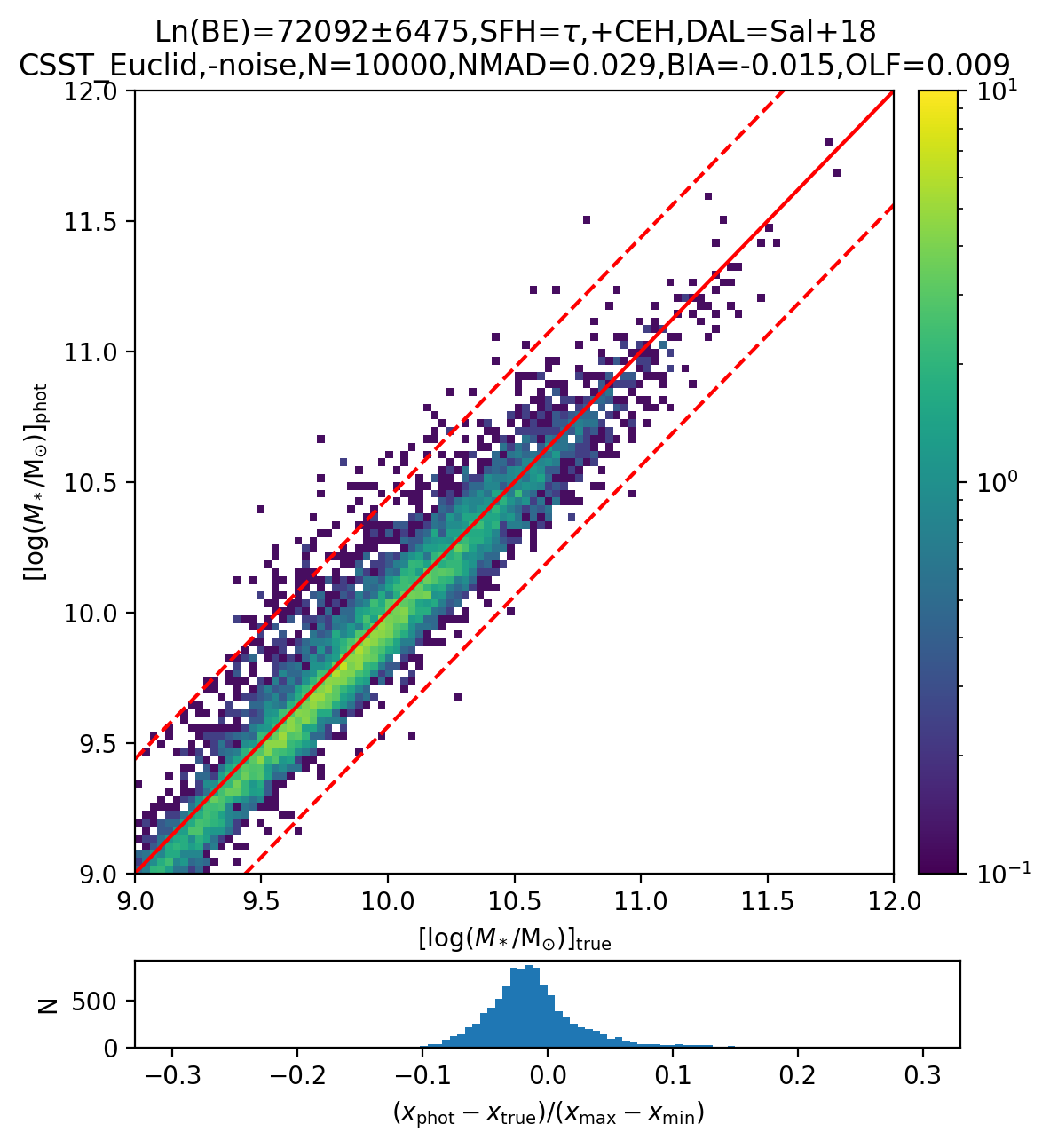}{0.48\textwidth}{(a)}
        \rightfig{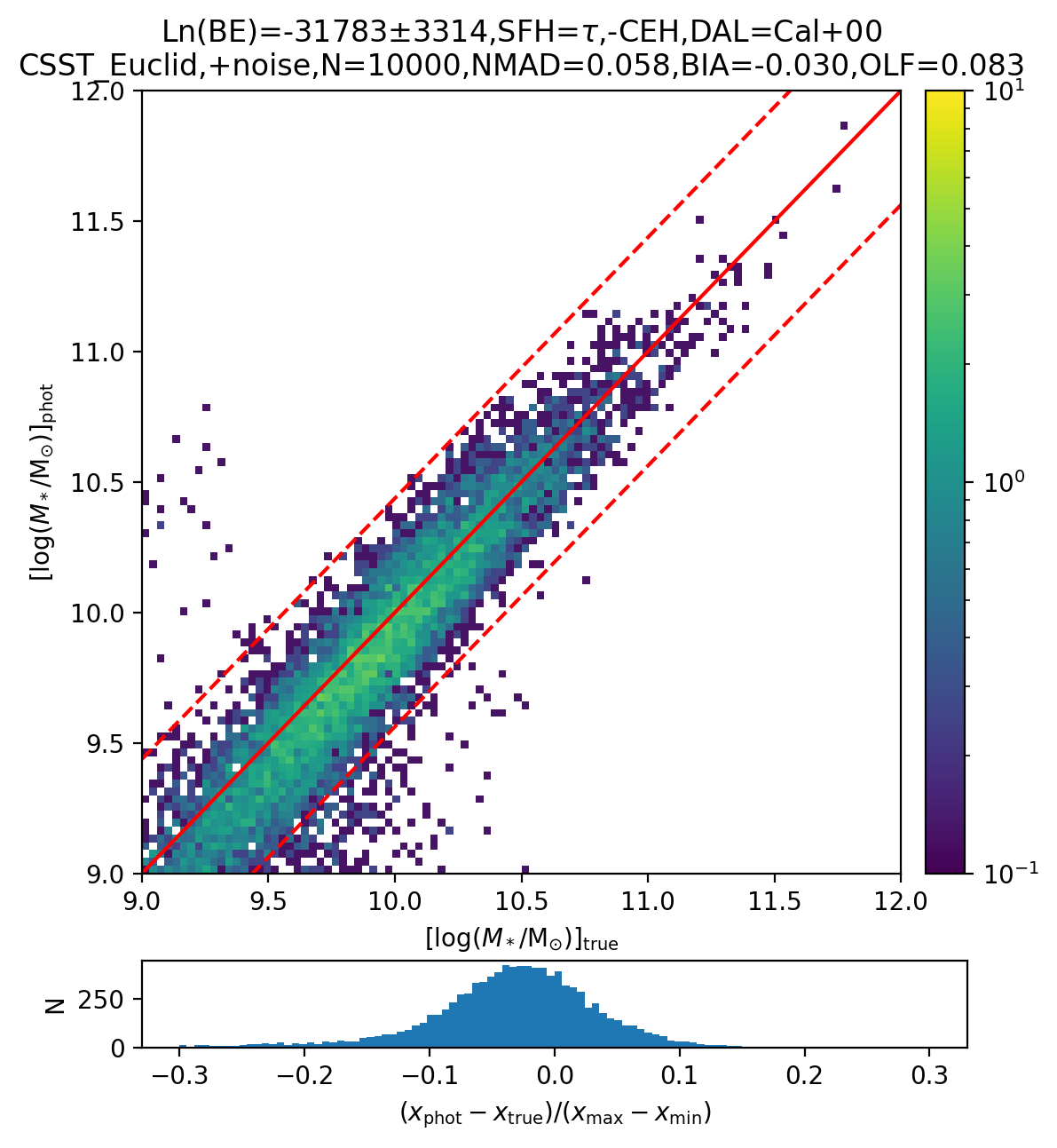}{0.48\textwidth}{(b)}
    }
    \gridline
    {
        \leftfig{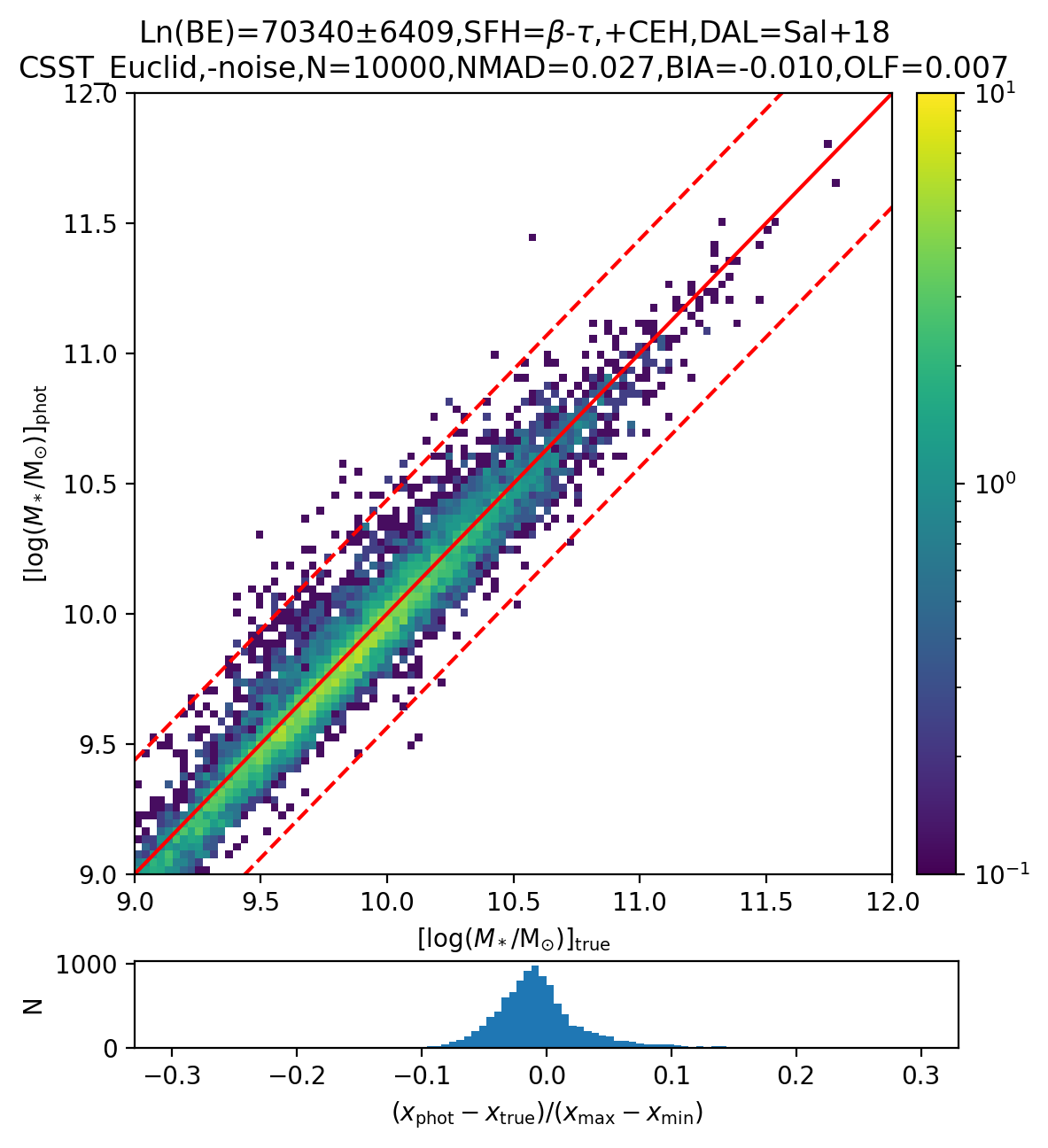}{0.48\textwidth}{(c)}
        \rightfig{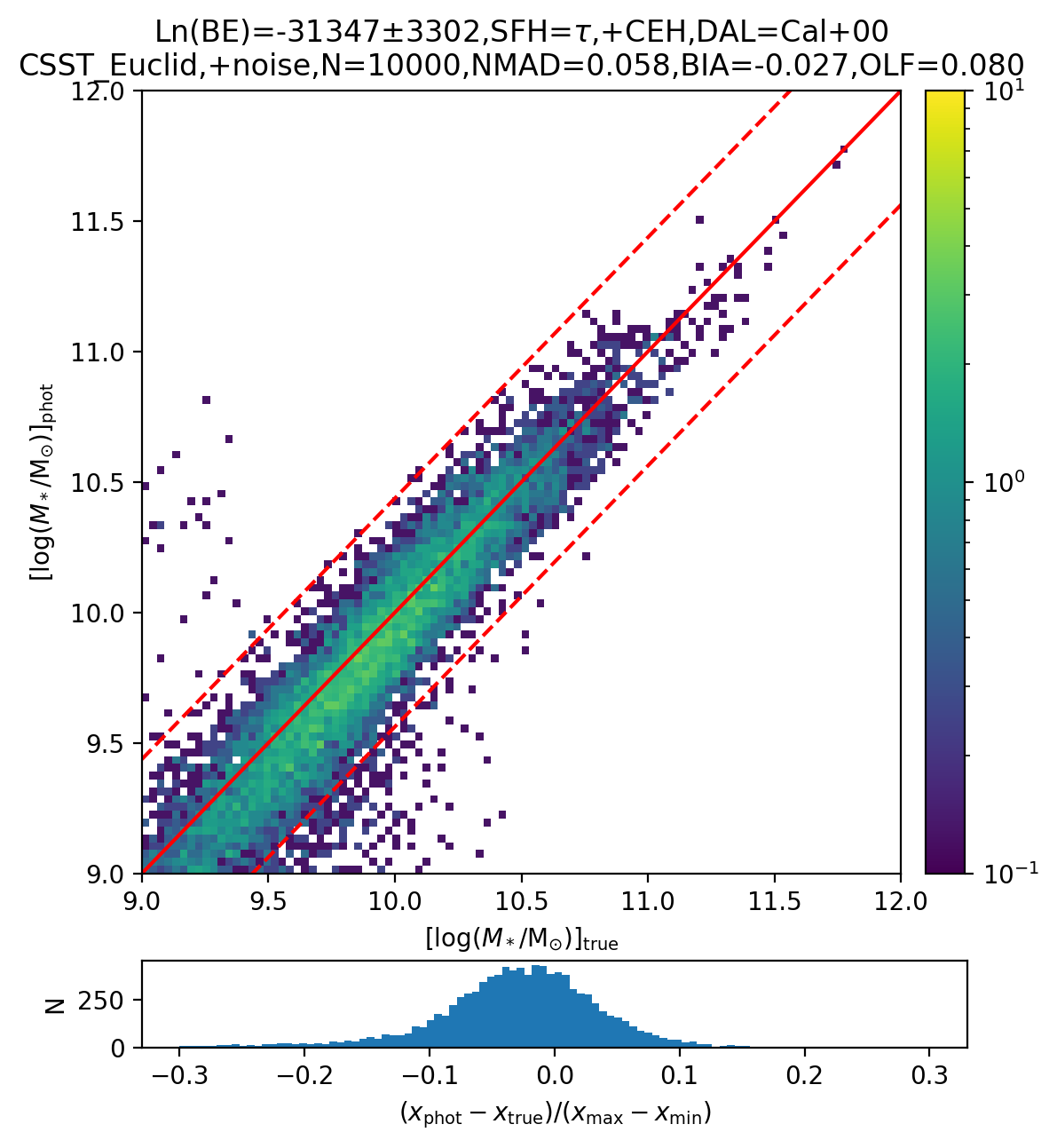}{0.48\textwidth}{(d)}
    }
    \caption
    {
        As in Figure \ref{fig:perf_phot_csst_euclid_z2}, but for the stellar  mass.
        By comparing with the results for CSST-like survey in Figure \ref{fig:perf_phot_csst_mass2}, it is clear that the quality of photometric stellar mass estimation has been significantly improved in both the cases with and without noise.
        Apparently, the inclusion of J, H and Y bands from Euclid is crucial for a more accurate estimation of stellar mass.
        Besides, with the more complicated $\beta$-$\tau$ form of SFH, the quality of photometric stellar mass estimation is only slightly improved.
    }
    \label{fig:perf_phot_csst_euclid_mass2}
\end{figure*}
The detailed results of photometric stellar mass estimation obtained by employing the two models with the largest Bayesian evidences are shown in Figure \ref{fig:perf_phot_csst_euclid_mass2}.
By comparing with the results for CSST-like survey in Figure \ref{fig:perf_phot_csst_mass2}, it is clear that the quality of photometric stellar mass estimation has been significantly improved in both the cases with and without noise.
Apparently, the inclusion of J, H and Y bands from Euclid is crucial for a more accurate estimation of stellar mass.
With the more complicated $\beta$-$\tau$ form of SFH, the quality of photometric stellar mass estimation is slightly improved.
However, as shown in Table \ref{tab:sum_CSST_Euclid} and Figure \ref{fig:model_lnE_metrics_CSST_Euclid}, the even more complicated forms of SFH is still not helpful for improving the quality of photometric stellar mass estimation.

\begin{figure*}
    \gridline
    {
        \leftfig{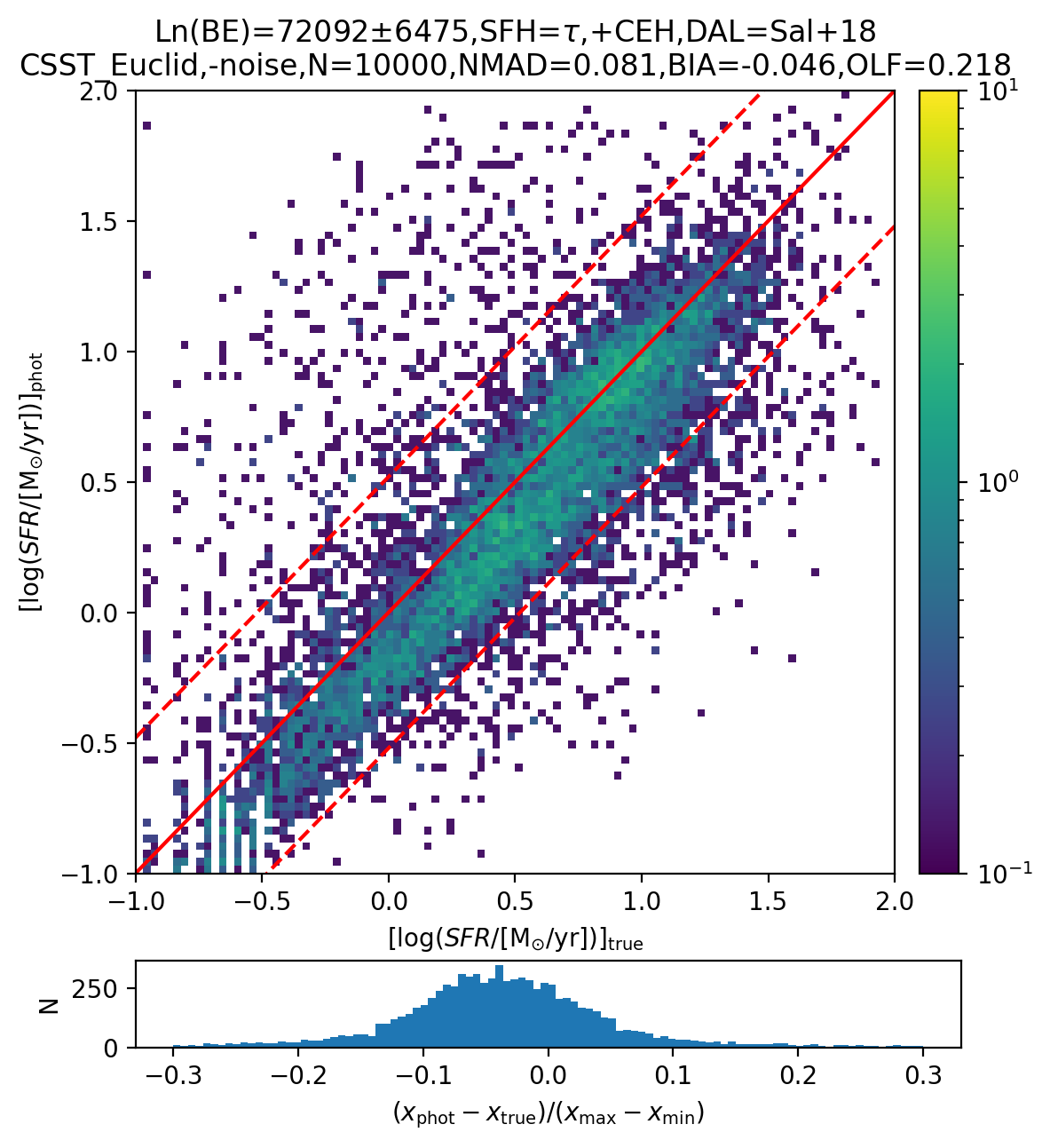}{0.48\textwidth}{(a)}
        \rightfig{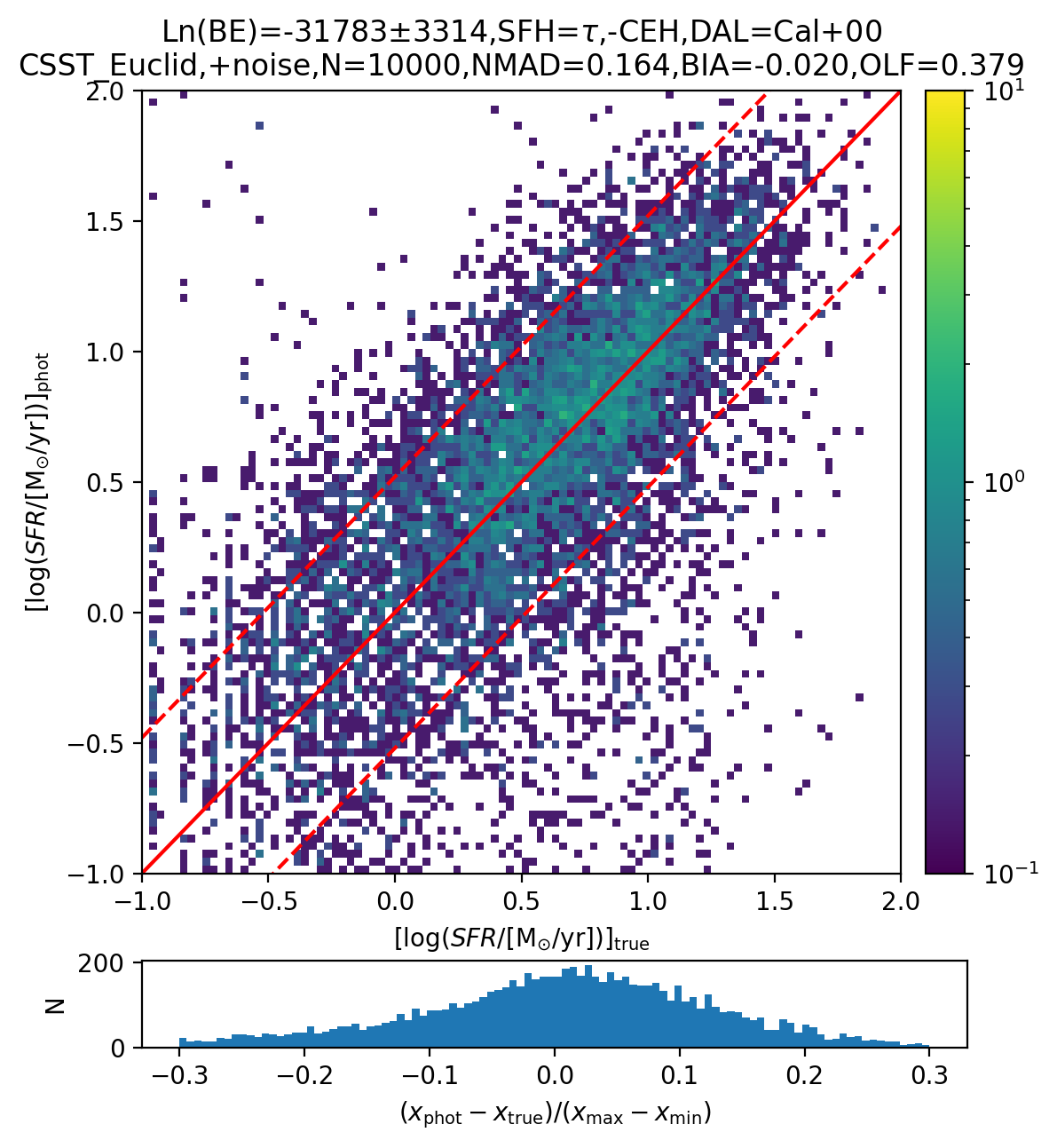}{0.48\textwidth}{(b)}
    }
    \gridline
    {
        \leftfig{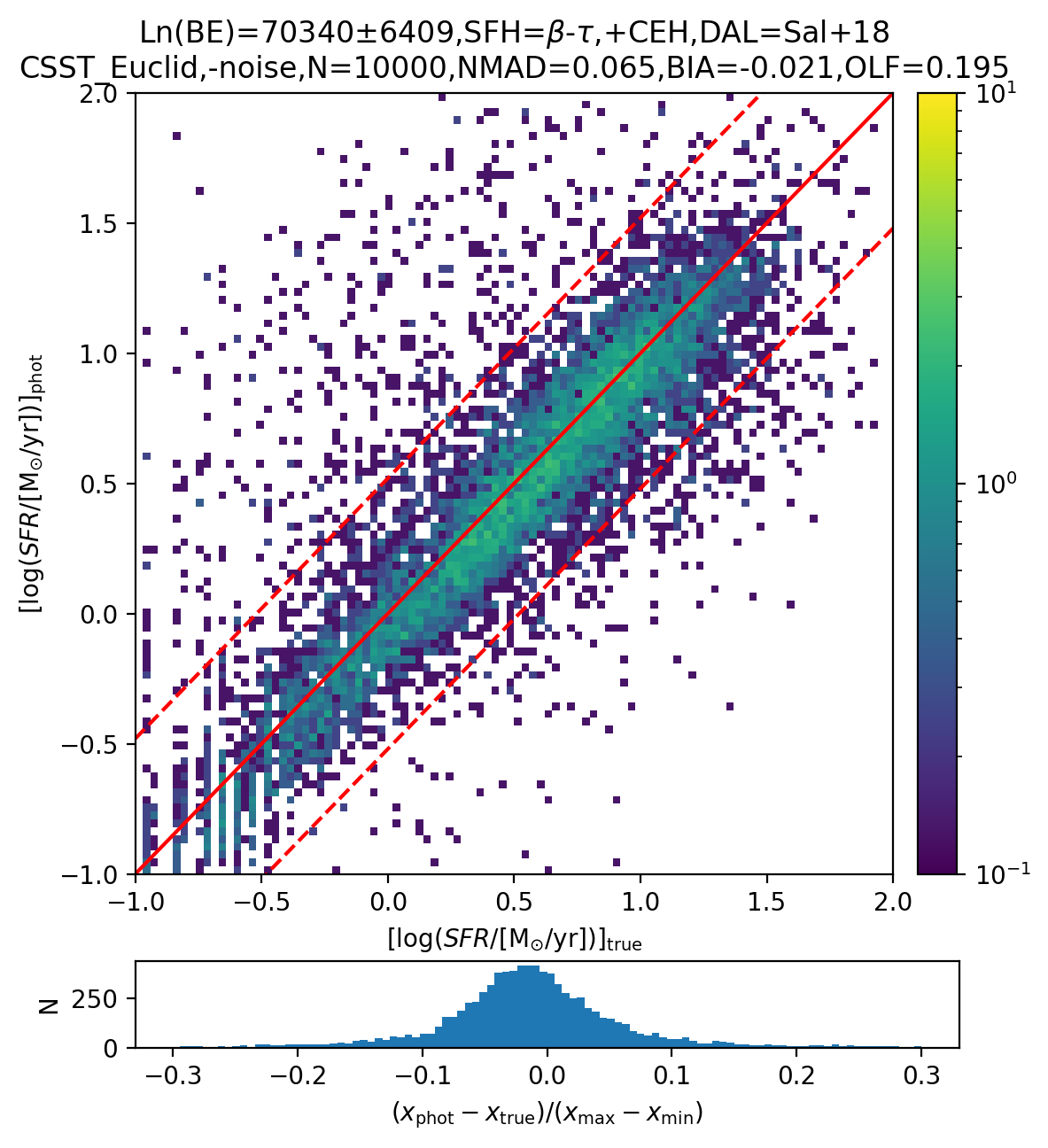}{0.48\textwidth}{(c)}
        \rightfig{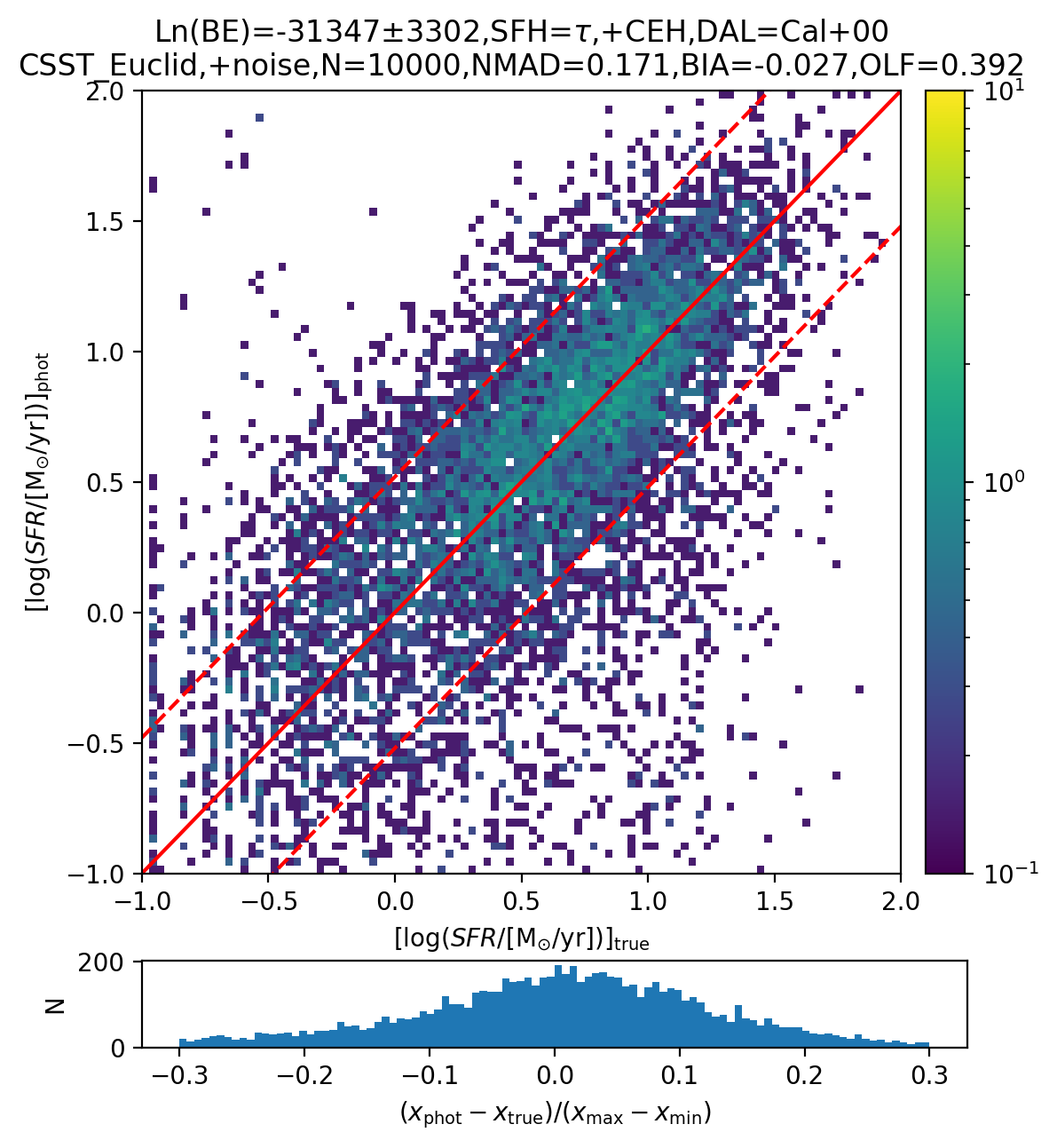}{0.48\textwidth}{(d)}
    }
    \caption
    {
        As in Figure \ref{fig:perf_phot_csst_euclid_z2}, but for the SFR.
        \textbf{(a, c)} The cases without noise. By comparing with the results for CSST-like survey in the left panes of Figure \ref{fig:perf_phot_csst_sfr2}, it is clear that the $\sigma_{\rm NMAD}$ and OLF of photometric SFR estimation are largely reduced, although the bias is slightly increased.
        Besides, with the more complicated $\beta$-$\tau$ form of SFH, the quality of photometric SFR estimation is improved.
        \textbf{(b, d)} The cases with noise. By comparing with the results for CSST-like survey in the right panes of Figure \ref{fig:perf_phot_csst_sfr2}, all of the $\sigma_{\rm NMAD}$, bias and OLF of photometric SFR estimation slightly increase.
        This suggest that, the inclusion of J, H and Y bands from Euclid is not very helpful for improving the photometric SFR estimation.
        Besides, with the additional consideration of metallicity evolution, the quality of photometric SFR estimation becomes slightly worse.
    }
    \label{fig:perf_phot_csst_euclid_sfr2}
\end{figure*}
The detailed results of photometric SFR estimation obtained by employing the two models with the largest Bayesian evidences are shown in Figure \ref{fig:perf_phot_csst_euclid_sfr2}.
In the case without noise, by comparing with the results for CSST-like survey in Figure \ref{fig:perf_phot_csst_sfr2}, it is clear that the $\sigma_{\rm NMAD}$ and OLF of photometric SFR estimation are largely reduced, although the bias is slightly increased.
However, in the case with noise, all of the $\sigma_{\rm NMAD}$, bias and OLF of photometric SFR estimation slightly increase.
So, the inclusion of J, H and Y bands from Euclid is not very helpful for improving the photometric SFR estimation.

\subsection{Effects of more flexible SFH and DAL for COSMOS-like survey} \label{ss:disc_cosmos}
The COSMOS-like survey covers many more bands than CSST-like data.
Although there is no NUV data, it extends to the longer wavelengths and includes some intermediate bands (IBs) \citep{LaigleC2016a}.
Since the COSMOS-like mock data has much stronger discriminative power than CSST-like and CSST+Euclid-like mock data, the Bayesian evidence of different SED models should show much larger difference, and the photometric redshift and stellar population parameter estimation should be better.

\begin{table*}
	\begin{center}
		\begin{tabular}{llllllllll}
			\toprule
			survey & noise & SFH & DAL & ln(BE) & ln(ML) & parameter & NMAD & BIA & OLF \\
\midrule
			 &  &  &  &  &  & $z_{\rm phot}$ & 0.007 & -0.002 & 0 \\
			COSMOS & - & $\tau$,-CEH & Cal+00 & -1710604$\pm$6811 & -1431790 & ${\rm log}(M_*)_{\rm phot}$ & 0.030 & -0.026 & 0.013 \\
			 &  &  &  &  &  & ${\rm log}(SFR/[\rm M_{\odot}/yr])_{\rm phot}$ & 0.110 & -0.062 & 0.290 \\
\midrule
			 &  &  &  &  &  & $z_{\rm phot}$ & 0.006 & -0.002 & 0 \\
			COSMOS & - & $\tau$,+CEH & Cal+00 & -1499618$\pm$6997 & -1207887 & ${\rm log}(M_*)_{\rm phot}$ & 0.032 & -0.017 & 0.010 \\
			 &  &  &  &  &  & ${\rm log}(SFR/[\rm M_{\odot}/yr])_{\rm phot}$ & 0.101 & -0.082 & 0.299 \\
\midrule
			 &  &  &  &  &  & $z_{\rm phot}$ & 0.002 & 0 & 0 \\
			COSMOS & - & $\tau$,+CEH & Sal+18 & 354581$\pm$7310 & 681554 & ${\rm log}(M_*)_{\rm phot}$ & 0.023 & -0.022 & 0.012 \\
			 &  &  &  &  &  & ${\rm log}(SFR/[\rm M_{\odot}/yr])_{\rm phot}$ & 0.069 & -0.068 & 0.124 \\
\midrule
			 &  &  &  &  &  & $z_{\rm phot}$ & 0.002 & 0 & 0 \\
			COSMOS & - & $\beta$-$\tau$,+CEH & Sal+18 & 387009$\pm$7273 & 717135 & ${\rm log}(M_*)_{\rm phot}$ & 0.020 & -0.011 & 0.006 \\
			 &  &  &  &  &  & ${\rm log}(SFR/[\rm M_{\odot}/yr])_{\rm phot}$ & 0.047 & -0.041 & 0.057 \\
\midrule
			 &  &  &  &  &  & $z_{\rm phot}$ & 0.003 & 0.001 & 0 \\
			COSMOS & - & $\beta$-$\tau$-r,+CEH & Sal+18 & 255069$\pm$7552 & 613256 & ${\rm log}(M_*)_{\rm phot}$ & 0.028 & -0.003 & 0.017 \\
			 &  &  &  &  &  & ${\rm log}(SFR/[\rm M_{\odot}/yr])_{\rm phot}$ & 0.070 & -0.001 & 0.124 \\
\midrule
			 &  &  &  &  &  & $z_{\rm phot}$ & 0.003 & 0.001 & 0 \\
			COSMOS & - & $\alpha$-$\beta$-$\tau$-r,+CEH & Sal+18 & 275129$\pm$7218 & 612401 & ${\rm log}(M_*)_{\rm phot}$ & 0.031 & -0.005 & 0.012 \\
			 &  &  &  &  &  & ${\rm log}(SFR/[\rm M_{\odot}/yr])_{\rm phot}$ & 0.067 & -0.006 & 0.121 \\
\midrule
			 &  &  &  &  &  & $z_{\rm phot}$ & 0.017 & -0.003 & 0.004 \\
			COSMOS & + & $\tau$,-CEH & Cal+00 & 99881$\pm$4881 & 259510 & ${\rm log}(M_*)_{\rm phot}$ & 0.044 & -0.020 & 0.006 \\
			 &  &  &  &  &  & ${\rm log}(SFR/[\rm M_{\odot}/yr])_{\rm phot}$ & 0.127 & -0.032 & 0.282 \\
\midrule
			 &  &  &  &  &  & $z_{\rm phot}$ & 0.017 & -0.004 & 0.003 \\
			COSMOS & + & $\tau$,+CEH & Cal+00 & 107009$\pm$4896 & 268492 & ${\rm log}(M_*)_{\rm phot}$ & 0.044 & -0.012 & 0.005 \\
			 &  &  &  &  &  & ${\rm log}(SFR/[\rm M_{\odot}/yr])_{\rm phot}$ & 0.126 & -0.044 & 0.288 \\
\midrule
			 &  &  &  &  &  & $z_{\rm phot}$ & 0.015 & 0.002 & 0.003 \\
			COSMOS & + & $\tau$,+CEH & Sal+18 & 131578$\pm$5115 & 313110 & ${\rm log}(M_*)_{\rm phot}$ & 0.042 & -0.017 & 0.036 \\
			 &  &  &  &  &  & ${\rm log}(SFR/[\rm M_{\odot}/yr])_{\rm phot}$ & 0.107 & -0.021 & 0.211 \\
\midrule
			 &  &  &  &  &  & $z_{\rm phot}$ & 0.015 & 0.002 & 0.003 \\
			COSMOS & + & $\beta$-$\tau$,+CEH & Sal+18 & 130766$\pm$5064 & 315205 & ${\rm log}(M_*)_{\rm phot}$ & 0.037 & -0.012 & 0.029 \\
			 &  &  &  &  &  & ${\rm log}(SFR/[\rm M_{\odot}/yr])_{\rm phot}$ & 0.090 & 0.002 & 0.176 \\
\midrule
			 &  &  &  &  &  & $z_{\rm phot}$ & 0.016 & 0.006 & 0.004 \\
			COSMOS & + & $\beta$-$\tau$-r,+CEH & Sal+18 & 109572$\pm$5116 & 305267 & ${\rm log}(M_*)_{\rm phot}$ & 0.044 & -0.021 & 0.052 \\
			 &  &  &  &  &  & ${\rm log}(SFR/[\rm M_{\odot}/yr])_{\rm phot}$ & 0.102 & 0.063 & 0.321 \\
\midrule
			 &  &  &  &  &  & $z_{\rm phot}$ & 0.015 & 0.005 & 0.003 \\
			COSMOS & + & $\alpha$-$\beta$-$\tau$-r,+CEH & Sal+18 & 110909$\pm$5006 & 306387 & ${\rm log}(M_*)_{\rm phot}$ & 0.045 & -0.020 & 0.066 \\
			 &  &  &  &  &  & ${\rm log}(SFR/[\rm M_{\odot}/yr])_{\rm phot}$ & 0.097 & 0.055 & 0.304 \\
			\bottomrule
		\end{tabular}
	\end{center}
	\caption
    {
        As in Table \ref{tab:sum_CSST}, but for COSMOS-like survey.
    }
	\label{tab:sum_COSMOS}
\end{table*}

\begin{figure*}
    \centering
    \includegraphics{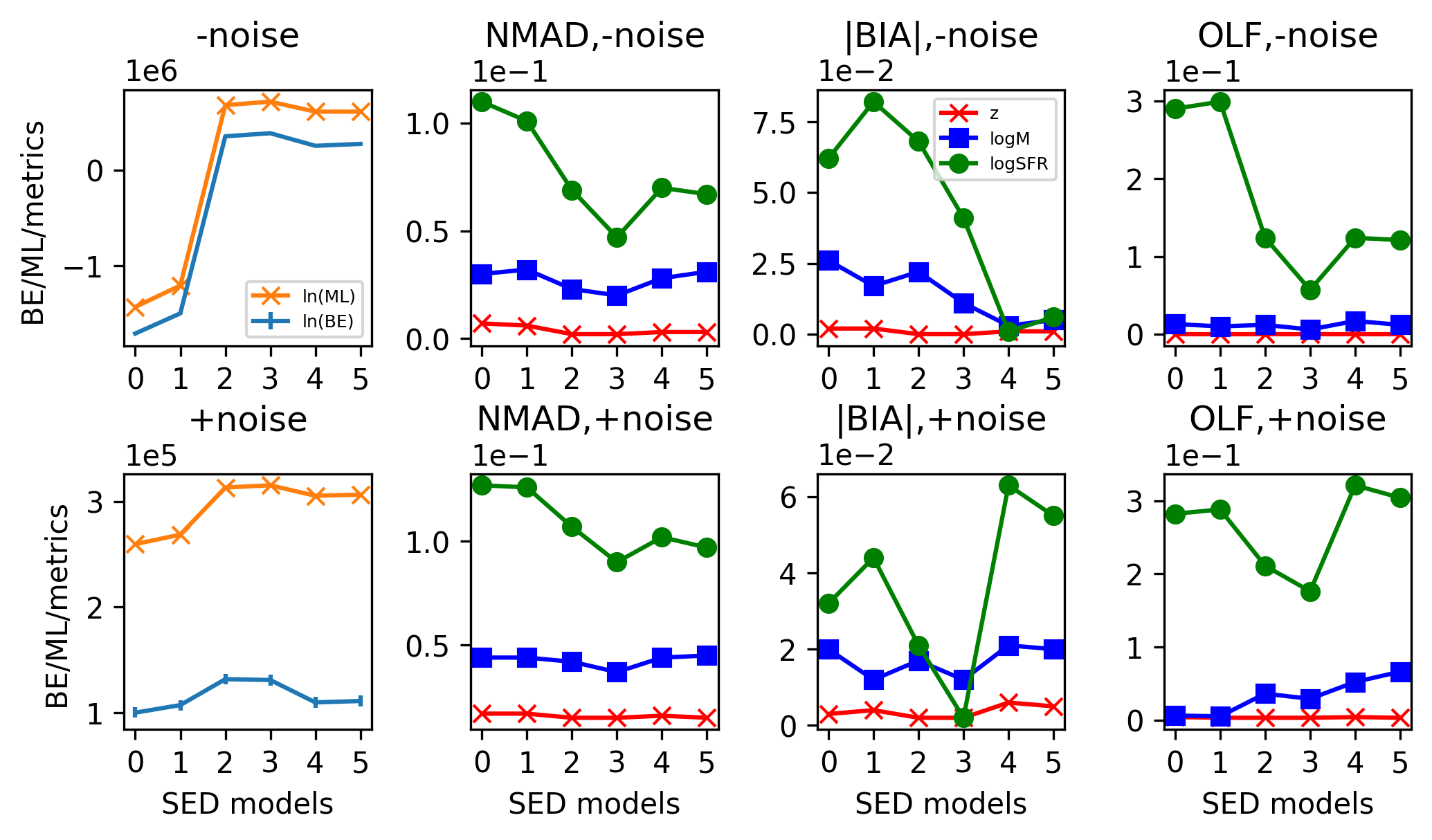}
    \caption
    {
        The Bayesian evidences (BE), maximum likelihood (ML) and metrics of photometric redshift (red star), stellar mass (blue square) and SFR (green circle) estimation from the Bayesian analysis of the hydrodynamical simulation-based mock galaxy sampe for COSMOS-like survey by employing six SED models (0:``SFH=$\tau$,-CEH,DAL=Cal+18'', 1:``SFH=$\tau$,+CEH,DAL=Cal+18'', 2:``SFH=$\tau$,+CEH,DAL=Sal+18'', 3:``SFH=$\beta$-$\tau$,+CEH,DAL=Sal+18'', 4:``SFH=$\beta$-$\tau$-r,+CEH,DAL=Sal+18'', 5:``SFH=$\alpha$-$\beta$-$\tau$-r,+CEH,DAL=Sal+18'') with increasing complexity in the forms of SFH and DAL , as well as for the cases with (bottom panels) and without noise (top panels).
        In the case without noise, the simplest model ``SFH=$\tau$,-CEH,DAL=Cal+18'' has the lowest Bayesian evidence of ${\rm ln}(BE)=-1710604\pm6811$.
        Meanwhile, the SED model ``SFH=$\beta$-$\tau$,+CEH,DAL=Sal+18'' which is neither the simplest nor the most complex models has the largest Bayesian evidence of ${\rm ln}(BE)=387009\pm7273$, and give the highest quality parameter estimates.
        As in the cases for CSST-like and CSST+Euclid-like surveys, the model selection with maximum likelihood (or equivalently minimum $\chi^2$) lead to similar results.
        In the case with noise, the simplest model ``SFH=$\tau$,-CEH,DAL=Cal+18'' still has the lowest Bayesian evidence of ${\rm ln}(BE)=99881\pm4881$.
        Meanwhile, the SED models ``SFH=$\tau$,+CEH,DAL=Sal+18'' and ``SFH=$\beta$-$\tau$,+CEH,DAL=Sal+18'' have the largest Bayesian evidences of ${\rm ln}(BE)=131578\pm5115$ and ${\rm ln}(BE)=130766\pm5064$ which are comparable within error bar.
        Unlike the cases for CSST-like and CSST+Euclid-like surveys, the model selection with maximum likelihood (or equivalently minimum $\chi^2$) lead to similar results.
        The two models with the largest Bayesian evidences (or maximum likelihoods) also give the highest quality parameter estimates.
        In general, in the cases with and without noise, the same and more clear results of model comparison are obtained.
        Meanwhile, in the more realistic case with noise, the more complicated SED models are more favored than in the cases for CSST-like and CSST+Euclid-like survey.
        All of these are the natural results of the much stronger discriminative power of the COSMOS-like survey than the CSST-like and CSST+Euclid-like surveys.
    }
    \label{fig:model_lnE_metrics_COSMOS}
\end{figure*}

In Table \ref{tab:sum_COSMOS}, we present a summary of the Bayesian evidences, maximum likelihoods and metrics of the quality of parameter estimation from the Bayesian analysis of the hydrodynamical simulation-based mock galaxy sampe for COSMOS-like survey by employing six different combinations of SFH and DAL with increasing complexity, as well as for the cases with and without noise.
The same results are also shown more clearly in Figure \ref{fig:model_lnE_metrics_COSMOS}.

\subsubsection{Model comparison} \label{ss:disc_cosmos_mod}
In the case without noise, as shown in the top left panel of Figure \ref{fig:model_lnE_metrics_COSMOS}, the simplest model ``SFH=$\tau$,-CEH,DAL=Cal+18'' has the lowest Bayesian evidence of ${\rm ln}(BE)=-1710604\pm6811$.
With the additional consideration of metallicity evolution, the Bayesian evidence of the model ``SFH=$\tau$,+CEH,DAL=Cal+18'' increases to ${\rm ln}(BE)=-1499618\pm6997$.
Then, with the adoption of the DAL of \cite{SalimS2018a}, the Bayesian evidence of the model ``SFH=$\tau$,+CEH,DAL=Sal+18'' increases significantly to ${\rm ln}(BE)=354581\pm7310$.
Apparently, the DAL of \cite{SalimS2018a} is also a much better choice than that of \cite{CalzettiD2000a} for the hydrodynamical simulation-based mock galaxy sampe in COSMOS-like survey.
Furthermore, by employing a more complicated $\beta$-$\tau$ form of SFH, the Bayesian evidence of the model ``SFH=$\beta$-$\tau$,+CEH,DAL=Sal+18'' increases further to ${\rm ln}(BE)=387009\pm7273$.
However, with a quenching (or rejuvenation) component added to the SFH, the Bayesian evidence of the model ``SFH=$\beta$-$\tau$-r,+CEH,DAL=Sal+18'' significantly decreases to ${\rm ln}(BE)=255069\pm7552$, which is similar to the case without noise.
Finally, by employing a even more flexible double power-law form of SFH, the Bayesian evidence of the model ``SFH=$\alpha$-$\beta$-$\tau$-r,+CEH,DAL=Sal+18'' increases to ${\rm ln}(BE)=275129\pm7218$.
As in the cases for CSST-like and CSST+Euclid-like surveys, the model selection with maximum likelihood (or equivalently minimum $\chi^2$) lead to similar results.

In the case with noise, as shown in the bottom left panel of Figure \ref{fig:model_lnE_metrics_COSMOS}, almost the same conclusions about SED model comparison are obtained, although the detailed values of Bayesian evidence are apparently different.
Unlike the cases for CSST-like and CSST+Euclid-like surveys, the model selection with Bayesian evidence and maximum likelihood (or equivalently minimum $\chi^2$) lead to similar results.
In general, for COSMOS-like survey, we can obtain more clear results of SED model comparison.
Meanwhile, in the more realistic case with noise, the more complicated SED models are more favored than in the cases for CSST-like and CSST+Euclid-like survey.
This is reasonable, since COSMOS-like survey has much stronger discriminative power than CSST-like and CSST+Euclid-like surveys.

\subsubsection{Parameter estimation} \label{ss:disc_cosmos_par}
In the right three panels of Figure \ref{fig:model_lnE_metrics_COSMOS}, we show the three metrics of the quality of photometric redshift, stellar mass and SFR estimation for different SED models, respectively.
In the cases with and without noise, the same SED model ``SFH=$\beta$-$\tau$,+CEH,DAL=Sal+18'' which is just the one with largest Bayesian evidence give the highest quality parameter estimates.
In the following, we discuss these results in more detail.

\begin{figure*}
    \gridline
    {
        \leftfig{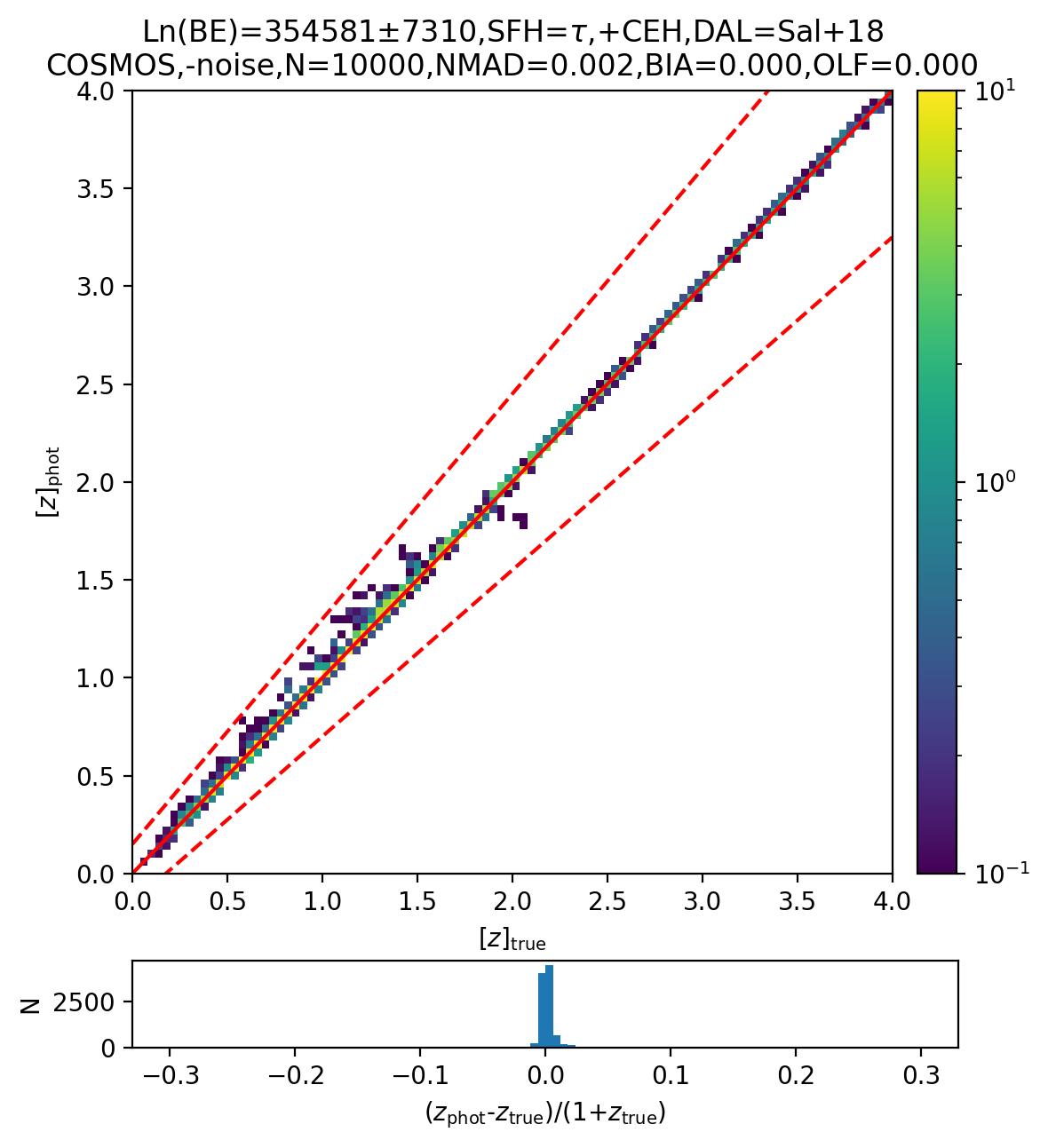}{0.48\textwidth}{(a)}
        \rightfig{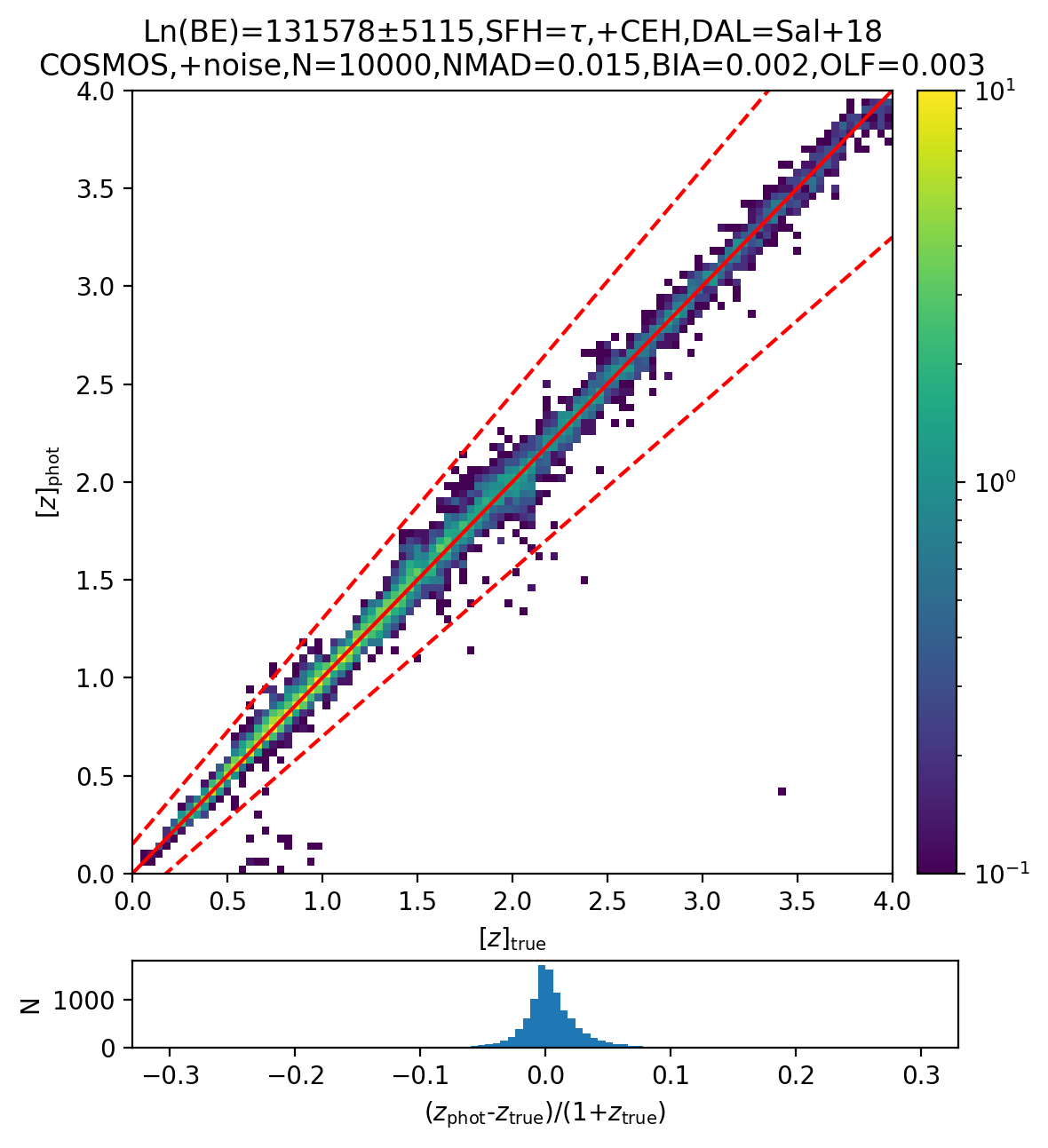}{0.48\textwidth}{(b)}
    }
    \gridline
    {
        \leftfig{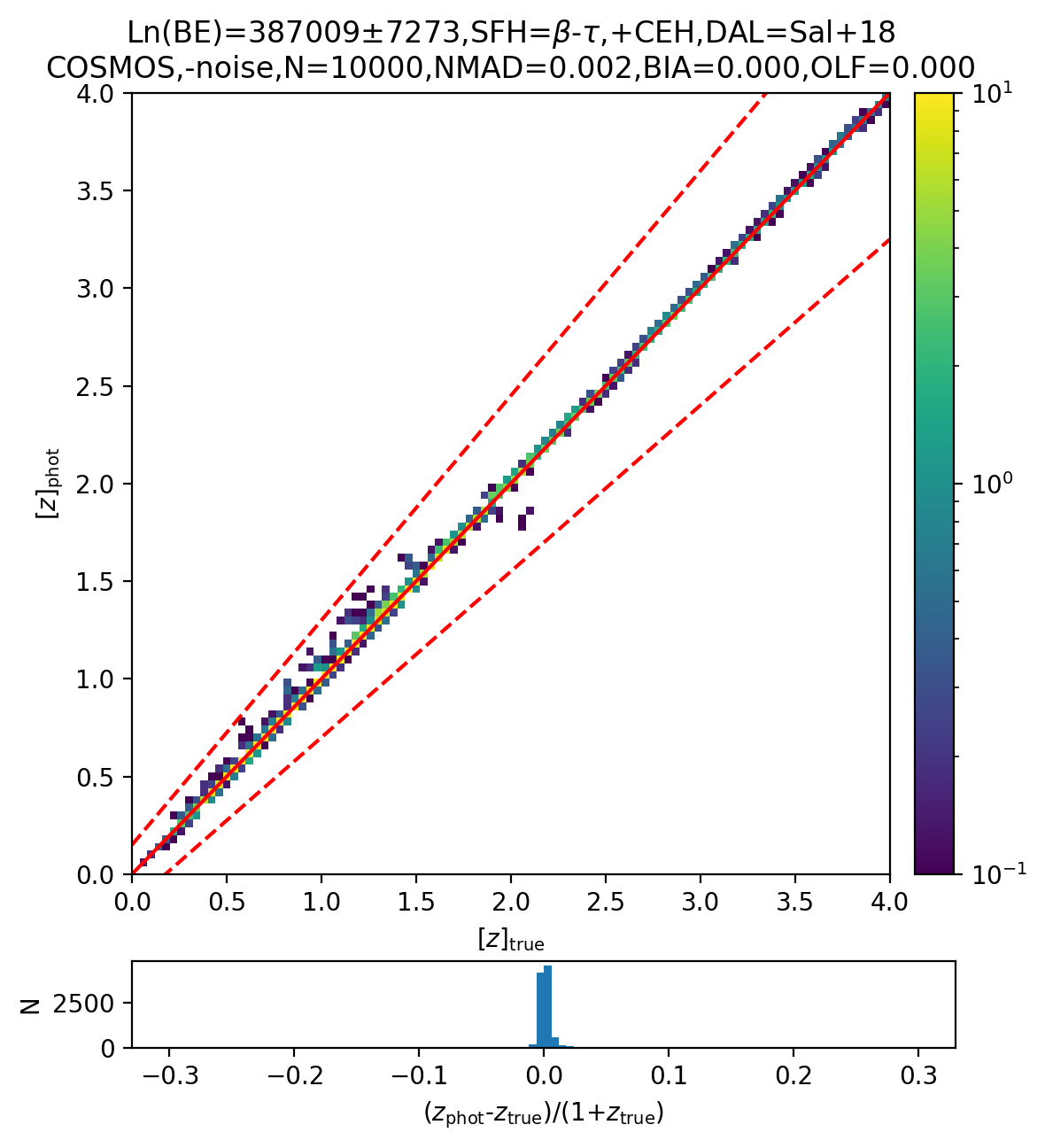}{0.48\textwidth}{(c)}
        \rightfig{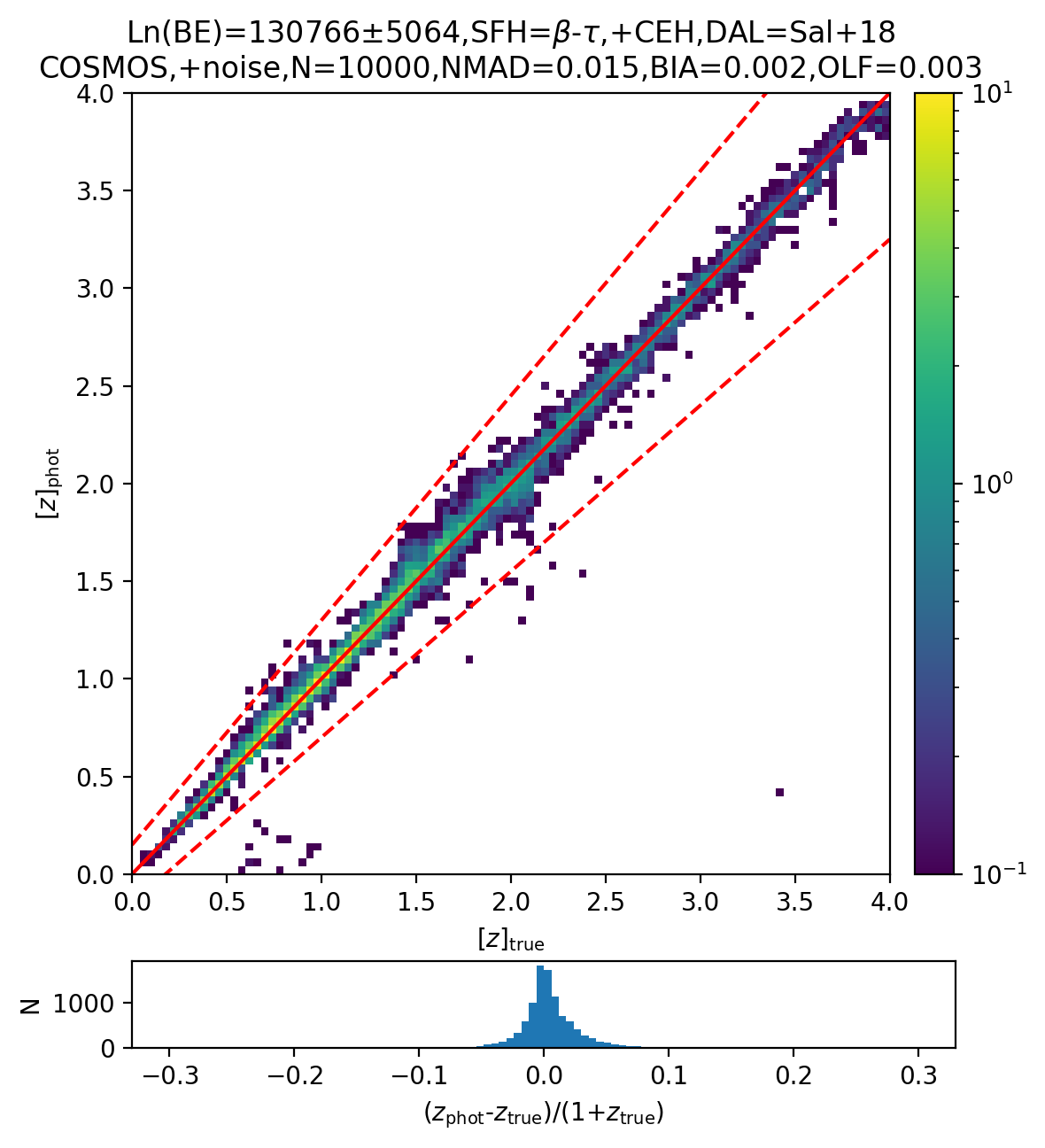}{0.48\textwidth}{(d)}
    }
    \caption
    {
        The results of photometric redshift estimation from the Bayesian analysis of the hydrodynamical simulation-based mock data for COSMOS-like survey by employing the two SED models with the largest Bayesian evidence.
        By comparing with the results for CSST-like survey in Figure \ref{fig:perf_phot_csst_z2} and that for for CSST+Euclid-like survey in Figure \ref{fig:perf_phot_csst_euclid_z2}, it is clear that the quality of photometric redshift estimation has been significantly increased in both the cases with and without noise.
        Meanwhile, in both cases, the best two SED models given identical quality of photometric redshift estimation.
        In the case without noise, the bias and OLF of photometric redshift estimation are almost zero while the $\sigma_{\rm NMAD}$ is only 0.002.
        Since the errors from parameter degeneracy and SED model error should be largely reduced, this suggest that the contribution of errors from the stochastic nature of MultiNest sampling algorithm and other potential errors in the BayeSED code should be less than 0.002.
        In the more realistic case with noise, the outliers caused by the mis-identification of Lyman and Balmer break features have been largely resolved.
    }
    \label{fig:perf_phot_cosmos_z2}
\end{figure*}
The detailed results of photometric redshift estimation obtained by employing the two models with the largest Bayesian evidences are shown in Figure \ref{fig:perf_phot_cosmos_z2}.
By comparing with the results for CSST-like survey in Figure \ref{fig:perf_phot_csst_z2} and that for for CSST+Euclid-like survey in Figure \ref{fig:perf_phot_csst_euclid_z2}, it is clear that the quality of photometric redshift estimation has been significantly increased in both the cases with and without noise.
Meanwhile, in both cases, the best two SED models given identical quality of photometric redshift estimation.
In the case without noise, the bias and OLF of photometric redshift estimation are almost zero while the $\sigma_{\rm NMAD}$ is only 0.002.
In this case, the errors from parameter degeneracy and SED model error should be largely reduced.
This suggest that the contribution of errors from the stochastic nature of MultiNest sampling algorithm and other potential errors in the BayeSED code should be less than 0.002.
In the more realistic case with noise, the outliers caused by the mis-identification of Lyman and Balmer break features have been largely resolved.
Finally, as in the cases for CSST-like and CSST+Euclid-like surveys, the more complicated SED models are not very helpful for improving the quality of photometric redshift estimation.

\begin{figure*}
    \gridline
    {
        \leftfig{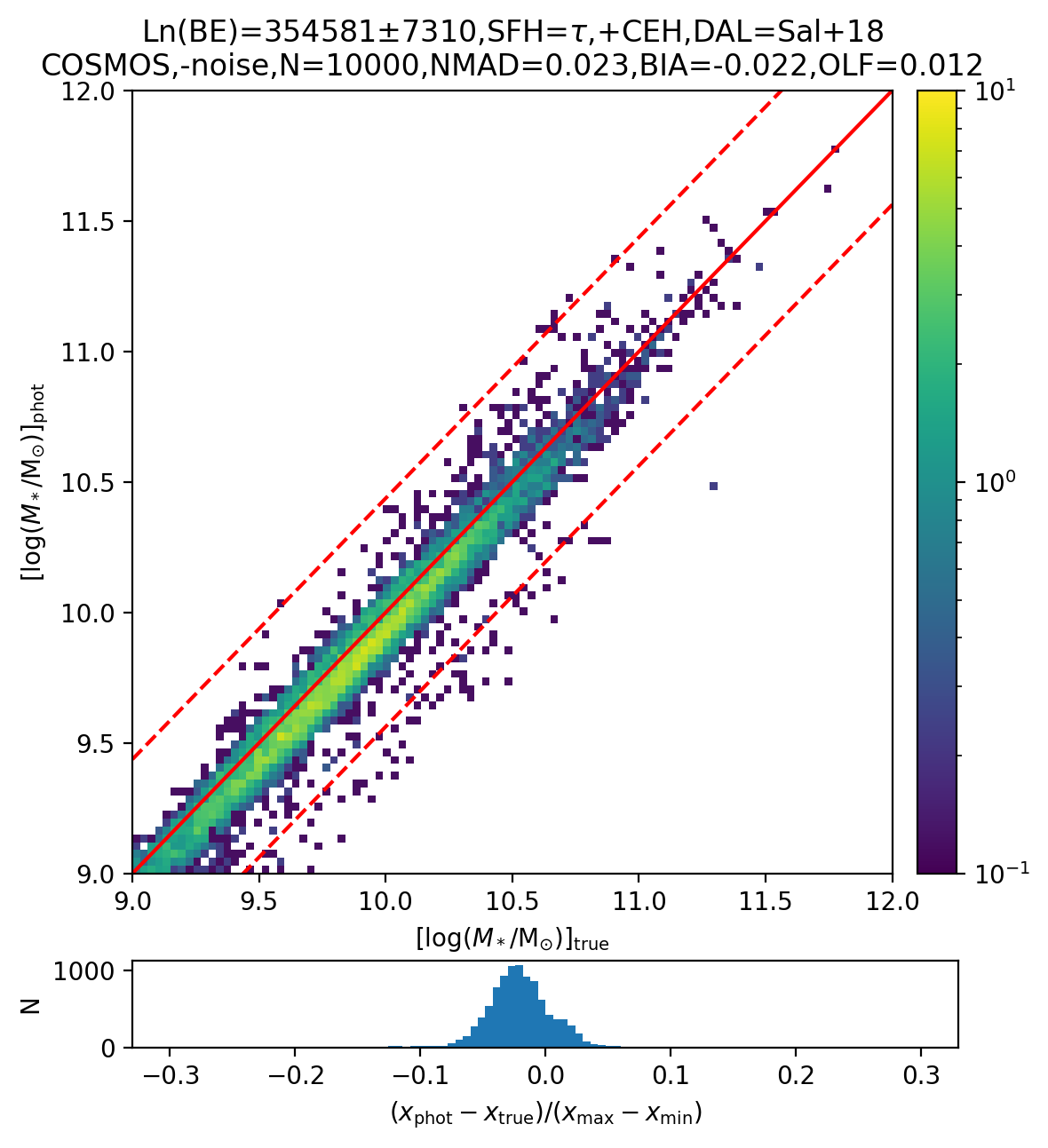}{0.48\textwidth}{(a)}
        \rightfig{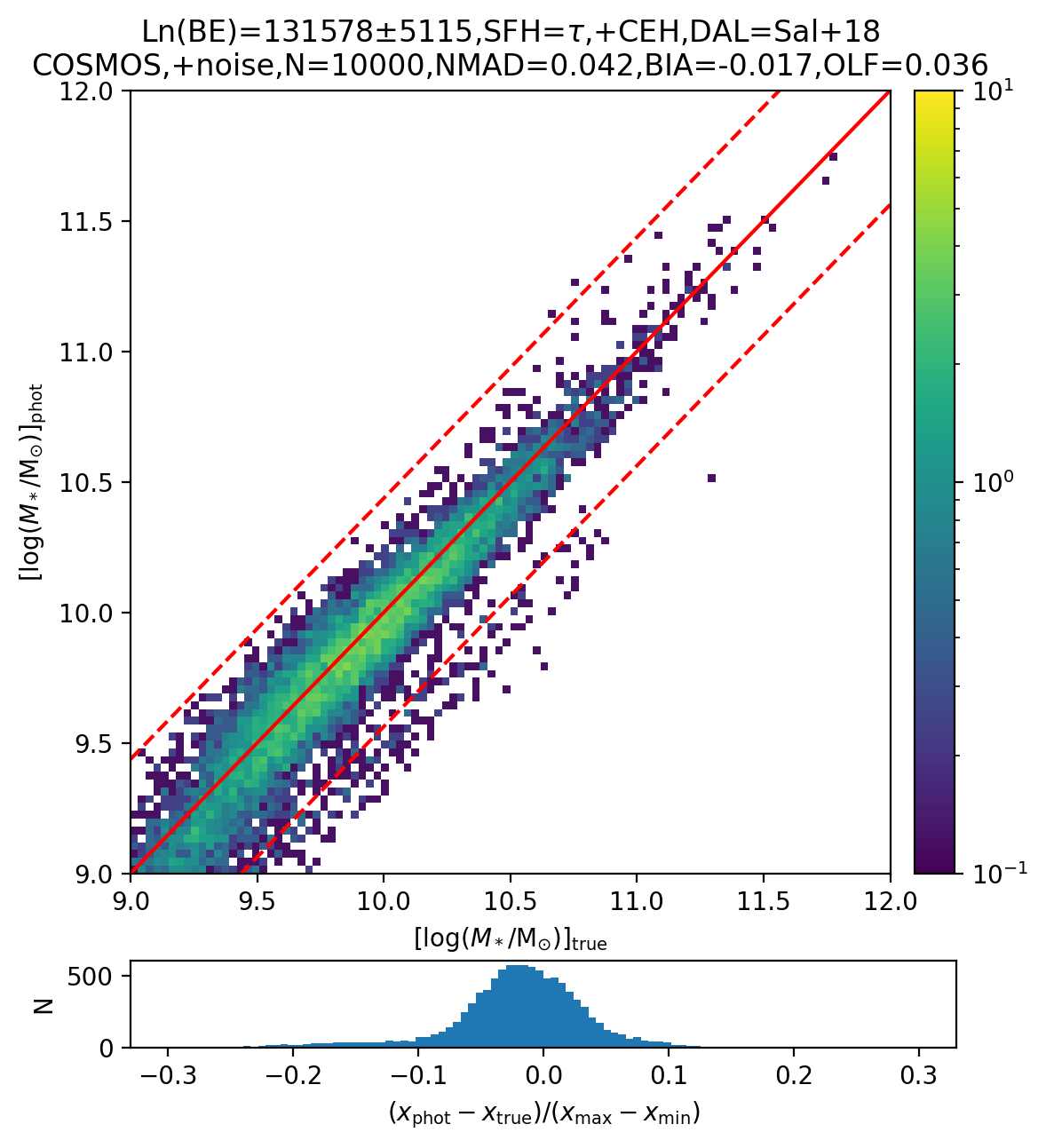}{0.48\textwidth}{(b)}
    }
    \gridline
    {
        \leftfig{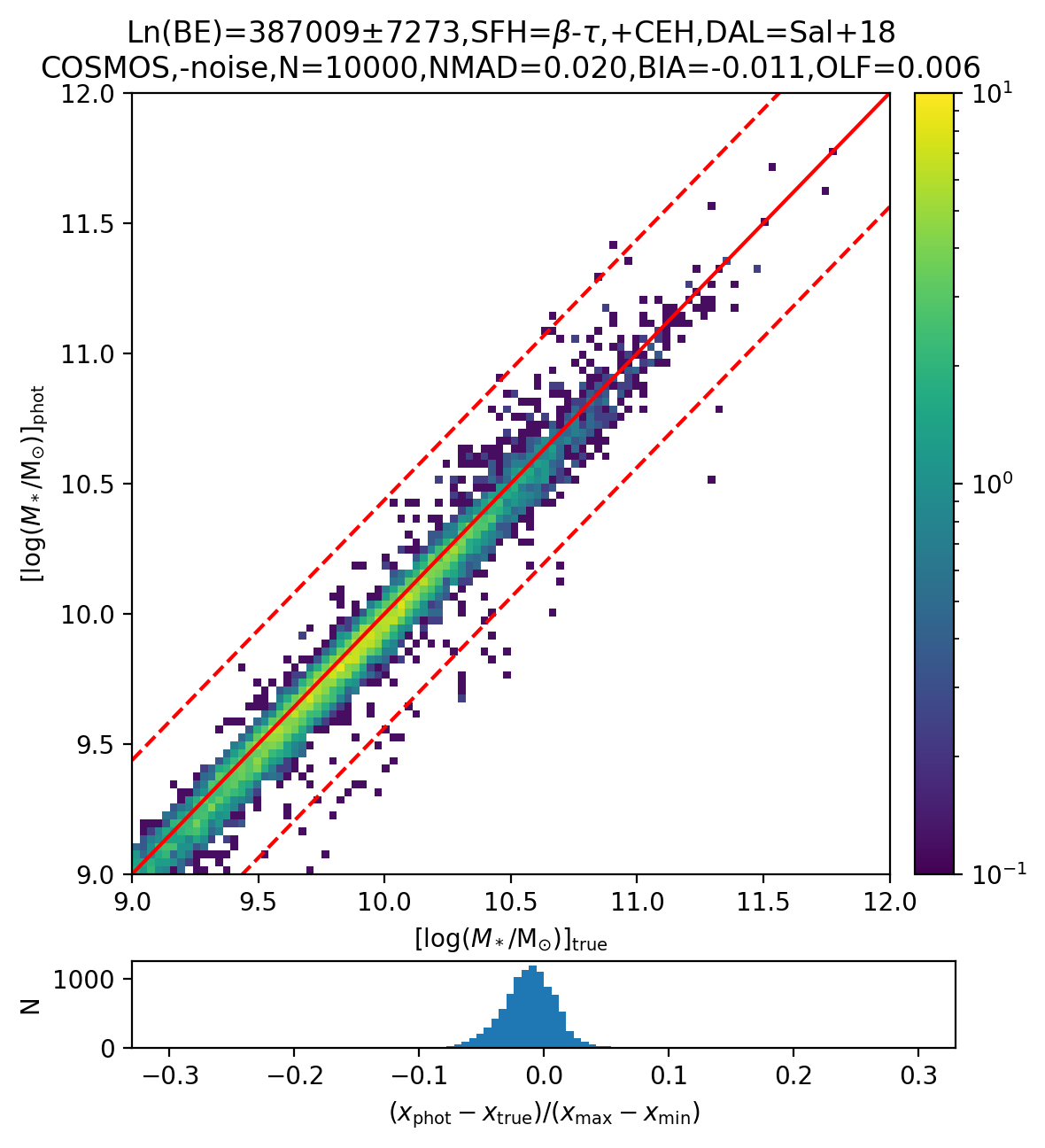}{0.48\textwidth}{(c)}
        \rightfig{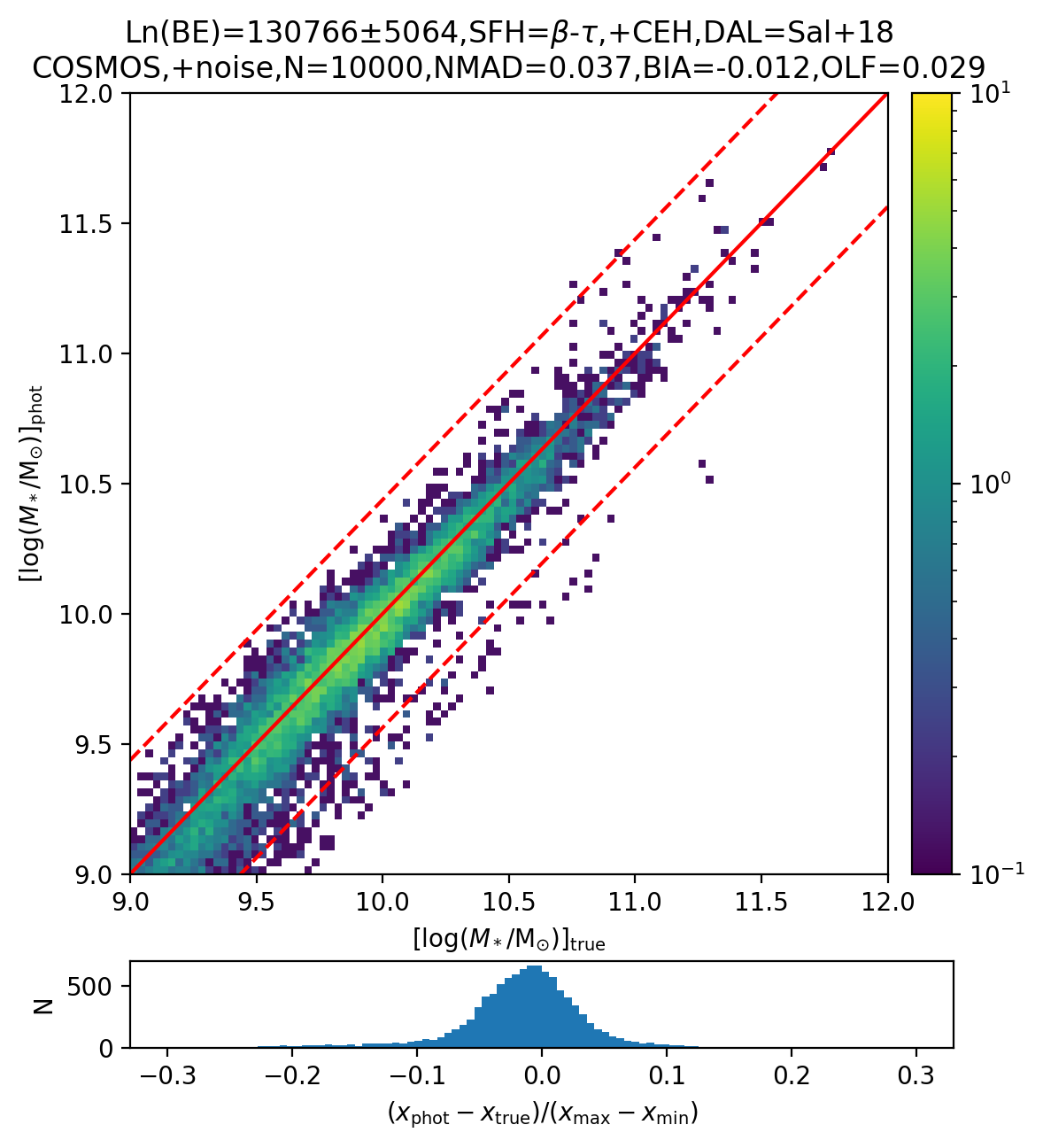}{0.48\textwidth}{(d)}
    }
    \caption
    {
        As in Figure \ref{fig:perf_phot_cosmos_z2}, but for the stellar mass.
        By comparing with the results for CSST+Euclid-like survey in Figure \ref{fig:perf_phot_csst_euclid_mass2}, it is clear that the quality of photometric stellar mass estimation has been improved further in both the cases with and without noise.
        Besides, the more complicated $\beta$-$\tau$ form of SFH makes the quality of photometric stellar mass estimation a little better.
    }
    \label{fig:perf_phot_cosmos_mass2}
\end{figure*}
The detailed results of photometric stellar mass estimation obtained by employing the two models with the largest Bayesian evidences are shown in Figure \ref{fig:perf_phot_cosmos_mass2}.
By comparing with the results for CSST+Euclid-like survey in Figure \ref{fig:perf_phot_csst_euclid_mass2}, it is clear that the quality of photometric stellar mass estimation has been improved further in both the cases with and without noise.
Besides, the more complicated $\beta$-$\tau$ form of SFH makes the quality of photometric stellar mass estimation a little better.
However, as shown in Table \ref{tab:sum_COSMOS} and Figure \ref{fig:model_lnE_metrics_CSST_Euclid}, the even more complicated forms of SFH make the quality of photometric stellar mass estimation worse.

\begin{figure*}
    \gridline
    {
        \leftfig{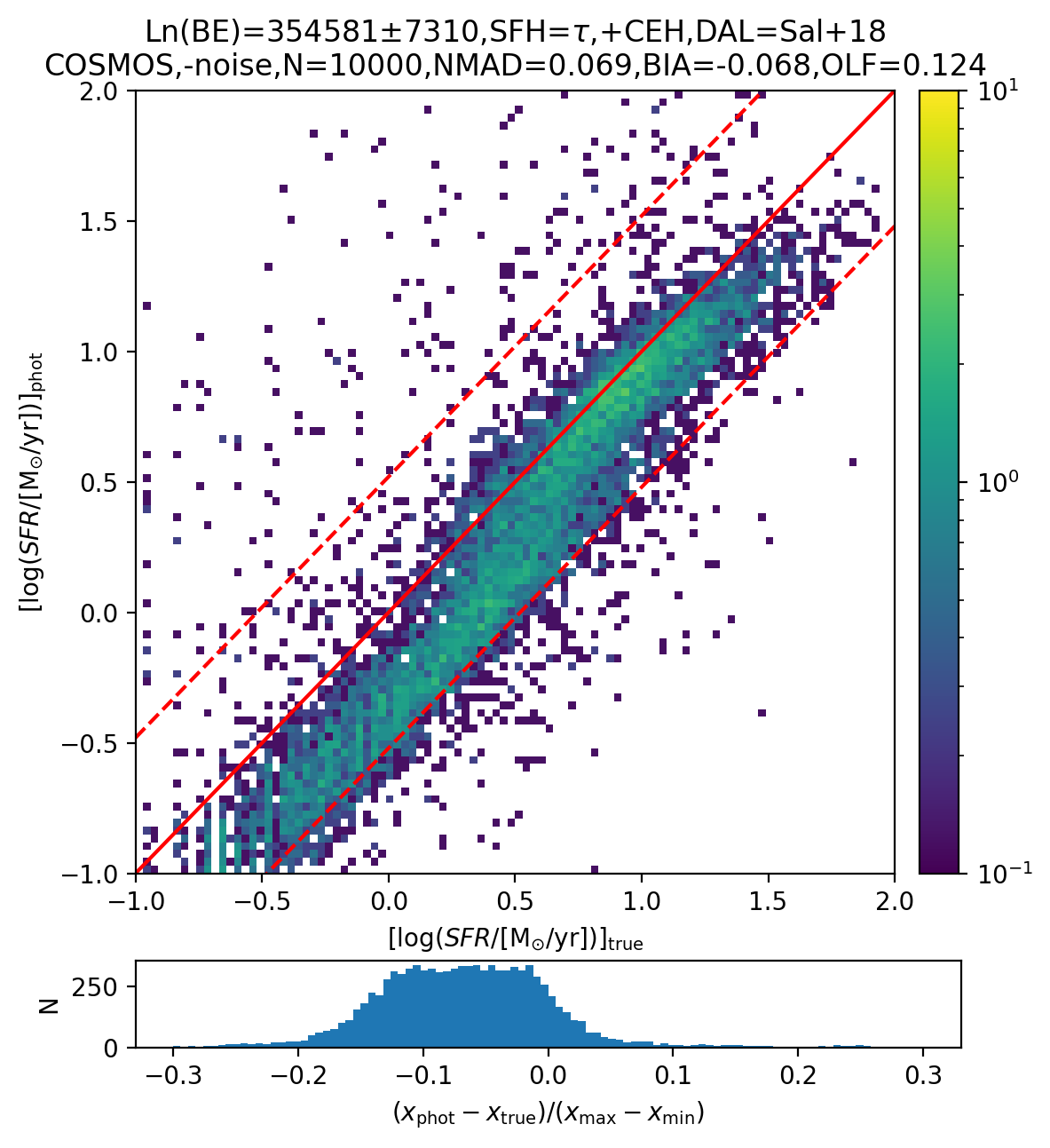}{0.48\textwidth}{(a)}
        \rightfig{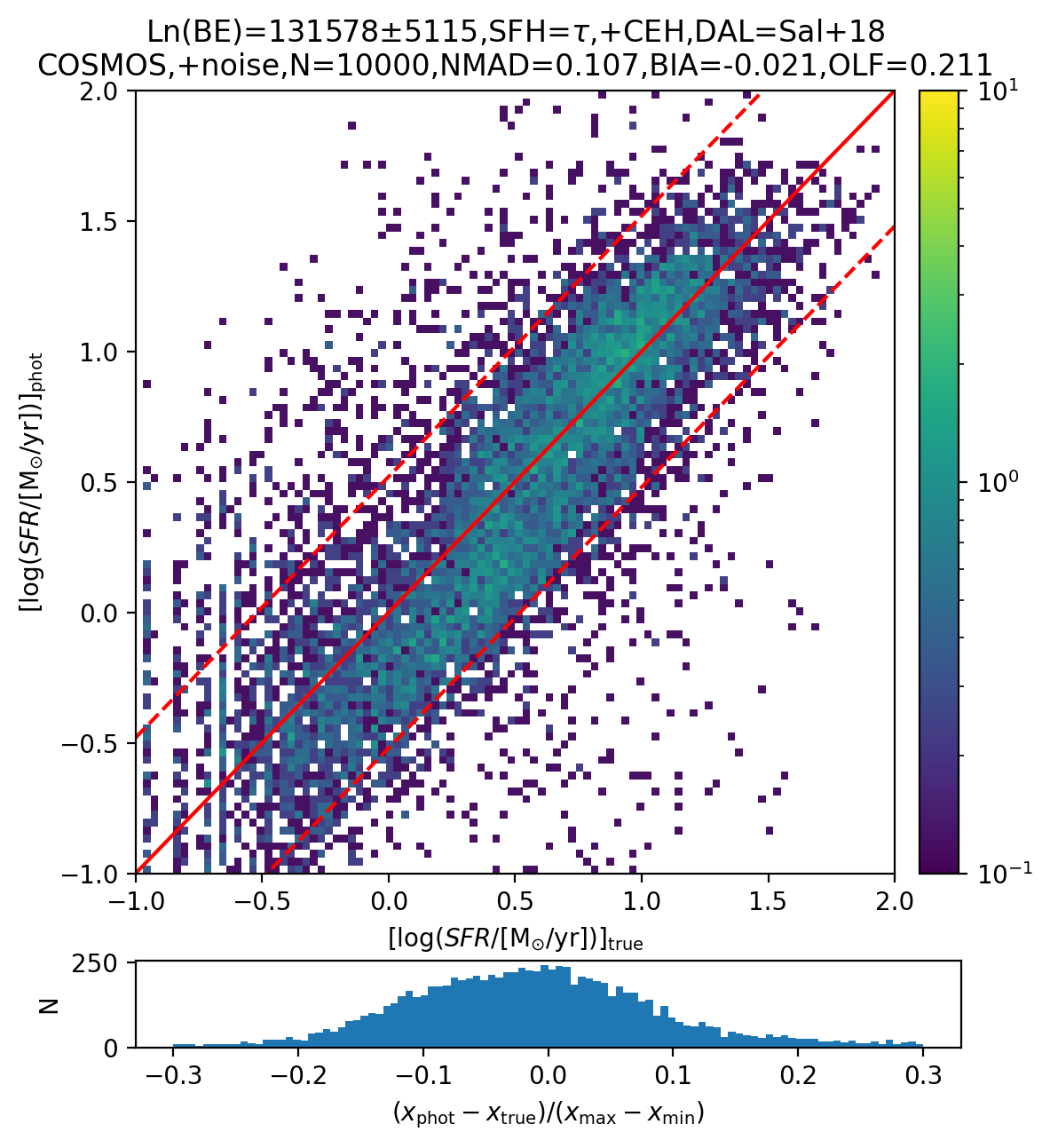}{0.48\textwidth}{(b)}
    }
    \gridline
    {
        \leftfig{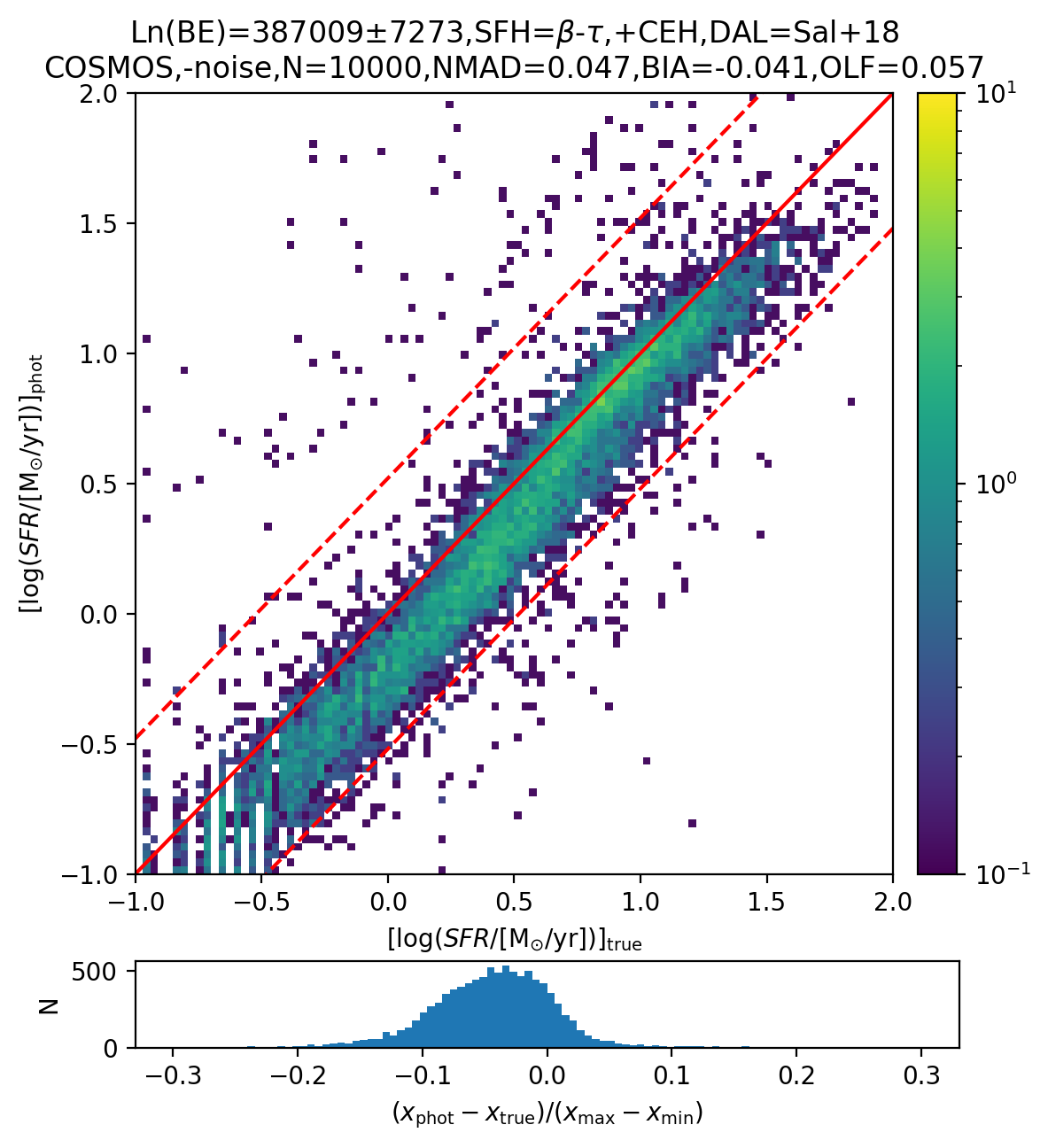}{0.48\textwidth}{(c)}
        \rightfig{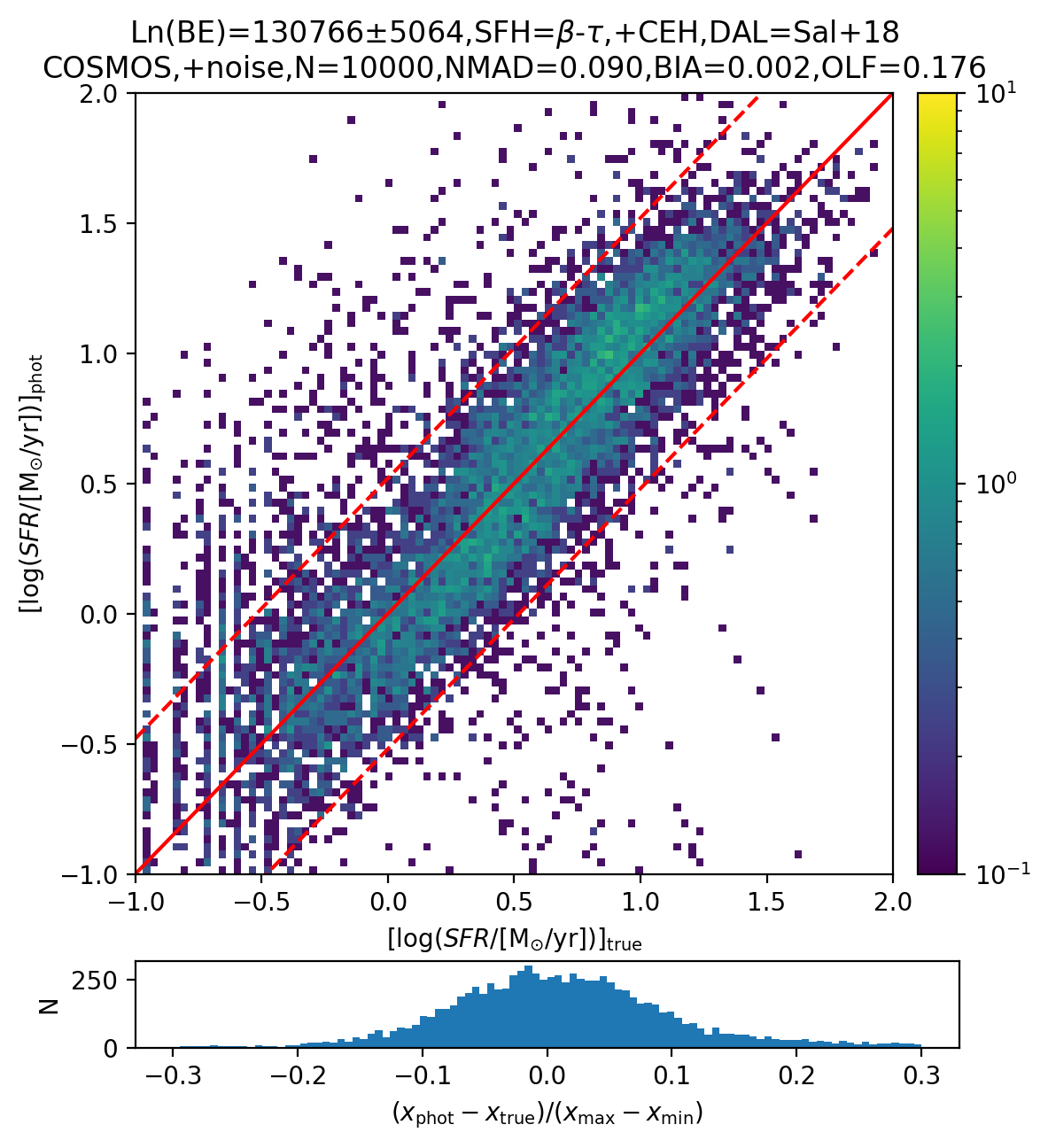}{0.48\textwidth}{(d)}
    }
    \caption
    {
        As in Figure \ref{fig:perf_phot_cosmos_z2}, but for the SFR.
        By comparing with the results for CSST+Euclid-like survey in Figure \ref{fig:perf_phot_csst_euclid_mass2}, it is clear that the quality of photometric SFR estimation has been significantly improved in both the cases with and without noise.
        Besides, the more complicated $\beta$-$\tau$ form of SFH makes the quality of photometric SFR estimation much better.
    }
    \label{fig:perf_phot_cosmos_sfr2}
\end{figure*}
The detailed results of photometric SFR estimation obtained by employing the two models with the largest Bayesian evidences are shown in Figure \ref{fig:perf_phot_cosmos_sfr2}.
By comparing with the results for CSST+Euclid-like survey in Figure \ref{fig:perf_phot_csst_euclid_mass2}, it is clear that the quality of photometric SFR estimation has been significantly improved in both the cases with and without noise.
Besides, the more complicated $\beta$-$\tau$ form of SFH makes the quality of photometric SFR estimation much better.
However, as shown in \ref{fig:model_lnE_metrics_COSMOS}, with a quenching (or rejuvenation) component added to the SFH, the quality of photometric SFR estimation becomes obviously worse.
Finally, by employing a even more flexible double power-law form of SFH, the quality of photometric SFR estimation becomes a little better.

\section{Summary and conclusion} \label{sec:summary}
In this work, based on the Bayesian SED synthesis and analysis techniques employed in the BayeSED-V3  code, we present a comprehensive and systematic test of its performance for simultaneous photometric redshift and stellar population parameter estimation of galaxies when combined with six SED models with increasing complexity in the form of SFH and DAL.
The main purpose is to make a systematic analysis of various factors affecting the simultaneous photometric redshift and stellar population parameter estimation of galaxies in the context of Bayesian SED fitting, so as to provide clues for further improvement.

To separate the different factors which could contribute to the errors of photometric redshift and stellar population parameter estimation of galaxies, the empirical statistics-based and hydrodynamical simulation-based approaches have been employed to generate mock photometric sample of galaxies with or without noise for CSST-like, COSMOS-like and CSST+Euclid-like surveys, respectively.
We compare the difference in performance of photometric parameter estimation with different run parameters of Bayesian analysis algorithm, different assumptions about the SFH and DAL of galaxies, and different observational datasets. Our main findings are as follows.

For the performance tests using empirical statistics-based mock galaxy sample with idealized SED modeling:
\begin{enumerate}
    \item The performance of photometric redshift and stellar population parameter estimation, in terms of speed and quality, is sensitive to the runtime parameters (the target sampling efficiency $efr$ and the number of live points $nlive$) of MultiNest algorithm.
    \item A good balance among the speed, quality of parameter estimation, and accuracy of model comparison can be achieved when adopting the MultiNest runtime parameters $efr$ equals to $0.1$ and $nlive$ equals to $50$.
    \item By employing the optimized runtime parameters of MultiNest and simplest SED modeling, a speed of $\sim2{\rm s/obj/cpu}$ ($\sim10{\rm s/obj/cpu}$) can be achieved for a detailed Bayesian analysis of photometries from CSST-like survey, which is sufficient for the analysis of massive photometric data.
    Meanwhile, a quality of photometric redshift estimation with $\sigma_{\rm NMAD}=0.056$, $BIA=-0.0025$, $OLF=0.215$, a quality of photometric stellar mass estimation with $\sigma_{\rm NMAD}=0.113$, $BIA=-0.025$, $OLF=0.285$, and a quality of photometric SFR estimation with $\sigma_{\rm NMAD}=0.08$, $BIA=-0.01$, $OLF=0.255$ can be achieved.
    \item With the optimized runtime parameters of MultiNest, the value of Bayesian evidence which is crucial for Bayesian model comparison can also be well estimated, although the error of Bayesian evidence tends to be overestimated, which may lead to a more conservative conclusion about model comparison.
    \item The random observational errors in photometries are more important sources of errors than the parameter degeneracies and Bayesian analysis method and tool.
    \item More complicated SED models apparently require longer running time.
    They also tend to overfit noisy photometries and lead to worse quality of photometric redshift, stellar mass and SFR estimation, which is likely due to more free parameters and more severe parameter degeneracies.
    \item The value of Bayesian evidence clearly decreases with the increasing of the complexity of the SED model in both of the cases with and without noise.
\end{enumerate}

For the performance tests using hydrodynamical simulation-based mock galaxy sample without idealized SED modeling:
\begin{enumerate}
    \item The commonly used simple assumptions about the SFH and DAL of galaxies have severe impact on the quality of photometric parameter estimation of galaxies, especially for CSST-like survey with only photometries from seven broad-bands.
    \item The performance of both Bayesian parameter estimation and model comparison highly depends on the discriminative power of the observational photometries. With more informative photometries, more clear results about SED model comparison and higher quality of photometric parameter estimation can be obtained.
    \item While the SED model comparison with Bayesian evidence may favor SED models with very different complexities when using photometries from different surveys, the maximum likelihood (or equivalently  minimum $\chi^2$) tend to favor more complex models.
    For photometries with strong enough discriminative power, the two methods lead to more consistent results.
    However, for photometries without strong enough discriminative power, the two methods may lead to contradictory results.
    In both cases, the results of model selection with Bayesian evidence are more consistent with the measurements of the quality of parameter estimation.
    \item In both of the cases with and without noise, the additional consideration of metallicity evolution helps to improve the quality of photometric redshift and stellar parameter estimation of galaxies, and increases the Bayesian evidence of corresponding SED model.
    \item In the case without noise, the DAL of \cite{SalimS2018a} is a much better choice than that of \cite{CalzettiD2000a} for the hydrodynamical simulation-based mock galaxy sampe in CSST-like, CSST+Euclid-like and COSMOS-like surveys.
    However, in the more realistic case with noise, it is only more favored in the COSMOS-like survey with Bayesian evidence-based model selection.
    With maximum likelihood (or equivalently minimum $\chi^2$)-based model selection, it could be more favored, but lead to worse parameter estimation.
    \item In the case without noise, the more flexible forms of SFH lead to better quality of parameter estimation and increase the Bayesian evidence of corresponding SED model.
    However, in the more realistic case with noise, they are only more favored in the COSMOS-like survey.
    \item With a quenching (or rejuvenation) component added to the SFH, the quality of parameter estimation and the Bayesian evidence of corresponding SED model decrease in all cases. Although the rejuvenation or rapid quenching events may happen in some galaxies, this additional component of SFH is not very effective for most of the galaxies in the hydrodynamical simulation-based mock galaxy sample.
    \item The quality of parameter estimation is closely related to the level of Bayesian evidence such that the SED model with largest Bayesian evidence tends to give the best quality of parameter estimation, which is more clear for photometries with larger discriminative power.
    By using photometries without strong enough discriminative power, the quality of parameter estimation, especially that of stellar mass and SFR estimation, tend to decrease with the increasing of SED model complexity
    \item Since the direct measurements of the quality of parameter estimation as indicated by NMAD, BIA and OLF are usually unavailable, the Bayesian model comparison with Bayesian evidence can be used to find the best SED model which is not only the most efficient but also give the best parameter estimation.
    \item For photometric redshift, stellar mass and SFR, their accurate estimation becomes increasingly difficult, while the latter two are also more sensitive to the selection of SED models. 
    \item For the photometric redshift estimation of galaxies in CSST-like survey, the observational noise is the more important source of error than the imperfect SED modeling.
    However, for the photometric stellar mass and SFR estimation of galaxies, the opposite is true.
    \item The combination of photometries from CSST-like and Euclid-like surveys is helpful for improving the quality of photometric redshift estimation and crucial for the more accurate stellar mass estimation, but not very useful for SFR estimation.
    \item With photometries in 26 bands from  COSMOS-like surveys, by employ the same SED model, BayeSED-V3 can achieve similar quality of photometric redshift, stellar mass and SFR estimation to previous works.
    Besides, with photometries in 26 bands from COSMOS-like surveys, more complicated SED models tend to be more favored, which is very different from the two cases with only photometries from CSST-like (7 bands) or CSST+Euclid-like (10 bands) surveys.
\end{enumerate}

We conclude that the latest version of BayeSED is capable of achieving a good balance among speed, the quality of simultaneous photometric redshift and stellar population parameter estimation of galaxies and the reliable SED model comparison.
This makes it suitable for the analysis of existing and forthcoming massive photometric data of galaxies in CSST wide-field multiband imaging survey and others.

Generally, the current main bottleneck that limits the performance of the Bayesian approach for the simultaneous photometric redshift and stellar population parameter estimation of galaxies is the reliability of the SED synthesis (or modeling) procedure.
Assuming a more flexible model is not a complete solution. 
We need a SED model that is not only more flexible but also more precisely accurate.
It can be achieved by gradually adding more informative priors and physical constraints to the SED synthesis (modeling) procedure of galaxies, which is the subject of future works.
The Bayesian model selection method with Bayesian evidence, a quantified Occam's razor, is very helpful to identify the best SED model which is not only the most efficient but also give the best parameter estimation.

The results about simultaneous photometric redshift and stellar population estimation presented in this work are not yet optimal, especially those about CSST.
The contributions of nebular lines and continuum emission to the SED, which may help break some parameter degeneracies, are still mising in this work.
It is also worth to mention that the results of Bayesian SED model comparison and the metrics (OLF, BIA or NMAD) of parameter estimation highly depend on the selected samples.
In this paper, we have chosen a relative broader sample to test the overall performance in the CSST wide-field imaging survey.
For a differently selected sample which is designed for answering a more specific scientific question, the results could be different.

Finally, in addition to multi-bands photometries, we may need more informations from other forms of data, such as slitless spectroscopy and morphology parameters from the imaging to break the severe parameter degeneracies.
More advanced methods may be able to take advantage of all information to give better redshift and/or stellar population parameter estimation of galaxies.
These will be the subjects of future works as well.

\acknowledgments 
{The authors gratefully acknowledge the ``PHOENIX Supercomputing Platform'' jointly operated by the Binary Population Synthesis Group and the Stellar Astrophysics Group at Yunnan Observatories, Chinese Academy of Sciences.
We warmly thank the Horizon-AGN team, especially C. Laigle, for making their photometric catalogs and spectra data publicly available.
We thank Fengshan Liu, Ye Cao, and Xinwen Shu for helpful discussion about SNR and magnitude limit.

We acknowledge support from the National Key R\&D Program of China (Nos. 2021YFA1600401 and 2021YFA1600400), the National Science Foundation of China (grant nos. 11773063, 12173037, 12233008, 12233005, 12073078, 12288102) the China Manned Space Project (grant nos. CMS-CSST-2021-A02, CMS-CSST-2021-A04, CMS-CSST-2021-A06, and CMS-CSST-2021-A07), the ``Light of West China'' Program of Chinese Academy of Sciences, the Yunnan Ten Thousand Talents Plan Young \& Elite Talents Project, the International Centre of Supernovae, Yunnan Key Laboratory (No. 202302AN360001).
LF also gratefully acknowledges the support of the CAS Project for Young Scientists in Basic Research (No. YSBR-092), the Fundamental Research Funds for the Central Universities (WK3440000006), and the Cyrus Chun Ying Tang Foundations.

This work made use of the following softwares: ASTROPY\citep{Astropy-Collaboration2013a,Astropy-Collaboration2018m}, MATPLOTLIB\citep{HunterJ2007x}, NUMPY\citep{WaltS2011s}, and H5PY\citep{ColletteA2023u}.
}

\bibliography{hanyk.bib}{}

\begin{thebibliography}{}
\expandafter\ifx\csname natexlab\endcsname\relax\def\natexlab#1{#1}\fi
\providecommand{\url}[1]{\href{#1}{#1}}
\providecommand{\dodoi}[1]{doi:~\href{http://doi.org/#1}{\nolinkurl{#1}}}
\providecommand{\doeprint}[1]{\href{http://ascl.net/#1}{\nolinkurl{http://ascl.net/#1}}}
\providecommand{\doarXiv}[1]{\href{https://arxiv.org/abs/#1}{\nolinkurl{https://arxiv.org/abs/#1}}}

\bibitem[{{Abdurro'uf} {et~al.}(2021){Abdurro'uf}, {Lin}, {Wu}, \&
  {Akiyama}}]{Abdurrouf2021j}
{Abdurro'uf}, {Lin}, Y.-T., {Wu}, P.-F., \& {Akiyama}, M. 2021, \apjs, 254, 15,
  \dodoi{10.3847/1538-4365/abebe2}

\bibitem[{{Acquaviva} {et~al.}(2011){Acquaviva}, {Gawiser}, \&
  {Guaita}}]{AcquavivaV2011a}
{Acquaviva}, V., {Gawiser}, E., \& {Guaita}, L. 2011, \apj, 737, 47,
  \dodoi{10.1088/0004-637X/737/2/47}

\bibitem[{{Acquaviva} {et~al.}(2015){Acquaviva}, {Raichoor}, \&
  {Gawiser}}]{AcquavivaV2015a}
{Acquaviva}, V., {Raichoor}, A., \& {Gawiser}, E. 2015, \apj, 804, 8,
  \dodoi{10.1088/0004-637X/804/1/8}

\bibitem[{{Alsing} {et~al.}(2023){Alsing}, {Peiris}, {Mortlock}, {Leja}, \&
  {Leistedt}}]{AlsingJ2023g}
{Alsing}, J., {Peiris}, H., {Mortlock}, D., {Leja}, J., \& {Leistedt}, B. 2023,
  \apjs, 264, 29, \dodoi{10.3847/1538-4365/ac9583}

\bibitem[{{Alsing} {et~al.}(2020){Alsing}, {Peiris}, {Leja}, {Hahn}, {Tojeiro},
  {Mortlock}, {Leistedt}, {Johnson}, \& {Conroy}}]{AlsingJ2020a}
{Alsing}, J., {Peiris}, H., {Leja}, J., {et~al.} 2020, \apjs, 249, 5,
  \dodoi{10.3847/1538-4365/ab917f}

\bibitem[{{Antonucci}(1993)}]{AntonucciR1993a}
{Antonucci}, R. 1993, \araa, 31, 473,
  \dodoi{10.1146/annurev.aa.31.090193.002353}

\bibitem[{{Antonucci}(2012)}]{AntonucciR2012l}
---. 2012, Astronomical and Astrophysical Transactions, 27, 557.
\newblock \doarXiv{1210.2716}

\bibitem[{{Arnouts} {et~al.}(2013){Arnouts}, {Le Floc'h}, {Chevallard},
  {Johnson}, {Ilbert}, {Treyer}, {Aussel}, {Capak}, {Sanders}, {Scoville},
  {McCracken}, {Milliard}, {Pozzetti}, \& {Salvato}}]{ArnoutsS2013a}
{Arnouts}, S., {Le Floc'h}, E., {Chevallard}, J., {et~al.} 2013, \aap, 558,
  A67, \dodoi{10.1051/0004-6361/201321768}

\bibitem[{Ashton {et~al.}(2022)Ashton, Bernstein, Buchner, Chen, Cs{\'a}nyi,
  Fowlie, Feroz, Griffiths, Handley, Habeck, Higson, Hobson, Lasenby,
  Parkinson, P{\'a}rtay, Pitkin, Schneider, Speagle, South, Veitch, Wacker,
  Wales, \& Yallup}]{AshtonG2022w}
Ashton, G., Bernstein, N., Buchner, J., {et~al.} 2022, Nature Reviews Methods
  Primers, 2, 39, \dodoi{10.1038/s43586-022-00121-x}

\bibitem[{{Astropy Collaboration} {et~al.}(2013){Astropy Collaboration},
  {Robitaille}, {Tollerud}, {Greenfield}, {Droettboom}, {Bray}, {Aldcroft},
  {Davis}, {Ginsburg}, {Price-Whelan}, {Kerzendorf}, {Conley}, {Crighton},
  {Barbary}, {Muna}, {Ferguson}, {Grollier}, {Parikh}, {Nair}, {Unther},
  {Deil}, {Woillez}, {Conseil}, {Kramer}, {Turner}, {Singer}, {Fox}, {Weaver},
  {Zabalza}, {Edwards}, {Azalee Bostroem}, {Burke}, {Casey}, {Crawford},
  {Dencheva}, {Ely}, {Jenness}, {Labrie}, {Lim}, {Pierfederici}, {Pontzen},
  {Ptak}, {Refsdal}, {Servillat}, \& {Streicher}}]{Astropy-Collaboration2013a}
{Astropy Collaboration}, {Robitaille}, T.~P., {Tollerud}, E.~J., {et~al.} 2013,
  \aap, 558, A33, \dodoi{10.1051/0004-6361/201322068}

\bibitem[{{Astropy Collaboration} {et~al.}(2018){Astropy Collaboration},
  {Price-Whelan}, {Sip{\H{o}}cz}, {G{\"u}nther}, {Lim}, {Crawford}, {Conseil},
  {Shupe}, {Craig}, {Dencheva}, {Ginsburg}, {VanderPlas}, {Bradley},
  {P{\'e}rez-Su{\'a}rez}, {de Val-Borro}, {Aldcroft}, {Cruz}, {Robitaille},
  {Tollerud}, {Ardelean}, {Babej}, {Bach}, {Bachetti}, {Bakanov}, {Bamford},
  {Barentsen}, {Barmby}, {Baumbach}, {Berry}, {Biscani}, {Boquien}, {Bostroem},
  {Bouma}, {Brammer}, {Bray}, {Breytenbach}, {Buddelmeijer}, {Burke},
  {Calderone}, {Cano Rodr{\'\i}guez}, {Cara}, {Cardoso}, {Cheedella}, {Copin},
  {Corrales}, {Crichton}, {D'Avella}, {Deil}, {Depagne}, {Dietrich}, {Donath},
  {Droettboom}, {Earl}, {Erben}, {Fabbro}, {Ferreira}, {Finethy}, {Fox},
  {Garrison}, {Gibbons}, {Goldstein}, {Gommers}, {Greco}, {Greenfield},
  {Groener}, {Grollier}, {Hagen}, {Hirst}, {Homeier}, {Horton}, {Hosseinzadeh},
  {Hu}, {Hunkeler}, {Ivezi{\'c}}, {Jain}, {Jenness}, {Kanarek}, {Kendrew},
  {Kern}, {Kerzendorf}, {Khvalko}, {King}, {Kirkby}, {Kulkarni}, {Kumar},
  {Lee}, {Lenz}, {Littlefair}, {Ma}, {Macleod}, {Mastropietro}, {McCully},
  {Montagnac}, {Morris}, {Mueller}, {Mumford}, {Muna}, {Murphy}, {Nelson},
  {Nguyen}, {Ninan}, {N{\"o}the}, {Ogaz}, {Oh}, {Parejko}, {Parley}, {Pascual},
  {Patil}, {Patil}, {Plunkett}, {Prochaska}, {Rastogi}, {Reddy Janga},
  {Sabater}, {Sakurikar}, {Seifert}, {Sherbert}, {Sherwood-Taylor}, {Shih},
  {Sick}, {Silbiger}, {Singanamalla}, {Singer}, {Sladen}, {Sooley},
  {Sornarajah}, {Streicher}, {Teuben}, {Thomas}, {Tremblay}, {Turner},
  {Terr{\'o}n}, {van Kerkwijk}, {de la Vega}, {Watkins}, {Weaver}, {Whitmore},
  {Woillez}, {Zabalza}, \& {Astropy Contributors}}]{Astropy-Collaboration2018m}
{Astropy Collaboration}, {Price-Whelan}, A.~M., {Sip{\H{o}}cz}, B.~M., {et~al.}
  2018, \aj, 156, 123, \dodoi{10.3847/1538-3881/aabc4f}

\bibitem[{{Aufort} {et~al.}(2020){Aufort}, {Ciesla}, {Pudlo}, \&
  {Buat}}]{AufortG2020a}
{Aufort}, G., {Ciesla}, L., {Pudlo}, P., \& {Buat}, V. 2020, \aap, 635, A136,
  \dodoi{10.1051/0004-6361/201936788}

\bibitem[{{Bastian} {et~al.}(2010){Bastian}, {Covey}, \&
  {Meyer}}]{BastianN2010a}
{Bastian}, N., {Covey}, K.~R., \& {Meyer}, M.~R. 2010, \araa, 48, 339,
  \dodoi{10.1146/annurev-astro-082708-101642}

\bibitem[{{Beckmann} {et~al.}(2017{\natexlab{a}}){Beckmann}, {Devriendt},
  {Slyz}, {Peirani}, {Richardson}, {Dubois}, {Pichon}, {Chisari}, {Kaviraj},
  {Laigle}, \& {Volonteri}}]{BeckmannR2017v}
{Beckmann}, R.~S., {Devriendt}, J., {Slyz}, A., {et~al.} 2017{\natexlab{a}},
  \mnras, 472, 949, \dodoi{10.1093/mnras/stx1831}

\bibitem[{{Beckmann} {et~al.}(2017{\natexlab{b}}){Beckmann}, {Devriendt},
  {Slyz}, {Peirani}, {Richardson}, {Dubois}, {Pichon}, {Chisari}, {Kaviraj},
  {Laigle}, \& {Volonteri}}]{BeckmannR2017a}
---. 2017{\natexlab{b}}, ArXiv e-prints.
\newblock \doarXiv{1701.07838}

\bibitem[{{Behroozi} {et~al.}(2013){Behroozi}, {Wechsler}, \&
  {Conroy}}]{BehrooziP2013a}
{Behroozi}, P.~S., {Wechsler}, R.~H., \& {Conroy}, C. 2013, \apj, 770, 57,
  \dodoi{10.1088/0004-637X/770/1/57}

\bibitem[{Beichman {et~al.}(2012)Beichman, Rieke, Eisenstein, Greene, Krist,
  McCarthy, Meyer, \& Stansberry}]{BeichmanC2012r}
Beichman, C.~A., Rieke, M., Eisenstein, D., {et~al.} 2012, in Space Telescopes
  and Instrumentation 2012: Optical, Infrared, and Millimeter Wave, ed. M.~C.
  Clampin, G.~G. Fazio, H.~A. MacEwen, \& J.~M.~O. Jr., Vol. 8442,
  International Society for Optics and Photonics (SPIE), 84422N,
  \dodoi{10.1117/12.925447}

\bibitem[{{Bertelli} {et~al.}(2008){Bertelli}, {Girardi}, {Marigo}, \&
  {Nasi}}]{BertelliG2008a}
{Bertelli}, G., {Girardi}, L., {Marigo}, P., \& {Nasi}, E. 2008, \aap, 484,
  815, \dodoi{10.1051/0004-6361:20079165}

\bibitem[{{Boquien} {et~al.}(2019){Boquien}, {Burgarella}, {Roehlly}, {Buat},
  {Ciesla}, {Corre}, {Inoue}, \& {Salas}}]{BoquienM2019a}
{Boquien}, M., {Burgarella}, D., {Roehlly}, Y., {et~al.} 2019, \aap, 622, A103,
  \dodoi{10.1051/0004-6361/201834156}

\bibitem[{{Bowman} {et~al.}(2020){Bowman}, {Zeimann}, {Nagaraj}, {Ciardullo},
  {Gronwall}, {McCarron}, {Weiss}, {Molina}, {Belles}, \&
  {Schneider}}]{BowmanW2020a}
{Bowman}, W.~P., {Zeimann}, G.~R., {Nagaraj}, G., {et~al.} 2020, \apj, 899, 7,
  \dodoi{10.3847/1538-4357/ab9f3c}

\bibitem[{{Breivik} {et~al.}(2022){Breivik}, {Connolly}, {Ford}, {Juri{\'c}},
  {Mandelbaum}, {Miller}, {Norman}, {Olsen}, {O'Mullane}, {Price-Whelan},
  {Sacco}, {Sokoloski}, {Villar}, {Acquaviva}, {Ahumada}, {AlSayyad}, {Alves},
  {Andreoni}, {Anguita}, {Best}, {Bianco}, {Bonito}, {Bradshaw}, {Burke},
  {Rodrigues de Campos}, {Cantiello}, {Caplar}, {Chandler}, {Chan}, {Nicolaci
  da Costa}, {Danieli}, {Davenport}, {Fabbian}, {Fagin}, {Gagliano}, {Gall},
  {Garavito Camargo}, {Gawiser}, {Gezari}, {Gomboc}, {Gonzalez-Morales},
  {Graham}, {Gschwend}, {Guy}, {Holman}, {Hsieh}, {Hundertmark}, {Ili{\'c}},
  {Ishida}, {Jurki{\'c}}, {Kannawadi}, {Kosakowski}, {Kova{\v{c}}evi{\'c}},
  {Kubica}, {Lanusse}, {Lazar}, {Levine}, {Li}, {Lu}, {Luna}, {Mahabal},
  {Malz}, {Mao}, {Medan}, {Moeyens}, {Nikoli{\'c}}, {Nikutta}, {O'Dowd},
  {Olsen}, {Pearson}, {Villicana Pedraza}, {Popinchalk}, {Popovi{\'c}},
  {Pritchard}, {Quint}, {Radovi{\'c}}, {Ragosta}, {Riccio}, {Riley},
  {Ro{\.z}ek}, {S{\'a}nchez-S{\'a}ez}, {Sarro}, {Saunders}, {Savi{\'c}},
  {Schmidt}, {Scott}, {Shirley}, {Smotherman}, {Stetzler}, {Storey-Fisher},
  {Street}, {Trilling}, {Tsapras}, {Ustamujic}, {van Velzen},
  {V{\'a}zquez-Mata}, {Venuti}, {Wyatt}, {Yu}, \& {Zabludoff}}]{BreivikK2022a}
{Breivik}, K., {Connolly}, A.~J., {Ford}, K.~E.~S., {et~al.} 2022, arXiv
  e-prints, arXiv:2208.02781.
\newblock \doarXiv{2208.02781}

\bibitem[{{Brott} {et~al.}(2011){Brott}, {de Mink}, {Cantiello}, {Langer}, {de
  Koter}, {Evans}, {Hunter}, {Trundle}, \& {Vink}}]{BrottI2011a}
{Brott}, I., {de Mink}, S.~E., {Cantiello}, M., {et~al.} 2011, \aap, 530, A115,
  \dodoi{10.1051/0004-6361/201016113}

\bibitem[{{Brown} {et~al.}(2019{\natexlab{a}}){Brown}, {Nayyeri}, {Cooray},
  {Ma}, {Hickox}, \& {Azadi}}]{BrownA2019a}
{Brown}, A., {Nayyeri}, H., {Cooray}, A., {et~al.} 2019{\natexlab{a}}, \apj,
  871, 87, \dodoi{10.3847/1538-4357/aaf73b}

\bibitem[{{Brown} {et~al.}(2019{\natexlab{b}}){Brown}, {Duncan}, {Landt},
  {Kirk}, {Ricci}, {Kamraj}, {Salvato}, \& {Ananna}}]{BrownM2019a}
{Brown}, M.~J.~I., {Duncan}, K.~J., {Landt}, H., {et~al.} 2019{\natexlab{b}},
  \mnras, 489, 3351, \dodoi{10.1093/mnras/stz2324}

\bibitem[{{Bruzual} \& {Charlot}(2003)}]{BruzualG2003a}
{Bruzual}, G., \& {Charlot}, S. 2003, \mnras, 344, 1000,
  \dodoi{10.1046/j.1365-8711.2003.06897.x}

\bibitem[{{Buchner}(2021)}]{BuchnerJ2021w}
{Buchner}, J. 2021, arXiv e-prints, arXiv:2101.09675.
\newblock \doarXiv{2101.09675}

\bibitem[{{Calzetti} {et~al.}(2000){Calzetti}, {Armus}, {Bohlin}, {Kinney},
  {Koornneef}, \& {Storchi-Bergmann}}]{CalzettiD2000a}
{Calzetti}, D., {Armus}, L., {Bohlin}, R.~C., {et~al.} 2000, \apj, 533, 682,
  \dodoi{10.1086/308692}

\bibitem[{Cameron \& Pettitt(2014)}]{CameronE2014f}
Cameron, E., \& Pettitt, A. 2014, Statistical Science, 29, 397 ,
  \dodoi{10.1214/13-STS465}

\bibitem[{{Cao} {et~al.}(2018){Cao}, {Gong}, {Meng}, {Xu}, {Chen}, {Guo}, {Li},
  {Liu}, {Xue}, {Cao}, {Fu}, {Zhang}, {Wang}, \& {Zhan}}]{CaoY2018a}
{Cao}, Y., {Gong}, Y., {Meng}, X.-M., {et~al.} 2018, \mnras, 480, 2178,
  \dodoi{10.1093/mnras/sty1980}

\bibitem[{{Cappellari} {et~al.}(2012){Cappellari}, {McDermid}, {Alatalo},
  {Blitz}, {Bois}, {Bournaud}, {Bureau}, {Crocker}, {Davies}, {Davis}, {de
  Zeeuw}, {Duc}, {Emsellem}, {Khochfar}, {Krajnovi{\'c}}, {Kuntschner},
  {Lablanche}, {Morganti}, {Naab}, {Oosterloo}, {Sarzi}, {Scott}, {Serra},
  {Weijmans}, \& {Young}}]{CappellariM2012b}
{Cappellari}, M., {McDermid}, R.~M., {Alatalo}, K., {et~al.} 2012, \nat, 484,
  485, \dodoi{10.1038/nature10972}

\bibitem[{{Caputi} {et~al.}(2015){Caputi}, {Ilbert}, {Laigle}, {McCracken}, {Le
  F{\`e}vre}, {Fynbo}, {Milvang-Jensen}, {Capak}, {Salvato}, \&
  {Taniguchi}}]{CaputiK2015f}
{Caputi}, K.~I., {Ilbert}, O., {Laigle}, C., {et~al.} 2015, \apj, 810, 73,
  \dodoi{10.1088/0004-637X/810/1/73}

\bibitem[{{Carnall} {et~al.}(2019){Carnall}, {Leja}, {Johnson}, {McLure},
  {Dunlop}, \& {Conroy}}]{CarnallA2019b}
{Carnall}, A.~C., {Leja}, J., {Johnson}, B.~D., {et~al.} 2019, \apj, 873, 44,
  \dodoi{10.3847/1538-4357/ab04a2}

\bibitem[{{Carnall} {et~al.}(2018){Carnall}, {McLure}, {Dunlop}, \&
  {Dav{\'e}}}]{CarnallA2018a}
{Carnall}, A.~C., {McLure}, R.~J., {Dunlop}, J.~S., \& {Dav{\'e}}, R. 2018,
  \mnras, 480, 4379, \dodoi{10.1093/mnras/sty2169}

\bibitem[{{Chabrier}(2003)}]{ChabrierG2003a}
{Chabrier}, G. 2003, \pasp, 115, 763, \dodoi{10.1086/376392}

\bibitem[{{Charlot} \& {Longhetti}(2001)}]{CharlotS2001a}
{Charlot}, S., \& {Longhetti}, M. 2001, \mnras, 323, 887,
  \dodoi{10.1046/j.1365-8711.2001.04260.x}

\bibitem[{{Chevallard} \& {Charlot}(2016)}]{ChevallardJ2016a}
{Chevallard}, J., \& {Charlot}, S. 2016, \mnras, 462, 1415,
  \dodoi{10.1093/mnras/stw1756}

\bibitem[{{Choi} {et~al.}(2019){Choi}, {Conroy}, \& {Johnson}}]{ChoiJ2019a}
{Choi}, J., {Conroy}, C., \& {Johnson}, B.~D. 2019, \apj, 872, 136,
  \dodoi{10.3847/1538-4357/aaff67}

\bibitem[{{Ciesla} {et~al.}(2017){Ciesla}, {Elbaz}, \& {Fensch}}]{CieslaL2017a}
{Ciesla}, L., {Elbaz}, D., \& {Fensch}, J. 2017, \aap, 608, A41,
  \dodoi{10.1051/0004-6361/201731036}

\bibitem[{{Ciesla} {et~al.}(2016){Ciesla}, {Boselli}, {Elbaz}, {Boissier},
  {Buat}, {Charmandaris}, {Schreiber}, {B{\'e}thermin}, {Baes}, {Boquien}, {De
  Looze}, {Fern{\'a}ndez-Ontiveros}, {Pappalardo}, {Spinoglio}, \&
  {Viaene}}]{CieslaL2016a}
{Ciesla}, L., {Boselli}, A., {Elbaz}, D., {et~al.} 2016, \aap, 585, A43,
  \dodoi{10.1051/0004-6361/201527107}

\bibitem[{{Coelho}(2009)}]{CoelhoP2009a}
{Coelho}, P. 2009, in American Institute of Physics Conference Series, Vol.
  1111, American Institute of Physics Conference Series, ed. G.~R. M.~L.
  L.~A.~A. N.~M. . E.~B. G.~Giobbi, A.~Tornambe, 67--74,
  \dodoi{10.1063/1.3141624}

\bibitem[{{Coelho} {et~al.}(2020){Coelho}, {Bruzual}, \&
  {Charlot}}]{CoelhoP2020k}
{Coelho}, P. R.~T., {Bruzual}, G., \& {Charlot}, S. 2020, \mnras, 491, 2025,
  \dodoi{10.1093/mnras/stz3023}

\bibitem[{Collette {et~al.}(2023)Collette, Kluyver, Caswell, Tocknell, Kieffer,
  Jelenak, Scopatz, Dale, Chen, VINCENT, Einhorn, payno, juliagarriga,
  Sciarelli, Valls, Ghosh, Pedersen, Kittisopikul, jakirkham, Raspaud,
  Danilevski, Abbasi, Readey, M{\"u}hlbauer, Paramonov, Chan, Schepper,
  Sol{\'e}, jialin, \& Guest}]{ColletteA2023u}
Collette, A., Kluyver, T., Caswell, T.~A., {et~al.} 2023, h5py/h5py:
  3.8.0-aarch64-wheels, 3.8.0-aarch64-wheels,  Zenodo,
  \dodoi{10.5281/zenodo.7568214}

\bibitem[{{Conroy}(2013)}]{ConroyC2013b}
{Conroy}, C. 2013, \araa, 51, 393, \dodoi{10.1146/annurev-astro-082812-141017}

\bibitem[{{Conroy} \& {Gunn}(2010)}]{ConroyC2010a}
{Conroy}, C., \& {Gunn}, J.~E. 2010, \apj, 712, 833,
  \dodoi{10.1088/0004-637X/712/2/833}

\bibitem[{{Conroy} {et~al.}(2009){Conroy}, {Gunn}, \& {White}}]{ConroyC2009a}
{Conroy}, C., {Gunn}, J.~E., \& {White}, M. 2009, \apj, 699, 486,
  \dodoi{10.1088/0004-637X/699/1/486}

\bibitem[{{Conroy} {et~al.}(2010){Conroy}, {White}, \& {Gunn}}]{ConroyC2010b}
{Conroy}, C., {White}, M., \& {Gunn}, J.~E. 2010, \apj, 708, 58,
  \dodoi{10.1088/0004-637X/708/1/58}

\bibitem[{{C{\^o}t{\'e}} {et~al.}(2016){C{\^o}t{\'e}}, {Ritter}, {O'Shea},
  {Herwig}, {Pignatari}, {Jones}, \& {Fryer}}]{CoteB2016c}
{C{\^o}t{\'e}}, B., {Ritter}, C., {O'Shea}, B.~W., {et~al.} 2016, \apj, 824,
  82, \dodoi{10.3847/0004-637X/824/2/82}

\bibitem[{{da Cunha} {et~al.}(2008){da Cunha}, {Charlot}, \&
  {Elbaz}}]{da-CunhaE2008a}
{da Cunha}, E., {Charlot}, S., \& {Elbaz}, D. 2008, \mnras, 388, 1595,
  \dodoi{10.1111/j.1365-2966.2008.13535.x}

\bibitem[{{Dahlen} {et~al.}(2013){Dahlen}, {Mobasher}, {Faber}, {Ferguson},
  {Barro}, {Finkelstein}, {Finlator}, {Fontana}, {Gruetzbauch}, {Johnson},
  {Pforr}, {Salvato}, {Wiklind}, {Wuyts}, {Acquaviva}, {Dickinson}, {Guo},
  {Huang}, {Huang}, {Newman}, {Bell}, {Conselice}, {Galametz}, {Gawiser},
  {Giavalisco}, {Grogin}, {Hathi}, {Kocevski}, {Koekemoer}, {Koo}, {Lee},
  {McGrath}, {Papovich}, {Peth}, {Ryan}, {Somerville}, {Weiner}, \&
  {Wilson}}]{DahlenT2013a}
{Dahlen}, T., {Mobasher}, B., {Faber}, S.~M., {et~al.} 2013, \apj, 775, 93,
  \dodoi{10.1088/0004-637X/775/2/93}

\bibitem[{{Davidzon} {et~al.}(2017){Davidzon}, {Ilbert}, {Laigle}, {Coupon},
  {McCracken}, {Delvecchio}, {Masters}, {Capak}, {Hsieh}, {Le F{\`e}vre},
  {Tresse}, {Bethermin}, {Chang}, {Faisst}, {Le Floc'h}, {Steinhardt}, {Toft},
  {Aussel}, {Dubois}, {Hasinger}, {Salvato}, {Sanders}, {Scoville}, \&
  {Silverman}}]{DavidzonI2017a}
{Davidzon}, I., {Ilbert}, O., {Laigle}, C., {et~al.} 2017, \aap, 605, A70,
  \dodoi{10.1051/0004-6361/201730419}

\bibitem[{{Davidzon} {et~al.}(2019){Davidzon}, {Laigle}, {Capak}, {Ilbert},
  {Masters}, {Hemmati}, {Apostolakos}, {Coupon}, {de la Torre}, {Devriendt},
  {Dubois}, {Kashino}, {Paltani}, \& {Pichon}}]{DavidzonI2019a}
{Davidzon}, I., {Laigle}, C., {Capak}, P.~L., {et~al.} 2019, \mnras, 489, 4817,
  \dodoi{10.1093/mnras/stz2486}

\bibitem[{{Debsarma} {et~al.}(2016){Debsarma}, {Chattopadhyay}, {Das}, \&
  {Pfenniger}}]{DebsarmaS2016a}
{Debsarma}, S., {Chattopadhyay}, T., {Das}, S., \& {Pfenniger}, D. 2016,
  \mnras, \dodoi{10.1093/mnras/stw1909}

\bibitem[{{Diemer} {et~al.}(2017){Diemer}, {Sparre}, {Abramson}, \&
  {Torrey}}]{DiemerB2017a}
{Diemer}, B., {Sparre}, M., {Abramson}, L.~E., \& {Torrey}, P. 2017, \apj, 839,
  26, \dodoi{10.3847/1538-4357/aa68e5}

\bibitem[{{Doore} {et~al.}(2023){Doore}, {Monson}, {Eufrasio}, {Lehmer},
  {Garofali}, \& {Basu-Zych}}]{DooreK2023a}
{Doore}, K., {Monson}, E.~B., {Eufrasio}, R.~T., {et~al.} 2023, \apjs, 266, 39,
  \dodoi{10.3847/1538-4365/accc29}

\bibitem[{{Dore} {et~al.}(2019){Dore}, {Hirata}, {Wang}, {Weinberg}, {Eifler},
  {Foley}, {Heinrich}, {Krause}, {Perlmutter}, {Pisani}, {Scolnic}, {Spergel},
  {Suntzeff}, {Aldering}, {Baltay}, {Capak}, {Choi}, {Dvorkin}, {Fall}, {Fang},
  {Fruchter}, {Galbany}, {Ho}, {Hounsell}, {Izard}, {Jain}, {Koekemoer},
  {Kruk}, {Leauthaud}, {Malhotra}, {Mandelbaum}, {Massara}, {Masters},
  {Miyatake}, {Plazas}, {Rhoads}, {Rhodes}, {Rose}, {Rubin}, {Sako},
  {Samushia}, {Shirasaki}, {Simet}, {Takada}, {Troxel}, {Wu}, {Yoshida}, \&
  {Zhai}}]{DoreO2019p}
{Dore}, O., {Hirata}, C., {Wang}, Y., {et~al.} 2019, \baas, 51, 341.
\newblock \doarXiv{1904.01174}

\bibitem[{Draine(2010)}]{DraineB2010a}
Draine, B. 2010, Physics of the Interstellar and Intergalactic Medium,
  Princeton Series in Astrophysics (Princeton University Press).
\newblock \url{https://books.google.co.kr/books?id=FycJvKHyiwsC}

\bibitem[{{Draine}(2003)}]{DraineB2003a}
{Draine}, B.~T. 2003, \araa, 41, 241,
  \dodoi{10.1146/annurev.astro.41.011802.094840}

\bibitem[{{Dries} {et~al.}(2016){Dries}, {Trager}, \& {Koopmans}}]{DriesM2016a}
{Dries}, M., {Trager}, S.~C., \& {Koopmans}, L.~V.~E. 2016, \mnras, 463, 886,
  \dodoi{10.1093/mnras/stw2049}

\bibitem[{{Dries} {et~al.}(2018){Dries}, {Trager}, {Koopmans}, {Popping}, \&
  {Somerville}}]{DriesM2018a}
{Dries}, M., {Trager}, S.~C., {Koopmans}, L.~V.~E., {Popping}, G., \&
  {Somerville}, R.~S. 2018, \mnras, 474, 3500, \dodoi{10.1093/mnras/stx2979}

\bibitem[{{Driver} {et~al.}(2013){Driver}, {Robotham}, {Bland-Hawthorn},
  {Brown}, {Hopkins}, {Liske}, {Phillipps}, \& {Wilkins}}]{DriverS2013a}
{Driver}, S.~P., {Robotham}, A.~S.~G., {Bland-Hawthorn}, J., {et~al.} 2013,
  \mnras, 430, 2622, \dodoi{10.1093/mnras/sts717}

\bibitem[{{Dubois} {et~al.}(2014){Dubois}, {Pichon}, {Welker}, {Le Borgne},
  {Devriendt}, {Laigle}, {Codis}, {Pogosyan}, {Arnouts}, {Benabed}, {Bertin},
  {Blaizot}, {Bouchet}, {Cardoso}, {Colombi}, {de Lapparent}, {Desjacques},
  {Gavazzi}, {Kassin}, {Kimm}, {McCracken}, {Milliard}, {Peirani}, {Prunet},
  {Rouberol}, {Silk}, {Slyz}, {Sousbie}, {Teyssier}, {Tresse}, {Treyer},
  {Vibert}, \& {Volonteri}}]{DuboisY2014a}
{Dubois}, Y., {Pichon}, C., {Welker}, C., {et~al.} 2014, \mnras, 444, 1453,
  \dodoi{10.1093/mnras/stu1227}

\bibitem[{{Eldridge} \& {Stanway}(2009)}]{EldridgeJ2009a}
{Eldridge}, J.~J., \& {Stanway}, E.~R. 2009, \mnras, 400, 1019,
  \dodoi{10.1111/j.1365-2966.2009.15514.x}

\bibitem[{{Feroz} \& {Hobson}(2008)}]{FerozF2008b}
{Feroz}, F., \& {Hobson}, M.~P. 2008, \mnras, 384, 449,
  \dodoi{10.1111/j.1365-2966.2007.12353.x}

\bibitem[{{Feroz} {et~al.}(2009){Feroz}, {Hobson}, \& {Bridges}}]{FerozF2009b}
{Feroz}, F., {Hobson}, M.~P., \& {Bridges}, M. 2009, \mnras, 398, 1601,
  \dodoi{10.1111/j.1365-2966.2009.14548.x}

\bibitem[{{Feroz} {et~al.}(2019){Feroz}, {Hobson}, {Cameron}, \&
  {Pettitt}}]{FerozF2019a}
{Feroz}, F., {Hobson}, M.~P., {Cameron}, E., \& {Pettitt}, A.~N. 2019, The Open
  Journal of Astrophysics, 2, 10, \dodoi{10.21105/astro.1306.2144}

\bibitem[{{Ferreras} {et~al.}(2013){Ferreras}, {La Barbera}, {de la Rosa},
  {Vazdekis}, {de Carvalho}, {Falc{\'o}n-Barroso}, \&
  {Ricciardelli}}]{FerrerasI2013a}
{Ferreras}, I., {La Barbera}, F., {de la Rosa}, I.~G., {et~al.} 2013, \mnras,
  429, L15, \dodoi{10.1093/mnrasl/sls014}

\bibitem[{{Fitzpatrick}(1986)}]{FitzpatrickE1986f}
{Fitzpatrick}, E.~L. 1986, \aj, 92, 1068, \dodoi{10.1086/114237}

\bibitem[{{Galliano} {et~al.}(2018){Galliano}, {Galametz}, \&
  {Jones}}]{GallianoF2018b}
{Galliano}, F., {Galametz}, M., \& {Jones}, A.~P. 2018, \araa, 56, 673,
  \dodoi{10.1146/annurev-astro-081817-051900}

\bibitem[{{Gardner} {et~al.}(2006){Gardner}, {Mather}, {Clampin}, {Doyon},
  {Greenhouse}, {Hammel}, {Hutchings}, {Jakobsen}, {Lilly}, {Long}, {Lunine},
  {McCaughrean}, {Mountain}, {Nella}, {Rieke}, {Rieke}, {Rix}, {Smith},
  {Sonneborn}, {Stiavelli}, {Stockman}, {Windhorst}, \&
  {Wright}}]{GardnerJ2006l}
{Gardner}, J.~P., {Mather}, J.~C., {Clampin}, M., {et~al.} 2006, \ssr, 123,
  485, \dodoi{10.1007/s11214-006-8315-7}

\bibitem[{{Gennaro} {et~al.}(2018){Gennaro}, {Tchernyshyov}, {Brown}, {Geha},
  {Avila}, {Guhathakurta}, {Kalirai}, {Kirby}, {Renzini}, {Simon}, {Tumlinson},
  \& {Vargas}}]{GennaroM2018d}
{Gennaro}, M., {Tchernyshyov}, K., {Brown}, T.~M., {et~al.} 2018, \apj, 855,
  20, \dodoi{10.3847/1538-4357/aaa973}

\bibitem[{{Gilda} {et~al.}(2021){Gilda}, {Lower}, \& {Narayanan}}]{GildaS2021m}
{Gilda}, S., {Lower}, S., \& {Narayanan}, D. 2021, \apj, 916, 43,
  \dodoi{10.3847/1538-4357/ac0058}

\bibitem[{{Gong} {et~al.}(2019){Gong}, {Liu}, {Cao}, {Chen}, {Fan}, {Li}, {Li},
  {Li}, {Zhang}, \& {Zhan}}]{GongY2019a}
{Gong}, Y., {Liu}, X., {Cao}, Y., {et~al.} 2019, \apj, 883, 203,
  \dodoi{10.3847/1538-4357/ab391e}

\bibitem[{{Green} {et~al.}(2012){Green}, {Schechter}, {Baltay}, {Bean},
  {Bennett}, {Brown}, {Conselice}, {Donahue}, {Fan}, {Gaudi}, {Hirata},
  {Kalirai}, {Lauer}, {Nichol}, {Padmanabhan}, {Perlmutter}, {Rauscher},
  {Rhodes}, {Roellig}, {Stern}, {Sumi}, {Tanner}, {Wang}, {Weinberg}, {Wright},
  {Gehrels}, {Sambruna}, {Traub}, {Anderson}, {Cook}, {Garnavich},
  {Hillenbrand}, {Ivezic}, {Kerins}, {Lunine}, {McDonald}, {Penny}, {Phillips},
  {Rieke}, {Riess}, {van der Marel}, {Barry}, {Cheng}, {Content}, {Cutri},
  {Goullioud}, {Grady}, {Helou}, {Jackson}, {Kruk}, {Melton}, {Peddie},
  {Rioux}, \& {Seiffert}}]{GreenJ2012c}
{Green}, J., {Schechter}, P., {Baltay}, C., {et~al.} 2012, arXiv e-prints,
  arXiv:1208.4012.
\newblock \doarXiv{1208.4012}

\bibitem[{{Hahn} \& {Melchior}(2022)}]{HahnC2022x}
{Hahn}, C., \& {Melchior}, P. 2022, \apj, 938, 11,
  \dodoi{10.3847/1538-4357/ac7b84}

\bibitem[{{Hahn} {et~al.}(2022){Hahn}, {Kwon}, {Tojeiro}, {Siudek}, {Canning},
  {Mezcua}, {Tinker}, {Brooks}, {Doel}, {Fanning}, {Gazta{\~n}aga}, {Kehoe},
  {Landriau}, {Meisner}, {Moustakas}, {Poppett}, {Tarle}, {Weiner}, \&
  {Zou}}]{HahnC2022d}
{Hahn}, C., {Kwon}, K.~J., {Tojeiro}, R., {et~al.} 2022, arXiv e-prints,
  arXiv:2202.01809.
\newblock \doarXiv{2202.01809}

\bibitem[{{Han} \& {Han}(2012)}]{HanY2012b}
{Han}, Y., \& {Han}, Z. 2012, \apj, 749, 123,
  \dodoi{10.1088/0004-637X/749/2/123}

\bibitem[{{Han} \& {Han}(2014)}]{HanY2014a}
---. 2014, \apjs, 215, 2, \dodoi{10.1088/0067-0049/215/1/2}

\bibitem[{{Han} \& {Han}(2019)}]{HanY2019a}
---. 2019, \apjs, 240, 3, \dodoi{10.3847/1538-4365/aaeffa}

\bibitem[{{Han} {et~al.}(2019){Han}, {Han}, \& {Fan}}]{HanY2019u}
{Han}, Y., {Han}, Z., \& {Fan}, L. 2019, in The Art of Measuring Galaxy
  Physical Properties, 5, \dodoi{10.5281/zenodo.3553688}

\bibitem[{{Han} {et~al.}(2020){Han}, {Han}, \& {Fan}}]{HanY2020a}
{Han}, Y., {Han}, Z., \& {Fan}, L. 2020, in IAU Symposium, Vol. 341, IAU
  Symposium, ed. M.~{Boquien}, E.~{Lusso}, C.~{Gruppioni}, \& P.~{Tissera},
  143--146, \dodoi{10.1017/S1743921319002746}

\bibitem[{{Han} {et~al.}(2007){Han}, {Podsiadlowski}, \&
  {Lynas-Gray}}]{HanZ2007a}
{Han}, Z., {Podsiadlowski}, P., \& {Lynas-Gray}, A.~E. 2007, \mnras, 380, 1098,
  \dodoi{10.1111/j.1365-2966.2007.12151.x}

\bibitem[{{Hern{\'a}ndez-P{\'e}rez} \& {Bruzual}(2013)}]{Hernandez-PerezF2013a}
{Hern{\'a}ndez-P{\'e}rez}, F., \& {Bruzual}, G. 2013, \mnras, 431, 2612,
  \dodoi{10.1093/mnras/stt368}

\bibitem[{{Hickox} \& {Alexander}(2018)}]{HickoxR2018i}
{Hickox}, R.~C., \& {Alexander}, D.~M. 2018, \araa, 56, 625,
  \dodoi{10.1146/annurev-astro-081817-051803}

\bibitem[{Higson {et~al.}(2019)Higson, Handley, Hobson, \&
  Lasenby}]{HigsonE2019b}
Higson, E., Handley, W., Hobson, M., \& Lasenby, A. 2019, Statistics and
  Computing, 29, 891, \dodoi{10.1007/s11222-018-9844-0}

\bibitem[{{Hogg} \& {Foreman-Mackey}(2018)}]{HoggD2018a}
{Hogg}, D.~W., \& {Foreman-Mackey}, D. 2018, \apjs, 236, 11,
  \dodoi{10.3847/1538-4365/aab76e}

\bibitem[{{Hopkins} {et~al.}(2014){Hopkins}, {Kere{\v{s}}}, {O{\~n}orbe},
  {Faucher-Gigu{\`e}re}, {Quataert}, {Murray}, \& {Bullock}}]{HopkinsP2014b}
{Hopkins}, P.~F., {Kere{\v{s}}}, D., {O{\~n}orbe}, J., {et~al.} 2014, \mnras,
  445, 581, \dodoi{10.1093/mnras/stu1738}

\bibitem[{{Hoversten} \& {Glazebrook}(2008)}]{HoverstenE2008a}
{Hoversten}, E.~A., \& {Glazebrook}, K. 2008, \apj, 675, 163,
  \dodoi{10.1086/524095}

\bibitem[{Hunter(2007)}]{HunterJ2007x}
Hunter, J.~D. 2007, Computing in Science \& Engineering, 9, 90,
  \dodoi{10.1109/MCSE.2007.55}

\bibitem[{{Ilbert} {et~al.}(2009){Ilbert}, {Capak}, {Salvato}, {Aussel},
  {McCracken}, {Sanders}, {Scoville}, {Kartaltepe}, {Arnouts}, {Le Floc'h},
  {Mobasher}, {Taniguchi}, {Lamareille}, {Leauthaud}, {Sasaki}, {Thompson},
  {Zamojski}, {Zamorani}, {Bardelli}, {Bolzonella}, {Bongiorno}, {Brusa},
  {Caputi}, {Carollo}, {Contini}, {Cook}, {Coppa}, {Cucciati}, {de la Torre},
  {de Ravel}, {Franzetti}, {Garilli}, {Hasinger}, {Iovino}, {Kampczyk},
  {Kneib}, {Knobel}, {Kovac}, {Le Borgne}, {Le Brun}, {F{\`e}vre}, {Lilly},
  {Looper}, {Maier}, {Mainieri}, {Mellier}, {Mignoli}, {Murayama}, {Pell{\`o}},
  {Peng}, {P{\'e}rez-Montero}, {Renzini}, {Ricciardelli}, {Schiminovich},
  {Scodeggio}, {Shioya}, {Silverman}, {Surace}, {Tanaka}, {Tasca}, {Tresse},
  {Vergani}, \& {Zucca}}]{IlbertO2009a}
{Ilbert}, O., {Capak}, P., {Salvato}, M., {et~al.} 2009, \apj, 690, 1236,
  \dodoi{10.1088/0004-637X/690/2/1236}

\bibitem[{{Ivezi{\'c}} {et~al.}(2019){Ivezi{\'c}}, {Kahn}, {Tyson}, {Abel},
  {Acosta}, {Allsman}, {Alonso}, {AlSayyad}, {Anderson}, {Andrew}, \&
  et~al.}]{IvezicZ2019a}
{Ivezi{\'c}}, {\v{Z}}., {Kahn}, S.~M., {Tyson}, J.~A., {et~al.} 2019, \apj,
  873, 111, \dodoi{10.3847/1538-4357/ab042c}

\bibitem[{{Iyer} \& {Gawiser}(2017)}]{IyerK2017a}
{Iyer}, K., \& {Gawiser}, E. 2017, \apj, 838, 127,
  \dodoi{10.3847/1538-4357/aa63f0}

\bibitem[{{Iyer} {et~al.}(2019){Iyer}, {Gawiser}, {Faber}, {Ferguson},
  {Kartaltepe}, {Koekemoer}, {Pacifici}, \& {Somerville}}]{IyerK2019a}
{Iyer}, K.~G., {Gawiser}, E., {Faber}, S.~M., {et~al.} 2019, \apj, 879, 116,
  \dodoi{10.3847/1538-4357/ab2052}

\bibitem[{{Iyer} {et~al.}(2020){Iyer}, {Tacchella}, {Genel}, {Hayward},
  {Hernquist}, {Brooks}, {Caplar}, {Dav{\'e}}, {Diemer}, {Forbes}, {Gawiser},
  {Somerville}, \& {Starkenburg}}]{IyerK2020a}
{Iyer}, K.~G., {Tacchella}, S., {Genel}, S., {et~al.} 2020, \mnras, 498, 430,
  \dodoi{10.1093/mnras/staa2150}

\bibitem[{{Joachimi}(2016)}]{JoachimiB2016b}
{Joachimi}, B. 2016, in Astronomical Society of the Pacific Conference Series,
  Vol. 507, Multi-Object Spectroscopy in the Next Decade: Big Questions, Large
  Surveys, and Wide Fields, ed. I.~{Skillen}, M.~{Balcells}, \& S.~{Trager},
  401

\bibitem[{{Kaviraj} {et~al.}(2017){Kaviraj}, {Laigle}, {Kimm}, {Devriendt},
  {Dubois}, {Pichon}, {Slyz}, {Chisari}, \& {Peirani}}]{KavirajS2017x}
{Kaviraj}, S., {Laigle}, C., {Kimm}, T., {et~al.} 2017, \mnras, 467, 4739,
  \dodoi{10.1093/mnras/stx126}

\bibitem[{{Kewley} {et~al.}(2019){Kewley}, {Nicholls}, \&
  {Sutherland}}]{KewleyL2019a}
{Kewley}, L.~J., {Nicholls}, D.~C., \& {Sutherland}, R.~S. 2019, \araa, 57,
  511, \dodoi{10.1146/annurev-astro-081817-051832}

\bibitem[{{Knowles} {et~al.}(2021){Knowles}, {Sansom}, {Allende Prieto}, \&
  {Vazdekis}}]{KnowlesA2021d}
{Knowles}, A.~T., {Sansom}, A.~E., {Allende Prieto}, C., \& {Vazdekis}, A.
  2021, \mnras, 504, 2286, \dodoi{10.1093/mnras/stab1001}

\bibitem[{{Knowles} {et~al.}(2019){Knowles}, {Sansom}, {Coelho}, {Allende
  Prieto}, {Conroy}, \& {Vazdekis}}]{KnowlesA2019t}
{Knowles}, A.~T., {Sansom}, A.~E., {Coelho}, P.~R.~T., {et~al.} 2019, \mnras,
  486, 1814, \dodoi{10.1093/mnras/stz754}

\bibitem[{{Kriek} \& {Conroy}(2013)}]{KriekM2013a}
{Kriek}, M., \& {Conroy}, C. 2013, \apjl, 775, L16,
  \dodoi{10.1088/2041-8205/775/1/L16}

\bibitem[{{Laigle} {et~al.}(2016){Laigle}, {McCracken}, {Ilbert}, {Hsieh},
  {Davidzon}, {Capak}, {Hasinger}, {Silverman}, {Pichon}, {Coupon}, {Aussel},
  {Le Borgne}, {Caputi}, {Cassata}, {Chang}, {Civano}, {Dunlop}, {Fynbo},
  {Kartaltepe}, {Koekemoer}, {Le F{\`e}vre}, {Le Floc'h}, {Leauthaud}, {Lilly},
  {Lin}, {Marchesi}, {Milvang-Jensen}, {Salvato}, {Sanders}, {Scoville},
  {Smolcic}, {Stockmann}, {Taniguchi}, {Tasca}, {Toft}, {Vaccari}, \&
  {Zabl}}]{LaigleC2016a}
{Laigle}, C., {McCracken}, H.~J., {Ilbert}, O., {et~al.} 2016, \apjs, 224, 24,
  \dodoi{10.3847/0067-0049/224/2/24}

\bibitem[{{Laigle} {et~al.}(2019){Laigle}, {Davidzon}, {Ilbert}, {Devriendt},
  {Kashino}, {Pichon}, {Capak}, {Arnouts}, {de la Torre}, {Dubois},
  {Gozaliasl}, {Le Borgne}, {Lilly}, {McCracken}, {Salvato}, \&
  {Slyz}}]{LaigleC2019a}
{Laigle}, C., {Davidzon}, I., {Ilbert}, O., {et~al.} 2019, \mnras, 486, 5104,
  \dodoi{10.1093/mnras/stz1054}

\bibitem[{{Laureijs} {et~al.}(2011){Laureijs}, {Amiaux}, {Arduini},
  {Augu{\`e}res}, {Brinchmann}, {Cole}, {Cropper}, {Dabin}, {Duvet}, {Ealet},
  {Garilli}, {Gondoin}, {Guzzo}, {Hoar}, {Hoekstra}, {Holmes}, {Kitching},
  {Maciaszek}, {Mellier}, {Pasian}, {Percival}, {Rhodes}, {Saavedra Criado},
  {Sauvage}, {Scaramella}, {Valenziano}, {Warren}, {Bender}, {Castander},
  {Cimatti}, {Le F{\`e}vre}, {Kurki-Suonio}, {Levi}, {Lilje}, {Meylan},
  {Nichol}, {Pedersen}, {Popa}, {Rebolo Lopez}, {Rix}, {Rottgering},
  {Zeilinger}, {Grupp}, {Hudelot}, {Massey}, {Meneghetti}, {Miller}, {Paltani},
  {Paulin-Henriksson}, {Pires}, {Saxton}, {Schrabback}, {Seidel}, {Walsh},
  {Aghanim}, {Amendola}, {Bartlett}, {Baccigalupi}, {Beaulieu}, {Benabed},
  {Cuby}, {Elbaz}, {Fosalba}, {Gavazzi}, {Helmi}, {Hook}, {Irwin}, {Kneib},
  {Kunz}, {Mannucci}, {Moscardini}, {Tao}, {Teyssier}, {Weller}, {Zamorani},
  {Zapatero Osorio}, {Boulade}, {Foumond}, {Di Giorgio}, {Guttridge}, {James},
  {Kemp}, {Martignac}, {Spencer}, {Walton}, {Bl{\"u}mchen}, {Bonoli},
  {Bortoletto}, {Cerna}, {Corcione}, {Fabron}, {Jahnke}, {Ligori}, {Madrid},
  {Martin}, {Morgante}, {Pamplona}, {Prieto}, {Riva}, {Toledo}, {Trifoglio},
  {Zerbi}, {Abdalla}, {Douspis}, {Grenet}, {Borgani}, {Bouwens}, {Courbin},
  {Delouis}, {Dubath}, {Fontana}, {Frailis}, {Grazian}, {Koppenh{\"o}fer},
  {Mansutti}, {Melchior}, {Mignoli}, {Mohr}, {Neissner}, {Noddle}, {Poncet},
  {Scodeggio}, {Serrano}, {Shane}, {Starck}, {Surace}, {Taylor},
  {Verdoes-Kleijn}, {Vuerli}, {Williams}, {Zacchei}, {Altieri}, {Escudero
  Sanz}, {Kohley}, {Oosterbroek}, {Astier}, {Bacon}, {Bardelli}, {Baugh},
  {Bellagamba}, {Benoist}, {Bianchi}, {Biviano}, {Branchini}, {Carbone},
  {Cardone}, {Clements}, {Colombi}, {Conselice}, {Cresci}, {Deacon}, {Dunlop},
  {Fedeli}, {Fontanot}, {Franzetti}, {Giocoli}, {Garcia-Bellido}, {Gow},
  {Heavens}, {Hewett}, {Heymans}, {Holland}, {Huang}, {Ilbert}, {Joachimi},
  {Jennins}, {Kerins}, {Kiessling}, {Kirk}, {Kotak}, {Krause}, {Lahav}, {van
  Leeuwen}, {Lesgourgues}, {Lombardi}, {Magliocchetti}, {Maguire}, {Majerotto},
  {Maoli}, {Marulli}, {Maurogordato}, {McCracken}, {McLure}, {Melchiorri},
  {Merson}, {Moresco}, {Nonino}, {Norberg}, {Peacock}, {Pello}, {Penny},
  {Pettorino}, {Di Porto}, {Pozzetti}, {Quercellini}, {Radovich}, {Rassat},
  {Roche}, {Ronayette}, {Rossetti}, {Sartoris}, {Schneider}, {Semboloni},
  {Serjeant}, {Simpson}, {Skordis}, {Smadja}, {Smartt}, {Spano}, {Spiro},
  {Sullivan}, {Tilquin}, {Trotta}, {Verde}, {Wang}, {Williger}, {Zhao},
  {Zoubian}, \& {Zucca}}]{LaureijsR2011a}
{Laureijs}, R., {Amiaux}, J., {Arduini}, S., {et~al.} 2011, ArXiv e-prints.
\newblock \doarXiv{1110.3193}

\bibitem[{{Lawler} \& {Acquaviva}(2021)}]{LawlerA2021n}
{Lawler}, A.~J., \& {Acquaviva}, V. 2021, \mnras, 502, 3993,
  \dodoi{10.1093/mnras/stab138}

\bibitem[{{Lee} {et~al.}(2010){Lee}, {Ferguson}, {Somerville}, {Wiklind}, \&
  {Giavalisco}}]{LeeS2010a}
{Lee}, S.-K., {Ferguson}, H.~C., {Somerville}, R.~S., {Wiklind}, T., \&
  {Giavalisco}, M. 2010, \apj, 725, 1644, \dodoi{10.1088/0004-637X/725/2/1644}

\bibitem[{{Lee} {et~al.}(2009){Lee}, {Idzi}, {Ferguson}, {Somerville},
  {Wiklind}, \& {Giavalisco}}]{LeeS2009a}
{Lee}, S.-K., {Idzi}, R., {Ferguson}, H.~C., {et~al.} 2009, \apjs, 184, 100,
  \dodoi{10.1088/0067-0049/184/1/100}

\bibitem[{{Leja} {et~al.}(2019){Leja}, {Carnall}, {Johnson}, {Conroy}, \&
  {Speagle}}]{LejaJ2019a}
{Leja}, J., {Carnall}, A.~C., {Johnson}, B.~D., {Conroy}, C., \& {Speagle},
  J.~S. 2019, \apj, 876, 3, \dodoi{10.3847/1538-4357/ab133c}

\bibitem[{{Leja} {et~al.}(2017){Leja}, {Johnson}, {Conroy}, {van Dokkum}, \&
  {Byler}}]{LejaJ2017a}
{Leja}, J., {Johnson}, B.~D., {Conroy}, C., {van Dokkum}, P.~G., \& {Byler}, N.
  2017, \apj, 837, 170, \dodoi{10.3847/1538-4357/aa5ffe}

\bibitem[{{Leja} {et~al.}(2020){Leja}, {Speagle}, {Johnson}, {Conroy}, {van
  Dokkum}, \& {Franx}}]{LejaJ2020a}
{Leja}, J., {Speagle}, J.~S., {Johnson}, B.~D., {et~al.} 2020, \apj, 893, 111,
  \dodoi{10.3847/1538-4357/ab7e27}

\bibitem[{{Lower} {et~al.}(2020){Lower}, {Narayanan}, {Leja}, {Johnson},
  {Conroy}, \& {Dav{\'e}}}]{LowerS2020u}
{Lower}, S., {Narayanan}, D., {Leja}, J., {et~al.} 2020, \apj, 904, 33,
  \dodoi{10.3847/1538-4357/abbfa7}

\bibitem[{{Lower} {et~al.}(2022){Lower}, {Narayanan}, {Leja}, {Johnson},
  {Conroy}, \& {Dav{\'e}}}]{LowerS2022g}
---. 2022, \apj, 931, 14, \dodoi{10.3847/1538-4357/ac6959}

\bibitem[{{Lyu} \& {Rieke}(2022)}]{LyuJ2022c}
{Lyu}, J., \& {Rieke}, G. 2022, Universe, 8, 304,
  \dodoi{10.3390/universe8060304}

\bibitem[{{Ma} {et~al.}(2016){Ma}, {Hopkins}, {Faucher-Gigu{\`e}re}, {Zolman},
  {Muratov}, {Kere{\v{s}}}, \& {Quataert}}]{MaX2016o}
{Ma}, X., {Hopkins}, P.~F., {Faucher-Gigu{\`e}re}, C.-A., {et~al.} 2016,
  \mnras, 456, 2140, \dodoi{10.1093/mnras/stv2659}

\bibitem[{{Madau}(1995)}]{MadauP1995h}
{Madau}, P. 1995, \apj, 441, 18, \dodoi{10.1086/175332}

\bibitem[{{Maiolino} \& {Mannucci}(2019)}]{MaiolinoR2019a}
{Maiolino}, R., \& {Mannucci}, F. 2019, \aapr, 27, 3,
  \dodoi{10.1007/s00159-018-0112-2}

\bibitem[{{Maraston}(2005)}]{MarastonC2005a}
{Maraston}, C. 2005, \mnras, 362, 799, \dodoi{10.1111/j.1365-2966.2005.09270.x}

\bibitem[{{Maraston} {et~al.}(2006){Maraston}, {Daddi}, {Renzini}, {Cimatti},
  {Dickinson}, {Papovich}, {Pasquali}, \& {Pirzkal}}]{MarastonC2006a}
{Maraston}, C., {Daddi}, E., {Renzini}, A., {et~al.} 2006, \apj, 652, 85,
  \dodoi{10.1086/508143}

\bibitem[{{Marigo} {et~al.}(2008){Marigo}, {Girardi}, {Bressan}, {Groenewegen},
  {Silva}, \& {Granato}}]{MarigoP2008a}
{Marigo}, P., {Girardi}, L., {Bressan}, A., {et~al.} 2008, \aap, 482, 883,
  \dodoi{10.1051/0004-6361:20078467}

\bibitem[{{Morishita}(2022)}]{MorishitaT2022a}
{Morishita}, T. 2022, {gsf: Grism SED Fitting package}, Astrophysics Source
  Code Library, record ascl:2211.012.
\newblock \doeprint{2211.012}

\bibitem[{{Narayanan} {et~al.}(2018){Narayanan}, {Conroy}, {Dav{\'e}},
  {Johnson}, \& {Popping}}]{NarayananD2018o}
{Narayanan}, D., {Conroy}, C., {Dav{\'e}}, R., {Johnson}, B.~D., \& {Popping},
  G. 2018, \apj, 869, 70, \dodoi{10.3847/1538-4357/aaed25}

\bibitem[{{National Academies of Sciences, Engineering, and
  Medicine}(2021)}]{National-Academies-of-SciencesE2021g}
{National Academies of Sciences, Engineering, and Medicine}. 2021, {Pathways to
  Discovery in Astronomy and Astrophysics for the 2020s} (The National
  Academies Press), \dodoi{10.17226/26141}

\bibitem[{{Netzer}(2015)}]{NetzerH2015a}
{Netzer}, H. 2015, \araa, 53, 365, \dodoi{10.1146/annurev-astro-082214-122302}

\bibitem[{{Noll} {et~al.}(2009){Noll}, {Burgarella}, {Giovannoli}, {Buat},
  {Marcillac}, \& {Mu{\~n}oz-Mateos}}]{NollS2009b}
{Noll}, S., {Burgarella}, D., {Giovannoli}, E., {et~al.} 2009, \aap, 507, 1793,
  \dodoi{10.1051/0004-6361/200912497}

\bibitem[{{Oke}(1974)}]{OkeJ1974a}
{Oke}, J.~B. 1974, \apjs, 27, 21, \dodoi{10.1086/190287}

\bibitem[{{Pacifici} {et~al.}(2012){Pacifici}, {Charlot}, {Blaizot}, \&
  {Brinchmann}}]{PacificiC2012a}
{Pacifici}, C., {Charlot}, S., {Blaizot}, J., \& {Brinchmann}, J. 2012, \mnras,
  421, 2002, \dodoi{10.1111/j.1365-2966.2012.20431.x}

\bibitem[{{Padoan} {et~al.}(1997){Padoan}, {Nordlund}, \&
  {Jones}}]{PadoanP1997a}
{Padoan}, P., {Nordlund}, A., \& {Jones}, B.~J.~T. 1997, \mnras, 288, 145

\bibitem[{{Pforr} {et~al.}(2012){Pforr}, {Maraston}, \& {Tonini}}]{PforrJ2012a}
{Pforr}, J., {Maraston}, C., \& {Tonini}, C. 2012, \mnras, 422, 3285,
  \dodoi{10.1111/j.1365-2966.2012.20848.x}

\bibitem[{{Pforr} {et~al.}(2013){Pforr}, {Maraston}, \& {Tonini}}]{PforrJ2013a}
---. 2013, \mnras, 435, 1389, \dodoi{10.1093/mnras/stt1382}

\bibitem[{{Qiu} \& {Kang}(2022)}]{QiuY2022x}
{Qiu}, Y., \& {Kang}, X. 2022, \apj, 930, 66, \dodoi{10.3847/1538-4357/ac63a1}

\bibitem[{{Reddy} {et~al.}(2012){Reddy}, {Pettini}, {Steidel}, {Shapley},
  {Erb}, \& {Law}}]{ReddyN2012a}
{Reddy}, N.~A., {Pettini}, M., {Steidel}, C.~C., {et~al.} 2012, \apj, 754, 25,
  \dodoi{10.1088/0004-637X/754/1/25}

\bibitem[{{Reddy} {et~al.}(2015){Reddy}, {Kriek}, {Shapley}, {Freeman},
  {Siana}, {Coil}, {Mobasher}, {Price}, {Sanders}, \& {Shivaei}}]{ReddyN2015a}
{Reddy}, N.~A., {Kriek}, M., {Shapley}, A.~E., {et~al.} 2015, \apj, 806, 259,
  \dodoi{10.1088/0004-637X/806/2/259}

\bibitem[{{Rieke} {et~al.}(2005){Rieke}, {Kelly}, \& {Horner}}]{RiekeM2005p}
{Rieke}, M.~J., {Kelly}, D., \& {Horner}, S. 2005, in Society of Photo-Optical
  Instrumentation Engineers (SPIE) Conference Series, Vol. 5904, Cryogenic
  Optical Systems and Instruments XI, ed. J.~B. {Heaney} \& L.~G. {Burriesci},
  1--8, \dodoi{10.1117/12.615554}

\bibitem[{{Robertson}(2022)}]{RobertsonB2022b}
{Robertson}, B.~E. 2022, \araa, 60, 121,
  \dodoi{10.1146/annurev-astro-120221-044656}

\bibitem[{{Robotham} {et~al.}(2020){Robotham}, {Bellstedt}, {Lagos}, {Thorne},
  {Davies}, {Driver}, \& {Bravo}}]{RobothamA2020a}
{Robotham}, A.~S.~G., {Bellstedt}, S., {Lagos}, C. d.~P., {et~al.} 2020,
  \mnras, 495, 905, \dodoi{10.1093/mnras/staa1116}

\bibitem[{{Salim} \& {Boquien}(2019)}]{SalimS2019a}
{Salim}, S., \& {Boquien}, M. 2019, \apj, 872, 23,
  \dodoi{10.3847/1538-4357/aaf88a}

\bibitem[{{Salim} {et~al.}(2018){Salim}, {Boquien}, \& {Lee}}]{SalimS2018a}
{Salim}, S., {Boquien}, M., \& {Lee}, J.~C. 2018, \apj, 859, 11,
  \dodoi{10.3847/1538-4357/aabf3c}

\bibitem[{{Salim} \& {Narayanan}(2020)}]{SalimS2020a}
{Salim}, S., \& {Narayanan}, D. 2020, \araa, 58, 529,
  \dodoi{10.1146/annurev-astro-032620-021933}

\bibitem[{{Salmon} {et~al.}(2016){Salmon}, {Papovich}, {Long}, {Willner},
  {Finkelstein}, {Ferguson}, {Dickinson}, {Duncan}, {Faber}, {Hathi},
  {Koekemoer}, {Kurczynski}, {Newman}, {Pacifici}, {P{\'e}rez-Gonz{\'a}lez}, \&
  {Pforr}}]{SalmonB2016a}
{Salmon}, B., {Papovich}, C., {Long}, J., {et~al.} 2016, \apj, 827, 20,
  \dodoi{10.3847/0004-637X/827/1/20}

\bibitem[{{Salvato} {et~al.}(2018){Salvato}, {Ilbert}, \&
  {Hoyle}}]{SalvatoM2018a}
{Salvato}, M., {Ilbert}, O., \& {Hoyle}, B. 2018, Nature Astronomy,
  \dodoi{10.1038/s41550-018-0478-0}

\bibitem[{{Schmidt}(1959)}]{SchmidtM1959a}
{Schmidt}, M. 1959, \apj, 129, 243, \dodoi{10.1086/146614}

\bibitem[{{Seon} \& {Draine}(2016)}]{SeonK2016c}
{Seon}, K.-I., \& {Draine}, B.~T. 2016, \apj, 833, 201,
  \dodoi{10.3847/1538-4357/833/2/201}

\bibitem[{{Sharma}(2017)}]{SharmaS2017b}
{Sharma}, S. 2017, \araa, 55, 213, \dodoi{10.1146/annurev-astro-082214-122339}

\bibitem[{{Shivaei} {et~al.}(2020){Shivaei}, {Reddy}, {Rieke}, {Shapley},
  {Kriek}, {Battisti}, {Mobasher}, {Sanders}, {Fetherolf}, {Azadi}, {Coil},
  {Freeman}, {de Groot}, {Leung}, {Price}, {Siana}, \& {Zick}}]{ShivaeiI2020k}
{Shivaei}, I., {Reddy}, N., {Rieke}, G., {et~al.} 2020, \apj, 899, 117,
  \dodoi{10.3847/1538-4357/aba35e}

\bibitem[{{Skilling}(2004)}]{SkillingJ2004a}
{Skilling}, J. 2004, in American Institute of Physics Conference Series, Vol.
  735, American Institute of Physics Conference Series, ed. .~U.~V.~T.
  R.~Fischer, R.~Preuss, 395--405, \dodoi{10.1063/1.1835238}

\bibitem[{Skilling(2006)}]{SkillingJ2006a}
Skilling, J. 2006, Bayesian Analysis, 1, 833

\bibitem[{{Speagle}(2020)}]{SpeagleJ2020a}
{Speagle}, J.~S. 2020, \mnras, 493, 3132, \dodoi{10.1093/mnras/staa278}

\bibitem[{Spergel {et~al.}(2003)Spergel, Verde, Peiris, Komatsu, Nolta,
  Bennett, Halpern, Hinshaw, Jarosik, Kogut, Limon, Meyer, Page, Tucker,
  Weiland, Wollack, \& Wright}]{SpergelD2003a}
Spergel, D.~N., Verde, L., Peiris, H.~V., {et~al.} 2003, \apjs, 148, 175,
  \dodoi{10.1086/377226}

\bibitem[{{Suess} {et~al.}(2022){Suess}, {Leja}, {Johnson}, {Bezanson},
  {Greene}, {Kriek}, {Lower}, {Narayanan}, {Setton}, \&
  {Spilker}}]{SuessK2022v}
{Suess}, K.~A., {Leja}, J., {Johnson}, B.~D., {et~al.} 2022, \apj, 935, 146,
  \dodoi{10.3847/1538-4357/ac82b0}

\bibitem[{{Tacconi} {et~al.}(2020){Tacconi}, {Genzel}, \&
  {Sternberg}}]{TacconiL2020b}
{Tacconi}, L.~J., {Genzel}, R., \& {Sternberg}, A. 2020, \araa, 58, 157,
  \dodoi{10.1146/annurev-astro-082812-141034}

\bibitem[{{Tanaka}(2015)}]{TanakaM2015a}
{Tanaka}, M. 2015, \apj, 801, 20, \dodoi{10.1088/0004-637X/801/1/20}

\bibitem[{{Thomas} \& {Maraston}(2003)}]{ThomasD2003a}
{Thomas}, D., \& {Maraston}, C. 2003, \aap, 401, 429,
  \dodoi{10.1051/0004-6361:20030153}

\bibitem[{{Thorne} {et~al.}(2021){Thorne}, {Robotham}, {Davies}, {Bellstedt},
  {Driver}, {Bravo}, {Bremer}, {Holwerda}, {Hopkins}, {Lagos}, {Phillipps},
  {Siudek}, {Taylor}, \& {Wright}}]{ThorneJ2021n}
{Thorne}, J.~E., {Robotham}, A. S.~G., {Davies}, L. J.~M., {et~al.} 2021,
  \mnras, 505, 540, \dodoi{10.1093/mnras/stab1294}

\bibitem[{{Tinsley}(1978)}]{TinsleyB1978a}
{Tinsley}, B.~M. 1978, \apj, 222, 14, \dodoi{10.1086/156116}

\bibitem[{{Tinsley}(1980)}]{TinsleyB1980n}
---. 1980, \fcp, 5, 287, \dodoi{10.48550/arXiv.2203.02041}

\bibitem[{{Tinsley} \& {Gunn}(1976)}]{TinsleyB1976a}
{Tinsley}, B.~M., \& {Gunn}, J.~E. 1976, \apj, 203, 52, \dodoi{10.1086/154046}

\bibitem[{{Valentini} {et~al.}(2019){Valentini}, {Borgani}, {Bressan},
  {Murante}, {Tornatore}, \& {Monaco}}]{ValentiniM2019a}
{Valentini}, M., {Borgani}, S., {Bressan}, A., {et~al.} 2019, \mnras, 485,
  1384, \dodoi{10.1093/mnras/stz492}

\bibitem[{van~der Walt {et~al.}(2011)van~der Walt, Colbert, \&
  Varoquaux}]{WaltS2011s}
van~der Walt, S., Colbert, S.~C., \& Varoquaux, G. 2011, Computing in Science
  \& Engineering, 13, 22, \dodoi{10.1109/MCSE.2011.37}

\bibitem[{{van Dokkum}(2008)}]{van-DokkumP2008a}
{van Dokkum}, P.~G. 2008, \apj, 674, 29, \dodoi{10.1086/525014}

\bibitem[{{Volonteri} {et~al.}(2016){Volonteri}, {Dubois}, {Pichon}, \&
  {Devriendt}}]{VolonteriM2016a}
{Volonteri}, M., {Dubois}, Y., {Pichon}, C., \& {Devriendt}, J. 2016, \mnras,
  460, 2979, \dodoi{10.1093/mnras/stw1123}

\bibitem[{{Walcher} {et~al.}(2011){Walcher}, {Groves}, {Budav{\'a}ri}, \&
  {Dale}}]{WalcherJ2011a}
{Walcher}, J., {Groves}, B., {Budav{\'a}ri}, T., \& {Dale}, D. 2011, \apss,
  331, 1, \dodoi{10.1007/s10509-010-0458-z}

\bibitem[{{Wang} \& {Lilly}(2020)}]{WangE2020a}
{Wang}, E., \& {Lilly}, S.~J. 2020, \apj, 892, 87,
  \dodoi{10.3847/1538-4357/ab7b7d}

\bibitem[{{Weingartner} \& {Draine}(2001)}]{WeingartnerJ2001a}
{Weingartner}, J.~C., \& {Draine}, B.~T. 2001, \apj, 548, 296,
  \dodoi{10.1086/318651}

\bibitem[{{Witt} \& {Gordon}(2000)}]{WittA2000a}
{Witt}, A.~N., \& {Gordon}, K.~D. 2000, \apj, 528, 799, \dodoi{10.1086/308197}

\bibitem[{{Yallup} {et~al.}(2022){Yallup}, {Jan{\ss}en}, {Schumann}, \&
  {Handley}}]{YallupD2022q}
{Yallup}, D., {Jan{\ss}en}, T., {Schumann}, S., \& {Handley}, W. 2022, European
  Physical Journal C, 82, 678, \dodoi{10.1140/epjc/s10052-022-10632-2}

\bibitem[{{Yan} {et~al.}(2019){Yan}, {Chen}, {Lazarz}, {Bizyaev}, {Maraston},
  {Stringfellow}, {McCarthy}, {Meneses-Goytia}, {Law}, {Thomas}, {Falcon
  Barroso}, {S{\'a}nchez-Gallego}, {Schlafly}, {Zheng}, {Argudo-Fern{\'a}ndez},
  {Beaton}, {Beers}, {Bershady}, {Blanton}, {Brownstein}, {Bundy}, {Chambers},
  {Cherinka}, {De Lee}, {Drory}, {Galbany}, {Holtzman}, {Imig}, {Kaiser},
  {Kinemuchi}, {Liu}, {Luo}, {Magnier}, {Majewski}, {Nair}, {Oravetz},
  {Oravetz}, {Pan}, {Sobeck}, {Stassun}, {Talbot}, {Tremonti}, {Waters},
  {Weijmans}, {Wilhelm}, {Zasowski}, {Zhao}, \& {Zhao}}]{YanR2019a}
{Yan}, R., {Chen}, Y., {Lazarz}, D., {et~al.} 2019, \apj, 883, 175,
  \dodoi{10.3847/1538-4357/ab3ebc}

\bibitem[{{Zhan}(2011)}]{ZhanH2011i}
{Zhan}, H. 2011, Scientia Sinica Physica, Mechanica \& Astronomica, 41, 1441,
  \dodoi{10.1360/132011-961}

\bibitem[{{Zhan}(2018)}]{ZhanH2018y}
{Zhan}, H. 2018, in 42nd COSPAR Scientific Assembly, Vol.~42, E1.16--4--18

\bibitem[{{Zhan}(2021)}]{ZhanH2021j}
---. 2021, Chinese Science Bulletin, 66, 1290

\bibitem[{{Zhang} {et~al.}(2005){Zhang}, {Han}, {Li}, \&
  {Hurley}}]{ZhangF2005c}
{Zhang}, F., {Han}, Z., {Li}, L., \& {Hurley}, J.~R. 2005, \mnras, 357, 1088,
  \dodoi{10.1111/j.1365-2966.2005.08739.x}

\bibitem[{{Zhou} {et~al.}(2022{\natexlab{a}}){Zhou}, {Gong}, {Meng}, {Chen},
  {Chen}, {Du}, {Fu}, \& {Luo}}]{ZhouX2022s}
{Zhou}, X., {Gong}, Y., {Meng}, X.-M., {et~al.} 2022{\natexlab{a}}, arXiv
  e-prints, arXiv:2206.13696.
\newblock \doarXiv{2206.13696}

\bibitem[{{Zhou} {et~al.}(2022{\natexlab{b}}){Zhou}, {Gong}, {Meng}, {Cao},
  {Chen}, {Chen}, {Du}, {Fu}, \& {Luo}}]{ZhouX2022v}
---. 2022{\natexlab{b}}, \mnras, 512, 4593, \dodoi{10.1093/mnras/stac786}

\end{thebibliography}
\bibliographystyle{aasjournal}

\end{document}